# Chem Soc Rev

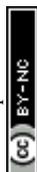

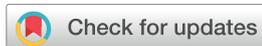
Check for updates

Cite this: DOI: 10.1039/d1cs00106j

## Solution-processed two-dimensional materials for next-generation photovoltaics

Sebastiano Bellani, [ab] Antonino Bartolotta, [c] Antonio Agresti, [d] Giuseppe Calogero, [c] Giulia Grancini, [e] Aldo Di Carlo, [df] Emmanuel Kymakis [g] and Francesco Bonaccorso [*ab]

In the ever-increasing energy demand scenario, the development of novel photovoltaic (PV) technologies is considered to be one of the key solutions to fulfil the energy request. In this context, graphene and related two-dimensional (2D) materials (GRMs), including nonlayered 2D materials and 2D perovskites, as well as their hybrid systems, are emerging as promising candidates to drive innovation in PV technologies. The mechanical, thermal, and optoelectronic properties of GRMs can be exploited in different active components of solar cells to design next-generation devices. These components include front (transparent) and back conductive electrodes, charge transporting layers, and interconnecting/recombination layers, as well as photoactive layers. The production and processing of GRMs in the liquid phase, coupled with the ability to ''on-demand'' tune their optoelectronic properties exploiting wet-chemical functionalization, enable their effective integration in advanced PV devices through scalable, reliable, and inexpensive printing/coating processes. Herein, we review the progresses in the use of solution-processed 2D materials in organic solar cells, dye-sensitized solar cells, perovskite solar cells, quantum dot solar cells, and organic–inorganic hybrid solar cells, as well as in tandem systems. We first provide a brief introduction on the properties of 2D materials and their production methods by solution-processing routes. Then, we discuss the functionality of 2D materials for electrodes, photoactive layer components/additives, charge transporting layers, and interconnecting layers through figures of merit, which allow the performance of solar cells to be determined and compared with the state-of-the-art values. We finally outline the roadmap for the further exploitation of solution-processed 2D materials to boost the performance of PV devices.

## 1. Introduction

Energy supply is one of the most pressing issues of the twenty-first century, having a harsh impact on the global economy and society.[1–3] Unending technological development in any human activity, ranging from transport to consumers electronics (*e.g.*, cell phones, laptops, *etc.*) and even stationary applications,[4] has led to a growing demand of cost-effective and environmentally

*a* BeDimensional S.p.A., Via Lungotorrente Secca 30R, 16163 Genova, Italy.
  E-mail: f.bonaccorso@bedimensional.it
*b* Istituto Italiano di Tecnologia, Graphene Labs, via Morego 30, 16163 Genova, Italy
*c* CNR-IPCF, Istituto per i Processi Chimico-Fisici, Via F. Stagno D'alcontres 37,
  98158 Messina, Italy
*d* CHOSE – Centre for Hybrid and Organic Solar Energy, University of Rome
  ''Tor Vergata'', via del Politecnico 1, 00133 Roma, Italy
*e* University of Pavia and INSTM, Via Taramelli 16, 27100 Pavia, Italy
*f* L.A.S.E. – Laboratory for Advanced Solar Energy, National University of Science
  and Technology ''MISiS'', 119049 Leninskiy Prosect 6, Moscow, Russia
*g* Department of Electrical & Computer Engineering, Hellenic Mediterranean
  University, Estavromenos 71410 Heraklion, Crete, Greece

friendly energy conversion and storage (ECS) devices.[5,6] In this context, photovoltaic (PV), or solar cell (SC), technology has been at the center of an ongoing research effort,[7–10] due to the direct exploitation of energy from sunlight, which can significantly contribute toward energy conversion in a sustainable and economical way.[11] Basically, SCs are electrical devices that use the PV effect to convert energy of light directly into electricity.[7–12] Thus, SCs require a light-harvesting material that absorbs photons and raises electrons from their molecular/atomic orbitals to generate free electron (e⁻)/hole (h⁺) pairs *via* the PV effect.[13,14] Once excited, charge carriers can either dissipate the energy as heat and recombine into their initial energy state or travel through the cell structure until they reach their respective electrodes.[15] In building the SC structure, a built-in potential barrier (ideally corresponding to the open circuit voltage $V_{OC}$) is typically created to act on the free charges, driving current through an external circuit, thereby powering desired loadings.[16,17]

The maximum theoretical solar-to-electrical energy conversion efficiency ($\eta_{th}$) of a SC for a single p–n junction ($\sim 33\%$ for 1 sun illumination) is determined by the Shockley–Queisser (S–Q) thermodynamic limit.[18] In agreement with the S–Q limit, the charge carriers generated by photons with energies ($E_{ph}$) larger than the semiconductor bandgap ($E_g$) lose their excess energy (= $E_{ph} - E_g$) as heat through the excitation of lattice vibrations.[18] Since the energy conversion efficiency ($\eta$), i.e., the fraction of incident power that is converted into electricity, remains one of the most critical parameters to optimize SCs for implementation, several approaches to overcome the S–Q limit have been proposed. Some examples include tandem cells (multiple p–n junctions),[19–21] hot-carrier SCs,[22,23] SCs generating multiple $e^-/h^+$ pairs for a single incident phonon,[24–26] and multi-band and impurity SCs.[27,28]

To improve commercially available SCs with respect to both performance and cost-effectiveness, several potential photo-active materials are under investigation. So far, doped forms of single- or polycrystalline Si (i.e., 1st-generation SCs) have comprised the lion's share of SCs in the PV market.[29,30] In fact, they achieved $\eta$ superior to 25%,[31,32] up to a record value of 26.7%.[33] The latter was demonstrated in a heterojunction with intrinsic thin-layer technology (HIT) based on thin amorphous Si (a-Si) passivating layers and on interdigitated back contacts on n-type Si wafers.[33] Subsequently, thin-film solar cells (TFSCs, i.e., 2nd-generation SCs), based on "thin" films having a thickness of ~1–2 μm, have played an important role in the

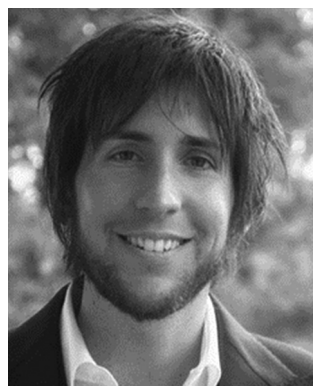

**Sebastiano Bellani**

*Sebastiano Bellani is a researcher at BeDimensional S.p.A. He received his PhD from Politecnico di Milano, while working at the Istituto Italiano di Tecnologia. Here, he investigated solid/liquid interfaces in organic semiconductor water-gated organic field-effect transistors, hybrid organic–inorganic photoelectrochemical cells, and biopolymer-based devices for optical cellular stimulation. Currently, he is participating at the European Commission's Future and Emerging Technology Graphene Flagship.* He has been the deputy leader of Graphene Flagship's Solar Farm Spearhead Project. His studies are focused on chemical-physical, spectroscopic, and photoelectrochemical characterizations of solution-processed two-dimensional materials and their energy-related applications including photovoltaics, (photo)-electrocatalysis, and energy storage systems.

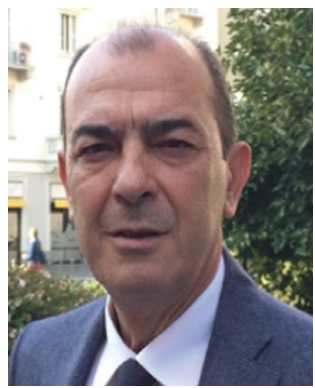

**Antonino Bartolotta**

*Antonino Bartolotta is a researcher at the Institute for the Chemical Physical Processes of the Italian National Research Council (CNR-IPCF). He gained his degree in chemistry from the University of Messina in Italy. His scientific activity started in the field of condensed matter physics, mainly devoted to glass transition and dynamical processes in disordered systems (glasses, polymers, and vitreous ionic conductors). Currently, his research is focused in the field of nanomaterials and energy conversion and storage devices.* He has been involved in several research projects, and he is the author/co-author of many publications in international peer-reviewed journals and book chapters.

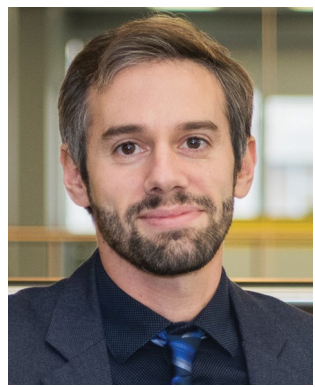

**Antonio Agresti**

*Antonio Agresti is an Assistant Professor at the Department of Electronic Engineering at the University of Rome Tor Vergata since 2016. His research activity mainly involves the design, engineering, fabrication, and electrical/spectroscopic character-ization of hybrid and organic solar cells; use of graphene and transition metal dichalcogenides and emerging two-dimensional materials such as MXenes for perovskite solar cells, tandem devices, large-area modules, and panels.* He has authored/co-authored more than 50 publications and has participated as an invited speaker to several conferences in the renewable energy field. He is currently the deputy leader of Horizon 2020 Spearhead 5—Graphene Core 3 project.

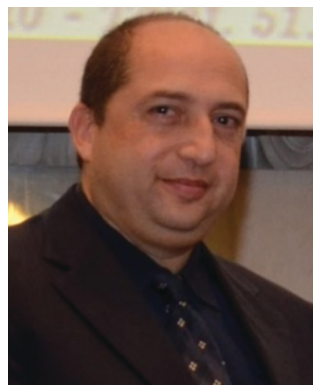

**Giuseppe Calogero**

*Giuseppe Calogero is a researcher at the Institute for the Chemical Physical Processes of the Italian National Research Council (CNR-IPCF), Messina, and he works in the field of energy and electron transfer processes. His research activity has mainly focused on the manufacturing, study, and characterization of dye-sensitized solar cells (DSSCs). He has synthetized supramolecular com-plexes based on polypyridine metal complexes and developed new nanostructured materials for DSSC applications, with focus on design, preparation, and processing.* He has authored over 60 articles in peer-reviewed international journals.

field of PV with regard to both $\eta$ ($>22\%$)[34,35] and cost-effectiveness.[36] Second-generation SCs are based on a large variety of semiconductor materials, including crystalline (c-Si)[37] and a-Si,[38] as well as GaAs[39] and metal chalcogenides, such as CdTe,[40] copper indium gallium diselenide (CuIn$_{1-x}$Ga$_x$Se$_2$ or CIGS),[41,42] copper indium gallium selenide sulfide Cu(In,Ga)(Se,S)$_2$ (CIGSSe),[34] CdTe/CdS or CdS/PbS heterojunctions,[43,44] and Cu$_2$ZnSnSe$_4$ (CZTSe).[45] Thin-film solar cells are characterized by some peculiar (opto)electronic features, such as nearly ideal $E_g$ for sunlight absorption ($\sim$1.4 eV, according to the S–Q limit for single-junction SCs[46]) and absorption coefficient ($\alpha$) ($\geq$10$^5$ cm$^{-1}$) over a wide spectral range (Fig. 1a and inset panel).[47,48] For example, CdTe has $E_g$ of 1.44 eV and $\alpha$ of $\sim$1.115 $\times$ 10$^6$ cm$^{-1}$, while CIGS has $E_g$ in the 1.0–1.6 eV range and $\alpha > 1 \times 10^5$ cm$^{-1}$.[9] In addition, both materials are direct-bandgap semiconductors, which implies that they can efficiently absorb above-$E_g$ light with a thin-film layer ($\sim$1–2 $\mu$m).[49] Based on the aforementioned

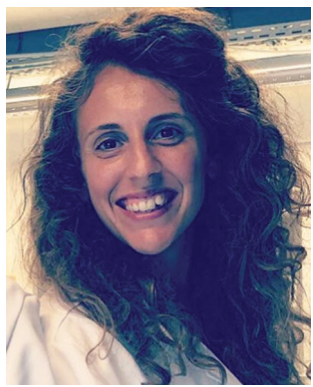

**Giulia Grancini**

*Giulia Grancini is an Associate Professor in Chemistry at the University of Pavia, leading the PVsquared2 team and the ERC Project "HYNANO" aiming at the development of advanced hybrid perovskites solar cells. She obtained her PhD in Physics from Politecnico di Milano in 2012 and worked as a Post-Doc Researcher at IIT, Milano. From 2015 to 2019, she joined the École Polytechnique Fédérale de Lausanne awarded by SNSF with the Ambizione Energy Grant. In 2020, she won the Journal of Materials Chemistry Lectureship. She is currently the USERN Ambassador for Italy and a board member of the Young Academy of Europe. In 2019 and 2020, she appeared among highly cited scientists. In 2020, she was listed in "100 Experts," which identified the top Italian women scientists in STEM. She is the author of 94 publications and owns 2 patents.*

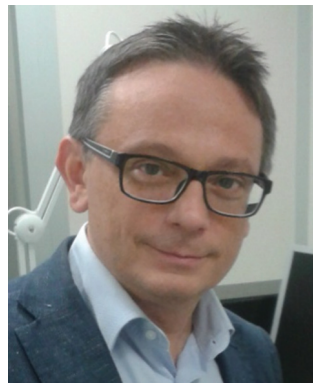

**Aldo Di Carlo**

*Aldo Di Carlo is a Full Professor at the University of Rome Tor Vergata and Director of the Institute of Structure of Matter of the National Research Council. His research activity mainly involves the design, fabrication, and characterization of solution-processed solar cells and other optoelectronic devices. He was the founder of the Centre for Hybrid and Organic Solar Energy (CHOSE), which involved more than 35 researchers. He was the CTO of Dyepower, a consortium for the industrialization of dye solar cells for façade applications. He has authored/co-authored more than 400 publications, review articles, and book chapters, and he owns 13 patents.*

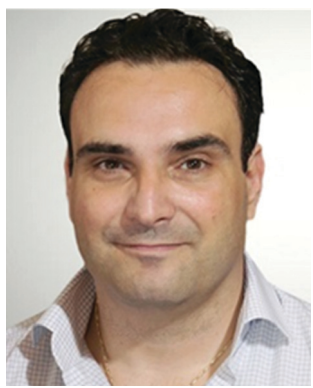

**Emmanuel Kymakis**

*Emmanuel Kymakis is a Full Professor at the Department of Electrical and Computer Engineering at the Hellenic Mediterranean University (HMU) and Vice President of the HMU Research Center. He obtained his PhD in engineering from the University of Cambridge in 2003. His research focuses on graphene and related 2D materials with regard to the interfacial engineering of emerging solar cells for improved performance and stability, as well as on the performance evaluation of PV systems. He was named as a ChemComm Emerging Investigator and served as a member of the founding GA of HFRI. He is the leader of the Energy Generation WP of the FET Flagship Initiative, Graphene Flagship.*

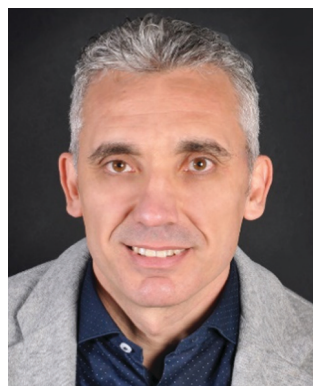

**Francesco Bonaccorso**

*Francesco Bonaccorso is the Founder and Scientific Director of BeDimensional S.p.A. and Visiting Scientist at the Istituto Italiano di Tecnologia. He gained his PhD from the University of Messina after working at the Italian National Research Council, the University of Cambridge, and the University of Vanderbilt. In 2009, he was awarded a Royal Society Newton International Fellowship at Cambridge University, and a Fellowship at Hughes Hall, Cambridge, receiving his MA. He was responsible in defining the ten years scientific and technological roadmap for the Graphene Flagship. He is now the deputy of the Innovation of Flagship. He was featured as 2016 Emerging Investigator by Journal of Materials Chemistry A and in 2019, by ChemPlusChem. His research interests encompass both fundamental understanding and solution processing of novel nanomaterials and their technological applications. He has authored/co-authored more than 160 publications and owns 13 patents.*

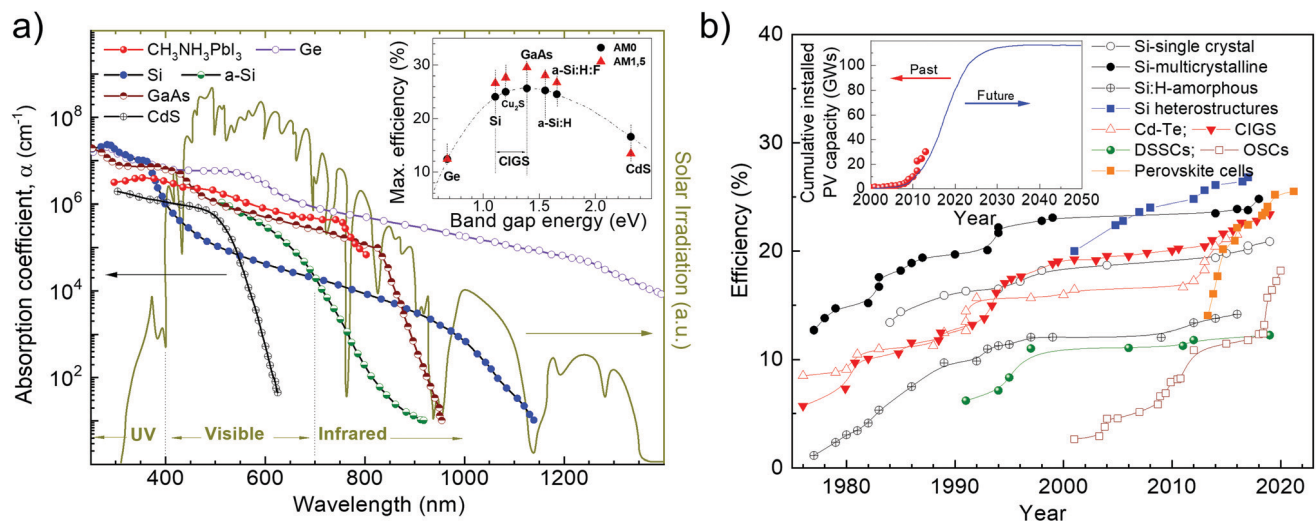

Fig. 1 (a) Light wavelength (λ) dependence of the absorption coefficient (α) at room temperature (RT) of some semiconductor materials used in PV technologies. The inset shows the maximum theoretical solar-to-energy conversion efficiency ($\eta_{th}$) of SCs under AM 1.5 light radiation determined by the S–Q limit. Adapted from: ref. 47 and 48. (b) Description of the development of laboratory SCs. Inset: Cumulative installed PV capacity and plausible projection in the near future. Adapted from: ref. 63, 64 and 65.

(opto)electronic properties, CdTe and CIGS TFSCs reached $\eta$ exceeding 19%[31] and 22%,[50] respectively, thus competing with mainstream c-Si-based technology. Beyond 1st- and 2nd-generation SCs, new potential PV technologies—most of which are based on thin films—have also emerged as the 3rd-generation SCs. These include organic solar cells (OSCs),[51] dye-sensitized solar cells (DSSCs),[52–54] quantum dot solar cells (QDSCs),[55–57] organic–inorganic hybrid SCs[58,59] and perovskite solar cells (PSCs).[60–62] Organic solar cells are based on conjugated polymers or small molecules for light absorption and charge transport,[51] DSSCs use an electrolyte as the charge transporting medium,[52–54] QDSCs exploit solution-processed nanocrystals (quantum dots (QDs)) as the light-harvesting material,[55–57] hybrid SCs mix both organic and inorganic materials as the photoactive component,[58,59] and PSCs are based on organic–inorganic halide perovskites material (e.g., $CH_3NH_3PbX_3$, where X = Cl, Br, I or their mixture) as the photosensitizer.[60–62]

The growth of the global market share of PV technology has been impressive and the demand for cumulative solar PV electricity generation is expected to move toward the scale of hundreds of gigawatts in the near future (Fig. 1b),[63–65] with $\eta$ of 2nd- and 3rd-generation SCs surpassing that of c-Si (Fig. 1b, inset panel).

Fundamentally, an ideal photoactive material for SCs based on thin films has to be a direct-bandgap semiconductor with an $E_g$ in the 1.0–2.0 eV range to absorb sunlight in a wide spectrum range.[46] Moreover, it should have high charge carrier mobility ($\mu$)[66] and should be compatible with one or the other material constituting the cell architecture to form reliable electrical connections.[67] Notably, the optical penetration depth ($\delta_p$), (i.e., the spatial region in which most of the incoming photons are absorbed to produce charge carrier pairs) of the photoactive material is crucial to determine its thickness ($t$). In fact, $\delta_p$ can be approximated to $\alpha^{-1}$, in agreement with the Lambert–Beer law: $T_r = (I/I_o)\cdot e^{-\alpha t}$, in which $T_r$ is the optical transmission, while $I$ and $I_o$ are the intensity of transmitted and incident light, respectively.[68] Consequently, the most appropriate $t$ value of the photoactive material should be as close as possible to the value of $\alpha^{-1}$, to facilitate charge transport toward the external circuitry, without significant charge recombination losses.

Following the research effort on graphene,[69,70] the development of other layered two-dimensional (2D) materials (named graphene-related materials (GRMs)),[71,72] as well as other 2D materials (e.g., nonlayered 2D materials and 2D perovskites), has burgeoned into the field of SCs and optoelectronic applications. In particular, graphene opens endless possibilities for new generations of SCs owing to its outstanding (opto)electronic properties (e.g., low sheet resistance ($R_s \approx 6.45$ kΩ □$^{-1}$),[73] excellent optical transparency in the UV-to-IR region ($T_r > 97.7\%$),[73] high intrinsic strength (~130 GPa), high Young modulus (~1 TPa), high electron mobility ($\mu_e$) ($>10^5$ cm$^2$ V$^{-1}$ s$^{-1}$),[71] large specific surface area (SSA) (~2630 m$^2$ g$^{-1}$),[74] and excellent chemical stability and catalytic activity toward photo(electro)chemical cell-related redox reaction.[75–77] Moreover, the (opto)electronic properties of graphene can be tuned via its chemical functionalization processes.[78–80] In this context, graphene oxide (GO) (i.e., graphene with C–O bonds and functionalities, such as –OH, C=O, and COO– groups)[81,82] or reduced graphene oxide (RGO),[83] as well as mono- and few-layered GRMs, exhibit electronic[84–86] and optical properties[87,88] that are complementary to the graphene ones. Among GRMs, transition metal dichalcogenides (TMDs) with the general stoichiometry of MX$_2$, where M is a transition element of groups IVB–VIIB and X is a chalcogen (i.e., S, Se, and Te) (Fig. 2), strongly emerged for their potential exploitation in the development of novel SCs due to their physical properties.[89–91] For example, together with graphene and graphene derivatives, TMDs are becoming attractive candidates as electron/hole transporting materials in several types of SCs[92] due to their electronic structure capable to optimize charge transport toward the current collectors.[93,94] Overall, the field of 2D materials is an ever-expanding research area, and new GRMs (*e.g.*, metallic group-V TMDs,[95,96] transition metal monochalcogenides (TMMs),[97,98] MXenes,[99,100] silicene,[101] phosphorene,[102] antimonene,[103] bismuthene,[104,105] arsenene,[106] and graphdiyne[107,108]) and even other types of synthetic 2D materials (not strictly belonging to the class of GRMs) are rapidly coming into the fray. Finally, the scenario of solution-processed 2D materials for PVs and, in general, optoelectronic applications, has been recently extended to both nonlayered 2D materials[109] and 2D perovskites.[110,111] In their review article,[112] Liu *et al.* outlined the advent of 2D materials for several PV technologies, showing the most important achievements up to 2015. It is now crucial to provide an update on the use of 2D materials in SCs, including OSCs, DSSCs, PSCs, QDSCs, organic–inorganic hybrid SCs, as well as tandem systems.

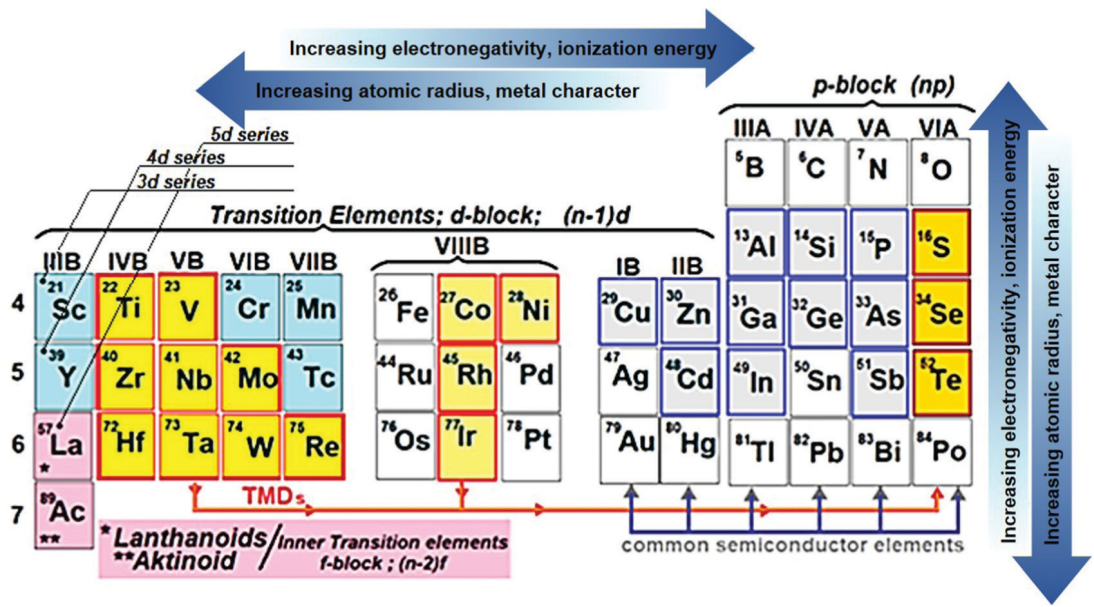

**Fig. 2** Selected elements across the periodic table that—as a single element (*e.g.*, Si, Ge), compounds (*e.g.*, GaAs, CdAs), or alloys (*e.g.*, $Si_xGe_{1-x}$, $Al_xGa_{1-x}$)—display semiconductor nature (into the blue frame) and the transition metals and three chalcogen (X) elements (enclosed by a red frame) that predominantly crystallize in layered TMDs. Partial highlights for Co, Rd, Ir, and Ni indicate that only some of their TMDs form layered structures.

The production of 2D materials by solution processing[83,113] represents an ideal platform for the advancement of PV technologies. In fact, liquid-dispersed 2D materials can be produced with on-demand morphological properties, *i.e.*, lateral size[114,115] and thickness[116–118] by exploiting sorting, or can be chemically modified to tuning the (opto)electronic properties.[119] Moreover, 2D materials produced by solution processing can be used for the realization of composites,[120] *i.e.*, blending with polymeric matrices, and the production of films by means of several coating techniques, such as inkjet[121,122] and screen[123,124] printing, drop[125] and dip[126] casting, and spin[127,128] and spray[129,130] coating.

The possibility to produce and process 2D materials and their heterostructures in the liquid phase represents a step forward toward the development of industrial-scale, reliable, inexpensive printing/coating processes, which can ultimately lead to a reduction in the levelized cost of energy (LCOE) of current PV technologies (less than 5 US cents kW h⁻¹)[131–133] to compete with fossil fuels.[134,135]

In this review, Section 2 provides an overview of the structural and (opto)electronic properties of 2D materials, highlighting the differences of GRMs compared to their bulk counterparts. The production and processing of GRMs in the liquid phase is thoroughly discussed in this section. A brief paragraph focuses on 2D nonlayered materials, while a specific discussion on 2D perovskites is provided in the section related to PSCs (*i.e.*, Section 6). In Section 3, we introduce the main figures of merit (FoM) of SCs and SC components to facilitate the discussion and understanding of subsequent sections. The use of solution-processed 2D materials as building blocks in OSCs, DSSCs, PSCs, and other types of SCs (*i.e.*, QDSCs and organic–inorganic hybrid SCs) is presented and critically discussed in Sections 4, 5, 6, and 7, respectively. Finally, Section 8 summarizes the key results of solution-processed 2D materials in PV technologies, providing the status, prospects, and challenges in this field.

## 2. Basic properties, production, and functionalization of 2D materials

### 2.1 Basic properties of GRMs

As depicted in Fig. 3a, graphene is a one-atom-thick layer of carbon atoms bonded together in a hexagonal honeycomb lattice.[136] Owing to its unique physical and chemical properties,[71] it became highly attractive for fabricating conductive and transparent thin films,[73] even though numerous other (opto)electronic[73] and ECS[5,137] applications exist. Graphene can be considered as the starting material for all fullerene allotropic dimensionalities,

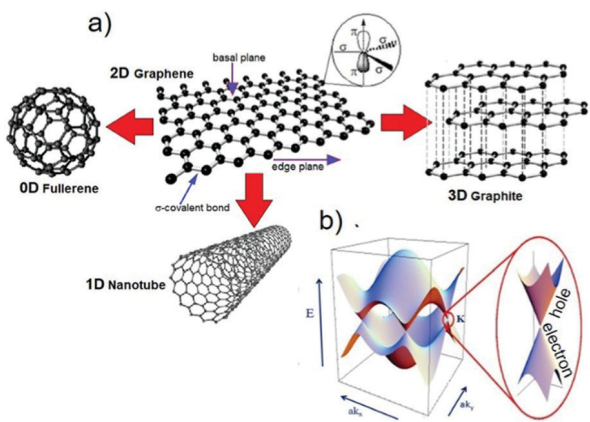

**Fig. 3** (a) Schematic of the honeycomb graphene network as formed by C atoms and bonded basal σ bonds perpendicular to π orbitals. The other graphene-derived allotropes of C are also shown. Adapted from ref. 139 and 140. (b) Band structure in the honeycomb lattice. In the enlarged picture, the energy bands close to one of the Dirac points are also sketched. Adapted from ref. 141.

including spherical buckyballs (zero-dimensional, 0D), one-dimensional (1D) carbon nanotubes (CNTs), further categorized in single- and multiwalled CNTs (SWCNTs and MWCNTs, respectively) depending on the number of graphene layers, as well as charcoal and graphite.[138]

Graphene nanoplatelets (GNPs), i.e., flakes of functionalized graphene with a thickness ranging from ∼2 to ∼15 nm and a lateral size ranging from the submicrometer scale to 100 μm,[137] and (R)GO, obtained by chemical/thermal processes,[82,83] are the most frequently used graphene derivatives for large-scale industrial applications, including composites[142–145] and ECS devices.[5,137,142,146,147] There is a large number of studies that detail the properties of GRMs.[71,137,148] Therefore, herein, we briefly focus only on the most peculiar properties of graphene, as well as those of other layered materials. Single-layer graphene (SLG) is a "zero-bandgap semiconductor" with the valence band (VB) and conduction band (CB) touching at the Dirac points (see Fig. 3b)[148,149] and charge carriers that can be regarded as massless electrons or Dirac fermions.[148] Electron mobilities exceeding $2 \times 10^5$ cm$^2$ V$^{-1}$ s$^{-1}$ at charge carrier densities of ∼$2 \times 10^{11}$ cm$^{-2}$ have been reported by Bolotin and co-workers by suspending SLG above a Si/SiO$_2$ gate electrode.[150] However, it has been shown that graphene on SiO$_2$ has a μ value that is limited by scattering from charged impurity states and impurities,[151–154] SiO$_2$ surface optical phonons,[153,154] and substrate surface roughness.[155–157] By searching for alternatives to SiO$_2$, it has been demonstrated that hexagonal boron nitride (h-BN), an insulating isomorph of graphite with B and N atoms and a small lattice mismatch (1.7%) relative to graphite,[158] represents an ideal, flat dielectric substrate for graphene.[159,160] Thus, graphene on h-BN can reach a $\mu_e$ value exceeding $6 \times 10^5$ cm$^2$ V$^{-1}$ s$^{-1}$,[159] which is 3 times higher than those shown on SiO$_2$. These results suggest graphene to be an ideal channel material for the fast transport of charge carriers in nanostructured and thin-film electrodes.[161,162] As a comparison, μ of graphene is ∼200 times higher than that of Si (∼1400 cm$^2$ V$^{-1}$ s$^{-1}$).[163]

Graphene, owing to its mechanical properties (i.e., flexibility and stretchability),[71] is an ideal material to fabricate flexible and ultralight devices.[164–167] It is important to highlight the dependence of the (opto)electronic properties (e.g., $R_s$ and $T_r$) on the number of graphene layers. In fact, by investigating the dependence of the $T_r$ value of graphene on the number of layers, Nair et al.[168] reported that the opacity of graphene increases by ∼2.3% for each added layer. Moreover, Li and co-workers[169] measured a $R_s$ value that varies from 2.1 kΩ □$^{-1}$ to 350 Ω □$^{-1}$, moving from SLG to 4-layer graphene, while $T_r$ is reduced to ∼90% (at λ = 550 nm) for 4-layer graphene (Fig. 4).

Overall, the aforementioned properties make graphene, as well its derivatives, a distinctive material for PV applications. In fact, low $R_s$, large SSA, and high μ and $T_r$ are essential requirements to be considered in the choice of material for the various building blocks of SCs. Beyond graphene, there is a plethora of other 2D materials that range from insulators (e.g., h-BN) to semiconductors (e.g., TMDs, such as MoS$_2$ and WS$_2$, and phosphorene), metals (e.g., TiS$_2$ and several group-5 TMDs, such as 2H- and 3R-TaS$_2$, 2H- and 3R-NbS$_2$, and 1T-VSe$_2$), and even superconductors (e.g., 2H-NbSe$_2$) and charge density wave materials (e.g., 1T and 2H-TaS$_2$, 1T- and 2H-TaSe$_2$, 2H-NbSe$_2$, 1T-VS$_2$, and 1T-VSe$_2$) at low temperatures.[170–173] In addition, 2D materials, such as Bi$_2$Te$_3$, Sb$_2$Te$_3$, Bi$_2$Se$_3$, SnTe, and even graphene, can display unique symmetry-protected helical metallic edge states with an insulating interior, yielding so-called topological insulators.[174,175] As for graphene, research on other GRMs, including TMDs,[176,177] TMMs,[178–181] transition metal oxides (TMOs),[182] monoelemental 2D materials (silicene, phosphorene, germanene, stanene, borophene, gallenene, arsenene, antimonene, bismuthene, plumbene, selenene, and tellurene),[183,184] and MXenes,[185] have provided evidences that the band structure of such materials drastically changes as they shrink from the bulk to the monolayer due to quantum confinement effects.[170,171,186] The abundance of GRMs and the ability to stack them in a layer-by-layer manner in desired sequences, eventually through solution-processed methods, offer the possibility to create novel three-dimensional (3D) architectures with entirely new functions,[187,188] which have been foreseen to design the next generation of PV devices.[189] In particular, the so-formed heterostructures are held together by van der Waals forces, as occurring in other layered materials. Since the family of GRMs is continuously expanding, the complexity of the heterostructures that can be created is nearly unlimited.[190] The stacking of different GRMs can lead to a series of synergistic effects, such as[190] (1) charge redistribution between closed materials (and even more distant materials) in the stack; (2) structural changes in the stacked materials, whose ultimate properties depend on their orientation relative to the neighboring materials. In addition to leading to the discovery and observation of novel thrilling physical phenomena, the enormous range of functionalities of 2D material heterostructures has yielded applications in PVs and optoelectronics. For example, photoactive semiconducting layers (e.g., TMDs) have been coupled with graphene as transparent electrodes to form photodetectors.[191,192] In addition, the combination of 2D materials with different work function ($\phi_W$) values (such as $MoS_2$ and $WSe_2$) enables photoexcited charges (electron and holes) to be accumulated in different layers, resulting in indirect excitons with long lifetimes and tunable binding energies.[193,194] At the atomic scale, p–n junctions (e.g., $GaTe/MoS_2$) have been created to provide highly efficient carrier separation, reaching external quantum efficiency (EQE) higher than 60%.[195] The exciting and vast topic of 2D material heterostructures is subject of relevant reviews, for which we refer to more in-depth discussions.[187,190,196–199] In the context of our work, it is worth pointing out that the realization of 2D material heterostructures with atomic precision and predetermined features by means of scalable solution-processing methods remains a formidable challenge. Nevertheless, the feasibility of printed heterostructures has been proven by several works,[200–204] paving the way toward their integration in advanced energy conversion devices, including SCs.

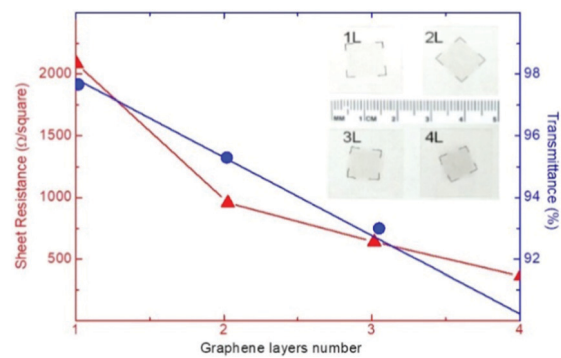

**Fig. 4** Sheet resistance ($R_s$) and transmittance ($T_r$) of a graphene (film) as a function of the number of stacked graphene layers. Adapted from ref. 169, Copyright 2008, American Chemical Society.

It is important to point out that GRM films formed by interconnected flakes are commonly proposed as functional components in several massive applications, including energy conversion and storage applications.[5] In fact, this approach can be realized by printing GRMs that are produced in the form of inks and pastes by means of scalable methods, such as liquid-phase exfoliation (LPE) (detailed discussion of solution-based syntheses and processing of GRMs is provided in Section 2.4).[205] Although, in principle, the properties of each flake can be preserved, the properties of whole films strongly depends on their morphology and structure, which are determined by the orientation and interconnection between the composing flakes. The same flakes can show different lateral sizes, thicknesses, and chemical compositions (in the case of heterogeneous films), resulting in different optoelectronic characteristics.[205] For the specific case of graphene flake films, contact resistance between the flakes and poor film compactness drastically decrease the conductivity of films compared to that of SLG and FLG ($> 10\,000$ S cm$^{-1}$).[69] Based on our experience, an as-deposited film of pristine graphene flakes typically shows a conductivity lower than 10 S cm$^{-1}$, together with poor mechanical properties. In order to strengthen the interconnection among graphene flakes, the incorporation of polymeric binders and other conductive additives, e.g., carbon black, is a common strategy that enables a low-temperature-processed film to achieve conductivity higher than 100 S cm$^{-1}$.[206] In particular, the use of carbon black nanoparticles (NPs) or other carbon NPs is effective to fill the voids of the as-deposited network of graphene flakes, which are consequently electrically bridged.[206,207] The application of pressure, as well as thermal treatments, can further increase the conductivity of graphene flake films.[208,209] In fact, compression treatments make the graphene-based films denser by decreasing the distance between flakes.[208] Meanwhile, thermal treatments can decompose or even evaporate the binders and/or surfactants, which limit the conductivity of the film.[209] Similar arguments apply to films composed of flakes of other GRMs besides graphene, although their functionalities can be different from those discussed above for graphene flake films.

## 2.2 Two-dimensional materials beyond "conventional" GRMs

### 2.2.1 Nonlayered materials.
Beyond the class of GRMs, nonlayered materials have been created in 2D forms, raising the research interest on either fundamental research or applications in the field of optoelectronics.[109,210–212] A comprehensive overview on the recent advancements of photoelectric devices based on 2D nonlayered materials is given in ref. 109, as well as in previous reviews.[210–213] In the context of PV applications, nonlayered materials display fascinating properties, which can complement those of GRMs.[214] In particular, the presence of structural distortions, surface dangling bonds, and coordinated–unsaturated surface atoms can promote rapid interfacial charge transfer,[214] thereby leading to efficient charge extraction in PV devices. In addition, their chemical reactivity can be used to create in situ interface engineering for the design/realization of novel concepts of charge extraction.[109,214] Example of 2D nonlayered materials are oxides/hydroxides (e.g., α-FeOOH,[215] CoOOH,[215] $TiO_2$,[215] γ-$Ga_2O_3$,[216] $Fe_2O_3$,[217] $Co_3O_4$,[217] $Mn_2O_3$,[217] and mixed oxides such as $ZnMn_2O_4$ (ZMO),[217] $ZnCo_2O_4$,[217] $NiCo_2O_4$,[217] and $CoFe_2O_4$[217]), sulfides (e.g., $Ga_2S_3$,[218] ZnS,[215] NiS,[215] $FeS_2$,[219] and $CuFeS_2$[220]), selenides (e.g., $In_2Se_3$[221] and ZnSe[222]), tellurides (e.g., ZnTe),[223] Ni-B oxide,[224] γ-CuBr,[225] CuI,[226] InI,[227] PbS,[215] carbonates (e.g., $CaCO_3$, $ZnCO_3$, $MnCO_3$, $FeCO_3$, and $PbCO_3$),[215] as well as elemental Se,[228] Bi,[229,230] Te,[231] and Se.[232] In this list, $In_2S_3$ is a direct-bandgap semiconductor in both monolayer and few-layer forms,[221] leading to a significantly different behavior compared to group-6 TMDs. Other 2D nonlayered materials, such as $Ga_2S_3$ and CuBr, exhibit bandgaps of around 3 eV,[225,233] which is between those of group-6 TMDs and h-BN. Therefore, they can be considered as photoactive materials in the UV spectrum, as well as advanced charge-selective layers. Meanwhile, materials such as elemental ones (e.g., Ge, Te, Se) or $CuFeS_2$ exhibit bandgaps of less than 1 eV,[220,228,231] bridging the optical properties of graphene and the most established TMDs. Moreover, such bandgap values are attractive for the development of near-infrared (NIR) to mid-infrared (MIR) photoabsorbers. Other essential features of 2D nonlayered materials are the high tunability of their optoelectronic properties by means of engineering their surface chemical properties (e.g., control of the number of vacancies in $In_2S_3$[234] and $Ga_2S_3$[218]), as well as theoretical high charge mobility (e.g., electron mobility up to 252 000 cm$^2$ V$^{-1}$ s$^{-1}$ for PbS).[235]

To date, 2D nonlayered materials have been demonstrated for UV-sensitive photodetectors, reaching a responsivity of up to 3.3 A W$^{-1}$ for Ga$_2$O$_3$,[236] 400 A W$^{-1}$ for $\alpha$-Bi$_2$O$_3$, and 3.17 A W$^{-1}$ for $\gamma$-CuBr.[225] In addition, visible-light photodetectors were successfully achieved using CdTe nanoflakes (responsivity of 0.6 mA W$^{-1}$),[237] ZnTe nanoflakes (responsivity as high as 453.9 A W$^{-1}$),[223] and $\alpha$-MnS (responsivity of 139 A W$^{-1}$).[238] PbS,[239] ZnSb,[240] and Te[231] nanoflakes were used for IR photo-detection, reaching responsivity of 1621 A W$^{-1}$, 89.2 A W$^{-1}$, and 13 A W$^{-1}$, respectively. Finally, CuGaSe$_2$,[241] $\alpha$-In$_2$Te$_3$,[242] Pb$_{1-x}$Sn$_x$Se,[243] Bi$_2$,[229,230] In$_2$S$_3$,[234] CuInSe$_2$,[244] Te,[245] and Ge[228] have also shown attractive properties for broadband photodetection.[109] Beyond the use of single 2D nonlayered materials, more complex photodetectors have been produced by coupling 2D materials, including layered and nonlayered ones, in the form of in-plane and out-of-plane heterostructures. Therefore, novel Schottky structures, p–n junctions, and phototransistors have been successfully proposed, as summarized in ref. 109. The application of 2D nonlayered materials has also been reported, although this technology is still in an early stage of development.[109] The progress in the control of unsaturated dangling bonds of 2D nonlayered materials is mandatory for the realization of high-quality PV devices. Recently, CdS/Cu$_2$S heterojunction with a clear PV effect was realized via the cation-exchange protocol, yielding an $\eta$ value of 2.1% (despite a cell thickness of only $\sim$30 nm).[246] Alternatively, GRMs have been used as an ideal interface for the growth of 2D nonlayered materials.[109] Based on this strategy, a PV device was fabricated by directly depositing a thin layer of MoO$_3$ onto MoS$_2$, reaching an $\eta$ value of 3.5%.[247] Despite these progresses, the application of 2D non-layered materials in prototypical SCs, including 1st-, 2nd-, and 3rd-generation SCs, has not been established yet, probably due to the difficulties in producing 2D nonlayered materials on a large scale.[109] Therefore, these materials will not be the specific subject of discussion in the present work. By achieving reproducibility in terms of thickness, crystallinity, and structural properties, their incorporation in practical PV devices could represent a key point to drive PV technologies beyond their current performances.

### 2.2.2 Two-dimensional conjugated metal–organic frameworks.

Metal–organic frameworks are crystalline coordination polymers that have emerged for various applications (e.g., energy conversion and storage systems, proton conduction membranes, and sensors) owing to their ultrahigh porosity (up to 90% free volume) and large surface area (even beyond 6000 m$^2$ g$^{-1}$).[248] The topic of MOFs is a research hotspot in materials science, as comprehensively reviewed by several recent literature works,[248–250] to which we specifically refer the reader of this work. As an evolution of MOFs, their 2D form, i.e., 2D c-MOFs, has also been developed to extend the properties of traditional MOFs. For example, the long-range $\pi$-conjugation in their 2D planes promotes the delocalization of charge carriers within the network, leading to high mobility and conductivity,[251,252] as well as providing additional possibility for multifunctional electronic devices for the recently called "MOFtronics."[253] In particular, 2D c-MOFs can exhibit high stability together with tunable optoelectronic properties, (photo)-electrochemical activity,[254,255] ferromagnetic ordering,[256] and topological states,[257] yielding a potential source for SCs, beyond their use in batteries[258,259] and supercapacitors.[260,261] In addition, their liquid-phase processability is particularly relevant for the realization of solution-processed SCs. For example, a thiol-functionalized 2D c-MOF has been recently used as an electron-extracting layer at the perovskite/cathode interface.[262] Meanwhile, a Te-based 2D c-MOF was introduced in PSCs to passivate the electron transporting layer (ETL) in TiO$_2$, while improving the morphology of the perovskite photoactive film.[263] Despite these promising results, the use of solution-processed 2D c-MOFs in SCs is still in its infancy, even though it is plausible that these materials can prospectively play a significant role in PV devices. In particular, we expect that the progresses in their synthesis, as well as the scaling-up of their synthesis strategies, will be crucial for the rational implementation of 2D c-MOFs in cutting-edge SC technologies.

### 2.2.3 Two-dimensional carbon nitrides.

By attempting to open the zero $E_g$ of graphene to provide intrinsic semiconductivity while maintaining a graphite-like atomic crystalline structure, bottom-up approaches with C-rich and N-rich precursors are successfully reported to produce 2D carbon nitrides (C$_x$N$_y$),[264] including the most prevalent ones such as graphitic carbon nitride (g-C$_3$N$_4$),[265,266] C$_2$N,[267–269] C$_3$N,[270,271] and C$_5$N$_2$.[272] Owing to their large surface area and tunable optoelectronic properties, such class of materials has been widely investigated for the realization of photocatalysts (and even photocatalyst supports)[273,274] and (electrochemical) energy storage systems.[275–277] Beyond these applications, 2D C$_x$N$_y$ also represents a promising class of solution-processable materials for SCs, as testified by their use in OSCs,[278,279] DSSCs,[280–282] and PSCs.[283–286] However, compared to GRMs, the rational engineering of most 2D C$_x$N$_y$ materials is still limited, and theoretical studies are needed to elucidate the influence of the number of layers, defects, and chemical modifications on their performance when used as functional components in SCs. Moreover, it should be noted that the precursors used for the synthesis of 2D C$_x$N$_y$ are often expensive, and the synthesis strategies are complex and require highly controlled experimental conditions. These aspects critically limit their use in massive applications, including SCs. Therefore, the present work will not focus on this class of materials, even though some results achieved with the most established C$_x$N$_y$ materials are mentioned in the discussion on PV technologies investigated here.

## 2.3 Classification of semiconductor 2D materials: n-type or p-type materials?

To provide some guidelines regarding their functional role in SCs, semiconductor 2D materials can be classified depending on their (opto)electronic properties. However, for the case of solution-processed 2D materials, such properties are strongly influenced by both structural and chemical characteristics. The possibility to on-demand tune the (opto)electronic properties by structural and chemical engineering is a key feature of solution-processed 2D materials, making them extremely versatile for application in PV devices. As discussed in the following sections (4, 5, 6, and 7), the structure of SCs is commonly engineered by introducing proper charge transporting layers (CTLs), which efficiently and selectively extract the photogenerated charges, improving the device performances. In this context, it is common to consider p-type and n-type materials to extract holes and electrons, respectively. However, the choice of CTLs can follow more complex rationales. In fact, the charge transporting properties are determined by the entire electronic structure of the materials, as well as by their chemical reactivity with the interfaced materials. As a striking example, $MoO_3$, which can also be found in the 2D form, is a typical n-type material that acts as an efficient hole transporting layer (HTL) due to its high $\phi_W$.[287,288] The latter can even be higher than 5 eV,[289] similar to that exhibited by common p-type materials used to extract photogenerated holes.[290] Therefore, $MoO_3$ can efficiently collect holes from its CB through an electron injection mechanism.[291] Furthermore, $MoO_3$ forms a highly p-type-doped interface with active materials having ionization energies lower than $\phi_W$ of $MoO_3$, favoring the hole extraction process.[291–293] Similar to $MoO_3$, 2D materials can go against the rules "p-type materials collect holes" and "n-type materials collect electrons"; therefore, they should be specifically examined to understand their functional role in the SC structure. Based on this consideration, semiconductor 2D materials will not be classified as n-type or p-type materials because there is no a clear one-to-one correspondence between 2D materials and their electronic properties, as well as between the electronic properties of solution-processed 2D materials and their functional role in PV devices.

## 2.4 Solution processing of 2D materials

The design, development, and production of (opto)electronic devices[73,86,294,295] inherently depend on the properties of the available materials.[83,296] Different methods have been reported for the production and processing of GRMs. The main approaches for the production of GRMs have been summarized in previous works.[83,296–298] Although proof-of-concept PV devices have been demonstrated for exploiting micromechanically cleaved materials,[299] the discovery of scalable methods to produce GRMs with "on-demand" tuned structural and (opto)electronic properties is a "must" for the realization of practical SCs. The production of large-area, high-quality GRMs is still one of the most urgent needs of this research area,[83,296,297] even though several progresses have been accomplished at the industrial level. The requirement to exercise control at the monolayer level needs the understanding of surface physics and chemistry, which has so far not been fully demonstrated in any multicomponent materials system. For example, progress is being made toward the production of large-area single crystals,[297,300–306] a key process for the development of high-quality thin films with both optical transparency and electrical conductivity.[307,308] Growth techniques reported in the literature for 2D materials, e.g., chemical vapor deposition (CVD), molecular beam epitaxy (MBE), and atomic layer deposition (ALD), have been conventionally used to create heterostructures based on graphene, other elemental 2D materials, TMDs, TMOs, h-BN, and oxide materials.[83] For example, significant progresses have been made in the growth of graphene on metals[309] and on silicon carbide (SiC).[83,310,311]

By carefully choosing the individual components, one can tune the growth/production parameters, creating GRMS "on demand" for the design and realization of van der Waals heterostructures with functional properties.[187,312,313] However, the growth of 2D materials by means of the aforementioned synthesis routes is challenging in the case of nonmetallic substrates.[314] In order to exploit the availability of high-quality synthetic 2D materials for practical devices, the transfer and alignment processes of 2D films on arbitrary substrates have to be developed. Several transfer processes classified as wet- or dry-transfer have been proposed and utilized so far.[83,297,305,315–319] In the wet-transfer process, the as-grown 2D material contacts the liquid during at least one step of the process.[83] This determines the occurrence of adsorbates that are trapped onto the 2D material surface, significantly influencing the interface quality. To avoid this drawback, dry-transfer processes have been established to create perfectly clean interfaces.[83,320] This has been a crucial step for the demonstration of the fundamental properties of 2D materials, which requires extremely low densities of interface traps and dangling bonds.[321]

Recent reports on the dry transfer of graphene using pick-and-place techniques[322] and exploiting h-BN as the 2D dielectric have successfully achieved extremely high $\mu$ (i.e., 350 000 cm$^2$ V$^{-1}$ s$^{-1}$) in graphene.[323] However, transfer processes intrinsically represent limitations for the integration of high-quality 2D materials in practical devices, in which direct material growth on ad hoc materials and/or solution-based processing are required for the realization of high-throughput device manufacturing chains. Recently, the direct growth of graphene on glass, creating the so-called "super graphene glass," has attracted enormous interest to circumvent transfer-process-related issues for practical applications,[324–327] including transparent conductors, smart windows, single heating devices, and SC electrodes. However, the CVD growth of high-quality graphene is still challenging, and "super graphene glasses" currently show (opto)electronic properties still far from those of CVD graphene grown on metallic substrates.[324] In fact, on a catalytically inert glass surface, one cannot expect yet to control the graphene growth as done onto a catalytically active metal surface.[83]

The direct exfoliation of bulk layered crystals by LPE[328–330] is an industrially relevant strategy for the scalable production and/or processing of GRMs. Herein, we will summarize the main methods for the production and processing of 2D materials in solution, while additional details can be found in recent literature reviews.[83,205,297,298,331] The LPE process enables the formulation of inks of GRMs in different solvents (Fig. 5a).[332–335] This is the starting point for the reliable production of devices based on printed technologies,[333] as well as for targeting the industrial fabrication of GRM-based devices, including SCs (Fig. 5b).[205]

Liquid phase exfoliation is a versatile technique and it has been established for the exfoliation of numerous layered materials,[328–330] including graphite, TMDs, TMMs, black phosphorus (BP), and h-BN, just to cite a few. As depicted in Fig. 6, the liquid-phase processing of bulk layered crystals generally

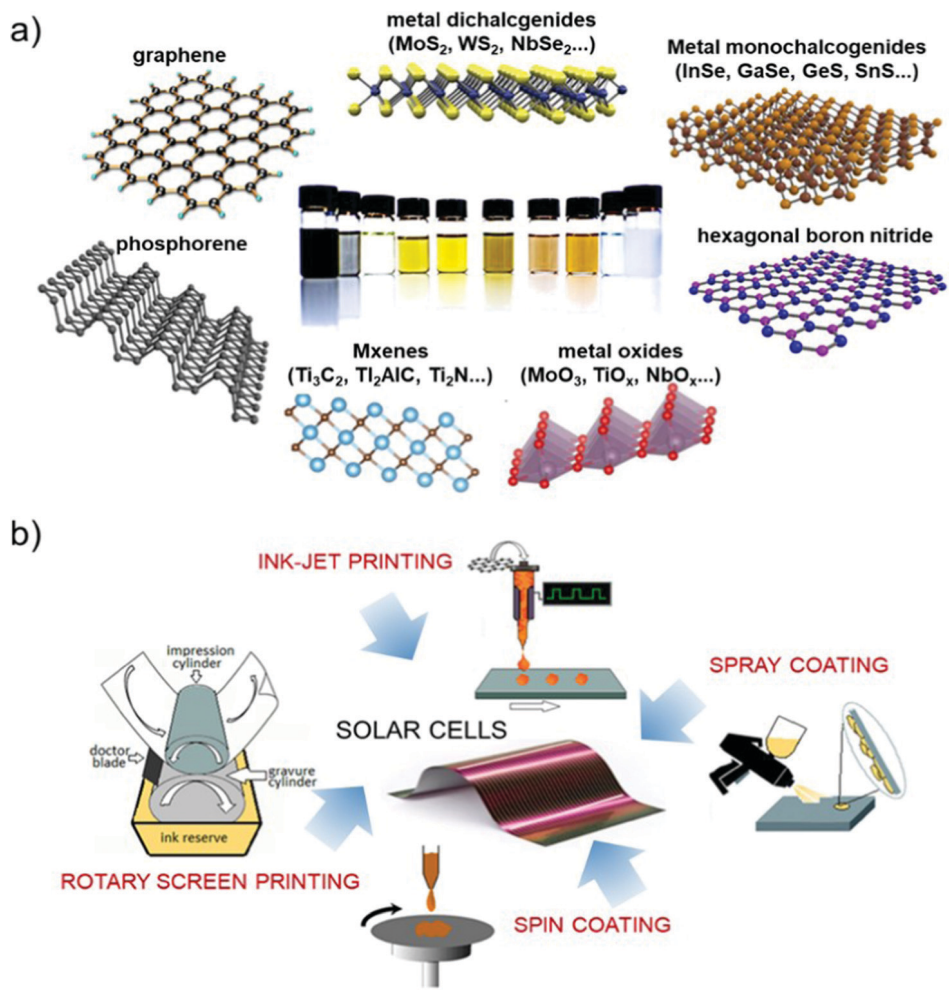

**Fig. 5** (a) Schematic of monolayer materials, *e.g.*, graphene, elemental 2D materials (phosphorene), metal dichalcogenides, metal monochalcogenides, MXenes, h-BN, and metal oxides, and their formulation in the form of ink. (b) Schematic of solution-processing methods of SCs, including relevant material deposition techniques (*e.g.*, spin coating, spray coating, inkjet printing, and rotary screen printing).

involves (1) the dispersion of bulk crystals in a solvent; (2) the exfoliation of bulk crystals through (acoustic) cavitation or shear forces (Fig. 6a); (3) the "sorting" (*e.g.*, by ultracentrifugation) of the material flake sizes (Fig. 6b).[83,205]

In general, the LPE process starts with the dispersion of bulk crystals either in organic solvents[330] or in aqueous solutions, the latter with the aid of surfactants[329,338,339,340] or polymers.[341,342] The exfoliation process is commonly carried out by exploiting cavitation[328–330] or shear forces[343] to produce single- and few-layer materials.[344] Ultrasonication-assisted exfoliation of bulk crystals is the prototypical LPE method.[328,329,345–349]

For the case of graphene, the ultrasonication process produces defect-free flakes (*i.e.*, no additional defects are introduced during exfoliation) as well as achieves concentrations of several grams per liter.[350] However, ultrasonication-assisted LPE is not a scalable process, since it is a time-consuming process requiring several hours.[205] Other approaches have also been proposed, such as ball milling,[351–353] shear exfoliation,[354,355] and microfluidization.[356–359] All these approaches have pros and cons compared with the ultrasonication method,[205] even though some of the apparatus can yield high-throughput production of 2D materials for industrial applications. Recently, Bonaccorso and co-workers presented a novel approach to exfoliate layered crystals, *i.e.*, graphite, h-BN, and TMDs, based on the high-pressure wet-jet milling (WJM) technique.[360] In detail, during the WJM process, a hydraulic piston applies a pressure between 180 and 250 MPa, forcing the solvent/layered-crystal mixture to pass through perforated disks with variable diameters (typically between 0.3 and 0.1 mm), called the nozzle. This process generates shear forces that promote the exfoliation of layered materials.[359,361] The key advantage of the WJM technique compared to other LPE methods is the small time required to process the sample, which is reduced to less than one second, instead of the several hours required during ultrasonication-assisted exfoliation[328,329,344–348] or shear exfoliation.[353,354] By means of the WJM method, a production rate higher than 2 L h$^{-1}$ of 2D crystal dispersion (concentration: 10 g L$^{-1}$) and an exfoliation yield (defined as the ratio between the weight of the exfoliated material and the weight of initial graphite) of 100% have been demonstrated with a single WJM apparatus.[359,362] The 2D crystals obtained through WJM have already been used for a wide range of ESC applications[363–371] and composites,[372] in which a large volume of material is needed for their industrial implementation. Another approach to upscale the 2D material production (beyond tens of grams per hour) is the electrochemical exfoliation process. In this method, a potential difference is applied between a layered anode/cathode in an electrolyte-containing medium.[373–375] In these experimental conditions, positive or negative charges can be imparted to the layered materials, promoting the intercalation of oppositely charged ions and facilitating the exfoliation process.[372–374] These processes can be broadly classified in two classes. The first one is the anodic exfoliation in the inorganic salts' aqueous solution, mineral acids, or mixture of water and ionic liquids. The second one is the cathodic exfoliation in organic solvents (*e.g.*, *N*-methyl-2-pyrrolidone, dimethyl sulfoxide, and propylene carbonate) in the presence of alkylammonium salts or Li.[374] Electrochemical methods are extremely attractive since they reduce the use of chemical oxidants as the driving force for intercalation or exfoliation, and the electromotive force is controllable for the creation of tunable-material intercalated compound.[372–374] In addition, the extensive capabilities of the electrochemical exfoliation method to modify materials enables the facile and direct synthesis of functionalized 2D materials with the desired properties for composites, electronics, and ECS applications.[372–374] Regardless of the LPE process used to produce 2D materials, a common key issue of the aforementioned methodologies is that the resulting samples are polydisperse in their dimension, typically showing broad distributions of flake thickness and size.[205,297] Thus, it is crucial to obtain a fine adjustment of the morphological properties by separating small from large flakes[337] and thin from thick ones.[339] This step is typically performed using ultracentrifugation protocols.[339,376–381] In this context, exfoliated GRMs can be sorted with respect to thickness and lateral size using techniques based on ultracentrifugation in uniform media (SBS)[382] or density gradient media (density gradient ultracentrifugation (DGU)).[381]

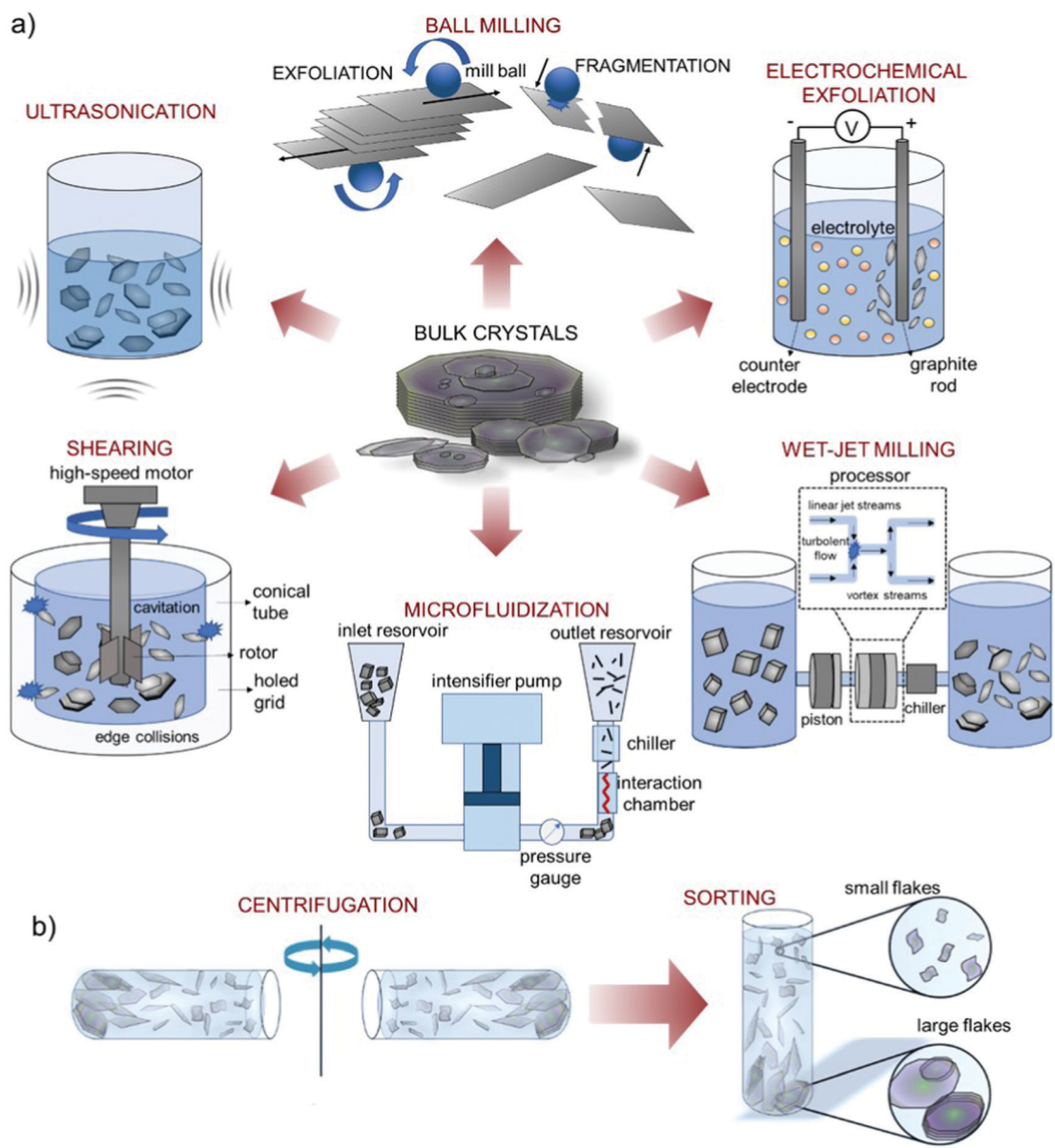

**Fig. 6** Schematic of the LPE processes. (a) Schematic of various LPE methods reported in the literature, including ultrasonication, shearing, wet-jet milling, microfluidization, ball milling, and electrochemical exfoliation. Schematic of the LPE methods adapted from ref. 336, 356, 360. (b) Dispersion purification by means of ultracentrifugation (sedimentation–based separation (SBS), and subsequent ''sorting'' of different material flake sizes. Adapted with permission from ref. 337, Copyright 2017, American Chemical Society.

Another important issue of 2D flakes produced in the form of ink through LPE methods is the re-aggregation of flakes after their deposition/coating. Therefore, flake re-aggregation might affect the electronic (i.e., $\mu$, contact resistance) and physical (i.e., roughness) properties of the resulting films. Therefore, suitable strategies must be developed to minimize flake reaggregation with regard to practical applications. For example, the addition of stabilizers (e.g., surfactants and polymers) physically hinders flakes' contact,[337–339] impeding their aggregation.[337–339] However, the effect of such stabilizers could affect the electrical performance of the assembled films.[297,383,384] In addition to the aforementioned issues, some layered materials, such as BP, are unstable under ambient conditions or in the presence of water and/or oxygen.[385,386] The instability issue, which might be valid for 2D materials either grown by the bottom-up approach (e.g., CVD)[83,387,388] or produced by micromechanical cleavage,[69,389,390] can be eliminated by introducing a solvent shell,[205] or residual surfactants/polymers adsorbed onto the surface of flakes. Importantly, solvents or surfactant residuals may imply an intrinsic doping of the flakes.[205] These effects can be advantageously used to attain controllable doping strategies. The LPE process can also be exploited for the exfoliation of bulk layered materials specifically prepared/synthetized with the desired chemical characteristics (e.g., doping), as for the case of graphite oxide[391] (i.e., to produce GO).[83,330] In particular, graphite oxide can be prepared by means of chemical processes that introduce functional groups both at the edges (e.g., COOH and C=O) and on the basal plane (e.g., OH or epoxide groups).[81,82] The occurrence of these functional groups is fundamental toward the production of GO using well-established methods, including thermal expansion,[392] ultrasonication,[393] and stirring[394] of graphite oxide. Moreover, the presence of the aforementioned functional groups introduces polarities[395–397] that facilitate the dispersion of exfoliated graphitic materials in aqueous solutions.[392,398] Although GO flakes can have lateral sizes up to several microns,[399] they exhibit a high density of structural defects,[393] which arise from the chemical treatment disrupting the $sp^2$-bonded network.[83] Thus, in order to restore both electrical and thermal conductivities of pristine graphene, various strategies have been developed to reduce GO flakes using either chemical[397] or physical[393,397,400] processes. These reduction processes are imperative to produce a sample of the quality approaching that of pristine graphene. Recently, tremendous progresses have been achieved in this direction, with the demonstration of $\mu$ exceeding 1000 cm$^2$ V$^{-1}$ s$^{-1}$ in field-effect transistors with microwave-reduced GO.[401]

Owing to its scalability and cost-effectiveness, LPE techniques can provide GRMs in massive quantities at an accessible price. Moreover, solution-processed 2D materials can be combined with polymeric materials, while being processed in the form of a coating on different substrates. In this context, progresses have been made on the large-scale placement of 2D material-based inks by means of different deposition/coating systems, such as Langmuir–Blodgett,[402] spin,[403–405] spray,[406–408] and rod[73] coating; vacuum filtration;[409–415] and inkjet,[332,333,416] gravure,[417] flexographic,[418] and screen[419] printing (including their roll-to-roll (R2R) configurations).[420] Advances in this area enabled the layer-by-layer printing of different 2D material-based films, as well as heterostructures and/or heterogeneous structures, on large areas (ranging from the scale of square centimeters to square meters).[205] However, beyond uniformity, the roughness of the deposited large-area films is a critical issue, which may degrade the (opto)-electronic properties expected from heterostructures produced through material transfer after micromechanical cleavage or direct growth.[187] However, different from the transfer approach, drop-on-demand printing could meet the large-scale fabrication requirements of practical devices.[205] For example, drop-on-demand inkjet printing has been demonstrated for the realization of all-printed, vertically stacked transistors with a graphene source, drain, and gate electrodes; a TMD channel; and an h-BN separator.[421] The proposed printed device, based on 2D material heterostructures, has shown a $\mu$ value of 0.22 cm$^2$ V$^{-1}$ s$^{-1}$.[420] Despite these important achievements, the obtained $\mu$ value is rather low, meaning that further insights are still needed into the assembly of such printed heterostructures.[187,332] Here, the challenges to be tackled are two-fold: (1) the optimization of ink formulation fulfilling the morphological (e.g., thickness and lateral dimension of the flakes) and rheological (e.g., viscosity and surface tension of the dispersions) property requirements; (2) the optimization of printing parameters for the deposition of uniform 2D material films with high-quality (i.e., clean) interfaces.[205] Noteworthily, the surfaces of 2D materials are strongly affected by both solvent and additive (i.e., stabilizers) residuals,[205] which, therefore, need to be minimized. Here, a balance must be found between the possibility to have a clean interface and the intrinsic doping (determined by the presence of solvent and additive residuals) on a case-to-case basis, depending on the final application. Overall, notwithstanding the scalable production of GRMs and their film deposition, understanding the precise determination of crystal structures and their crystallographic relationships is of utmost importance for the design and realization of any (opto)electronic device, including PV ones that are discussed here. Further, chemical doping and functionalization are pivotal to properly tune the (opto)electronic properties of the structures.[344,422–425] Both covalent and noncovalent functionalizations introduce a systematic modification of 2D material properties to control their solubility/processability, the prevention of flake re-aggregation, and their (opto)electronic characteristics (e.g., $E_g$).[344,421–423] The chemical modification/functionalization also allows the properties of 2D materials to be combined with the property portfolio of other compounds.[344,420] Overall, a thorough understanding of the charge transport and transfer properties, defects (including edge terminations, dopants, point defects, and grain boundaries), environmental contaminants (*e.g.*, surfactants and adsorbates), and chemical reactivity is crucial for the design of practical GRM-based devices.

### 2.5 Functional roles of solution-processed 2D materials

The understanding of "*how to use 2D materials in SCs*" is not trivial, since their versatility resulting from the immense portfolio of their (tunable) properties can lead to apparently contradictory experimental results. In fact, there exist solution-processed 2D materials that have been applied to collect either photogenerated holes or electrons, while being used as buffer layers to stabilize the interfaces between the materials comprising the SCs, or even as catalysts for the redox reactions involved at the counter electrodes (CEs) in DSSCs, or as electrically conductive materials for current collectors. This aspect is so surprising to the extent that it could even be disappointing, albeit it reveals the easiness to incorporate 2D materials in SC structures to improve their performances. Scheme 1 reports a sketch of the various functional roles of material components in SCs, as they will be detailed for each type of technology in the subsequent sections. Clearly, solution-processed 2D materials have been applied almost everywhere, most of them for more than one functionality. The most representative example material class, namely, "graphene and its derivatives," has been used for all the functional roles identified here, indicating the importance to specify the structural, morphological, and chemical properties for each material, thereby using a "case-by-case approach." In addition, this point implicitly stresses the importance of providing a full set of experimental characterizations of 2D materials when used in SCs, so that it is possible to uniquely correlate their functional role to their intrinsic attributes. Even though it is common to refer to electronic structures of 2D materials in ideal stoichiometry to explain their functional role in SCs, it is recommended to provide experimental measurements (beyond those which are used for the characterization of SCs) to confirm the absence of a relevant discrepancy compared to such ideal cases. In fact, defects, surface oxidation, chemical functionalization, and even the simple morphology of 2D materials can result in optoelectronic properties that are completely different from those of their ideally stoichiometric structures. Examples of effective characterizations are absorbance/reflectance measurements coupled with ultraviolet photoelectron spectroscopy and Kelvin probe measurements to provide the first sketch of the energy-band edge positions and WF values of the materials used in the different components of SCs. Possible discrepancies should be explained by investigating the chemistry of the material surface through X-ray photoelectron spectroscopy (XPS). The impact of 2D morphology on the functional role of 2D materials should be supported by proper lateral and thickness analyses through transmission electron microscopy (TEM), atomic force microscopy (AFM), and surface area measurement techniques (*e.g.*, physisorption characterizations), while the structural properties of 2D materials can be rapidly assessed by both Raman spectroscopy and X-ray diffraction (XRD) characterizations. Electrical and photoelectrical properties, such as (photo)resistivity/(photo)conductivity, of 2D materials could be accessed by realizing and characterizing complementary devices, such as field-effect transistors, as well as a simple four-probe method. These considerations indicate the key importance of providing reliable insights into the nanomaterials, devoted to improve the performance of entire PV systems, which must be carefully rationalized through in-depth experimental characterization. In this context, the efforts recently made to standardize the sequence of methods for characterizing the structural properties of graphene, bilayer graphene, and graphene nanoplatelets in both powder and liquid (*i.e.*, dispersion) forms are noteworthy. The need of such a standard, namely, ISO/TS 21356-1:2021, emerged from the confusion around the terminology of "graphene" used to label commercially available materials. In conjunction with the international ISO/IEC terminology, the ISO/TS 80004-13:2017 standard represents a step forward to the use of (solution-processed) 2D materials with well-defined properties in both laboratory and commercial applications, including SCs.

## 3. Figures of Merit of Solar Cells

For facilitating comparison, SCs are often ranked in terms of the following FoM:[53]

(i) EQE, which represents the ratio between the number of charge carriers collected by the cell and that of photon flux (of a given energy) that strikes the cell, *i.e.*,

$$\text{EQE}(\lambda) = \frac{\left(\dfrac{I}{e^-}\right)}{\left(\dfrac{P_{\text{in}}}{hc/\lambda}\right)} \quad (3.1)$$

where $I$ is the electrical current given by the SC, $e$ is the elementary charge ($1.6021766208 \times 10^{-19}$ C), $P_{\text{in}}$ is the power of incident light, $h$ is the Planck constant, $c$ is the speed of light in a vacuum, and $\lambda$ is the photon wavelength.

(ii) Internal quantum efficiency (IQE), *i.e.*, the fraction of absorbed photons converted in $I$, *i.e.*,

$$\text{IQE}(\lambda) = \frac{\left(\dfrac{I}{e}\right)}{\left(\dfrac{P_{\text{in}}}{hc/\lambda}\right) \times (1 - R)} \quad (3.2)$$

(iii) the overall $\eta$, defined as the ratio between the maximum output electrical power ($P_{\text{max}}$) of the cell, and $P_{\text{in}}$, *i.e.*,

$$\eta = \frac{P_{\text{max}}}{P_{\text{in}}} = (V_{\text{OC}} \times I_{\text{SC}} \times \text{FF}) \quad (3.3)$$

where $V_{\text{OC}}$ is the maximum open-circuit voltage, $I_{\text{SC}}$ is the short-circuit current, and FF is the fill factor. Here,

$$\text{FF} = \frac{(V_{\text{MPP}} \times I_{\text{MPP}})}{(V_{\text{OC}} \times I_{\text{SC}})} \quad (3.4)$$

| ETLs |
|---|
| • Graphene & derivatives |
| • Transition metal dichalcogenides (*e.g.*, $MoS_2$, $TiS_2$, $SnS_2$) |
| • MXenes (*e.g.*, $Ti_3C_2T_x$) |
| • Elemental 2D materials (*e.g.*, black phosphorous/phosphorene) |

| HTLs |
|---|
| • Graphene & derivatives |
| • Transition metal dichalcogenides (*e.g.*, $MoS_2$, $MoSe_2$, $WS_2$, $WSe_2$, $TaS_2$, $NbSe_2$ |
| • Group-15 metal calchogenides (e.g., $Bi_2Se_3$) |
| • MXenes (*e.g.*, $Ti_3C_2T_x$ |
| • III-VI compounds (*e.g.*, $In_2Se_3$) |
| • Elemental 2D materials (*e.g*,. black phosphorous |

| Dopants for CTLs |
|---|
| • Graphene & derivatives |
| • Transition metal dichalcogenides (*e.g.*, $MoS_2$, $WS_2$) |
| • Elemental 2D materials (*e.g.*, black phosphorous, Antimonene |
| • $g$-$C_3N_4$ |
| • Group-15 metal calchogenides (*e.g.*, $Bi_2Te_3$) |

| Electrodes (current collector) |
|---|
| • Graphene & derivatives |
| • MXenes (*e.g.*, $Ti_3C_2T_x$) |

| Catalysts for CEs (in DSSCs) |
|---|
| • Graphene & derivatives |
| • Transition metal dichalcogenides (*e.g.*, $MoS_2$, $WS_2$, $WSe_2$, $NbSe_2$, $NiS_2$, $NiSe_2$, $TiS_2$ and $TaSe_2$) |
| • Group-15 metal calchogenides (e.g., $Bi_2Se_3$) |

| Additives for photoactives |
|---|
| • Graphene & derivatives |
| • Transition metal dichalcogenides (*e.g.*, $WSe_2$) |
| • Elemental 2D materials (*e.g.*, black phosphorous/phosphorene) |
| • MXenes (*e.g.*, $Ti_3C_2T_x$) |
| • Metal oxysulfides (*e.g.*, $Bi_2OS_2$) |
| • MOFs |
| • 2D perovskites (*e.g.*, $(PEA)_2PbI_4$) |
| • $g$-$C_3N_4$ |

| Buffer layers (protective layers and passivating layers) |
|---|
| • Graphene & derivatives |
| • Transition metal dichalcogenides (*e.g.*, $MoS_2$ and $TiS_2$) |
| • Elemental 2D materials (*e.g.*, black phosphorous) |
| • Group-15 metal chalcogenides (*e.g.*, $Bi_2Te_3$) |
| • $g$-$C_3N_4$ |
| • 2D c-MOFs |
| • 2D perovskites |

| Interconnecting layers |
|---|
| • Graphene & derivatives |
| • Transition metal dichalcogenides (*e.g.*, $MoS_2$) |

| Encapsulants |
|---|
| • h-BN |

**Scheme 1** Functional components of SCs and the corresponding 2D materials reported in the literature for such a role. The 2D materials listed here correspond to those reported in the subsequent sections for each type of SC technology reviewed in this work.

where $V_{MPP}$ and $I_{MPP}$ are the voltage and current, respectively, at the maximum power point (MPPT), defined as the voltage at which $d(IV)/dV = 0$.

Since the application of solution-processed GRMs as transparent conductive electrode (TCEs) for SCs will be examined here, the FoM determining the quality of TCEs are also reported and discussed. The quality of TCEs is mainly assessed through two crucial parameters: $R_s$ and $T_r$, which should be $<10\ \Omega\ \square^{-1}$ and $>90\%$, respectively.[73] Moreover, a trade-off between $R_s$ and $T_r$ is unavoidable for TCEs.

To evaluate TCEs, the following semiempirical FoM has been proposed:[426]

$$\Phi_{H} = \frac{T_r^x}{R_s} \quad (3.5)$$

where exponent $x$ determines the required $T_r$ value for a specific purpose.

Notably, $R_s$ depends on both charge carrier density ($N_d$) and $\mu$ (cm$^2$ V$^{-1}$ s$^{-1}$),[427] as expressed by the following equation:

$$R_s = \frac{1}{\mu N_d t} \quad (3.6)$$

where $t$ is the thickness of the TCE film.

In order to describe the frequency dependence of the $T_r$ losses in TCEs,[428–430] as well as the critical reflection at the air/film/substrate interfaces,[431] the following equation for $T_r$ has been proposed (for thickness $\ll \lambda/2\cdot\pi\cdot n_{film}$, where $n_{film}$ is the refractive index of the film):

$$T_r(\lambda) = \frac{16 n_{sub}^2}{R_s(1 + n_{sub})^4} \times \frac{1}{\left(1 + \dfrac{Z_0}{R_s} \times \dfrac{1}{(1 + n_{sub})} \times \dfrac{\sigma_{opt}}{\sigma_{dc}}\right)^2} \quad (3.7)$$

In eqn (3.7), $Z_0$ is the vacuum impedance (377 $\Omega$);[427] $\sigma_{opt}$ and $\sigma_{dc}$ are the optical and electrical dc conductivities (also simply referred to as $\sigma$) of the material, respectively; and $n_{sub}$ is the refractive index of the substrate. In eqn (3.5), the relationship between $T_r$ and $R_s$ strongly depends on the ratio $\sigma_{dc}/\sigma_{op}$, which can be used as another FoM.[432] A high value of $\sigma_{dc}/\sigma_{op}$ implies high $T_r$ (>90%) and low $R_s$ (<10 $\Omega$ $\square^{-1}$), which are the desired properties for a TCE.[73] In order to achieve commercial TCE performance ($R_s \leq 100 \Omega \square^{-1}$ and $T_r \geq 90\%$ in the visible frequency range), an ideal value of $\sigma_{dc}/\sigma_{opt} \geq 35$ is typically required. It is noteworthy that $\alpha$ in the visible spectrum ($\alpha_{vis}$) arises from the tail of the free-carrier absorption, as described by Drude's theory[433] and is determined by

$$\alpha_{vis} = \frac{e^3}{4\pi^2 \varepsilon_0 c^3} \frac{N_d \lambda^2}{\sqrt{\varepsilon_\infty} \mu m_{eff}^2} \quad (3.8)$$

The latter equation shows the direct proportionality between $\alpha_{vis}$ and $N_d/\mu$. This evidences that $\alpha_{vis}$ can be reduced by decreasing $N_d$ and increasing $\mu$, showing a strategy commonly adopted to design effective TCEs.

In addition to $R_s$ and $T_r$, environmental stability and abrasion resistance are also decisive factors to select TCE materials.

For the specific case of DSSCs, for example, the transport of charge carriers from the photoanode to CE is hindered by several resistances.[434–437] The latter include the series resistance comprising $R_s$ of TCE and contact resistance of the cell; the transport resistance of electrons in the TiO$_2$ film ($R_{TiO_2}$); the resistance at the TCO/TiO$_2$ contact ($R_{TCO-TiO_2}$); the charge transfer resistance of charge recombination between the electrons in the TiO$_2$ film and ions in the electrolyte ($R_{rec}$); the charge transfer resistance at the CE/electrolyte interface ($R_{CT}$); the charge transfer resistance at the exposed TCO/electrolyte interface ($R_{TCO-electr.}$); and the Warburg parameter, which describes the Nernst diffusion of active ions in the electrolyte ($Z_d$).

Typically, $R_{CT}$ is often dominant among multiple charge transfer resistances. However, in large-area DSSCs, $R_s$ also significantly determines FF losses.[438] The smaller the $R_s$, the higher is the FF, resulting in higher $\eta$.[439–441] Concerning the electrocatalytic activity of CE, $R_{CT}$ can be explained in terms of current density ($J$), as expressed by the following equation:

$$R_{CT} = \frac{RT}{nFJ} \quad (3.9)$$

where $R$, $T$, $n$, and $F$ are the gas constant, temperature, number of electrons transferred in the elementary electrode reaction ($n = 2$), and Faraday constant, respectively.[442]

# 4. OSCs

OSCs hold remarkable potential for low-cost, flexible PVs, presenting both compatibility with R2R large-area fabrication[443–445] and impressive short-energy pay-back times.[446] Bulk-heterojunction (BHJ) OSCs, exploiting blends of p-type polymer (or organic small-molecule) donor/n-type fullerene (or other kind of organic small-molecule) acceptor materials dissolved in a common solvent, have opened an avenue for promising research activity to improve the $\eta$ value of SCs,[447] as well as the overall performance of photoelectrochemical cells.[448–455] The BHJ configuration maximizes the donor/acceptor interfacial area, facilitating exciton dissociation and charge transfer by forming a bicontinuous interpenetrated charge transport network in the photoactive layer.[456,457] In addition, the incorporation of layers with hole and electron transporting (or blocking) properties between the donor/acceptor active layer and anode/cathode promotes and balances the extraction/collection of photogenerated charges.[458] All these properties make the BHJ concept a landmark in OSC development, as well as a plethora of other applications (e.g., photodetectors[459] and biosensing devices[460–462]). Historically, the development of low-bandgap polymers, interfacial engineering, and fabrication techniques allowed BHJ to achieve $\eta$ exceeding 9% for single-junction cells[463–467] and 10% for tandem cells.[468] More recently, non-fullerene acceptors (NFAs) have dominated the OSC field due to significant performance and stability improvements.[469] Compared with their fullerene-based counterparts, NFAs exhibit tunable bandgaps that extend their light absorption in the NIR region.[468] In addition, their tunable energy levels can adjust the energy-level alignments between the constituent layers in OSCs to minimize the energy offsets, increasing the $V_{OC}$.[468] Lastly, their crystallinity can be easily tuned to finely control the photoactive-layer morphology, improving the device stability.[470–472] Nowadays, state-of-the-art OSCs exhibit $\eta$ values over 17% for both single-junction cells and two-terminal tandem cells, mainly due to the rapid developments of NFAs, as well as advanced device engineering.[473–477] In particular, the combination of low-bandgap donors and NFA-enabled OSCs has resulted in the achievement of record efficiency of 18.3%.[478]

In a typical OSC structure, GRMs can be incorporated either as additional components or to replace traditional materials, aiming at both performance and stability enhancement. In this context, GRMs have been used to fulfil several functions (Fig. 7) such as (a) transparent front electrode (i.e., TCE)[479–481] or back electrode;[482–484] (b) electron acceptors in binary OSCs or additives in ternary OSCs in the form of nanoflakes dispersed in donor–acceptor matrices;[336,485–487] (c) $\phi_W$-tuned HTLs/ETLs[488–490] or interfacial layers in tandem OSCs.[491–493] In the following subsections, the application of solution-processed GRMs into OSC structures will be examined for each functional device component.

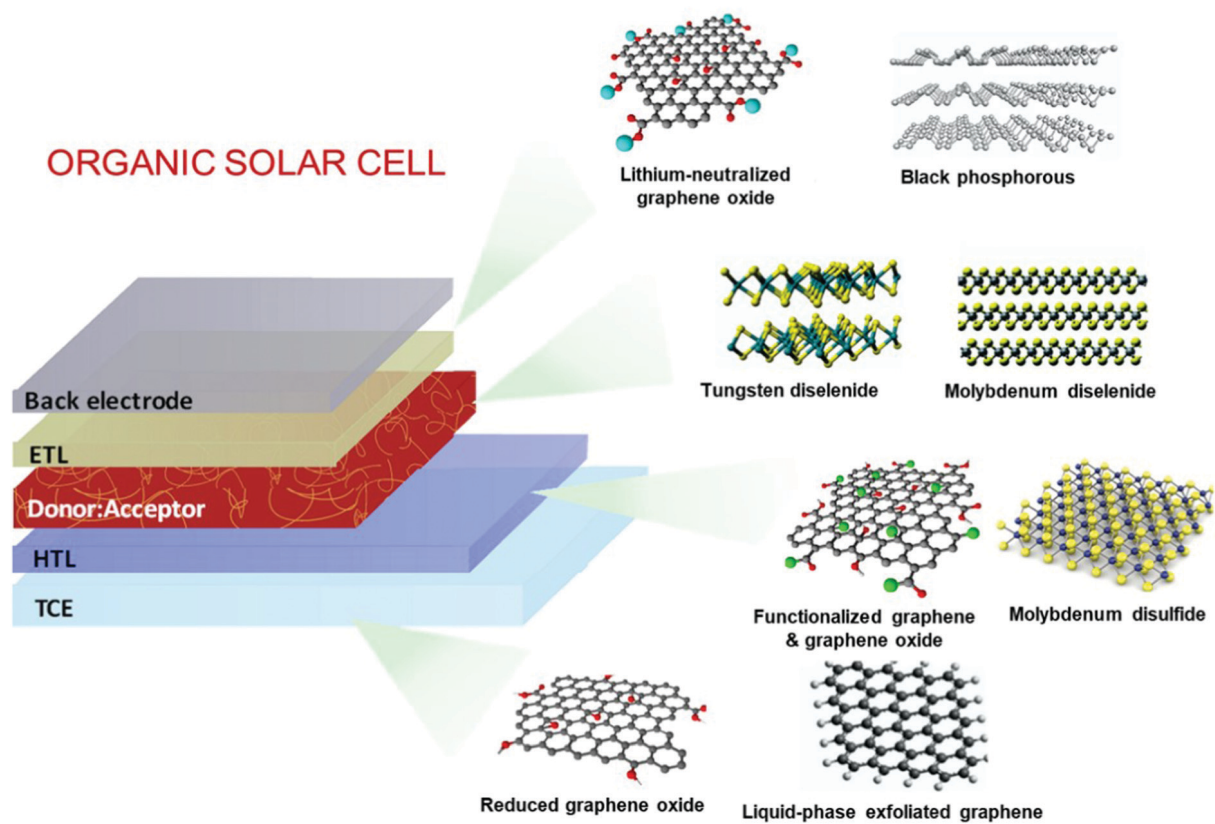

**Fig. 7** Two-dimensional materials used as OSC components, including electrodes, CTLs/buffer layers, and photoactive layers. RGO and electrochemically exfoliated graphene have been used as the TCEs. Functionalized graphene molecules, GO, and TMDs such as MoS$_2$ and WS$_2$ have been investigated as HTLs. Graphene-based molecules, WSe$_2$, and MoSe$_2$ have been used as electron acceptors in binary blends or additives in ternary OSCs. Lastly, OSCs including functionalized graphene or BP as ETLs or interlayers have also been reported.

### 4.1. TCEs

Graphene has been largely investigated as the TCE in OSCs to replace traditional ITO electrodes. Actually, ITO is currently the most established TCE material for rigid OSCs due to its excellent conductivity (i.e., $R_s <$ 10 Ω □$^{-1}$ for 100 nm-thick films)[495] and high $T_r$ (>80%) in the visible spectrum. However, some drawbacks, including the scarcity of In, expensiveness of the sputter deposition processes, and its polycrystalline structure, makes the ITO films brittle when they are repeatedly bent or stretched,[496] nullifying their use in flexible OSCs. In addition, it is recognized that ITO elements diffuse through the photoactive layer, leading to a significant decrease in the OSC performance.[497–499] Alternative TCEs based on CNTs,[500,501] metallic nanowires,[502] and conductive polymers[503] have been proposed and used in OSCs.

However, these TCEs exhibit high surface roughness and/or large $R_s$, which reduce the reproducibility rate of the devices.[499–502]

Alternatively, TCEs based on graphene rapidly emerged, and several approaches have been implemented to decrease the $R_s$ values of graphene-based TCEs toward commercially competitive values.[73] For example, Wang et al. reported poly(3-hexylthiophene): [6,6]-phenyl C61-butyric acid methyl ester (P3HT:PC$_{61}$BM)-based OSCs incorporating a TCE comprising 4-layer HNO$_3$-doped graphene prepared by a layer-by-layer transfer method.[504] An $\eta$ value of 2.5% was obtained by the additional evaporation of a thin layer of MoO$_3$ over the TCE in order to improve its hydrophilicity and to tune its $\phi_W$ from 4.36 to 5.37 eV.[503] Currently, the highest $\eta$ values of 6.1% and 7.1% reported for flexible conventional and inverted OSCs, respectively, have been achieved using graphene-based TCEs produced through the CVD method.[505] Notably, these results have been achieved by applying a MoO$_3$ buffer layer onto graphene-based TCEs.[504] More recently, $\eta$ as high as 8.48% has been achieved in tandem OSCs, which combine a wide-bandgap small molecule with low-bandgap polymer using Au-doped single-layer graphene nanoribbons (GNRs) as the TCE.[506] Although the CVD is an efficient approach to produce effective graphene-based TCEs,[507–509] the transfer process of the as-grown graphene films onto a target substrate is still critical, negatively impacting the manufacturing time and cost. In this regard, the chemical exfoliation of GO through ultrasonication or rapid thermal expansion, followed by reduction with chemical[510] or photo-assisted routes,[511] is a reliable low-cost top-down alternative approach, compatible with R2R mass production.[83] As discussed in Section 2, RGO can be easily produced in bulk quantities in the form of ink, taking advantage of its solubility in common solvents,[512] including water. Consequently, there has been an extensive research effort on the use of RGO as TCE in OSCs.[513–515] Flexible OSCs based on a RGO film as the TCE were firstly fabricated using P3HT:PC$_{61}$BM.[514] The RGO TCE was produced by spin coating GO flakes over a rigid SiO$_2$/Si substrate. The resulting GO film was then reduced by thermal annealing and transferred onto a polyethylene terephthalate (PET) substrate, yielding the RGO TCE. However, the constructed devices (area: 1 mm$^2$) have shown a low $\eta$ value ($\sim$0.78%), which was attributed to the low $T_r$ (65%) and high $R_s$ of the RGO films ($\sim$3.2 kΩ $\square^{-1}$) compared with those of the ITO reference (90% and 15 Ω $\square^{-1}$, respectively).

Geng et al.[516] realized graphene-based TCEs using chemically converted graphene (CCG). This was produced by the chemical reduction of GO produced in the form of dispersion without the need of dispersants.[515] The reduction was accomplished by annealing GO under a vacuum in a furnace tube. This treatment resulted in the recovery of the sp$^2$-carbon networks of the graphene sheets. The CCG films exhibited $R_s$ = 103 Ω $\square^{-1}$ and $T_r$ = 50%. The P3HT:PC$_{61}$BM-based OSC with CCG-based TCE yielded an $\eta$ value of 1.01%, which was approximately half that reached by the reference OSC based on ITO.

In the same framework, an efficient reduction method based on laser illumination was demonstrated by Kymakis et al.[517] for the fabrication of flexible graphene-based TCEs, which can be spin cast on substrates that are sensitive to temperature. Femtosecond laser-treated RGO (LRGO) films exhibited $T_r$ of 70% and $R_s$ of 1.6 kΩ $\square^{-1}$ and were subsequently incorporated as the TCE in P3HT:PC$_{61}$BM-based OSCs, yielding an $\eta$ value of $\sim$1.1%.[516] Additionally, the as-produced graphene-based OSCs were bent to angles up to 135° without $\eta$ deterioration, which is different from ITO-based OSCs that failed completely at bending angles greater than 65°.[518,519]

In order to improve the trade-off between $T_r$ and $R_s$, the use of a mesh structure with periodic lines, as exploited for Cu-based[520] and Si-based[521] electrodes appeared to be an eye-catching strategy even for graphene-based TCEs. Following this strategy, $R_s$ and $T_r$ of TCEs can be controlled by varying the grid width, spacing, and thickness of the mesh structure.[522] Konios et al.[523] demonstrated a scalable one-step patterning of RGO films on PET or glass substrates based on femtosecond laser irradiation treatments. The authors proved an accurate control of RGO micromesh (RGOMM) features on both rigid (glass) and flexible (PET) substrates.[522] In particular, they obtained a RGO electrode with $T_r$ varying from $\sim$20% to $\sim$85% without deteriorating $R_s$.[522] The as-produced RGOMM was then used as TCE in small- and large-area OSCs based on poly[N-9′-heptadecanyl-2,7-carbazole-alt-5,5-(4′,7′-di-2-thienyl-2′,1′,3′benzothiadiazole)]:[6,6]-phenyl-C71-butyric acid methyl ester (PCDTBT:PC$_{71}$BM), achieving an $\eta$ value of 3.67% and 3.05% on glass and flexible substrates, respectively.[522]

More recently, electrochemically exfoliated graphene (e-graphene) was used as an alternative to RGO for TCE, avoiding the need for the harsh conditions necessary for the graphite oxidation step.[524] TCEs based on e-graphene were then formed by spray coating e-graphene dispersion.[523] The as-produced films exhibited $R_s$ between 520 Ω $\square^{-1}$ (at $T_r$ of 70%) and 180 Ω $\square^{-1}$ (at $T_r$ of 55%).[525] The as-produced TCEs were used in a thieno[3,4-b]-thiophene/benzodithiophene:phenyl-C71-butyric acid methyl ester (PTB7:PCB$_{71}$M)-based OSCs, which reached an $\eta$ value of 4.23%.[523] Subsequently, a mixed-dimensional TCE using silver nanowires (AgNWs) and e-graphene was also demonstrated, achieving an $\eta$ value of 6.57%.[525] The addition of e-graphene on the AgNW network led to a decrease in $R_s$ from 78 to 13.7 Ω $\square^{-1}$ and a reduction in film roughness from 16.4 to 4.6 nm.[526]

Recently, a graphene-based TCE prepared by stacking polyimide on graphene led to an ultraclean graphene surface, allowing the flexible device to reach a record high $\eta$ value of 15.2% for flexible OSCs.[527] Alternative to the use of high-quality graphene, benzimidazole-doped graphene was also proposed to achieve a trade-off between $T_r$ and $R_s$, enabling the realization of flexible OSC based on a 3-layer benzimidazole-doped graphene-based anode, with an $\eta$ value of 6.85%.[528]

Lastly, both coupling of graphene with metallic grids[529,530] and graphene/metal hybridization[531–534] are currently prevalent strategies used to achieve an optimal balance between $T_r$ ($>$90%) and $R_s$ ($<$100 Ω $\square^{-1}$). Table 1 summarizes the main experimental results achieved with OSCs using graphene-based TCEs.

### 4.2 Active layer components

**Electron acceptors.** The photoactive layer of an OSC typically comprises a bicontinuous interpenetrating network of electron donor and acceptor materials at the nanometer scale, which is referred to as the BHJ.[535–538] Traditionally, electron donors are mainly based on conjugated polymers,[539–541] while typical electron acceptors are fullerene derivatives.[534–537,542] Although fullerenes exhibit high electron $\mu$ ($\mu_e$) and high exciton diffusion length,[543] their low light absorption within the solar spectrum restricts the maximum attainable $V_{OC}$ in OSCs.[544–548] Therefore, alternative acceptors based on both graphene derivatives[549–552] and small molecules[553–559] have been successfully proposed to provide an "on-demand" tuning of the LUMO level. Among the 2D materials, functionalized GO and graphene QDs (GQDs) have been largely investigated as solution-processed electron acceptors in OSCs.[560–562] Liu et al.[563] functionalized GO with phenyl isocyanate to be used as an electron acceptor in OSCs. The resulting OSCs based on poly(3-octylthiophene-2,5-diyl) (P3OT) as the polymer donor exhibited an $\eta$ value of 1.4%.[562] Functionalized GO was also blended with P3HT, achieving an $\eta$ value of $\sim$1.1%.[548] When P3HT was blended with GQDs functionalized with aniline (ANIGQDs), the resulting OSCs reached an $\eta$ value of $\sim$1.14%.[564]

By a simple lithiation synthesis, Yu et al.[565] covalently joined C$_{60}$ onto a GO surface. Thus, they obtained a GO:C$_{60}$ hybrid that was used as an electron acceptor in P3HT-based OSCs, providing an $\eta$ value of 1.22% (2.5 times higher than the $\eta$ value measured for GO-free device ($\eta = 0.47\%$). This performance enhancement was attributed to the optimal percolation networks for electron transport through the GO flakes.

**Table 1** Summary of the PV performance of OSCs using graphene-based TCEs

| Material | Usage | Device structure | Cell performance | | | | Ref. |
|---|---|---|---|---|---|---|---|
| | | | $J_{SC}$ (mA cm$^{-2}$) | $V_{OC}$ (V) | FF(−) | $\eta$ (%) | |
| Chemical and thermal RGO | TCE | PET/RGO/PEDOT:PSS/P3HT:PC$_{61}$BM/LiF/Al | 1.18 | 0.46 | 0.24 | 0.13 | 512 |
| Chemical and thermal RGO | TCE | Glass/RGO/PEDOT:PSS/P3HT:PC$_{61}$BM/Al | 1.84 | 0.44 | 0.25 | 0.2 | 513 |
| Chemical RGO | TCE | PET/RGO/PEDOT:PSS/P3HT:PC$_{61}$BM/TiO$_2$/Al | 4.39 | 0.56 | 0.32 | 0.78 | 514 |
| Laser RGO (LRGO) | TCE | PET/LRGO/PEDOT:PSS/P3HT:PC$_{61}$BM/Al | 5.62 | 0.57 | 0.34 | 1.1 | 516 |
| RGO micromesh (RGOMM) | TCE | PET/RGOMMs/PEDOT:PSS/PCDTBT:PC$_{71}$BM/TiO$_x$/Al | 7.81 | 0.85 | 0.46 | 3.05 | 522 |
| Electrochemical exfoliated graphene (EG) | TCE | PEN/EG/PEDOT:PSS/PTB7:PC$_{61}$BM/Ba/Al | 9.97 | 0.71 | 0.59 | 4.23 | 524 |
| EG-AgNWs | TCE | PEN/EG-AgNWs/PEDOT:PSS/PTB7:PC$_{71}$BM/Ba/Al | 15.5 | 0.73 | 0.58 | 6.57 | 525 |
| PEDOT-doped graphene | TCE | PEDOT-doped graphene/PEDOT:PSS/P3HT:PC$_{60}$BM/ bathocuproine (BCP)/Al | 9.07 | 0.55 | 0.49 | 2.45 | 494 |
| PEDOT-doped graphene | TCE | PEDOT-doped graphene/PEDOT:PSS/PBDTTT-C-T: PC$_{70}$BM/BCP/Al | 14.57 | 0.70 | 0.45 | 4.64 | 493 |
| Polyimide/graphene | TCE | Polyimide/graphene/PEDOT:PSS/PM6:Y6/PDINO/Al | 25.8 | 0.84 | 0.70 | 15.2 | 526 |
| Cu/graphene hybrid | TCE | Graphene/Cu/PEIE$^+$Blm$_4^-$/PC$_{71}$BM:PTB7/MoO$_3$/Ag | 13.01 | 0.73 | 0.46 | 4.38 | 530 |
| Cu/graphene hybrid | TCE | Graphene/Cu/PEDOT:PSS/PTB7:PC$_{71}$BM/PEIE$^+$Blm$_4^-$/aL | 12.99 | 0.58 | 0.42 | 3.16 | 530 |
| Ag grid/graphene | TCE | Ag grid/graphene/PEDOT:PSS/P3HT:PCBM/Ca/Al | 7.64 | 0.57 | 0.58 | 2.55 | 528 |
| Graphene quantum dots (GQDs)-mixed Ag nanowires (NWs)/graphene | TCE | GQDs-mixed Ag NWs/graphene/PEDOT:PSS/P3HT: PCBM/Al | 10.43 | 0.59 | 0.59 | 3.66 | 531 |
| Ag NWs/GO | TCE | Ag NWs/GO/PEDOT:PSS/P3HT:PC$_{70}$BM/Al | 9.53 | 0.59 | 0.57 | 3.26 | 532 |
| Ag NWs/GO | TCE | Ag NWs/GO/PEDOT:PSS/ptb7:PC$_{70}$BM/LiF/Al | 19.84 | 0.68 | 0.57 | 7.62 | 532 |
| Graphene:Ag NWs composite | TCE | Graphene:Ag NWs/PH1000/PEDOT:PSS/PM6:Y6/pdino/Al | 23.2 | 0.83 | 0.70 | 13.44 | 533 |

Stylianakis *et al.*[566] functionalized GO flakes by linking them via peptide bonds to acylated groups (GO-COCl), as well as to 3,5-dinitrobenzoyl chloride with the amino groups of ethylene-diamine (GO-EDNB). The resulting GO-EDNB was used as an electron acceptor material in P3HT-based OSCs, which achieved an $\eta$ value of 0.96%.[565] However, it is noteworthy that the LUMO level of GO-EDNB was 3.4 eV, which means that it is able to provide an energetic offset for exciton dissociation only with P3HT (LUMO$_{P3HT}$ = 3 eV).[565] This condition, which is not met by the state-of-the-art polymer donors, prevents the use of GO-EDNB as a universal electron acceptor.[565] These results evinced the need of exploring alternative functionalization routes for graphene derivatives to improve the distribution of flakes in the polymer matrix, while tuning their electronic structure (*i.e.*, achieving an ideal energy offset between the LUMO levels of the polymers and graphene derivatives). Based on this consideration, a photochemical functionalization of GO through laser-induced covalent grafting of GO nanosheets with EDNB molecules (LGO-EDNB) was subsequently demonstrated to tune the GO energy levels.[567] The as-produced LGO-EDNB has shown excellent processability in organic solvents commonly used for prototypical polymer donors.[566] The HOMO/LUMO levels of LGO-EDNB were tuned by adjusting the laser irradiation parameters.[566] The optimized LGO-EDNB displayed an $E_g$ value of 1.7 eV and LUMO level of 4.1 eV. Thus, it was used as an electron acceptor in PCDTBT-based OSCs, achieving a $V_{OC}$ value of 1.17 V and an $\eta$ value of 2.41%.[566]

Pristine RGO sheets were also incorporated in the nano-architecture of TiO$_2$ nanorod (NR)–ZnO NP/P3HT hybrid OSCs,[568] and $\eta$ of ~3.8% was achieved for a 900 nm-thick TiO$_2$ NR array. According to the authors, the RGO behaves as an energy-matched auxiliary electron acceptor in the hybrid structure, connecting the electron transport pathways provided by the 3D ZnO network and TiO$_2$ NR array to the fluorine-doped tin oxide (FTO) substrate.[567] In addition, it was concluded that the incorporation of RGO with low C-to-O atomic ratio stabilizes the active layer infiltrated in the interstices of the TiO$_2$ NR array.[567]

Beyond more conventional 2D materials, 2D-conjugated polymers have been commonly proposed as potential donor materials for high-performance OSCs. In particular, 2D-conjugated polymers based on bithienylbenzodithiophene-*alt*-benzotriazole backbone bearing different conjugated side chains, commonly named J-series polymers, enabled the realization of OSCs with $\eta$ approaching values obtained from state-of-the-art materials.[569]

**Additives in ternary OSCs.** An effective way to enhance the performance of BHJ OSCs relies on the addition of a third component into the polymer–fullerene binary blend, generating a ternary OSC.[570]

In principle, the ternary structure can address most of the deficiencies of the BHJ binary blend. In particular, the absorption spectral window of the polymer donor can be extended and the exciton dissociation and charge transport can be enhanced owing to the introduction of additional interfaces, and the morphological properties of the photoactive layer can also be tuned for favorable cell operation. However, it is crucial that the LUMO and HOMO levels of the additive component must lie between the LUMO and HOMO levels of fullerene and the polymer, respectively, so that suitable energy offsets are present at the material interfaces. In this regard, indene-C$_{60}$ bisadduct (ICBA), whose energy levels lie between the polymer donor and fullerene acceptor, has been successfully used as the third component in ternary blends.[571] As an alternative to ICBA, solution-processed graphene derivatives can be ideal additives in ternary OSCs, since a remarkable $\mu$ value in the device is expected to be achieved via graphene addition. In addition, graphene plays a relevant role in charge transfer processes,[572] increasing the exciton separation efficiency. Consequently, pristine

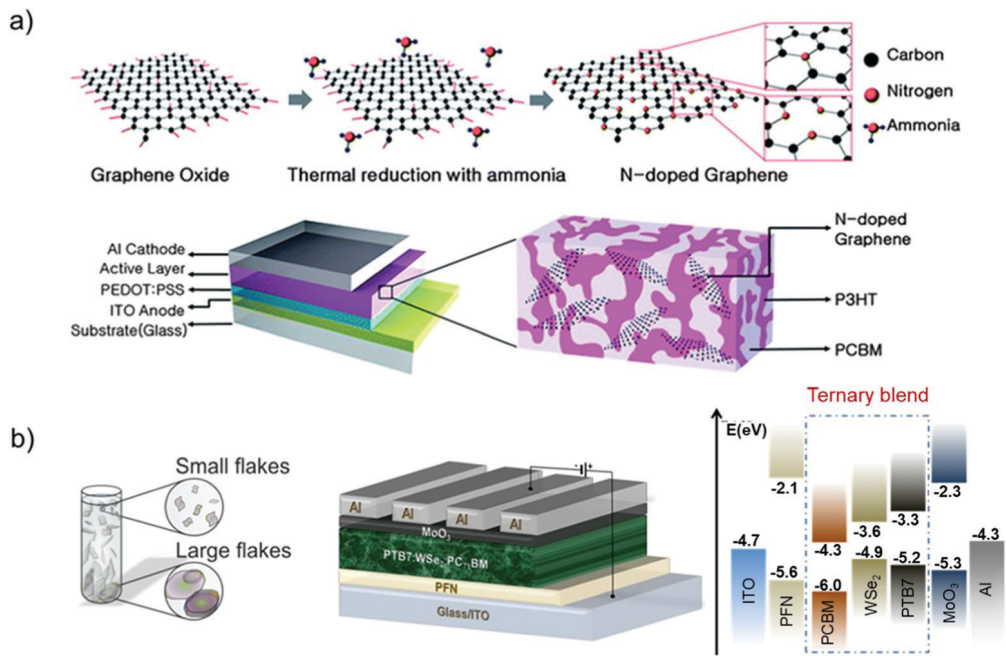

a)

Graphene Oxide — Thermal reduction with ammonia — N-doped Graphene

Carbon
Nitrogen
Ammonia

Al Cathode
Active Layer
PEDOT:PSS
ITO Anode
Substrate(Glass)

N-doped Graphene
P3HT
PCBM

b)

Small flakes
Large flakes

Ternary blend

E (eV)

Fig. 8 (a) Schematic of the N-doping process of RGO and BHJ OSC using N-doped graphene/P3HT:PC$_{61}$BM as the active layer. Adapted from ref. 572. (b) Schematic of the WSe$_2$ flake production through LPE, and schematic of the device structure and energy levels. Adapted with permission from ref. 336, Copyright 2017, American Chemical Society.

graphene flakes and RGO have been investigated as additives in ternary OSCs to increase their PV performance. For example, Jun et al.[573] used RGO flakes n-doped with N (NRGO) as the additive material in P3HT:PC$_{61}$BM-based OSCs, which exhibited a ~40% increase in $\eta$ (4.39%) compared to that of a binary OSC. The beneficial effect of NRGO addition was associated to the enhancement of $\mu_e$ in the photoactive layer (from 3.1 × 10$^{-7}$ to 5.4 × 10$^{-7}$ m$^2$ V$^{-1}$ s$^{-1}$) (Fig. 8a). However, because of the absence of an appropriate bandgap, the flakes act as carrier traps in the BHJ. Therefore, NRGO was not an energy cascade material, but it only provided additional charge transport pathways.

Similarly, Robaeys et al.[574] used solution-processed graphene flakes, produced by the LPE of pristine graphite, as an additive in P3HT:PC$_{61}$BM-based OSCs. It was shown that graphene addition determines the formation of a continuous active film with an interpenetrating structure by improving the crystallinity of P3HT. Nevertheless, like NRGO, solution-processed graphene flakes cannot be considered as an energy cascade component in a ternary BHJ OSC due to lack of a bandgap and therefore appropriate energy level matching. Contrarily, solution-processed graphene flakes can be considered as an additive to improve the crystallization and morphology of P3HT, beyond the improvement of charge transport properties. Consequently, graphene flakes can favor better balancing between $\mu_e$ and $\mu_h$ compared to the reference cell.

Graphene nanoflakes with controlled lateral size and functionalized with EDNB (EDNB-GNFs) were demonstrated as a ternary compound acting as an efficient electron-cascade acceptor material in air-processed PCDTBT:PC$_{71}$BM-based OSCs.[484] The functionalization process allowed the HOMO and LUMO levels of GNFs to be matched with the HOMO and LUMO levels of the hosting polymer and fullerene components, respectively. Furthermore, EDNB-GNFs acted as a highly conductive bridge between polymer chains and fullerene balls, thus offering two additional interfaces for exciton dissociation, as well as multiple routes for charge transfer at the donor/acceptor interfaces. The as-prepared ternary OSCs achieved an $\eta$ value of 6.59%, which was ~18% higher than that of the binary reference ($\eta$ = 5.59%). The same group investigated the role of GO covalently linked with porphyrin moieties (GO-TPP) into the active layer of PCDTBT:PC$_{71}$BM and PTB7:PC$_{71}$BM,[575] showing that the addition of GO-TPP induces favorable energy alignment between the energy levels of the donor and acceptor, facilitating the electron-cascade effect. The optimized ternary PTB7-based OSCs, containing 0.3% GO-TPP, exhibited a remarkable $\eta$ of 8.81%, which was ~16% higher than the binary reference one.[574]

Kim et al.[576] incorporated GO-QDs in PTB7:PC$_{71}$BM-based OSCs and investigated the effect of reduction of GO on the PV performance. It was found that the addition of partially reduced GO-QDs (RGO-QDs) in the active layer enhanced the $\eta$ value from 6.7% up to 7.6% because of the ideal balance between optical absorption and conductivity of QDs.[575] Most recently, RGO-Sb$_2$S$_3$ hybrid flakes have been used as additives in PCDTBT:PC$_{71}$BM-based OSCs.[577] Hybrid RGO-Sb$_2$S$_3$ combines the advantages of individual materials in which Sb$_2$S$_3$—acting as a secondary light-harvesting antenna in the visible spectrum—enhances the light absorption of the device, while RGO flakes offer highly conductive multiple charge-transfer percolation paths, suitable for ballistic electron transport to the LUMO of PC$_{71}$BM.[576] Moreover, the RGO sheets accelerate charge transfer, hindering the recombination phenomena in inorganic nanocrystals.[576] Therefore, the resulting cells exhibited a significant $\eta$ of 7%, corresponding to an enhancement of ∼23% compared to the reference device.[576] Kim *et al.* recently reported the utilization of size-selected GO flakes as the third component in PTB7: PC$_{71}$BM-based OSCs.[578] GO nanosheets with lateral sizes ranging from nanometers to micrometers were fabricated by a physical sonication process.[577] The physical size of the GO flakes affects the GO dispersion stability and morphological aggregation of the ternary blend.[577] In particular, it was found that the use of GO with lateral sizes of 500–750 nm maximizes both hole and electron mobilities of ternary OSCs.[577] Consequently, the non-geminate recombination was reduced. The corresponding ternary OSCs reached an $\eta$ value as high as 9.21% by increasing the FF to 69.4% in inverted devices, while the reference binary OSCs exhibited an $\eta$ value of 7.94%.[577]

Other 2D materials have been exploited as additives in OSCs. Bruno *et al.*[579] used WS$_2$ nanotubes in P3HT-QDs devices (which can also be classified as an organic–inorganic hybrid SC; see Section 7) as additives. *In situ* laser-induced anchoring of noble-metal NPs onto the surface of thin GO, WS$_2$, MoS$_2$, and BN have been developed to design special additives for OSCs.[580] In particular, WS$_2$ nanosheet-Au NP assemblies added in PCDTBT:PC$_{71}$BM allowed the corresponding cells to achieve an ∼13% enhancement in $\eta$ compared to the binary reference.[579] This effect was attributed to the efficient synergy of plasmon-enhanced absorption of Au NPs and superior charge transport into WS$_2$ nanosheets, as well as energy-level matching between the polymer and intermediate WS$_2$ nanosheets.[579] WSe$_2$ nanoflakes of different sizes were also used as the third component in ternary PTB7:PC$_{71}$BM-based OSCs (Fig. 8b).[336]

Three WSe$_2$ samples, with different average lateral sizes (below 20 nm, between 30 and 50 nm, and above 50 nm) were investigated.[336] Upon the introduction of medium-sized flakes, an $\eta$ value of 9.45% was measured, which is one of highest reported for OSCs based on PTB7 as the polymer donor.[336] The observed enhancement was attributed to the synergistic effect of absorption and charge transfer processes.[336] Notably, only medium-sized WSe$_2$ flakes contributed to $\eta$ enhancement.[336] This was linked with the similar size of WSe$_2$ flakes and PC$_{71}$BM domains in the ternary blend.[336] Therefore, the insertion of such nanoflakes introduces additional percolation pathways in the photoactive blend, promoting electron extraction and therefore collection.[336] These results highlighted the importance to match the morphological properties of 2D materials with the photoactive components of OSC blends.[336] Lately, Yang *et al.* incorporated LPE-produced black phosphorus nanoflakes (BPNFs) with an average size of 46 nm in poly([2,6′-4,8-di(5-ethylhexylthienyl)benzo[1,2-b;3,3-b]dithiophene]{3-fluoro-2[(2-ethyl-hexyl)carbonyl]thieno[3,4-b]thiophenediyl}):low-bandgap NFA (PTB7-Th:IEICO-4F)-based OSCs as the third component.[583] BPNFs were used as the morphology modifier to improve the performance of fullerene-free OSCs.[582] The incorporation of BPFNs promotes molecular ordering and higher phase purity of the ternary blend, contributing to lowering the charge transport resistance and suppressing charge recombination compared to the binary blend without BPFNs.[582] As a result, ternary OSCs exhibited an $\eta$ value of 12.2%, whilst the $\eta$ value of binary OSCs was 11.4%.[582] Moreover, the ternary OSC with BPNFs retains 73% of its initial $\eta$ after thermal treatment at 150 °C in a N$_2$ atmosphere for over 3 h, while the binary OSC retains only 60% of its initial $\eta$ under the same condition.[582] The improvement in stability was ascribed to the retarding of phase mixing in BHJ during the aging period as a consequence of the space confinement effect induced by BPFNs.[582]

The effect of hydrogenation on MoSe$_2$ nanosheets, used as additives in PTB7-Th:PC$_{71}$BM OSCs, was also investigated in ternary devices.[584] The OSCs exhibited an $\eta$ value of 10.44%, which represent a 16% increment compared to the reference binary OSCs.[583] The obtained results were associated with the establishment of optimized percolation pathways in the active layer.[583] Furthermore, the ternary OSC maintained 70% of its initial $\eta$ value after continuous heating at 100 °C for approximately 1 h.[583] The improvement in performance, compared to the reference OSC, was attributed to enhanced exciton generation and dissociation at the MoSe$_2$–fullerene interfaces and balanced $\mu_e$ and $\mu_h$.[583] Very recently, chlorine-functionalized graphdiyne has been successfully applied as a multifunctional solid additive to fine-tune the morphology and improve the efficiency and reproductivity of NFA-based OSCs, which reached an $\eta$ value of 17.3% (certified $\eta$ of 17.1%).[585]

Table 2 summarizes the main results achieved with OSCs using GRMs as the active layer components.

## 4.3 CTLs

The most successful application of GRMs for OSCs is in CTLs as either ETLs or HTLs. To design high-efficiency OSCs, ETL/HTL are positioned between the photoactive layer and anode/cathode, to reduce the potential barriers at both the interfaces and suppress the current leakage and/or charge recombination.[586] Preferably, to ensure ohmic contacts at both interfaces, the $\phi_W$ value of an ETL should match the LUMO level of the acceptor, while the $\phi_W$ value of the HTL should match the HOMO level of the donor.[587]

A large number of HTL materials for OSCs have been investigated, including transition metal oxides (*e.g.*, V$_2$O$_5$, NiO$_x$)[588,589] and self-assembled organic molecules, *e.g.*, poly(3,4-ethylene-dioxythiophene) polystyrene sulfonate (PEDOT:PSS).[590,591] With regard to ETLs, the most efficient materials currently used are n-type inorganic (*e.g.*, TiO$_x$ and ZnO)[592] and organic semi-conductors.[593] However, there are several issues related to the most established CTLs. In particular, the main issues are related to the strong acidic and hygroscopic character of PEDOT:PSS and the sensitiveness of sol–gel-prepared TiO$_x$ to moisture. Therefore, costly manufacturing in a controlled atmosphere is often required.[594] In addition, charge transporting materials do not always allow $\phi_W$ tuning and/or solution processability.[595] In this context, graphene derivatives[596] and other 2D materials[597] have been extensively investigated as buffer layers in order to fully exploit their features, including solution processability, low-cost fabrication, environmental stability, and $\phi_W$ tunability *via* functionalization methods.

**HTLs.** PEDOT:PSS as well as metal oxides (*e.g.*, V$_2$O$_5$, VO$_x$, MoO$_x$, and NiO) have been widely used as HTL components in

**Table 2** Summary of the PV performance of OSCs using GRMs as the active layer components

| Material | Usage | Device structure | Cell performance | | | | |
|---|---|---|---|---|---|---|---|
| | | | $J_{SC}$ (mA cm$^{-2}$) | $V_{OC}$ (V) | FF (−) | $\eta$ (%) | Ref. |
| Fullerene-grafted graphene | Electron acceptor | ITO/PEDOT:PSS/P3HT:C$_{60}$-G/Al | 4.45 | 0.56 | 0.49 | 1.22 | 564 |
| Chemically synthesized GO-ethylene-dinitro-benzoyl (GO-EDNB) | Electron acceptor | ITO/PEDOT:PSS/P3HT:GO-EDNB/Al | 3.32 | 0.72 | 0.4 | 0.96 | 565 |
| Laser produced GO-ethylene-dinitro-benzoyl (LGO-EDNB) | Electron acceptor | ITO/PEDOT:PSS/PCDTBT:LGO-EDNB/TiO$_x$/Al | 5.29 | 1.17 | 0.39 | 2.41 | 566 |
| RGO | Electron acceptor | FTO/TiO$_2$ NR-ZnO NP/RGO/P3HT/PEDOT:PSS/Au | 10.78 | 0.68 | 0.52 | 3.79 | 567 |
| Nitrogen doped graphene (N-RGO) | Additive | ITO/PEDOT:PSS/P3HT:PC$_{61}$BM:N-RGO/Ca/Al | 14.90 | 0.6 | 0.49 | 4.50 | 572 |
| Graphene flakes | Additive | ITO/PEDOT:PSS/P3HT:PC$_{61}$BM:graphene/Ca/Al | 8.00 | 0.6 | 0.66 | 3.17 | 573 |
| Functionalized graphene nanoflakes (GNFs) | Additive | ITO/PEDOT:PSS/PCDTBT/PC$_{71}$BM:GNF-EDNB60/TiO$_x$/Al | 12.56 | 0.89 | 0.57 | 6.41 | 484 |
| Graphene-based porphyrin molecule (GO-TPP) | Additive | ITO/PEDOT:PSS/PTB7:PC$_{71}$BM:GO-TPP/TiO$_x$/Al | 17.98 | 0.77 | 0.61 | 8.58 | 574 |
| Graphene oxide quantum dots (GOQDs) | Additive | ITO/PEDOT:PSS/PTB7:PC$_{71}$BM:GOQD/TiO$_x$/Al | 15.20 | 0.74 | 0.68 | 7.60 | 575 |
| RGO-antimony sulfide (RGO-Sb$_2$S$_3$) hybrid nanosheets | Additive | ITO/PEDOT:PSS/PCDTBT:PC$_{71}$BM:RGO-Sb$_2$S$_3$/TiO$_x$/Al | 13.47 | 0.92 | 0.55 | 6.81 | 576 |
| Medium sized GO | Additive | ITO/ZnO/PTB7:PC$_{71}$BM:MGO/MoO$_3$/Al | 18.00 | 0.74 | 0.69 | 9.09 | 577 |
| WS$_2$ decorated with Au NPs | Additive | ITO/PEDOT:PSS/PCDTBT:PC$_{71}$BM:PCDTBT:WS$_2$-Au/TiO$_x$/Al | 12.3 | 0.89 | 0.58 | 6.30 | 579 |
| WSe$_2$ | Additive | ITO/PFN/PTB7-WSe$_2$-PC$_{71}$BM/MoO$_3$/Al | 17.84 | 0.73 | 0.72 | 9.45 | 336 |
| BPFNs | Additive | ITO/PEDOT-Th:IEICO-4F:BPNFs/MoO$_3$/Ag | 23.44 | 0.71 | 0.73 | 12.20 | 582 |
| Hydrogen plasma–treated MoSe$_2$ | Additive | ITO/ZnO/PTB7-TH:PC$_{71}$BM/MoO$_3$/Al | 18.69 | 0.78 | 0.70 | 10.2 | 583 |
| Chlorine-functionalized graphdiyne | Additive | ITO/PEDOT:PSS/PM6:Y6/PFN-Br/Al | 26.09 | 0.84 | 0.79 | 17.32 | 584 |
| Zn–porphyrin based metal-organic framework nanosheets (Zn$_2$(ZnTCPP)) | Additive | ITO/PEDOT:PSS/PM6:Y6:Zn$_2$(ZnTCPP):BCP/Al | 10.80 | 0.69 | 0.69 | 5.2 | 581 |
| Bi$_2$OS$_2$ nanosheets | Additive | ITO/ZnO/ITIC:Bi$_2$OS$_2$:PBDB-T/MoO$_3$/Ag | 18.61 | 0.94 | 0.71 | 12.31 | 582 |
| g-C$_3$N$_4$ QDs | Additive | ITO/ZnO/g-C$_3$N$_4$:P3HT:PC$_{61}$BM/PEDOT:PSS/Ag | 11.44 | 0.61 | 0.60 | 4.23 | 278 |
| | | ITO/ZnO/g-C$_3$N$_4$:PBDTTT-C: PC$_{71}$BM/PEDOT:PSS/Ag | 15.9 | 0.70 | 0.57 | 6.62 | |
| | | ITO/ZnO/g-C$_3$N$_4$:PTB7-Th:PC$_{71}$BM/PEDOT:PSS/Ag | 16.74 | 0.78 | 0.70 | 9.2 | |

order to block electrons as well as transport holes, thus minimizing carrier recombination in OSCs.[598] Unfortunately, the highly acidic nature of PEDOT:PSS,[599,600] as well as the high cost of vacuum processes (e.g., ALD) used to deposit inorganic oxide (e.g., ZnO, VO$_x$) films[601–603] or the insufficient performance of solution-processed metal oxide films (compared to the organic reference),[604] pushed research toward the search for solution-processed alternatives. In this context, GO and RGO were found to be effective materials for replacing both PEDOT:PSS and inorganic oxides. In this context, Li et al.[605] reported graphene-based HTL using spin-coated 2 nm-thick GO film to replace PEDOT:PSS in P3HT:PC$_{61}$BM-based OSCs. The devices with GO exhibited a slower recombination rate and better stability than PEDOT:PSS-based OSCs.[604] In addition, the PEDOT:PSS-GO composite was investigated as the HTL in PTB7:PC$_{71}$BM-based OSCs.[606] The composite layer improved the $\mu_h$ value in the presence of benzoid–quinoid transitions, which also provided $\phi_W$ alignment between GO and PEDOT:PSS.[605] Consequently, PEDOT:PSS-GO-based OSCs achieved an $\eta$ value of 8.21%, which was 12% higher than that achieved by PEDOT:PSS-based OSCs.[605] In ref. 607, two layers of GO and vanadium oxide (VO$_x$) were subsequently spin coated to yield a hybrid film used as the HTL in PTh4FBT:PC$_{71}$BM-based OSCs, reaching an $\eta$ value of 6.7%. The authors demonstrated that thin films of graphene derivatives can improve the electron-blocking properties of the metal-oxide-based HTLs, while offering a barrier against the penetration of metal oxide films into organic active layers.[606] Despite the promising results on GO as the HTL, its insulating nature leads to severe limitations for efficient hole transport. Therefore, as a general strategy to improve the HTL performance, GO-based HTLs were modified to a partially reduced GO (pRGO) via thermal annealing and chemical and photoreduction processes.[608] For example, Yun et al.[609] prepared RGO by a novel p-toluenesulfonyl hydrazide (p-TosNHNH$_2$) reductant to be used as the HTL in P3HT:PC$_{61}$BM-based OSCs, reaching an $\eta$ value of 3.6% (similar to that of the PEDOT:PSS-based reference). Furthermore, the RGO-based OSC exhibited a lifetime significantly longer than that of the PEDOT:PSS-based device.[608] Similarly, Murray et al.[610] photoreduced GO with UV irradiation to obtain a HTL with $\phi_W$ aligned with the HOMO level of the PTB7 donor (Fig. 9a). It was demonstrated that the resulting pRGO HTL positively influenced the PTB7 π-stacking orientation, promoting the hole extraction process. In addition, although the $\eta$ value of pRGO-based OSCs (7.5%) was comparable to that of the PEDOT:PSS-based reference, the prolonged lifetime in air highlighted the key advantage of pRGO as the HTL, which is in agreement with other related works.[608,609] By following a different reduction method, Yeo et al.[611] produced a RGO HTL by functionalizing GO with p-hydrazinobenzene sulfonic acid hemihydrate as the reducing agent. The resulting sRGO has shown both high dispersion concentration in water (without the need of surfactants) and high electrical conductivity (3.18 S cm$^{-1}$).[610] Moreover, sRGO exhibited a higher $\phi_W$ value

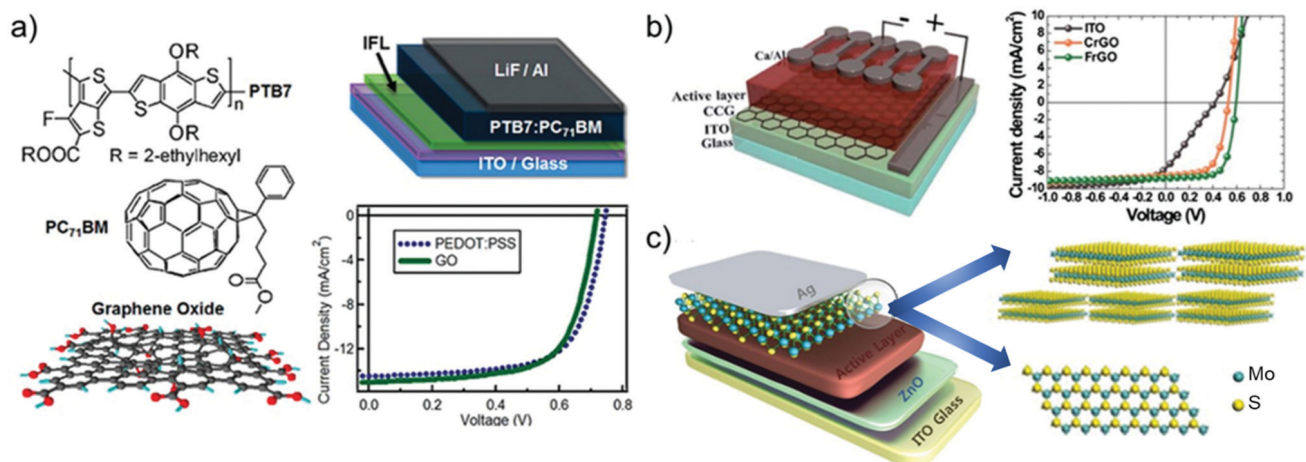

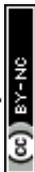

Fig. 9 (a) Chemical structures of the PTB7 donor polymer, PC$_{71}$BM acceptor, and GO HTL. Schematic of a standard OSC indicating the location of CTLs. Comparative PV performance of PTB7:PC$_{71}$BM-based OSC with PEDOT:PSS or GO HTLs. Adapted with permission from ref. 609, Copyright 2011, American Chemical Society. (b) Schematic of the OSC structure. I–V curves of OSCs based on different HTLs. Adapted from ref. 616. (c) Schematic of the inverted-type OSC incorporating a MoS$_2$ HTL. Schematic of the structure of the thin-layer MoS$_2$ buffer layer (side view) and schematic of the monolayer flake of MoS$_2$ along the 0001 direction (top view). Adapted from ref. 620.

(i.e., 5.04 eV) compared to that of RGO. Therefore, sRGO was compatible with the HOMO level of conventional donor polymers.[610] Further, sRGO was successfully applied in OSCs based on P3HT, poly[1-(6-{4,8-bis[(2-ethylhexyl)oxy]-6-methylbenzo[1,2-b:4,5-b']dithiophen-2-yl}-3-fluoro-4 methylthieno [3,4-b]thiophen-2-yl)-1-octanone] (PBDTTT-CF), and PTB7 as polymer donors.[610] In particular, an $\eta$ value higher than 7% was achieved for sRGO-based OSCs, which also exhibited device stability superior to that of the PEDOT:PSS-based reference.[610] Liu et al.[612] produced a sulfated RGO by introducing -OSO$_3$H groups into the basal plane of GO (RGO-OSO$_3$H). The corresponding RGO-OSO$_3$H HTL displayed a conductivity as high as 1.3 S m$^{-1}$ and $\phi_W$ aligned with the HOMO level of P3HT.[611] The corresponding RGO-OSO$_3$H-based OSC achieved an $\eta$ value of ~4.37%, which was similar to that obtained for PEDOT:PSS-based reference (4.39%).[611]

An alternative way to increase the electrical conductivity of GO relies on its mixing with SWCNTs.[613] This approach allowed P3HT:PC$_{61}$BM-based OSCs to reach an $\eta$ value of 4.1%.[612] Furthermore, surfactant-free Au NPs were incorporated between the photoactive layers and GO HTL, leading to an $\eta$ value increase of ~30% compared to the PEDOT:PSS-based reference.[614] In addition, the GO-based devices retained 50% of their initial $\eta$ after 45 h of continuous illumination, while the reference devices based on PEDOT:PSS completely degraded after 20 h.[613] The $\eta$ improvement was attributed to an increase in the exciton generation rate caused by Au NP-induced plasmon absorption enhancement. Meanwhile, the stability performance was ascribed to the suppression, in presence of GO, of oxygen and/or In diffusion from ITO toward the P3HT:PC$_{61}$BM.[613]

Li et al.[615] investigated deposited GQDs as the HTL material in DR$_3$TBDT:PC$_{71}$BM- and P3HT:PC$_{61}$BM-based OSCs (DR$_3$TBDT is a small-molecule donor based on the benzo[1,2-b:4,5-b']dithiophene unit).[616] GQD films exhibited homogenous morphology and high conductivity, yielding P3HT:PC$_{61}$BM-based OSCs with an $\eta$ value of 3.51%.[615] This value was similar to that measured for

PEDOT:PSS-based OSCs ($\eta$ = 3.52%).[615] In addition, GQD-based OSCs exhibited longer lifetime and more reproducible $\eta$ compared to the reference device.[615]

An effective approach to enhance the performance of GO-based HTLs is to tune their $\phi_W$ through functionalization routes.[112] In ref. 617, a fluorinated RGO (FRGO) was synthesized with a $\phi_W$ value of 4.9 eV using a F-containing phenylhydrazine-based reductant. The as-produced FRGO was then used as the HTL in PTB7:PC$_{71}$BM- and P3HT:ICBA-based OSCs (Fig. 9b).[616] The functionalization process detached oxygen functional groups from GO flakes, while concomitantly doping the edges and basal planes of the flakes themselves with F.[616] Due to the $\phi_W$ increase, the FRGO-based OSCs exhibited similar performance and higher stability compared to those of the PEDOT:PSS-based reference.[616] A series of GOs with tuned oxidation (pr-GO) were synthesized by Li et al.[618] strictly by controlling the preoxidation steps, oxidation time, and oxidant content, leading to $\phi_W$ values between 4.74 and 5.06 eV. By precisely controlling the oxidation time, a P3HT:PC$_{61}$BM-based OSC using pr-GO HTL reached an $\eta$ value of 3.74%, which was ~3.60% higher than that reported for the PEDOT:PSS-based reference.[617] Stratakis et al.[619] demonstrated that GO $\phi_W$ can be effectively tuned by UV laser irradiation in the presence of Cl gas. In particular, by irradiating ultrathin GO films with a pulsed laser in the presence of a dopant Cl precursor gas, a simultaneous reduction and Cl doping of GO lattice was achieved.[618] Following the irradiation process, Cl atoms were linked to both basal planes and edges of GO. The $\phi_W$ value of GO was tuned by controlling the laser exposure time.[618] In particular, the $\phi_W$ value of chlorinated GO (GO-Cl) was adjusted from 4.9 eV in GO to a maximum of 5.23 eV in GO-Cl by increasing the laser exposure level up to 60 laser pulses (pulse duration = 20 ns; wavelength = 248 nm; power of 50 mW; beam profile = 20 × 10 mm$^2$).[618] The induced polar character of C–Cl bonds is responsible for the downward shift in $E_F$ in the VB of GO-Cl and the subsequent increase in $\phi_W$ compared to pristine GO.[618]

This $\phi_W$ tuning determined the energy matching between GO-Cl and the PCDTBT donor, allowing the resulting OSC to reach an $\eta$ value higher than that of PEDOT:PSS-based reference.[618] Phosphorylated GO was recently used as HTL in PTB7:PC$_{71}$BM-, PBDTTT-C:PC$_{71}$BM-, and P3HT:PC$_{61}$BM-based OSCs, enhancing their $\eta$ from 6.28%, 5.07%, and 2.78% (in pristine GO-based devices) to 7.90%, 6.59%, and 3.85%, respectively.[620] The proposed phosphate ester modification increased the GO film roughness and hydrophobicity, while the p-doping of the GO increased $\phi_W$ from 4.24 to 4.70 eV, providing better matching with the HOMO level of the polymer donor.[619]

In addition to graphene-based materials, solution-processed TMDs have also been widely investigated as HTL materials. For example, Gu et al.[621] exploited a film of MoS$_2$ flakes, produced by the chemical Li intercalation method, as the HTL in P3HT:PC$_{61}$BM- and PTB7:PC$_{71}$BM-based OSCs. The resulting MoS$_2$-based OSCs achieved $\eta$ values of 4.02% and 8.11% for P3HT:PC$_{61}$BM and PTB7:PC$_{71}$BM active layers, respectively (Fig. 9c).[620] These $\eta$ values were higher than those measured for the reference OSCs using thermally evaporated MoO$_3$ HTLs.[620]

The superior HTL performance of MoS$_2$ compared to that of vacuum-evaporated MoO$_3$ was attributed to the inferior trap density compared to the MoO$_3$ reference, providing higher hole concentration at $V_{OC}$ (i.e., $\sim 10^{16}$ cm$^{-3}$ in MoS$_2$ vs. $\sim 10^{16}$ cm$^{-3}$ in MoO$_3$).[620] In addition, at the MoS$_2$/P3HT interface, the presence of a surface dipole with the negative charge end pointing toward the active film electrode and positive charge end pointing toward the Ag electrode reinforces the actual built-in potential across the device, suppressing charge recombination and leading to a more effective charge extraction capability.[620] Likewise, Yun et al. prepared a p-type MoS$_2$ (p-MoS$_2$) layer by HAuCl$_4$·3H$_2$O doping.[621] This process increased the MoS$_2$ $\phi_W$ value from 4.52 to 4.76 eV.[621] As a result, P3HT:PC$_{61}$BM-based OSCs using p-MoS$_2$ HTL exhibited an $\eta$ value of 3.4%, which was higher than that of pristine MoS$_2$-based OSCs ($\eta = 2.8\%$), owing to the better energy-level matching between the P3HT HOMO level and HTL $\phi_W$.[621] In the research activity of energy-level optimization of HTL, Le et al. further increased the $\phi_W$ value of MoS$_2$ up to 4.9 eV by UV/ozone (UVO) treatment, providing excellent matching with the HOMO level of P3HT ($\sim 5$ eV).[623] The resulting MoS$_2$-based OSCs achieved an $\eta$ value of 2.44%, which was similar to that of the PEDOT:PSS-based reference ($\eta = 2.81\%$).[622] Moreover, the use of MoS$_2$ HTL extended the device stability in air by protecting the ITO surface from the hygroscopic nature of PEDOT:PSS.[622] An increase in MoS$_2$ $\phi_W$ was also achieved by introducing O atoms inside the lattice of MoS$_2$ flakes (O-MoS$_2$) via UVO post-treatment.[624] The optimized O-MoS$_2$ flakes were used as HTLs in PTB7:PC$_{71}$BM-based OSCs, which displayed an $\eta$ value of 7.64%—53% higher than that of the cell using pristine MoS$_2$ and comparable to that obtained using PEDOT:PSS (7.6%).[623] In addition, the $R_s$ value of the device with O-MoS$_2$ was considerably lower (1.88 Ω □$^{-1}$) than that obtained using MoS$_2$ (4.03 Ω □$^{-1}$).[623] The incorporation of O atoms into the MoS$_2$ lattice can act as a type of doping or alloy, reducing structural defects by the filling of vacancies, as well as increasing $\phi_W$ (up to 4.93 eV) to match the HOMO level of P3HT.[623] Liu et al. proposed a further surface modification pathway of MoS$_2$ with a hydrophilic surfactant via electrostatic interaction.[625] Subsequently, they fabricated PTB7:PC$_{71}$BM-based OSCs with a modified MoS$_2$ HTL, achieving $\eta > 7\%$.[624] Yang et al. decorated MoS$_2$ flakes with Au NPs in order to create localized surface plasmon resonance effects to boost $\eta$.[626] In fact, Au NPs act as plasmonic near-field antennas,[627,628] increasing the absorption cross-section of the photoactive layer.[625] As a result, PTB7:PC$_{71}$BM-based OSCs using the MoS$_2$-Au hybrid as the HTL, exhibited an $\eta$ value of 7.25%, which represents a 17.3% increase compared to that of pristine MoS$_2$ HTL-based devices ($\eta = 6.18\%$).[625] Zheng et al. proposed a graphene–MoS$_2$ heterostructure (GMo) as an interlayer between the ITO and PEDOT:PSS HTL in OSCs based on a binary PTB7-Th:PC$_{71}$BM system.[629] GMo was hydrothermally synthesized using thiourea/glycerol, LPE-produced graphene, and phosphomolybdic acid as the precursors.[628] The few layers of oxygen-incorporated MoS$_2$ contained both 2H and 1T phases.[628] GMo-based OSCs reached $\eta = 9.5\%$, while retaining more than 93% of the initial $\eta$ over 1000 h.[628] Beyond MoS$_2$, other TMDs have been investigated as HTLs. Kwon et al. used WS$_2$ treated with UVO as the HTL in P3HT:PC$_{61}$BM-based OSCs.[630] The UVO treatment modified the $\phi_W$ value of WS$_2$ from 4.75 to 4.95 eV, improving the alignment with the LUMO level of P3HT in addition to the removal of surface contaminants.[629] The combination of these effects allowed the achievement of $\eta = 3.08\%$ (comparable to that of the PEDOT:PSS-based reference (3.23%)).[629] UVO treatment was also used for TaS$_2$ nanosheets, used both as the HTL and ETL in P3HT:PC$_{61}$BM-based OSCs.[631] The $\phi_W$ value of TaS$_2$ changed from 4.4 eV to 5.1 eV and $\eta = \sim 3.06\%$ could be achieved. This value was similar to that measured for the PEDOT:PSS-based OSCs as the reference (3.28%).[629] Gu et al. introduced NbSe$_2$ HTL in inverted PTB7:PC$_{71}$BM- and P3HT:PC$_{61}$BM-based OSCs, reaching $\eta$ of $\sim 8.10\%$ and $\sim 3.05\%$, respectively.[632] These $\eta$ values were higher than those of OSCs based on vacuum-deposited MoO$_3$ (7.54%) and spin-coated PEDOT:PSS (2.7%).[631]

The enhancement of $\eta$ was attributed to the flake-like 2D structure, which exhibits a lower trap density, as well as to the existence of surface dipoles, which promote charge extraction processes. Lastly, layered bismuth selenide nanoplatelets (L-Bi$_2$Se$_3$) were implemented as the HTL in inverted P3HT:PC$_{61}$BM-based OSCs.[633] The corresponding OSCs reached $\eta = 4.37\%$, which was higher than the $\eta$ value of OSCs based on evaporated MoO$_3$ HTL (3.91%).[632] The $\eta$ improvement was ascribed to the high conductivity of L-Bi$_2$Se$_3$.[632] Moreover, the L-Bi$_2$Se$_3$ $\phi_W$ was found to increase with aging under the ambient conditions due to O-induced p-doping, resulting in improved $V_{OC}$ and FF.[632] More recently, Li et al. demonstrated the use of LPE-produced few-layer WS$_2$ and MoS$_2$ nanosheets as the HTL materials for high-efficiency NFA-based OSCs (Fig. 10).[634] The cells used Y6[471] or IT-4F[635] as small-molecule NFAs and PBDB-T-SF[636] or PBDB-T-2F[634] as the polymer donors. Binary PBDB-T-SF:IT-4F and ternary PBDB-T-2F:Y6:PC$_{71}$BM OSCs based on WS$_2$ as the HTL exhibited an $\eta$ value of 15.8% and 17.0%, respectively, which were higher than the corresponding reference OSCs based on PEDOT:PSS, i.e., $\eta$ of 13.5% and 16.4%, respectively.[633] The observed performance enhancement was attributed to a reduction in bimolecular recombination losses (i.e., losses determined by the recombination of an

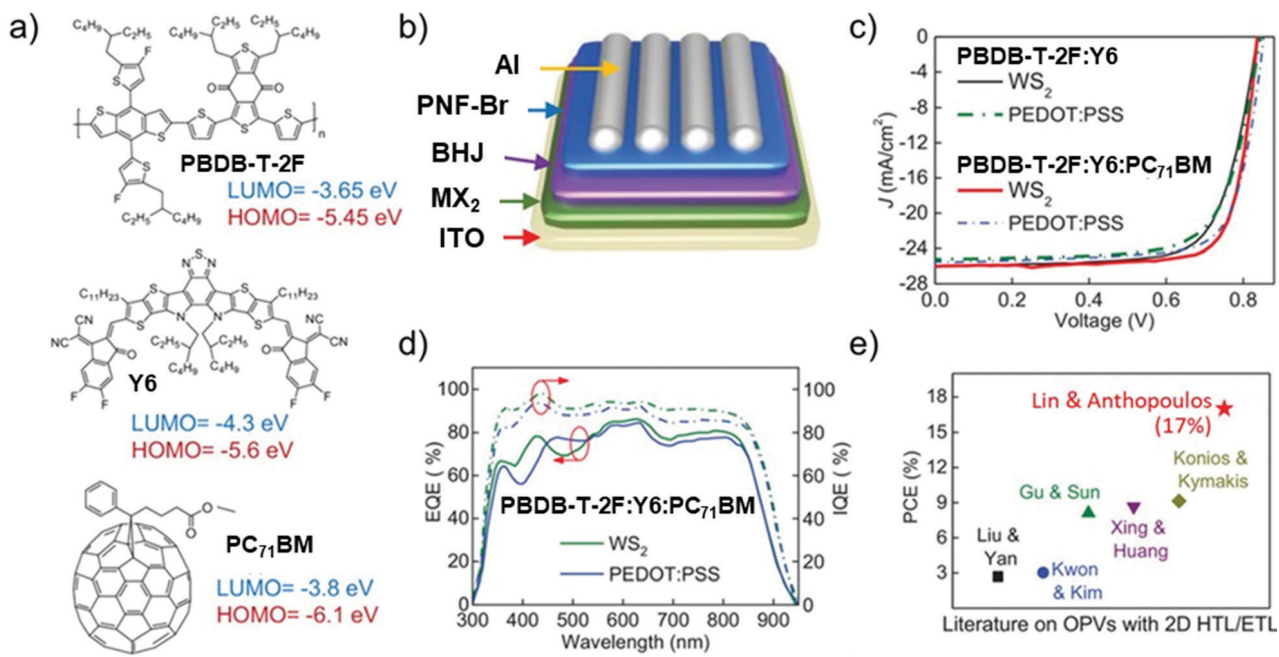

Fig. 10 (a) Chemical structure of PBDB-T-2F, Y6, and PC$_{71}$BM and the corresponding LUMO and HOMO energies. (b) Schematic of the standard OSC architectures employed. (c) J–V curves of OSCs based on PBDB-T-2F:Y6 and PBDB-T-2F:Y6:PC$_{71}$BM with different HTLs. (d) EQE curves of OSCs based on a PBDB-T-2F:Y6:PC$_{71}$BM active layer for different HTLs. (e) A comparison of the performances of previously reported OSCs with 2D material interfaces. Adapted from ref. 633.

electron with a hole, thus directly depending on both electron and hole concentrations) compared to PEDOT:PSS-based OSCs.[633] The lower bimolecular recombination in WS$_2$-based devices compared to MoS$_2$ and PEDOT:PSS devices was ascribed to the deeper $\phi_W$ value of WS$_2$ on ITO (i.e., 5.5 eV vs. 5.4 eV and 4.8 eV for MoS$_2$ and PEDOT:PSS on ITO, respectively).[633] The optimal WS$_2$ $\phi_W$ allowed charge collection to be improved and surface energy to be reduced, leading to quasi-ideal phase separation.[633]

Table 3 summarizes the main results achieved with OSCs using HTLs based on GRMs.

**ETLs.** For the use of GRMs as ETLs, proper functionalization routes have been used to decrease the $\phi_W$ values close to that of the HOMO level of fullerene acceptors, aiming to facilitate efficient electron transport from the electron acceptor to the ETL. Liu et al. first reported GRM-based ETLs based on Cs$_2$CO$_3$-functionalized GO in P3HT:PC$_{61}$BM-based OSCs.[637] By replacing the -COOH groups of GO with -COOCs groups through charge neutralization, $\phi_W$ was decreased from 4.7 to 4.0 eV. Consequently, $\phi_W$ of Cs$_2$CO$_3$-functionalized GO matched the LUMO level of PC$_{61}$BM, thereby facilitating electron collection.[636] An $\eta$ value of 3.67% and 2.97% were obtained using normal and inverted OSC structures, respectively.[636] The PV performance were similar to those measured for the reference cell using LiF as the standard ETL.[636]

Similar Cs$_2$CO$_3$-based functionalization was applied to GQDs (GQDs-Cs$_2$CO$_3$), which were then used as the ETL in inverted P3HT:PC$_{61}$BM-based OSCs.[638] The OSCs exhibited an $\eta$ value of 3.23%, which was 56% higher than that of OSCs using pristine Cs$_2$CO$_3$ HTL.[637] In addition, while GQDs-Cs$_2$CO$_3$-based devices retained 50% of their original $\eta$ under ambient conditions (exposition for 1200 h), the $\eta$ value of pristine Cs$_2$CO$_3$-based device decreased to 17% of its initial value.[637] The high $\eta$ and stability of GQDs-Cs$_2$CO$_3$-based OSCs were attributed to both optimal electron extraction and suppression of leakage current, as well as the immobilization of Cs$^+$ ions on GQDs in the HTL, delaying their diffusion into the P3HT.[637] Meanwhile, n-doped GO was produced through chemical Li intercalation, leading to functionalized GO-Li with $\phi_W$ of 4.3 eV.[639]

The low $\phi_W$ value of GO-Li was ascribed to the presence of Li atoms with low electronegativity.[638] In detail, Li atoms bonded to GO release their valence electrons to GO, leading to the formation of an electric dipole induced by Li$^+$.[638] The charge transfer from Li to GO plane shifts the Fermi level toward the vacuum, inducing a difference between the Fermi level of the two materials of 0.67 eV, explaining the decrease of $\phi_W$.[638] In PCDTBT:PC$_{71}$BM-based OSCs, the GO-Li layer, which is inserted between TiO$_x$ and photoactive blend, acts as an interfacial engineering material, increasing $\eta$ up to 6.29% (4.89% in GO-based OSCs and 5.51% in interlayer-free OSCs).[638] These results prove the bifunctional role of GO-Li acting as (1) an interfacial engineering material that improves the ohmic contact between the cathode and the ETL, while increasing the internal electric field amplitude;[638] (2) a protection layer against humidity and oxygen, enhancing the device stability during prolonged illumination.[638]

Beyond the intercalation of alkali metals in GO, an alternative n-doping strategy of RGO was developed by producing RGO-ZnO and RGO-TiO$_2$ nanocomposites, which were then used as the ETL in inverted PTB7:PC$_{71}$BM-based OSCs.[644] The RGO-ZnO- and RGO-TiO$_2$-based OSCs achieved $\eta$ values of 7.50% and 7.46%, respectively.[643] These values were comparable to those obtained using pristine ZnO (7.39%) and TiO$_2$ (7.22%).[643] The authors also compared their RGO-metal oxide (MO)-based OSCs with devices containing thermally evaporated bathocuproine (or 2,9-dimethyl-4,7-diphenyl-1,10-phenanthroline) (BCP) as ETLs, obtaining fairly comparable $\eta$ (7.47%) due to the capability of RGO to balance hole and electron mobilities of the devices.[643] Subsequently, RGO-MO ETLs were also exploited in PCDTBT:PC$_{71}$BM-based,[645] P3HT:PC$_{61}$BM-based,[646] and low-bandgap quinoxaline-based D-A copolymer:PCBM-based[647] OSCs.

**Table 3** Summary of the PV performance of OSCs using GRMs as the HTL

| Material | Usage | Device structure | Cell performance | | | | Ref. |
|---|---|---|---|---|---|---|---|
| | | | $J_{SC}$ (mA cm$^{-2}$) | $V_{OC}$ (V) | FF (−) | $\eta$ (%) | |
| GO | HTL | ITO/ZnO/PTh4FBT:PC$_{71}$BM/VO$_x$/GO/Ag | 13.2 | 0.76 | 0.67 | 6.7 | 606 |
| Partial reduced GO | HTL | ITO/pr-GO/PCDTBT:PC$_{71}$BM/TiO$_x$/Al | 11.18 | 0.89 | 0.59 | 5.96 | 607 |
| UV-O$_3$ treated GO | HTL | ITO/UV-O$_3$ treated GO/PTB7:PC$_{71}$BM/LiF/Al | 15.21 | 0.72 | 0.68 | 7.39 | 609 |
| Sulfonic acid-functionalized RGO | HTL | ITO/sr-GO/PTB7:PC$_{71}$BM/Ca/Al | 15.3 | 0.75 | 0.63 | 7.18 | 610 |
| GO-SWCNT | HTL | ITO/GO-SWCNT/P3HT:PC$_{61}$BM/Ca/Al | 10.82 | 0.6 | 0.63 | 4.1 | 612 |
| GQDs | HTL | ITO/GQD/P3HT:PC$_{61}$BM/Ca/Al | 10.20 | 0.52 | 0.66 | 3.51 | 614 |
| Fluorine-functionalized RGO | HTL | ITO/FRGO/P3HT:PC$_{61}$BM/Ca/Al | 8.78 | 0.6 | 0.7 | 3.64 | 616 |
| Partial oxidized GO (pr-GO) | HTL | ITO/pr-GO/P3HT:PC$_{61}$BM/LiF/Al | 10.40 | 0.61 | 0.59 | 3.74 | 617 |
| Photochlorinated GO (GO-Cl) | HTL | ITO/GO-Cl/PCDTBT:PC$_{71}$BM/TiO$_x$/Al | 13.65 | 0.88 | 0.55 | 6.56 | 618 |
| Phosphorylated GO | HTL | ITO/P-GO/PTB7:PC$_{71}$BM/Ca/Al | 16.12 | 0.71 | 0.68 | 7.9 | 619 |
| MoS$_2$ | HTL | ITO/ZnO/PTB7:PC$_{71}$BM/MoS$_2$/Ag | 15.9 | 0.72 | 0.71 | 8.11 | 620 |
| p-Doped MoS$_2$ | HTL | ITO/p-doped MoS$_2$/P3HT:PC$_{61}$BM/Ca/Al | 8.62 | 0.59 | 0.66 | 3.38 | 621 |
| UV/ozone-treated MoS2 (UVO MoS$_2$) | HTL | ITO/UVO MoS$_2$/P3HT:PC$_{61}$BM/LiF/Al | 7.97 | 0.52 | 0.68 | 2.81 | 622 |
| Oxygen-incorporated chemical exfoliated MoS$_2$ (O-ceMoS$_2$) | HTL | ITO/O-ceMoS$_2$/PTB7:PC$_{71}$BM/PFN/Al | 14.98 | 0.73 | 0.7 | 7.64 | 623 |
| Modified ce-MoS$_2$ (m-MoS$_2$) | HTL | ITO/m-MoS$_2$/PTB7:PC$_{71}$BM/PFN/Al | 14.71 | 0.73 | 0.67 | 7.26 | 624 |
| MoS$_2$ decorated with Au NPs (MoS$_2$@Au) | HTL | ITO/MoS$_2$@Au/PTB7:PC$_{71}$BM/PFN/Al | 15.44 | 0.72 | 0.65 | 7.25 | 625 |
| Graphene-MoS$_2$/PEDOT:PSS | HTL | ITO/Graphene-MoS$_2$/PEDOT:PSS/PTB7-Th:PC$_{71}$BM/Ca/Ag | 17.2 | 0.77 | 0.72 | 9.4 | 628 |
| UV-O$_3$ treated MoS$_2$ | HTL | ITO/UV-O$_3$ treated MoS$_2$/P3HT:PC$_{61}$BM/LiF/Al | 7.81 | 0.6 | 0.63 | 2.96 | 629 |
| UV-O$_3$ treated WS$_2$ | HTL | ITO/UV-O$_3$ treated WS$_2$/P3HT:PC$_{61}$BM/LiF/Al | 7.87 | 0.61 | 0.64 | 3.08 | 629 |
| UV-O$_3$ treated TaS$_2$ | HTL | ITO/TaS$_2$/P3HT:PC$_{61}$BM/LiF/Al | 7.87 | 0.61 | 0.64 | 3.06 | 630 |
| NbSe$_2$ nanosheets | HTL | ITO/ZnO/PTB7:PC$_{71}$BM/NbSe$_2$/Ag | 16.04 | 0.72 | 0.7 | 8.1 | 631 |
| Layered bismuth selenide (ι-Bi$_2$Se$_3$) nanoplates | HTL | ITO/ZnO/P3HT:PC$_{61}$BM/L-Bi$_2$Se$_3$/Ag | 9.91 | 0.65 | 0.68 | 4.37 | 632 |
| WS$_2$ | HTL | ITO/WS$_2$/PBDB-T-SF:IT-4F/PFN-Br/Al | 20.6 | 0.88 | 0.74 | 13.5 | 633 |
| WS$_2$ | HTL | ITO/WS$_2$/PBDB-T-2F: Y6: PC$_{71}$BM/PFN-Br/Al | 26 | 0.84 | 0.78 | 17 | 633 |
| UV treated Ti$_3$C$_2$T$_x$ | HTL | ITO/UV-MXene/PBDB-T:ITIC/Ca/Al | 15.98 | 0.89 | 0.64 | 9.02 | 651 |
| GO | HTL | ITO/GO/P3HT:PC$_{61}$BM/GOCs/Al | 10.30 | 0.61 | 0.59 | 3.67 | 636 |
| GO-Cl | HTL | ITO/GO-Cl/PTB7:PC$_{71}$BM/GO-Li/TiO$_x$/Al | 19.59 | 0.76 | 0.62 | 9.14 | 650 |
| PEDOT:PPS-GO | HTL | ITO/PEDOT:PPS-GO/PM6:Y6/PDINO–G/Al | 25.65 | 0.85 | 0.76 | 16.5 | 652 |
| PEDOT:PSS:GO | HTL | ITO/PEDOT:PSS:SPGO/PTB7:PC$_{71}$BM/Al | 17.3 | 0.67 | 0.41 | 4.82 | 640 |
| g-C$_3$N$_4$-doped PEDOT:PSS | HTL | ITO/g-C$_3$N$_4$-doped PEDOT:PSS/PM6:Y6/PFN-Br/Ag | 26.71 | 0.84 | 0.73 | 16.38 | 641 |
| α-In$_2$Se$_3$ nanosheets | HTL | ITO/α-In$_2$Se$_3$/PBDB-T:ITIC/Ca/Al | 16.69 | 0.88 | 0.65 | 9.58 | 642 |
| WS$_2$ nanosheets | HTL | ITO/WS$_2$/PBDB-T-2F:Y6:PC$_{71}$BM/PFN-Br/Ag | 26.0 | 0.83 | 0.72 | 15.6 | 643 |
| MoS$_2$ nanosheets | HTL | ITO/WS$_2$/PBDB-T-2F:Y6:PC$_{71}$BM/PFN-Br/Ag | 25.3 | 0.81 | 0.71 | 14.9 | 642 |

A RGO-PC$_{61}$BM composite was produced by Qu et al. by anchoring PC$_{61}$BM onto GO through a pyridine moiety to be used as the ETL in P3HT:PC$_{71}$BM-based OSCs.[648] The RGO-PC$_{61}$BM nanocomposite exhibits higher solubility compared to RGO and a low $\phi_W$ value of 4.4 eV, which matched the LUMO level of the electron acceptor.[647] Therefore, the modified PC$_{61}$BM OSCs significantly improved the $\eta$ value (3.89%) compared to OSCs using pristine RGO or pyrene-PC$_{61}$BM ETLs.[647]

Hu et al. used GQDs functionalized with ammonium iodide at the edge as a thickness-insensitive ETL with high optical transparency.[649] PCDTBT:PC$_{71}$BM-based OSCs using functionalized

GQDs exhibited an $\eta$ value of 7.49%, which was significantly higher than that of the reference cells using calcium as the ETL.[648] Importantly, the performance of OSCs was insensitive to the thickness of the GQD layer (i.e., 2–22 m).[648]

Solution-processed BP flakes in ethanol were also recently demonstrated as an effective interfacial layer between the ZnO ETL and PTB7:PC$_{71}$BM active layer in inverted OSCs.[650] The addition of the BP interlayer enhanced the $\eta$ value by 11%, reaching the maximum value of 8.25%.[649] The improvement of $\eta$ was attributed to the formation of a cascaded band structure between PC$_{71}$BM, ZnO, and BP flakes, which facilitates the electron transport and suppresses the carrier recombination near the cathode.[649] Furthermore, the BP-incorporated OSC has shown superior air stability, exhibiting a degradation of 5.82% after two days, compared to the reference device, which exhibited a degradation of 9.29% in the same timeframe.[649] Konios et al. demonstrated the simultaneous use of $\phi_W$-tuned functionalized GO derivatives as both HTL and ETL in PCDTBT:PC$_{71}$BM- and PTB7:PC$_{71}$BM-based OSCs.[651] The $\phi_W$ tuning of GO took place by either photochlorination[618] or Li neutralization[638] for $\phi_W$ increase or decrease, respectively.[651] Consequently, it was possible to match the GO-Cl $\phi_W$ with the HOMO level of both PCDTBT and PTB7 donor, as well as GO-Li $\phi_W$ with the fullerene LUMO level, enabling the balance between $\mu_e$ and $\mu_h$.[650] As a result, both graphene-based OSCs significantly outperformed the reference ones, leading to $\eta$ improvement of 30% and 19% for PCDTBT- and PTB7-based devices, respectively.[650] In particular, the champion device exhibited an $\eta$ value of 9.14%, which was a record-high value for OSCs using a graphene-based buffer layer.[650] In the same context, Yu et al. demonstrated the use of $\phi_W$-tuned MXenes, particularly $Ti_3C_2T_x$, as both HTL and ETL in NFA-based OSCs using PBDB-T:ITIC as the active layer.[652] The $\phi_W$ tuning took place through UVO or hydrazine treatments for $\phi_W$ increase and decrease, respectively.[651] Therefore, the $\phi_W$ value was tuned in the 4.08–4.95 eV range.[651] The $\phi_W$ modification mechanism was ascribed to the oxidation or reduction of the C element of $Ti_3C_2T_x$ by UVO or $N_2H_4$, respectively.[651] The UVO-and $N_2H_4$-treated MXenes were used as the HTLs in conventional OSCs and as the ETL in inverted OSCs, respectively.[651] The resulting cells exhibited an $\eta$ value of 9.02% or 9.06% respectively, both comparable to the performance achieved with PEDOT:PSS-based references.[651] Pan et al. developed an n-doped graphene ETL for OSCs by adding micromechanically exfoliated single-layer graphene to (N,N-dimethyl-ammonium N-oxide)propyl perylene diimide (PDINO).[653] The conductivity of graphene was increased by n-doping with the nitroxide radical of N-oxide in PDINO.[652] The resultant n-doped graphene (PDINO-G) possessed increased conductivity, lower $\phi_W$, reduced charge recombination, and increased charge extraction rate compared to pristine PDINO.[652] The OSCs based on PTQ10:IDIC-2F with PDINO-G as the ETL exhibited an $\eta$ value of 13.01%, which was superior to that achieved by OSCs without graphene.[652] Furthermore, PM6:Y6-based OSCs using PEDOT:PPS-GO as the HTL and PDINO–G as the ETL displayed an $\eta$ value as high as 16.52%, significantly higher than that for OSCs without GO and graphene (15.1%).[652] The observed performance enhancement was attributed to the higher (by two orders of magnitude) conductivity of graphene-based ETL compared to that of graphene-free ETL, suitable $\phi_W$, and optimal charge extraction.[652]

More recently, Lee et al. used $MoS_2$ nanoflakes, with an average diameter of 27 nm, as an effective electron transporting interlayer between polyethylenimine ethoxylated (PEIE) and the photoactive layer in OSCs.[654]

$MoS_2$ nanoflakes acted not only as an ETL but also as a sub-photosensitizer (i.e., additional light-absorbing layer), enhancing $\eta$ by 27%, 11%, and 15% compared to P3HT:$PC_{60}BM$-, PTB7: $PC_{71}BM$-, and PTB7-Th:$PC_{71}BM$-based reference cells, respectively.[653] The observed performance enhancement was attributed to effective electron transport via $MoS_2$ nanoflakes supported by an increased Förster resonance energy transfer[375,655,656] efficiency of 67% from PTB7:$PC_{71}BM$ to $MoS_2$ nanoflakes.[653]

Table 4 summarizes the main results achieved with OSCs using ETLs based on GRMs.

**Interconnection layers (ICLs).** Tandem OSCs stack two or more single-junction sub-cells (with complementary $E_g$ values) to harvest light from the entire solar spectrum.[657] Ideally, the $V_{OC}$ value of the tandem devices is the sum of the $V_{OC}$ values of the sub-cells, while the $I_{SC}$ value is the lowest $I_{SC}$ of the two sub-cells, the latter being in series.

An ICL collects electron and holes from the respective sub-cells, acting as a recombination site between them.[658,659] Therefore, ICL is a critical component in tandem architectures. In addition, an optimal ICL should be uniform, transparent, highly conductive, and resistant to solvents.[660] So far, PEDOT:PSS/ $TiO_2$,[661] PEDOT:PSS:ZnO,[662] and LiF/Al/Au/PEDOT:PSS[663] have been the most established ICLs, despite the well-known drawbacks attributable to the acidic and aqueous nature of PEDOT: PSS, which have a huge impact on OSC stability.[664] In the development of ICL, GRMs have been used both to improve the stability of PEDOT:PSS and in combination with other materials as an alternative to PEDOT:PSS.

Tung et al. used GO:PEDOT:PSS nanocomposite as the ICL in a tandem OSCs consisting of two identical P3HT:$PC_{61}BM$-based sub-cells.[665] The tandem OSCs were fabricated by a direct adhesive lamination process enabled by the sticky GO:PEDOT film.[664] An $\eta$ value of 4.14 and $V_{OC}$ of 0.94 V (~84% of the sum of the $V_{OC}$ of the two sub-cells) were reported.[664] Surprisingly, the presence of GO in the composite increased the PEDOT:PSS electrical conductivity by altering its chain conformation and

---

**Table 4** Summary of the PV performance of OSCs using GRMs as the ETLs

| Material | Usage | Device structure | Cell performance | | | | Ref. |
|---|---|---|---|---|---|---|---|
| | | | $J_{SC}$ (mA cm$^{-2}$) | $V_{OC}$ (V) | FF (—) | $\eta$ (%) | |
| GO & Cs-neutralized GO (GOCs) | ETL | ITO/GO/P3HT:$PC_{61}BM$/GOCs/Al | 10.30 | 0.61 | 0.59 | 3.67 | 650 |
| $Cs_2CO_3$ functionalized GQDs-$Cs_2CO_3$) | ETL | ITO/GQDs-$Cs_2CO_3$/P3HT:$PC_{61}BM$/V_2O_5/Au | 9.18 | 0.58 | 0.61 | 3.23 | 637 |
| Lithium-neutralized GO (GO-Li) | ETL | ITO/PEDOT:PSS/PCDTBT:$PC_{71}BM$/GO-Li/$TiO_x$/Al | 12.51 | 0.89 | 0.57 | 6.29 | 638 |
| ZnO-RGO hybrids | ETL | ITO/PEDOT:PSS/PTB7:$PC71BM$/ZnO-RGO/Al | 15.19 | 0.72 | 0.69 | 7.5 | 643 |
| $TiO_2$-RGO hybrids | ETL | ITO/PEDOT:PSS/PTB7:$PC_{71}BM$/$TiO_2$-RGO/Al | 14.99 | 0.74 | 0.67 | 7.46 | 643 |
| $TiO_x$:RGO composites | ETL | ITO/rGO:$TiO_x$/P3HT:$PC_{61}BM$/MoO_3$/Ag | 9.85 | 0.64 | 0.61 | 3.82 | 646 |
| RGO-pyrene-$PC_{61}BM$ | ETL | ITO/PEDOT:PSS/P3HT:$PC_{61}BM$/RGO-pyrene-$PC_{61}BM$/Al | 9.78 | 0.64 | 0.62 | 3.89 | 647 |
| GQDs functionalized with ammonium iodide (GQD-NI) | ETL | ITO/PEDOT:PSS/PCDTBT:$PC_{71}BM$/GQD-NI/Al | 10.98 | 0.93 | 0.73 | 7.49 | 648 |
| n-Doped $MoS_2$ | ETL | ITO/n-doped $MoS_2$/P3HT:$PC_{61}BM$/PEDOT:PSS/Ag | 8.16 | 0.59 | 0.55 | 2.73 | 621 |
| Black phosphorus (BP) | ETL | ITO/ZnO/BP/PTB7:$PC_{71}BM$/MoO_3$/Ag | 18.78 | 0.72 | 0.61 | 8.25 | 649 |
| $N_2H_4$ treated $Ti_3C_2T_x$ | ETL | ITO/$N_2H_4$-$Ti_3C_2T_x$/PBDB-T:ITIC/MoO3/Al | 17.36 | 0.87 | 0.6 | 9.06 | 651 |
| Small sized $MoS_2$ | ETL | ITO/PEIE/$MoS_2$/PTB7-Th:$PC_{71}BM$/MoO_3$/Ag | 17.02 | 0.8 | 0.66 | 9.08 | 653 |
| GO-Li | ETL | ITO/GO-Cl/PTB7:$PC_{71}BM$/GO-Li/$TiO_x$/Al | 19.59 | 0.76 | 0.62 | 9.14 | 650 |
| PDINO-G | ETL | ITO/PEDOT:PSS-GO/PM6:Y6/PDINO–G/Al | 25.65 | 0.85 | 0.76 | 16.5 | 652 |

| Material | Usage | Device structure | Cell performance | | | | Ref. |
|---|---|---|---|---|---|---|---|
| | | | $J_{SC}$ (mA cm$^{-2}$) | $V_{OC}$ (V) | FF (−) | $\eta$ (%) | |
| GO:PEDOT:PSS composite | ICL | Glass/ITO/PEDOT:PSS/P3HT:PC$_{61}$BM/ZnO/GO:PEDOT:PSS/P3HT:PC$_{61}$BM/Ca/Al | 7.2 | 1 | 0.58 | 4.14 | 490 |
| GO | ICL | Glass/ITO/GO/PSEHTT:ICBA/TiO$_2$/GO/PSBTBT:PC$_{71}$BM/ZnO/Al | 8.23 | 1.62 | 0.63 | 8.4 | 491 |
| Cesium neutralized GO (GO-Cs) & GO | ICL | Glass/ITO/GO/PSS/PCDTBT:PC$_{71}$BM/GO-Cs/Al/GO/MoO$_3$/PCDTBT:PC$_{71}$BM/Ca/Al | 5.03 | 1.69 | 0.46 | 3.91 | 492 |

morphology. Moreover, GO increased the PEDOT:PSS dispersion viscosity, leading to a beneficial effect on the adhesion properties.[664] Overall, the addition of GO effectively improved charge extraction at the interface between the HTL and active layer.[664]

Yusoff *et al.* incorporated a GO/TiO$_2$ recombination layer into a tandem OSC.[491] The overall $V_{OC}$ (1.62 V) was approximately the sum of those of the individual sub-cells (0.94 V and 0.68 V).[491] This result indicated that the incorporation of GO in traditional recombination layers can allow the realization of ideal resistance-free interconnection between the front and rear cell, while preserving the optical transparency of the same recombination layers (*e.g.*, TiO$_2$).[491] Notably, all the tandem OSCs were solution-processed and stable.[491] Finally, Cs-functionalized GO was used in GO-Cs/Al/GO/MoO$_3$ ICL between two PCDTBT-based sub-cells.[666] The ICL based on GO promoted recombination between the electrons and holes generated from the front and rear cells, owing to the energy-level matching of the interfaced materials.[665] In fact, after MoO$_3$ modification, the $\phi_W$ value of GO increased up to 5.3 eV, matching the HOMO level of PCDTBT, while that of Al-modified GO-Cs decreased to 4 eV, matching the LUMO level of PCBM.[665] The resulting $\eta$ and $V_{OC}$ values were 3.91% and 1.69 V, respectively.[665] The $V_{OC}$ was almost equal to the sum of the two sub-cells, proving the beneficial role of GRM-based ICLs.[665]

Table 5 summarizes the main results achieved with tandem OSCs using ICLs based on GRMs.

#### 4.4 Summary and outlook

The effort for OSC commercialization has recently seen a renaissance after the development of BHJ single-junction devices based on low-bandgap polymer donors and NFAs,[474–476] which reached $\eta$ values exceeding 17%,[474–476] up to the record value of 18.3%.[477] Importantly, the LCOE for organic solar modules with an $\eta$ value of 10% in a 20 year range has recently been estimated to be between 0.185 and 0.486 ¥ kW h$^{-1}$ (*i.e.*, between 2.7 and 7.3 US cent kW h$^{-1}$),[667] which is competitive with the LCOEs afforded by current PV technologies (less than 5 US cents kW h$^{-1}$)[131–133] and fossil fuels.[134,135] In this context, the incorporation of solution-processed 2D materials enabled the OSCs to further increase their performance up to $\eta$ values of more than 17%.[584,633] In ref. 633, this achievement was attained by replacing a traditional HTL, *i.e.*, hygroscopic PEDOT:PSS, with solution-processed WS$_2$ flakes. More generally, several solution-processed GRMs have been proven to combine all the key attributes required by ideal CTLs and/or buffer layers for OSCs. In particular, the energy levels of 2D materials can be tuned on demand to conceive advanced interface engineering at the device heterojunctions, enhancing the exciton dissociation, while providing optical transparency and high carrier mobilities for efficient charge transport toward the electrodes. Meanwhile, solution-processed 2D materials have tunable energy levels to act as an additive in ternary blends together with NFAs and low-bandgap polymer donors. In addition, solution-processed 2D materials have been demonstrated to regulate the morphology of the active layer, improving device $\eta$ (up to 17.3%) as well as reproducibility.[584] Therefore, we are looking forward to seeing the implementation of 2D materials into the most efficient reported OSC architectures to achieve $\eta$ over 20% in the near future.[668]

Despite the progress seen in 2D materials as CTLs, additives for photoactive layers, and buffer layers, the printing of 2D materials over a large scale in air using either R2R or sheet-to-sheet (S2S) processes is still unreported for the practical realization of large-area ($>1$ cm$^2$) OSCs and modules.[669–673] The establishment of scalable LPE methods for the production of 2D materials with controlled size and thickness can stimulate research toward the realization of commercially competitive large-area OSC technologies, including flexible devices, with $\eta$ exceeding 14%.[674–678]

Moreover, OSCs based on solution-processed 2D materials can find applications for solution-processed tandem PVs.[679] For example, the latter can be fabricated using a PSC and OSC as the sub-cells, both produced through R2R methods. The solution processability of efficient tandem SCs is a very attractive alternative to perovskite/Si tandem SCs,[678] which recently gained enormous interest for off-grid power generation.[680] In the scenario involving flexible and bifacial OSCs, (semi)transparent and flexible electrodes based on solution-processed graphene and graphene derivatives are still not competitive with commercially established TCO-based technologies (see additional discussion in Section 8). Lastly, OSCs are attractive for indoor applications,[681–683] such as ideal power sources for indoor IoT devices. In this context, $\eta$ over 21% has been reached,[680,682] and the exploitation of 2D materials may rapidly contribute toward further improving these performances.

Overall, we believe that the use of cost-effective solution-processed 2D materials as CTLs, buffer layers, and additives in OSCs, coupled with the development of low-bandgap donors and NFAs, can be the key to unlock their spread in both large-scale and niche (*i.e.*, indoor) applications.

## 5. DSSCs

DSSCs are an intriguing alternative to the more conventional Si-based PV technology, owing to their potential low-cost production and compatibility with flexible design.[684–686] In a typical DSSC, a dye—sensitizing nanocrystalline $TiO_2$ on a TCO glass—absorbs the solar energy.[52,687] The photoexcitation of the dye promotes electron injection toward the CB of nanocrystalline $TiO_2$.[52,686] The electrons flow in the direction of the transparent electrode where they are collected for powering a load.[52,686] After flowing through the external circuit, the electrons are reintroduced into the cell on a metal electrode on the back, called CE, where they are transferred to an electrolyte (also called the mediator).[52,686] Then, the electrolyte transports the electrons back to the dye molecules.[52,686] Thus, the original state of the dye is recovered. The mediator can be either an electrolyte containing a redox couple (such as $I^-/I_3^{-[688]}$ and $Co^{2+}/Co^{3+}$,[689] for liquid-state DSSCs) or a HTL such as 2,2′,7,7′-tetrakis[$N$,$N$-di(4-methoxyphenyl)amino]-9,9′-spirobifluorene (spiro-OMeTAD) for solid-state DSSCs.[690–692]

The theoretical maximum $V_{OC}$ value of a DSSC is regulated by the energy difference between the Fermi level of the metal oxide (typically $TiO_2$) semiconductor on the photoanode and the redox potential of the mediator.[52,686] However, at a nonzero current, the output voltage is inferior to $V_{OC}$. In detail, the overall overpotential of the CE determines a voltage loss related to the delivery of current through the electrolyte/CE interface (kinetic overpotential or charge transfer overpotential) and through the electrolyte (mass transfer overpotential).[693] The mass transfer overpotential is mainly affected by the ionic conductivity of the electrolyte and the transport of mediator species from the CE to the photoanode.[694,695] Instead, the catalytic activity of the CE for the mediator reduction reaction defines the magnitude of charge transfer overpotential.[693,694] Detailed descriptions of each component of a DSSC, as well as recent advances in DSSCs, can be found in several reviews.[686,696,697]

To improve the performance and reducing the cost of DSSCs, the incorporation of new materials as well as the development of solution-processing techniques are actively pursued. In this context, GRMs have been extensively exploited in different DSSC components.[698–702] In detail, they were first used as a transparent electrode to replace FTO at the photoanode.[701] Subsequently, they have been used as light absorbers,[703] additives for improving charge transport through both $TiO_2$[704–710] and electrolyte,[711,712] and CE material for Pt replacement.[699,713,714] Herein, we will overview the use of solution-processed GRMs in different components of DSSCs.

### 5.1 TCEs and photoanodes

TCEs, as well as entire photoanodes, are the key components in DSSCs. Traditionally, ITO ($R_S \sim 10$–$30 \ \Omega \ \square^{-1}$, $T_r$ (550 nm) > 90%, $\phi_W \sim 4.8$ eV)[715] and FTO ($R_S \sim 15$–$20 \ \Omega \ \square^{-1}$, $T_r \sim 85\%$ (300 < $\lambda$ <550 nm), $\phi_W \sim 4.4$–5 eV)[716–719] have been the most frequently used TCEs. However, ITO- and FTO-based electrodes have economic and technical issues, such as scarcity of In and high production/processing costs,[720] as well as their crack susceptibility under tensile stress and structural defects.[715,721] In this framework, pristine graphene and its derivatives promise to be ideal alternatives to traditional TCEs[507,722] due to their high $\sigma$ and $T_r$.[73] In particular, the preparation and processability of GRMs, with controlled size,[723] thickness,[118] and chemical functionalities,[119] by solution-based methods[130,333] have provided simple and scalable ways to fabricate TCEs, compatible with high-throughput printing processes.[333,724] In order to design DSSC photoanodes, several strategies have been pursued to minimize competition between light absorption, electron transport, and charge recombination processes: (i) the formulation of a composite of metal oxides with appropriate $E_g$ to guarantee the ideal $T_r$ value, allowing light to be effectively absorbed by the dye;[725,726] (ii) the design of 1D metal oxides such as nanowires, nanorods, and nanotubes for (1) reducing the electron transport pathway (i.e., increasing the electron diffusion coefficient ($D_n$)); (2) increasing the electron lifetime ($\tau$) (i.e., reducing the charge recombination losses);[727–731] (iii) engineering 3D-metal-oxide-based light-trapping scaffolds for enhancing the light absorption, while reducing the photoanode thickness.[732–734]

The first attempt to exploit graphene-based materials to replace conventional TCEs in DSSCs can be tracked back to the pioneering work of Wang et al.[701] The authors used transparent and conductive ultrathin (10 nm) graphene-based films, produced through graphite oxide exfoliation followed by a thermal reduction treatment.[701] However, poor PV performance was achieved as a consequence of both $R_{series}$ and electronic interfacial changes introduced by defected edges in graphene platelets, which limited the hopping mechanism of electrons during their transport.[735–739] Therefore, large graphene sheets are beneficial to reduce the number of boundaries and thus the contact resistances through the entire graphene film. These conclusions were very similar to those drawn in the field of OSCs. In the context of DSSCs, major efforts focusing on the incorporation of GRMs into DSSC photoanodes were aimed to form 2D bridges within mesoporous/nanostructured electrodes and to improve the electron collection efficiency.[702,703,708,740–748] In fact, graphene and its derivatives can hold energy levels between those of photoanode metal oxide (typically $TiO_2$ or ZnO) CBs and $\phi_W$ of FTO (Fig. 11a).[749] Consequently, graphene acts as an extended current collector (or electron transfer channel) for a rapid collection/transfer of the photogenerated electrons to the conductive substrate before the electrons recombine by interacting with the dye and/or the redox species (Fig. 11b).[750] In addition, graphene insertion into photoanode metal oxides introduces hierarchical structures (i.e., structures with a multiscale nanostructural ordering), which enhance light scattering.[751,752] Thus, graphene effectively acts as light-capture centers to improve the overall $\eta$ of DSSCs.[747,753] Based on the aforementioned considerations, Yang and co-workers reported the use of $TiO_2$/graphene front electrode for enhancing DSCC $\eta$ and $J_{SC}$ by approximately 39% and 45%, respectively, compared to the reference DSSC using nanocrystalline $TiO_2$.[703] Similar effects have also been reported by Gao and co-workers.[739] In particular, Nafion ($C_7HF_{13}O_5S \cdot C_2F_4$)-functionalized graphene dispersion and commercial $TiO_2$ NPs (P25) were used to prepare a graphene/$TiO_2$ nanocomposite-based TCE with a continuous 2D conductive network.[739] The hydrophobic fluorine backbones of Nafion avoided the agglomeration of graphene flakes and conferred stability through electrostatic interactions between

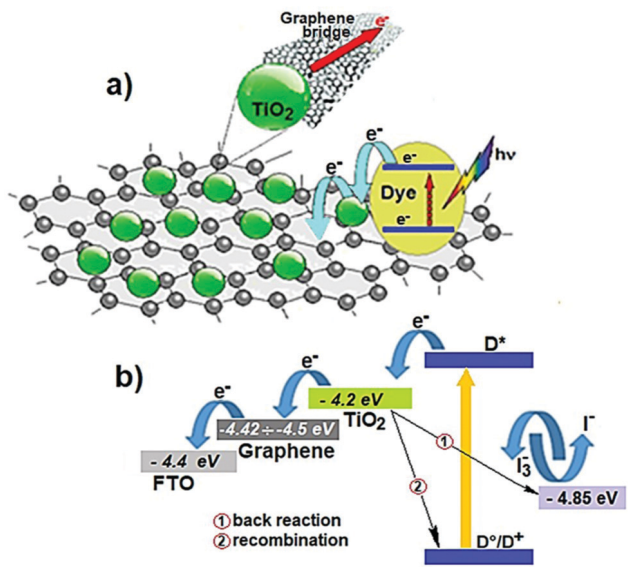

**Fig. 11** (a) Schematic of light capture and electron transfer pathway from TiO$_2$ to the FTO, in TiO$_2$/graphene-based DSSC photoanodes. (b) Schematic of the electron cascade route in DSSC using TiO$_2$/graphene-based photoanodes.

the composite materials.[754] It was proposed that this TCE morphology strongly influenced dye absorption, modulating the incident light harvesting, as well as the number of photo-induced electrons injected from the excited dye to the TiO$_2$ CB.[753] Consequently, the DSSC incorporating 0.5 wt% graphene in the TiO$_2$ photoanode demonstrated an $\eta$ value of 4.28%, which was 59% higher than that of the reference DSSC.[753] Tsai et al. demonstrated that graphene-incorporated TiO$_2$ photoanode, prepared by spin coating, suppresses electron recombination, as well as enhances dye absorption onto the electrode surface.[755] By using 1 wt% graphene, a 15% improvement in $\eta$ (from 5.98% to 6.86%) was demonstrated.[754]

Xu et al. developed a highly conductive graphene scaffold incorporated into ZnO hierarchically structured NPs (HSN) capable of capturing and transporting photogenerated electrons.[752] The DSSCs, with a 1.2 wt% of graphene into the ZnO photoanode, exhibited an $\eta$ value of ~5.86%, higher than that of DSSCs without graphene.[752] Performance improvement due to graphene addition could be attributed to the combination of fast electron transport and long electron lifetime, which reduced the electron recombination losses (Fig. 12).[752]

Kusumawati et al. prepared a composite TiO$_2$/RGO porous photoanode in order to investigate the influence of RGO content (0.6, 1.2, and 3 wt%) on dye (N719) loading, as well as on the charge extraction transport properties of TiO$_2$ NPs.[740] The authors found that RGO incorporation increased the film SSA, thereby promoting dye loading.[740] This effect improved the light absorption, increasing $J_{SC}$ and $\eta$ (by ~12%) as compared to those of the reference DSSC.[740] The $\eta$ enhancement was also ascribed to the optimized electron transport (~60% increase in $\sigma$ compared to bare TiO$_2$) in the TiO$_2$/RGO (1.2 wt%) composite photoelectrodes film.[740] Hayashi et al. fabricated a multistep electron transfer system based on organic–inorganic ternary composites of Zn–porphyrin (ZnP), ZnO NPs, and RGO onto a FTO/SnO$_2$ electrode.[756] The RGO flakes randomly distributed in the composite film acting as a 2D network, which assists the electron flow from ZnO-NP/ZnP composite to the FTO/SnO$_2$ electrode.[755] This effect limited the charge recombination at the electrolyte interface and improved the photocurrent generation, resulting in an IPCE value of ~70% over the absorbed wavelength.[755] The authors assessed that RGO can act as electron acceptor from ZnO-NP/ZnP composite, as well as a medium to store and shuttle electrons within the composite film.[755] Fang et al. introduced different GO contents in TiO$_2$ NPs by ball milling and reported an $\eta$ value of 5.09%, which was remarkably higher than that of the reference DSSC ($\eta$ = 4.43%).[757] Sun and co-workers reported DSSCs based on graphene–TiO$_2$ composite photoanodes, achieving an $\eta$ value of 4.28%, which was 59% higher than the reference DSSC.[739] It was found that the incorporation of graphene caused both increased dye adsorption and significantly longer electron lifetime compared to the graphene-free case.[739] Chemically exfoliated graphene sheets (GS), incorporated by the grafting method in TiO$_2$ NP films, were synthesized by Tang *et al.* and used as the photoanode scaffold.[758] Both high $\sigma$ of reduced GS and optimum attachment of TiO$_2$ NPs on the GS were achieved by regulating the oxidation time during the chemical exfoliation process.[757] Uniform films of GS/TiO$_2$ composite with large SSA were prepared on a conductive glass by electrophoretic deposition, and the integration of GS substantially increased the $\sigma$ value of the film of TiO$_2$ NPs by more than two orders of magnitude.[757] In addition, DSSCs based on GS/TiO$_2$ composite films reached $\eta$ that is 5 times higher than that based on TiO$_2$ alone.[757] The better PV performance of GS/TiO$_2$-based DSSC was also attributed to the higher dye loading of the GS/TiO$_2$ film compared to the GS-free case.[757] Durantini *et al.* reported graphene–TiO$_2$ composite electrodes fabricated by hydrothermal synthesis or simple deposition by the spin-coating technique of TiO$_2$ paste onto a graphene layer.[709] In both cases, no significant morphological differences were observed in the electrodes prepared with and without graphene.[709] The DSSCs containing the graphene composite or layered films, enhanced both the $J_{SC}$ (from 11.6 mA cm$^{-2}$ to 14.0 mA cm$^{-2}$) and $\eta$ (from 5.8% to 7.3%) values compared to the reference devices.[709] Similar results have been reported by Chen *et al.* who developed a TiO$_2$/graphene/TiO$_2$ sandwich structure used as the photoactive layer in DSSCs, reaching $\eta$ that is ~60% higher than the reference cell.[759]

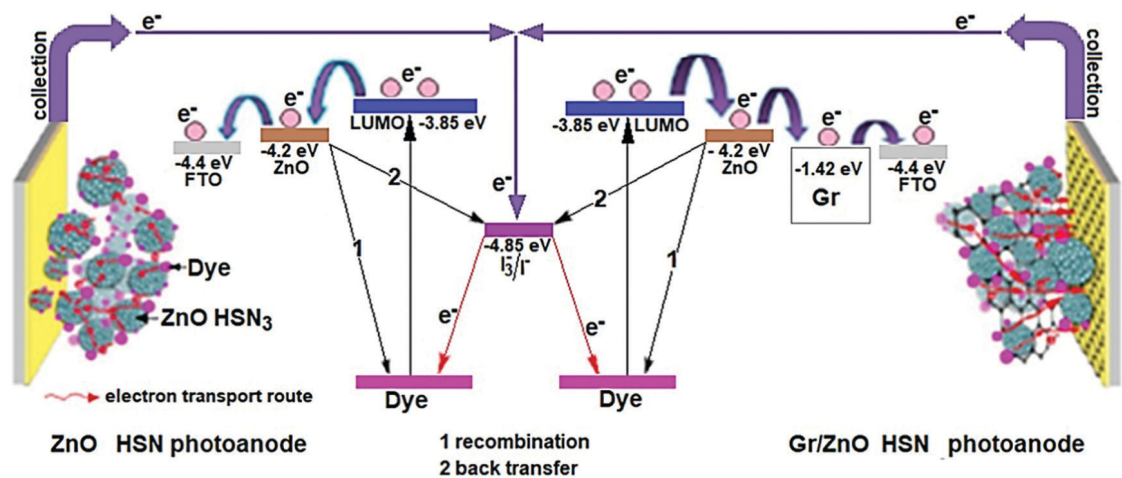

**Fig. 12** Operating principle and energy diagram of the DSSC using ZnOHSN (left) or Gr/ZnOHSN (right.). Electron injection from the excited dye into the nanostructured ZnO semiconductor, electron transport to the collection electrode, and the recombination (1) and back transfer (2) pathways are also shown. Reprinted with permission from ref. 752, Copyright 2013, American Chemical Society.

It was speculated that electrons from the photoexcited dye are rapidly and efficiently transported to the CB of TiO$_2$ through the graphene-layer bridge, which both enhances the $\sigma$ value of the photoelectrode and reduces charge recombination and back-reaction processes compared to the reference photoanode.[703] In addition, the sandwich structure allowed light to be absorbed over a wide spectral range, enhancing the $V_{OC}$ of DSSCs from 0.55 V to 0.6 V.[703] Xiang *et al.* fabricated DSSCs based on TiO$_2$ photoanodes modified by GO and N-RGO, revealing better PV performance for N-RGO TiO$_2$ photoanode compared to the case based on GO.[760] The DSSCs using N-RGO TiO$_2$ photoanode reached a 13.23% higher $\eta$ compared to that of conventional TiO$_2$-based DSSCs.[759] In particular, the $V_{OC}$ value increased with N-RGO addition due to the suppression of electron recombination, while $J_{SC}$ exhibited its maximum value at N-RGO content of 0.2 wt% owing to the synergistic effects of electron transfer efficiency, light scattering, and dye adsorption.[759]

Ding *et al.* produced RGO-TiO$_2$ composite films by mixing TiO$_2$ NPs with flakes of GO and ascorbic acid (vitamin C).[742] The latter enabled GO to be reduced at ambient temperature.[742] After treatment in a TiCl$_4$/H$_2$O solution followed by sintering at 450 °C, the RGO-TiO$_2$ NPs were sensitized by N719 dye and used as the photoanode in DSSCs.[742] The influence of RGO on the DSSC PV performances was evaluated at different RGO contents, varying from 0.25 to 0.75 wt%.[742] For a content of 0.75 wt%, the DSSCs reached the best PV performance with an $\eta$ value of 7.89%, which was ~30% higher compared to that of its RGO-free device (6.06%).[742] This performance improvement was attributed to the remarkable electric transport properties of RGO,[761] which captures and transports electrons, decreasing the overall charge recombination rate.[742] Notably, an excessive content of RGO (>0.75 wt%) caused the restacking of flakes, which were then ineffective in covering the TiO$_2$ NPs, eliminating their beneficial effects on the PV performance of DSSCs.[742] Mehmood *et al.* also studied the dependence of the $\eta$ value of DSSC by GNP content in TiO$_2$/graphene composite-based DSSCs.[762] In particular, they fabricated photoanodes by adding GNPs into TiO$_2$ NP paste, obtaining the highest $\eta$ of 4.03% with a GNP content of 0.16 wt%.[761] Higher GNP content negatively affected the DSSC performance.[761] This was attributed to the reduced $T_r$ value of the TiO$_2$/graphene film, as well as to the presence of graphene aggregates inside the TiO$_2$ matrix, which can act as charge-trapping sites.[761] Sacco *et al.* investigated the charge transport and recombination properties in GO/TiO$_2$ composite-based DSSCs.[741] Impedance spectroscopy analysis revealed that GO incorporation into TiO$_2$ led to an increase in both $D_n$ and $\tau$, limiting the charge recombination processes and increasing the $V_{OC}$.[739,763] He *et al.* designed and prepared a DSSC photoanode based on a RGO-TiO$_2$ heterostructure by cetyl-trimethyl-ammonium-bromide (CTAB)-assisted hydrothermal method in order to wrap and anchor RGO with high-density TiO$_2$ NPs, resulting in a high-SSA (~83 cm$^2$ g$^{-1}$) composite.[764] The inner RGO flakes ensured a rapid charge carrier transport route for effective charge collection at the conductive substrate, while the closely packed TiO$_2$ NPs limited direct contact between the RGO surface (rich in e$^-$) and electrolyte (rich in h$^+$), preventing charge recombination processes.[763] Because of these multiple effects, DSSCs based on RGO have shown an $\eta$ value enhancement of ~40% compared to the reference one.[763] Ranganathan *et al.* exploited N-doped graphene@nickel oxide (NG/NiO) nanocomposite-doped TiO$_2$, deposited onto FTO substrates by screen printing, as the photoanode.[765] The corresponding DSSCs have shown $\eta$ up to 9.75%, which was higher than those of DSSCs using GO/TiO$_2$-, TiO$_2$-, and NiO/TiO$_2$-based photoanodes (8.55, 8.69, and 9.11%, respectively).[764]

Graphene was also used in QD-based DSSCs (QDDSSCs),[766] also named QD-sensitized solar cells (QDSSCs) (see additional discussion in Section 6), to realize graphene–TiO$_2$ hybrid photoanodes. For example, Zu *et al.* fabricated CdS–QDDSSCs based on a graphene–TiO$_2$ film photoanode (0.8 wt% of graphene), improving the $\eta$ value by ~55% compared to a reference DSSC with a pristine TiO$_2$-based photoanode.[767]

Yan *et al.* synthesized—via stepwise solution chemistry—large, soluble graphene QDs with 1,3,5-trialkyl-substituted phenyl moieties covalently attached at the edge of graphene QDs and used as sensitizers in DSSCs.[702] However, despite the higher molar extinction coefficient ($\kappa$) (~1 × 10$^5$ M$^{-1}$ cm$^{-1}$) of GQDs compared to that of N719 dye (~1.5 × 10$^4$ M$^{-1}$ cm$^{-1}$),[702] the fabricated QDDSSC exhibited suitable $V_{OC}$ (0.58) and FF (0.48 V) values, but small $J_{SC}$ (0.2 mA cm$^{-2}$) due to low affinity between GQDs and TiO$_2$ surface.[702] Subsequently, the incorporation of both 3D graphene structure and GSs into TiO$_2$ was evaluated to clarify their influence on charge transport through the graphene/TiO$_2$ interface in QDDSSCs using CdS/CSse QDs.[768] From the I–V

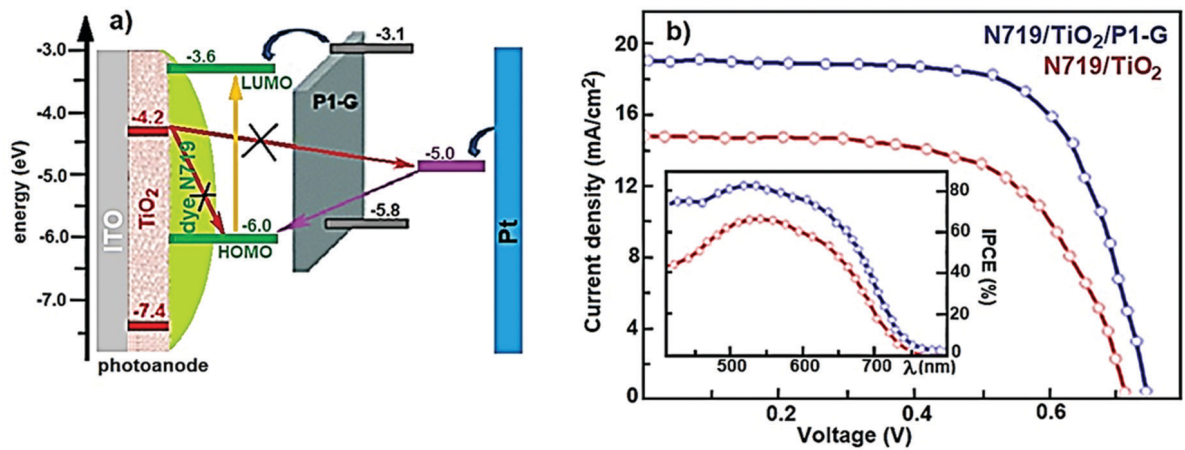

**Fig. 13** (a) Schematic of the energy-level diagram and electron transfer process in P1-graphene-based DSSCs. (b) J–V curves and IPCE spectra for DSSCs with and without a P1-G layer. Adapted with data from ref. 769.

curve analysis, as a function of GQD content in photoanode formulation, it was observed that $\eta$ and $J_{SC}$ reached the maximum value at a graphene content of 1.5 wt%, and then decreasing at higher contents.[767] On the contrary, both FF and $V_{OC}$ have shown no correlation with the GQD concentration.[767] These features highlighted the strong correlation between $J_{SC}$ and $\eta$[739,741,766,767] with graphene content, while it was suggested that the composite semiconductor $E_F$[52] and device series resistance[769] are not affected by the incorporation of graphene.

In a recent work, a novel approach based on graphene has been used to fabricate a DSSC with an $\eta$ value of 10.4%, representing an ~28% improvement compared to the reference cell based on conventional TiO₂-based photoanodes ($\eta$ = 7.5%).[770] In detail, graphene dispersed in o-dichlorobenzene was uniformly incorporated in a semiconducting polymer with commensurate band edges (P1) (see chemical structure in ref. 771) and the resulting composite (P1-graphene) solution was spin coated over the cell photoanode to act as a barrier layer limiting the back-transfer process of electrons (Fig. 13).[769] At a graphene concentration of 0.9 wt%, experimental data proved the favorable influence of the P1-G barrier layer in improving the dye regeneration ability.[769] In detail, graphene effectively acts as a scavenger for electrons at P1, directing the electrons to the HOMO of the dye for the regeneration process.[769] The P1-G-based device has shown higher recombination resistance ($R_{rec}$) compared to that of the reference device.[769] In addition, P1-G-based device displayed a $\tau$ value of 113 ms at the photoanode, more than double that of the reference device (i.e., 45 ms).[769] Finally, cyclic voltammetry (CV) and photoluminescence measurements revealed that the use of P1-G resulted in charge injection from the redox electrolyte to the HOMO level of the dye that was higher than that exhibited by the standard device (TiO₂ surface). Moreover, the addition of graphene in P1 decreased the photoluminescence intensity from $9.3 \times 10^4$ in the P1 film to $5.2 \times 10^4$ counts in P1-G films. This last feature further testified that graphene acts as a scavenger for electrons at P1, leading the electrons to the HOMO of the dye for the regeneration process.

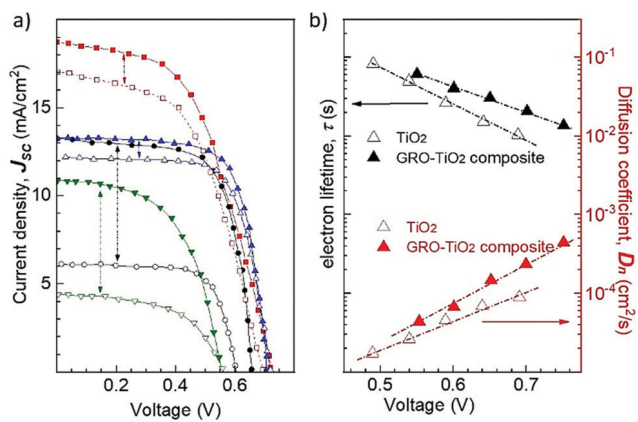

**Fig. 14** (a) Typical J–V curves of DSSCs based on different GRM-based photoactive layers (line and full symbols) compared to those of the corresponding traditional TiO₂-based photoelectrode. (b) Electron diffusion coefficient ($D_n$) and lifetime ($\tau$) dependence on the applied voltage for DSSCs based on TiO₂ photoactive layers with and without GRMs. Adapted from ref. 741,742,758,759.

Fig. 14 summarizes the J–V curves of representative DSSCs using TiO₂-GRMs composite-based photoanodes, compared to those of the equivalent TiO₂-based cells, and the corresponding $D_n$ and $\tau$ data.

Table 6 lists the experimental results achieved in DSSCs using solution-processed GRMs as the photoanode material. For each case, the $\eta$ value is compared with that of the TiO₂-based photoanode reference.

### 5.2 CEs

**5.2.1 Graphene and graphene derivatives.** In DSSCs, the CE collects the photogenerated electrons from the external circuit and catalyzes the oxidized electrolyte regeneration.[52] The reduction in the overall overpotential of CE (i.e., decrease of $R_{CT}$), attributable to the delivery of current through the electrolyte/CE interface, is crucial for limiting the voltage loss within the DSSCs.[52] An effective CE should have high $\sigma$ and exhibit high electrocatalytic activity toward the mediator reduction reaction. For the case of liquid-state DSSCs, the reduction reaction involves the redox couple $I^-/I_3^-$ and is $I_3^- + 2e \rightarrow 3I^-$. Thus, in order to minimize the charge transfer overpotential, high $\sigma$ for charge transport and electrocatalytic activity for reducing the redox couple, as well as electrochemical stability, are fundamental requirements for CE materials.[792]

**Table 6** Summary of the PV performance of DSSCs using 2D material-based photoanodes

| Photoanode structure | Dye | Cell performance | | | | Δη (%) | Ref. |
|---|---|---|---|---|---|---|---|
| | | $J_{SC}$ (mA cm$^{-2}$) | $V_{OC}$ (V) | FF (−) | η (%) | | |
| TiO₂ | N3 | 11.26 | 0.69 | 64.5 | 5.01 | 39.1 | 703 |
| TiO₂/GO (0.6 wt%) | | 16.29 | 0.69 | 62.0 | 6.97 | | |
| TiO₂ | N719 | 18.83 | 0.684 | 46.48 | 5.98 | 14.7 | 754 |
| TiO₂/graphene (1 wt%) | | 19.92 | 0.704 | 48.86 | 6.86 | | |
| TiO₂ | N719 | 11.0 | 0.71 | 74.1 | 5.78 | 29.6 | 740 |
| TiO₂/graphene (1.2 wt%) | | 14.4 | 0.68 | 76.80 | 7.49 | | |
| TiO₂ nanofibers | N719 | 13.9 | 0.71 | 63 | 6.3 | 20.6 | 772 |
| TiO₂ nanofibers/graphene (0.7 wt%) | | 16.2 | 0.71 | 66 | 7.6 | | |
| TiO₂ | N719 | 8.69 | 0.77 | 66 | 4.42 | 36.9 | 773 |
| TiO₂/graphene (1.0 wt %) | | 12.89 | 0.68 | 69 | 6.05 | | |
| TiO₂ | N3 | 8.787 | 0.606 | 65.97 | 4.43 | 14.9 | 756 |
| TiO₂/GO | | 10.284 | 0.616 | 63.75 | 5.09 | | |
| TiO₂ | N3 | 9.58 | 0.82 | 62 | 4.89 | 32.7 | 707 |
| TiO₂/graphene (0.5 wt%) | | 12.78 | 0.82 | 62 | 6.49 | | |
| TiO₂ | N719 | 10.99 | 0.68 | 71.3 | 5.3 | 34 | 774 |
| TiO₂/graphene (0.2 wt%) | | 13.93 | 0.70 | 73.4 | 7.1 | | |
| TiO₂ | N719 | 13.2 | 0.691 | 52.4 | 4.78 | 60.1 | 775 |
| TiO₂/RGO (1.6 wt%) | | 18.39 | 0.682 | 61.2 | 7.68 | | |
| TiO₂ | N719 | 10.75 | 0.686 | 56.6 | 4.2 | 31.0 | 776 |
| TiO₂/RGO (0.75 wt%) | | 12.16 | 0.668 | 67.7 | 5.5 | | |
| TiO₂ | N719 | 6.18 | 0.606 | 71 | 2.67 | 110.5 | 741 |
| TiO₂/RGO (0.25 wt%) | | 13.04 | 0.645 | 67 | 5.62 | | |
| TiO₂ | N719 | 12.5 | 0.669 | 66.0 | 5.52 | 17.6 | 777 |
| TiO₂/graphene (0.5 wt%) | | 13.7 | 0.685 | 69.2 | 6.49 | | |
| TiO₂ | N719 | 16.13 | 0.62 | 65.3 | 6.57 | | 778 |
| N-Doped TiO₂ | | 16.71 | 0.74 | 61.6 | 7.64 | 16.3 | |
| N-Doped TiO₂/GO (0.1 wt%) | | 19.65 | 0.74 | 64.70 | 9.32 | 41.9 | |
| TiO₂ | N719 | 10.30 | 0.64 | 73 | 4.81 | 29.3 | 779 |
| TiO₂/GQDs | | 11.72 | 0.68 | 78 | 6.22 | | |
| TiO₂ | N719 | 12.59 | 0.704 | 65.07 | 5.77 | | 780 |
| TiO₂/RGO (0.6 wt%) | | 14.52 | 0.697 | 68.22 | 6.91 | 19.8 | |
| TiO₂/graphene sheets (0.6 wt%) | | 17.31 | 0.690 | 69.04 | 8.24 | 42.8 | |
| Pristine TiO₂ | N719 | 11.1 | 0.693 | 67.7 | 5.21 | 45.1 | 781 |
| TiO₂/graphene (0.03 wt%) | | 16.5 | 0.703 | 65.2 | 7.56 | | |
| TiO₂ | N719 | 10.7 | 0.75 | 76.82 | 6.13 | 11.7 | 782 |
| TiO₂ RGO (3.12 wt%) | | 12.9 | 0.76 | 69.20 | 6.85 | | |
| TiO₂ nanotubes (TT) | N719 | 9.19 | 0.71 | 61.3 | 4.00 | 33.3 | 783 |
| TT/RGO (2 wt%) | | 10.7 | 0.78 | 63.9 | 5.33 | | |
| TiO₂ | N719 | 15.5 | 0.71 | 0.68 | 7.51 | 38.9 | 769 |
| TiO₂/P1-graphene | | 19.8 | 0.74 | 0.71 | 10.43 | | |
| TiO₂ | N719 | 17.46 | 0.75 | 0.66 | 8.69 | | 764 |
| TiO₂/GO | | 16.70 | 0.74 | 0.68 | 8.55 | −1.6 | |
| TiO₂/NiO/NGE | | 19.04 | 0.76 | 0.67 | 9.75 | 12.2 | |
| TiO₂ | N719 | 11.51 | 0.72 | 0.66 | 5.52 | 37.1 | 784 |
| TiO₂/RGO | | 16.75 | 0.74 | 0.65 | 7.57 | | |
| ZnO | N719 | 3.0 | 0.45 | — | 3.7 | 76.2 | 785 |
| ZnO/B-doped GQDs | | 7.5 | 0.43 | — | 2.1 | | |
| TiO₂ | N719 | 9.30 | 0.68 | 0.48 | 3.0 | 173.3 | 786 |
| TiO₂/graphene | | 27.49 | 0.67 | 0.45 | 8.2 | | |
| TiO₂ nanofibers | N719 | 14.0 | 0.73 | 0.72 | 7.3 | 21.9 | 787 |
| TiO₂ nanofibers/graphene | | 18.0 | 0.73 | 0.68 | 8.9 | | |
| TiO₂ | N719 | 13.3 | 0.82 | 0.58 | 6.32 | 36.4 | 788 |
| TiO₂/RGO | | 20.6 | 0.79 | 0.53 | 8.62 | | |
| TiO₂ | Mimosa pudica | | | | 0.059 | 69.5 | 789 |
| TiO₂/RGO | | 16.085 | 0.248 | — | 0.1 | | |
| SnO₂:TiO₂ | N719 | 7.76 | 0.67 | 0.56 | 2.91 | 15.8 | 790 |
| Graphene-doped SnO₂:TiO₂ | | 9.03 | 0.65 | 0.58 | 3.37 | | |
| MoS₂ | N719 | 9.32 | 0.67 | 0.52 | 3.36 | 165 | 791 |
| MoS₂/graphene nanocomposite | | 15.82 | 0.82 | 0.71 | 8.92 | | |

Noble metals, (e.g., Pt, Au, and Ag) have been largely used as CE materials, with Pt representing the most popular one.[793] However, noble metals are expensive and their corrosion in DSSC liquid electrolytes is a critical shortcoming that hinders the commercialization of DSSC technology. Therefore, research activities have been focused on the development of metal-free

CEs for low-cost DSSCs. In this framework, carbon-based materials (e.g., CNTs,[794–796] GNPs,[713,797–799] functionalized graphene sheets (FGSs),[712] hybrid structures, GNRs/RGO/ CNT,[700,800] graphene/SWNT composite,[801] and N-doped GNPs (NGNPs)[802,803]) and scalable solution-based processes for electrode fabrication methods (e.g., spin coating,[804] flexographic printing[798] electrophoretic deposition (EpD)[805]) have been developed and exploited with the aim to replace traditional, expensive Pt-based CEs. Graphene-based materials have been used as the CE in the DSSC ambience since 2008[806] and some investigations have shown that combinations of two carbon materials (e.g., CNTs/ graphene[807] or porous carbon/CNTs[808]) can enhance the electrochemical activity of CEs. In particular, CE based on micrometer-thick graphene films can show $R_{CT}$ approaching to those of Pt-based CEs, which can even be inferior to 1 $\Omega$ cm² for the reduction of $I_3^-$.[809–811] For example, Kaniyoor and Ramaprabhu used thermally exfoliated graphene (TEGr) as a novel electrocatalyst material for $I_3^-$ reduction.[812] Their TEGr-based CE has shown a $R_{CT}$ of ~11.7 $\Omega$ cm² that, although higher than that of their Pt-based CE (~6.5 $\Omega$ cm²), was lower than that reported for graphite-based CE.[813] Counter electrodes based on thin-film GRMs have been demonstrated to be extremely effective for DSSCs using mediators different than traditional $I^-/I_3^-$, particularly the Co(bpy)$_3^{2+/3+}$ redox couple.[697] For example, thermally reduced graphene oxide (TRGO)-based CEs allow flexible DSSCs to be realized with an $\eta$ value of ~5%, comparable to that obtained using platinized FTO (5.5%), suggesting that the functional groups and defects of graphene can play an important role in catalysis.[712] More recently, a printable graphene-based ink obtained by the LPE of graphite has been spray-coated onto a TCO substrate to replace Pt in a large area ($\geq$90 cm²) semitransparent ($T_r$ = 44%) CE.[405] A large-area DSSC module (43.2 cm² active area) using the constructed CE achieved an $\eta$ value of 3.5%, $V_{OC}$ of 711 mV, $J_{SC}$ of 14.8 mA cm⁻², and FF of 34.7%.[405] Yen et al.[814] prepared a composite dispersion of graphene/metal NPs via the H$_2$O/(CH$_2$OH)$_2$ synthesis method.[815] Then, they fabricated a graphene/Pt NP-based CE on the FTO substrate.[813] The Pt incorporation improved the graphene reduction degree, and the composite materials synergistically enhanced the electrocatalytic properties of the CE.[813] The DSSC based on the as-produced CE afforded an $\eta$ value of 6.35% ($J_{SC}$ of ~12 mA cm⁻², $V_{OC}$ of ~0.8 V, and FF of ~0.7), which is ~20% higher compared to the reference DSSCs.[813] Ju and co-workers developed heteroatom-doped graphene nanomaterials for DSSC CEs.[801] In detail, they used N-doped GNPs, sprayed onto FTO/glass substrates through electrospray coating, as the CEs in a DSSC using a Co(bpy)$_3^{3+/2+}$ redox couple and JK225 organic dye sensitizer containing bis-dimethylfluorenyl amino group as the electron donor and cyanoacrylic acid as the electron acceptor and bridged by an indeno[1,2-b]thiophene unit.[816] N-doping induced a structural deformation in the hexagonal lattice of the graphene layer of the GNPs via local strains, as well as produced additional electron/ion pair with electrocatalytic activity.[817,818] Therefore, the DSSCs based on N-GNP CE reached an $\eta$ value of 9.05%, which outperformed that obtained using Pt-based CE (8.43%).[801] By means of N-GNP CE, the measured $R_{CT}$

(1.73 $\Omega$ cm²) was significantly lower than that obtained with Pt-based CE (3.15 $\Omega$ cm²).[801]

Functionalized graphene sheets (FGSs), containing lattice defects and oxygen functional groups (e.g., OH, C=O, and epoxides), were thermally treated at a temperature of 1000– 1500 °C to heal the lattice defects and tune the C/O ratio and used as the CEs.[712] FGS-based inks were directly cast on a nonconductive plastic substrate without the need of a conductive substrate.[712] The corresponding DSSCs based on FGS/surfactant/ polymer network produced via the thermlysis process displayed an $\eta$ value of 4.99%, which was close to that of Pt-based DSSCs (5.48%).[712] Electrochemical impedance spectroscopy (EIS) and CV measurements suggested that $R_{CT}$ increases with decreasing O-containing functional groups, lowering the PV performance.[712] Therefore, CE with C/O ratio of <7 was not sufficiently conductive for use as a CE material.[712]

In addition, it was proved that a low C/O ratio gives rise to a coarse film structure because of increased agglomeration.[712] All these results proved that the optimization of functionalization/ morphology is crucial to control the $R_{CT}$, realizing low-cost and flexible CEs. By applying graphene-based CEs (namely, GNP-based CES) to the most advanced DSSC configurations, Kakiuge et al. reached the current record $\eta$ value for DSSCs, i.e., 14.7%.[830] Noteworthily, the previous record $\eta$ of 13.0% was also achieved using graphene-based CEs.[831]

Table 7 summarizes the PV parameters of selected DSSC configurations using solution-processed graphene-based CEs.

**5.2.2 Two-dimensional TMDs.** Among the GRMs, TMDs such as MoS$_2$, MoSe$_2$, WS$_2$, and WSe$_2$ have been the most studied 2D materials to be used as CEs for low-cost Pt-free DSSCs.[832,833] Wu et al. proposed MoS$_2$ and WS$_2$ as the CE materials in $I_3^-/I^-$, $T_2/T^-$, and $S_2/S^-$-based DSSCs.[834] CV measurements revealed that both MoS$_2$- and WS$_2$-based CEs hold a catalytic activity, for the redox couple regeneration, comparable to that of the Pt-based CE.[833] Consequently, the DSSCs using MoS$_2$- and WS$_2$-based CEs have shown an $\eta$ value of 7.59% and 7.73%, respectively, which were comparable to that of the reference DSSC using a Pt-based CE.[833] EIS analysis estimated $R_{CT}$ values of 0.5 and 0.3 $\Omega$ for MoS$_2$- and WS$_2$-based CE, respectively, together with large electrode capacitances (134 and 198 μF, respectively) (i.e., large SSA).[833] These values outperformed those of Pt-based CE ($R_{CT}$ of 3 $\Omega$ and capacitance of 2.1 μF).[833]

Freitas et al. prepared MoS$_2$ through a hydrothermal route to be used as the CE material using $I^-/I_3^-$ as the redox couple.[835] MoS$_2$-based DSSCs reached an $\eta$ value of 2.9%, while DSSCs based on Pt CE has shown $\eta$ = 5.2%.[834] Although the $\eta$ value of MoS$_2$-based DSSCs was lower than that of Pt-based DSSCs, the possibility to dry MoS$_2$-based CE at a temperature of 120 °C was considered to be an advantage for the manufacturing of low-cost Pt-free DSSCs.[834] Next, Al-Mamun et al. grew ultrathin MoS$_2$-nanostructured films onto the FTO substrate through a facile one-pot hydrothermal method.[836] It was demonstrated that the temperature of the hydrothermal reaction and the molar ratio of reaction precursors have a relevant effect on the structure of the resulting MoS$_2$.[835] An ultrathin MoS$_2$ film was obtained through a hydrothermal process with a reaction solution comprising $NH_2CSNH_2$ and $(NH_4)_6Mo_7O_{24}\cdot4H_2O$ with a molar ratio of $28:1$ at 150 °C for 24 h.[835] After calcination at 400 °C in Ar, the resulting $MoS_2$ films were used as CEs for DSSCs, reaching an $\eta$ value of 7.41%.[835] This value was superior to that measured for Pt-based DSSCs (7.13%).[835] Meanwhile, homogeneous CNTs-$MoS_2$-C with ultrathin, uniform lamellar structure of $MoS_2$ was synthesized by Liu et al. via wet impregnation and calcination method.[837] This work indicated that the addition of a nonionic surfactant (PEG400) promotes the dispersion of $(NH_4)_2MoS_4$ onto the surface of CNTs, inhibiting the formation of independent particles of $MoS_2$. The DSSCs using CNTs-$MoS_2$-C composite-based CE achieved an $\eta$ value of 7.23%, which was higher than that of Pt-based DSSCs (6.19%). Yue et al. synthetized flower-like structure complexes of $MoS_2$/SWCNTs with glucose and PEDOT:PSS-assisted in situ hydrothermal route.[838]

**Table 7** Summary of the PV performance of DSSCs using graphene-based CEs

| Materials | Device structure | Cell performance | | | | | Ref. |
| | | $R_{CT}$ ($\Omega$ cm$^2$) | $J_{SC}$ (mA cm$^2$) | $V_{OC}$ (V) | FF (−) | $\eta$ (%) | |
|---|---|---|---|---|---|---|---|
| Graphene | FTO/TiO$_2$/N719/I$^-$/I$_3^-$/graphene | 38.0 | 14.3 | 0.54 | 0.653 | 5.69 | 819 |
| TiN/graphene | FTO/TiO$_2$/dye/I$^-$/I$_3^-$/TiN-graphene | 5.67 | 12.34 | 0.73 | 0.643 | 5.78 | 802 |
| Pt/graphene | FTO/TiO$_2$/N719/I$^-$/I$_3^-$/Pt-graphene | 0.67 | 12.06 | 0.79 | 0.67 | 6.35 | 813 |
| GNPs | FTO/TiO$_2$/Y123/Co$^{2+/3+}$/GNPs | 0.70 | 12.70 | 1.03 | 0.70 | 9.30 | 796 |
| Pt-RGO | FTO/TiO$_2$/dye/I$^-$/I$_3^-$//Pt-RGO | — | 14.10 | 0.72 | 0.67 | 6.77 | 820 |
| GNSs/AC | FTO/TiO$_2$/N719/I$^-$/I$_3^-$/GNSs-AC | 0.5 ÷ 0.8 | 13.30 | 0.76 | 0.738 | 7.50 | 821 |
| NGNPs | FTO/TiO$_2$/dye/Co(III/II)/NGNPs | 1.73 | 13.83 | 0.883 | 0.742 | 9.05 | 801 |
| GNTs/graphene-Rib | FTO/TiO$_2$/N719/I$^-$/I$_3^-$/GNT/graphene-Rib | — | 16.73 | 0.730 | 0.670 | 8.23 | 806 |
| GQD-PPy | FTO/TiO$_2$/N719/I$^-$/I$_3^-$/GQD-doped PPy | — | 14.36 | 0.723 | 0.580 | 5.27 | 822 |
| Au/GNP | FTO/TiO$_2$/ADEKA-1/LEG4/Co$^{2+/3+}$/Au/GNP | — | 19.55 | 0.995 | 0.776 | 14.7 | 829 |
| NGNPs | FTO/TiO$_2$/YD2-o-C8/Co$^{2+/3+}$/NGNPs | 0.45 | 13.33 | 0.870 | 0.720 | 8.30 | 823 |
| aGNP | FTO/TiO$_2$/N719/I$^-$/I$_3^-$/aGNP | 2.68 | 22.54 | 0.73 | 0.47 | 7.7 | 824 |
| Pt/GONF | FTO/TiO$_2$/Y123/Cu +/Cu + +/Pt/GONF | 1.1 | 14.01 | 1.02 | 0.665 | 9.5 | 825 |
| PEDOT/RGO | FTO/TiO$_2$/N719/I$^-$/I$_3^-$/PEDOT-RGO | 18.17 | 15.82 | 0.73 | 0.67 | 7.79 | 826 |
| 3D graphene networks/RGO | FTO/TiO$_2$:RGO/N719/I$^-$/I$_3^-$/3D graphene networks/RGO | 9.61 | 21.0 | 0.75 | 0.661 | 9.79 | 827 |
| RGO QDs | FTO/TiO$_2$/N719/I$^-$/I$_3^-$/RGO QDs | 64.8 | 6.1 | 0.784 | 0.52 | 2.5 | 828 |
| GO | FTO/TiO$_2$/N719/I$^-$/I$_3^-$/GO | 52.26 | 11.83 | 0.66 | 0.715 | 5.6 | 829 |
| CVD graphene/FLG | FTO/TiO$_2$/Y123/Co$^{2+/3+}$/FLG/CVD graphene | — | 11.23 | 0.958 | 0.47 | 5.09 | 406 |
| GNPs | FTO/TiO$_2$/SM315Co(bpy)$_3$]$^{2+/3+}$/GNPs | — | 18.1 | 0.91 | 0.78 | 13.0 | 830 |

Dye-sensitized solar cells based on $MoS_2$/SWCNTs as the CE exhibited an $\eta$ value of 8.14%, superior to that of Pt-based DSSCs (7.78%).[837] Kim et al. used atomically thin 2D $MoS_2$ nanoflakes, produced by a simple intercalation/exfoliation process, for fabricating transparent CEs via spin coating of the $MoS_2$ dispersion followed by thermal treatment.[839] The authors found that DSSCs based on $MoS_2$ thermally treated at 100 °C exhibited an $\eta$ value of 7.35%, which was comparable to that of the reference one, i.e., Pt-based DSSC ($\eta$ = 7.53%).[838,840] Solution-processed mesoporous $WO_3$ films with 3D, rough, and high-curvature surfaces followed by a rapid sulfurization process to prepare an edge-oriented $WS_2$ thin film was presented (Fig. 15). The maximized active edge sites on the high-curvature surface and electron transfer via continuous $WS_2$ building blocks enhanced the catalytic activity toward the $I_3^-$ reduction reaction in $WS_2$-based CEs.[839] This feature allowed $WS_2$-based DSSCs to reach an $\eta$ value of 8.85%, i.e., superior to that of the Pt-based

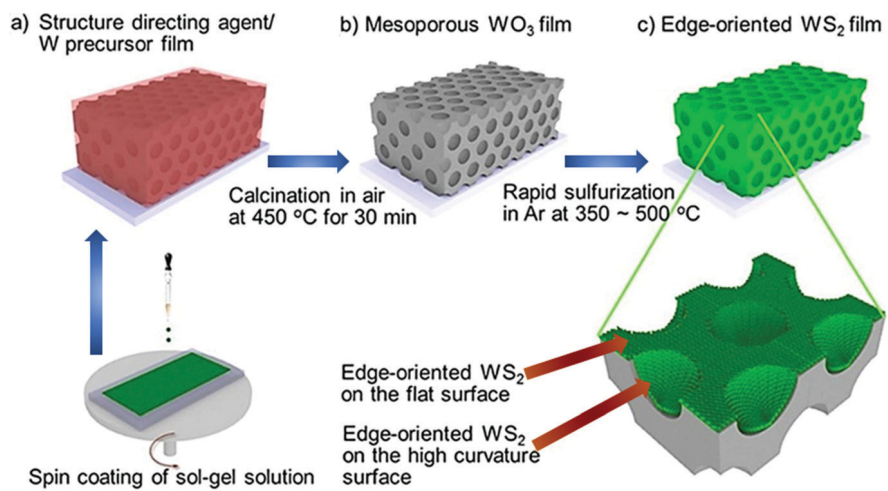

a) Structure directing agent/ W precursor film

b) Mesoporous $WO_3$ film

c) Edge-oriented $WS_2$ film

Calcination in air at 450 °C for 30 min

Rapid sulfurization in Ar at 350 ~ 500 °C

Spin coating of sol-gel solution

Edge-oriented $WS_2$ on the flat surface

Edge-oriented $WS_2$ on the high curvature surface reference (7.20%).[839] More recently, Vikraman *et al.* proposed a synthesis route to fabricate $MoS_2/FTO$ CE *via* a simple chemical bath (pH ≈ 10) deposition process by means of thiourea ($CH_4N_2S$, 0.5 M) as the sulfur source and ammonium-heptamolybdate-tetrahydrate (($NH_4)_6Mo_7O_{24}\cdot4H_2O$, 0.01–0.03 M), followed by annealing (450 °C for 60′) in a S environment to obtain crystalline $MoS_2$.[841] Under these conditions, the following reactions have been suggested: (1) $CH_4N_2S + 2H_2O \rightarrow CO_2\uparrow + 2NH_3\uparrow + H_2S\uparrow$; (2) $(NH_4)_6Mo_7O_{24}\cdot4H_2O + 28H_2S + 8NH_3 \rightarrow 7(NH_4)_2MoS_4 + 28H_2O$; (3) $(NH_4)_2MoS_4 + 2N_2H_4 \rightarrow MoS_2 + N_2\uparrow + 2(NH_4)_2S$.[840]

**Fig. 15** Schematic of the preparation of solution-processed mesoporous $WO_3$ thin film and its conversion to edge-oriented $WS_2$ thin film. Adapted from ref. 839.

Raman spectroscopy, XPS, and scanning electron microscopy (SEM) analyses evidenced the presence of tri- and tetra-layered $MoS_2$ and agglomeration effects, depending on the molybdate concentration and deposition time, respectively.[840] Mo precursor concentration of 0.01 M and bath temperature and deposition time of 90 °C and 5′, respectively, led to the optimal morphology of $MoS_2$ layers with spherical-shaped grains and absence of agglomeration effects.[840] In addition, agglomerations of uniform spherical grains provide large SSA with numerous edge sites, thereby promoting the electrocatalytic activity.[840] In this context, a deposition time of 30 min resulted in the optimal performance of DSSCs using the $MoS_2/$ FTO CE.[840] Under longer deposition times (~45 min), $MoS_2$ changed from the layered to bulk structure, showing a phase transformation from $MoS_2$ to $Mo_2S_3$.[840] These effects lowered the catalytic activity of the CE due to the presence of an insufficient number of sulfur active sites in the material structure.[840] In addition, $MoS_2/FTO$ CE has shown an electrocatalytic activity toward $I_3^-$ reduction, corresponding to $R_{CT}$ (~8.3 Ω $\square^{-1}$) comparable to that of Pt/FTO (~7.2 Ω $\square^{-1}$).[840] The PV performance of optimized $MoS_2/FTO$-based DSSCs ($J_{SC}$ = 15.92 mA cm$^{-2}$, $V_{OC}$ = 0.73 V, FF = 0.61, and η = 7.14%) almost reached that of the reference one based on Pt/FTO CE ($J_{SC}$ = 17.84 mA cm$^{-2}$, $V_{OC}$ = 0.71 V, FF = 0.69, and η = 8.73%).[840] Fig. 16 compares the *J–V* curves of the $MoS_2/FTO$ and Pt/FTO CE-based DSSCs under the same illumination conditions.[840] The anodic and cathodic branches of the Tafel polarization curves (Fig. 16, inset) indicate that the catalytic activity of $MoS_2/FTO$ and Pt/FTO CEs is comparable.[840] In fact, larger the slope in the anodic/cathodic branch, higher are the exchange current densities on the electrode, which then shows higher electrocatalytic activity and lower $R_{CT}$ at the electrolyte/CE interface (see eqn (3.5)).[842]

Recently, $MoSe_2/WS_2$ heterostructures on FTO have been proposed as the DSSC anode, providing a simple route to optimize interfacial transport for enhancing the electrocatalytic properties of DSSC anodes. By optimizing the thickness of the $WS_2$ layer, a high η value of 9.92% was reached, proving the potential of combining TMDs in advanced functional structured anodes.[843]

Table 8 summarizes the PV performance obtained in DSSCs using solution-processed TMD-based CEs.

### 5.2.3 Two-dimensional material-based hybrids.
The hybridization of different materials is, in principle, an effective method to produce advanced CE composites with enhanced synergistic electrocatalytic activity in comparison to the single counterparts. In fact, Wen *et al.* developed a metal-nitride/ graphene nanohybrid (*i.e.*, TiN-decorated N-doped graphene) to be used as a CE material for DSSCs.[802] The latter exhibited higher η (5.78%) than that of the Pt-based reference (5.03%), demonstrating the potential role of these hybrid structures to replace Pt-based CEs, with the added value of cost reduction and easy cell fabrication.[802] Tjoa and co-workers developed a low-temperature route to synthesize hybrid GO/Pt NP composites by light-assisted spontaneous co-reduction of GO and chloroplatinic acid.[819] The hybrid composites were used as CE materials in DSSCs, achieving an η value of 6.77%, which was higher than that of Pt-based references (6.29%).[819] In addition, the hybrid materials were compatible with flexible plastic substrates, yielding flexible DSSCs with an η value of 4.05%.[819] Lin *et al.* produced hybrid $MoS_2/GNS$ through EpD onto a FTO substrate to be used as the CE material in DSSCs.[844] The resulting DSSCs achieved an η value of 5.81% and low $R_{CT}$ (2.34 Ω cm$^2$), which was the result of a synergic effect derived by the combination of the single material components.[843] In fact, DSSCs using only GNS or $MoS_2$ as the CE exhibited poor η (4.68% and 4.15%, respectively) and higher $R_{CT}$ (6.24 and 3.65 Ω cm$^2$, respectively), testifying the occurrence of synergistic catalytic effects in the hybrids.[843] $MoS_2/$graphene hybrid as the CE for DSSCs have also been reported by Yue *et al.* showing performance comparable to that of Pt CE.[845] The hybrid electrodes were more efficient than those based on $MoS_2$, with $MoS_2/$ graphene-based DSSCs reaching an η value of 5.98%, which is similar to that of the Pt-based reference (6.23%).[844]

Liu *et al.* also reported $MoS_2/$graphene hybrid as the CE material.[864] The hybrids were synthesized by mixing GO nanosheets with ammonium tetrathiomolybdate and converting the solid intermediate into $MoS_2/RGO$ hybrid by flowing $H_2$ at 650 °C.[863] The DSSCs using $MoS_2/RGO$ hybrid CE have shown excellent electrocatalytic activity toward $I_3^-$ reduction, together with optimal electrochemical stability.[863] CV and EIS measurements evidenced the superior electrocatalytic activity and lower $R_{CT}$ (0.57 Ω cm$^2$) of the $MoS_2/RGO$-based CE compared to the CEs based on RGO, $MoS_2$, and sputtered Pt.[863] The DSSCs assembled with MoS₂/RGO CEs have shown PV characteristics ($\eta$ = 6.0%, $J_{SC}$ = 12.51 mA cm⁻², $V_{OC}$ = 0.73 V, and FF = 0.66) comparable to those of Pt-based DSSCs ($J_{SC}$ = 13.42 mA cm⁻², $V_{OC}$ = 0.72 V, FF = 0.66, and $\eta$ = 6.38%).[863] Li et al. prepared RGO-NiS₂ NP hybrids to develop CEs with excellent electrocatalytic activity toward I₃⁻ reduction.[865] The fabricated DSSCs using the NiS₂/RGO-based CE exhibited an $\eta$ value of ∼8.55%, which was higher than that of the DSSC using either NiS₂-based (∼7%) or RGO-based (∼3.14) CE, as well as Pt-based CE (∼8.15%).[864]

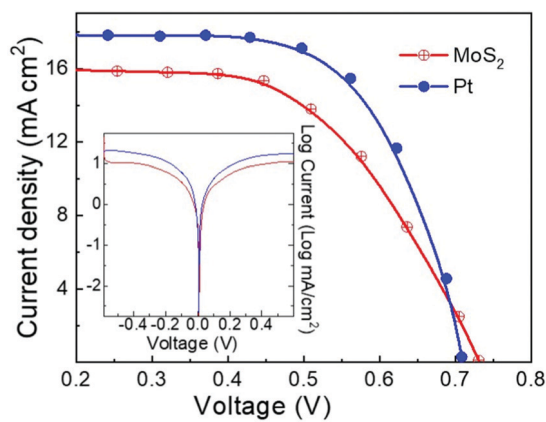

**Fig. 16** *J–V* curves of DSSCs with $MoS_2/FTO$ and Pt/FTO CEs. Inset shows the Tafel polarization curves of symmetrical cells with $MoS_2/FTO$ and Pt/FTO CEs. Adapted from ref. 840.

**Table 8**  Summary of the PV performance of DSSCs using TMD-based CEs

| Material | Device structure | Cell performance | | | | | Ref. |
|---|---|---|---|---|---|---|---|
| | | $R_{CT}$ ($\Omega$ cm²) | $J_{SC}$ (mA cm⁻²) | $V_{OC}$ (V) | FF (—) | $\eta$ (%) | |
| MoS₂ catalyst | FTO/TiO₂/N719/I⁻/I₃⁻/MoS₂/FTO | 0.50 | 13.84 | 0.76 | 0.73 | 7.59 | 833 |
| WS₂ catalyst | FTO/TiO₂/N719/I⁻/I₃⁻/WS₂/FTO | 0.30 | 14.13 | 0.78 | 0.70 | 7.73 | 833 |
| Carbon-coated WS₂ | FTO/TiO₂/N719/I⁻/I₃⁻/WS₂/FTO | 5.0 | 13.10 | 0.67 | 0.62 | 5.5 | 846 |
| NbSe₂ nanorods | FTO/TiO₂/N719/I⁻/I₃⁻/NbSe₂/FTO | 6.21 | 13.94 | 0.76 | 0.64 | 6.78 | 846 |
| NbSe₂ nanosheets | FTO/TiO₂/N719/I⁻/I₃⁻/NbSe₂/FTO | 2.59 | 15.04 | 0.77 | 0.63 | 7.34 | 847 |
| NbSe₂ nanosheets | FTO/TiO₂/N719/I⁻/I₃⁻/NbSe₂/FTO | — | 16.85 | 0.74 | 0.62 | 7.73 | 848 |
| NbSe₂ nanorods | FTO/TiO₂/N719/I⁻/I₃⁻/NbSe/FTO | — | 14.85 | 0.74 | 0.46 | 5.05 | 847 |
| NbSe₂ nanoparticles | FTO/TiO₂/N719/I⁻/I₃⁻/NbSe₂/FTO | — | 14.93 | 0.75 | 0.55 | 6.27 | 847 |
| Porous MoS₂ sheets | FTO/TiO₂/N719/I⁻/I₃⁻/MoS₂/FTO | 1.73 | 15.40 | 0.763 | 0.53 | 6.35 | 849 |
| MoS₂ nanoparticles | FTO/TiO₂/N719/I⁻/I₃⁻/MoS₂/FTO | 93.0 | 14.72 | 0.745 | 0.490 | 5.41 | 850 |
| MoS₂ | FTO/TiO₂/N719/I⁻/I₃⁻/MoS₂/FTO | 18.50 | 18.46 | 0.680 | 0.580 | 7.01 | 851 |
| MoS₂ nanosheets | FTO/TiO₂/N719/I⁻/I₃⁻/MoS₂/FTO | 0.619 | 18.37 | 0.698 | 0.578 | 7.41 | 835 |
| MoSe₂ | FTO/TiO₂/N719/I⁻/I₃⁻/MoSe₂/FTO | 229.8 | 13.0 | 0.67 | 0.68 | 5.90 | 852 |
| NiS₂ | FTO/TiO₂/N719/I⁻/I₃⁻/NiS₂ | 50.40 | 14.70 | 0.72 | 0.52 | 5.50 | 851 |
| NiSe₂ | FTO/TiO₂/N719/I⁻/I₃⁻/NiSe₂ | 45.00 | 14.30 | 0.75 | 0.68 | 7.30 | 851 |
| CoSe₂ | FTO/TiO₂/N719/I⁻/I₃⁻/CoSe₂ | 10.70 | 13.50 | 0.72 | 0.68 | 6.60 | 851 |
| MoSe₂ | FTO/TiO₂/N719/I⁻/I₃⁻/MoSe₂ | 2.43 | 14.11 | 0.73 | 0.65 | 6.70 | 853 |
| WSe₂ | FTO/TiO₂/N719/I⁻/I₃⁻/WSe₂ | 0.78 | 15.50 | 0.73 | 0.66 | 7.48 | 852 |
| TaSe₂ | FTO/TiO₂/N719/I⁻/I₃⁻/TaSe₂ | 1.89 | 15.81 | 0.73 | 0.64 | 7.32 | 852 |
| MoS₂ | FTO/TiO₂/N719/I⁻/I₃⁻/MoS₂/FTO | 30.98 | 16.90 | 0.727 | 0.517 | 6.35 | 854 |
| MoS₂ | FTO/TiO₂/N719/I⁻/I₃⁻/1 T-MoS₂/FTO | 19.0 | 8.76 | 0.730 | 0.520 | 7.08 | 855 |
| MoS₂ | FTO/TiO₂/N719/I⁻/I₃⁻/MoS₂/FTO | 15.29 | 14.94 | 0.718 | 0.67 | 7.19 | 856 |
| MoS₂ | FTO/TiO₂/N719/I⁻/I₃⁻/MoS₂/FTO | 12.9 | 16.96 | 0.74 | 0.66 | 8.28 | 857 |
| MoS₂ | FTO/TiO₂/N719/I⁻/I₃⁻/1T-MoS₂/FTO | 19.60 | 11.54 | 0.80 | 0.65 | 6.0 | 858 |
| TiS₂ | FTO/TiO₂/N719/I⁻/I₃⁻/TiS₂/FTO | 0.63 | 17.48 | 0.73 | 0.603 | 7.66 | 859 |
| MoS₂ | FTO/TiO₂/N719/I⁻/I₃⁻/MoS₂/graphite paper | 2.16 | 13.34 | 0.696 | 0.698 | 6.48 | 860 |
| MoS₂ | FTO/TiO₂/N719/I⁻/I₃⁻/MoS₂/FTO | 2.77 | 15.68 | 0.72 | 0.634 | 7.16 | 861 |
| MoS₂ | FTO/TiO₂/N719/I⁻/I₃⁻/MoS₂/FTO | 25.77 | 15.92 | 0.73 | 0.615 | 7.14 | 862 |
| MoS₂ | FTO/TiO₂/N719/I⁻/I₃⁻/MoS₂/FTO | 2.86 | 19.6 | 0.795 | 0.36 | 6.6 | 863 |
| MoSe₂/WS₂ | FTO/TiO₂/N719/I⁻/I₃⁻/MoSe₂/WS₂/FTO | 18.3 | 23.1 | 0.69 | 0.651 | 9.92 | 842 |

The DSSCs using CE based on NiO NPs/RGO hybrids have shown an $\eta$ value of ∼7.42%, which was comparable to that of a conventional Pt-based DSSC (∼8.18%).[866]

The NiO NPs/RGO hybrids exhibited lower $R_{ct}$ value (1.93 $\Omega$ cm²) compared to those based on NiO-based (44.39 $\Omega$ cm²) and GO-based (12.19 $\Omega$ cm²) CEs.[865] Experimental investigations indicated that the synergic effects of two different low-dimensional carbon materials, such as CNT/graphene nanoribbons (CNT/graph-nRib), can be used to further amplify the CE catalytic activity of the single nanomaterial.[806]

More recently, Zhai et al. obtained an $\eta$ value of ∼8.3% in porphyrin-sensitized DSSCs using N-doped GNP-based CE [Co-(bpy)₃]³⁺/²⁺ redox complexes.[822] The obtained result was the consequence of better electrocatalytic activity in comparison to that of the Pt-based CE, whose corresponding cell reached an $\eta$ value of 7.95%.[822] The performance increase was ascribed to a higher number of catalytic sites, due to the introduction of pyridinic and pyrrolic N into the carbon-conjugated lattice, compared to the GNPs.[822]

Shen et al. synthesized NiS/RGO-based CE with a material mass ratio varying from 0.2 up to 0.6 through a low-temperature hydrothermal method.[867] The NiS/RGO hybrid-based CE exhibited the best catalytic property for the NiS:RGO mass ratio of 0.4%, yielding a DSSC with an $\eta$ value of 8.26%, a value much higher than that of pristine RGO-based or NiS-based references (1.56% and 7.41%, respectively).[866] The obtained results demonstrated that a correct load of NiS hinders the agglomeration of RGO flakes, favoring the diffusion of electrolyte into the NiS/RGO network.[866]

Zhou et al. synthesized graphene-wrapped CuInS₂ hybrids to be used as the CE material.[868] The DSSCs based on CuInS₂/RGO as CE achieved an $\eta$ value of 6.4%, which was comparable to that of Pt-based CE (6.9%).[867] Huo et al. developed a sponge-like CoS/RGO-based CE with a low $R_{CT}$ value of 3.59 $\Omega$ cm².[869] The corresponding DSSCs have shown an $\eta$ value of 9.39%.[868]

Li et al. synthesized a nanostructured architecture of 3D bismuth sulfide (Bi₂S₃) microspheres on RGO through a solvo-thermal route.[877] The as-produced architecture was used as the CE for DSSCs.[876] By combining the characteristics of direct-bandgap Bi₂S₃ semiconductor (low bandgap of 1.7 eV and absorption coefficient of the order of 10⁴–10⁵ cm⁻¹) with out-standing carrier transfer properties of RGO, an $\eta$ value ∼3 times greater than that of DSSCs with 3D Bi₂S₃ without RGO (1.9%) was achieved.

Chen et al. prepared GQDs through a chemical oxidation approach to dope conductive polymers (polypyrrole (PPy)) on FTO glass as the CE for DSSCs.[821] The as-prepared DSSCs displayed an $\eta$ value (i.e., 5.27%) lower than that of Pt-based reference devices ($\eta \approx 6.02\%$), but ~20% higher compared to that of DSSCs based on PPY as the CE (4.46%).[821]

**Table 9** Summary of the PV performance of DSSCs using 2D material-based hybrid CEs

| Material | Device structure | $R_{CT}$ ($\Omega$ cm$^2$) | $J_{SC}$ (mA cm$^{-2}$) | $V_{OC}$ (V) | FF ($-$) | $\eta$ (%) | Ref. |
|---|---|---|---|---|---|---|---|
| MoS$_2$/graphene | FTO/TiO$_2$/N719/I$^-$/I$_3^-$/MoS$_2$/graphene | 24.42 | 12.41 | 0.71 | 0.68 | 5.98 | 844 |
| GNS | FTO/TiO$_2$/dye/I$^-$/I$_3^-$/GNS | 6.24 | 11.99 | 0.754 | 0.30 | 2.68 | 843 |
| MoS$_2$/GNS | FTO/TiO$_2$/dye/I$^-$/I$_3^-$/MoS$_2$/GNS | 2.34 | 12.79 | 0.773 | 0.59 | 5.81 | |
| RGO/NiS$_2$ | FTO/TiO$_2$/N719/I$^-$/I$_3^-$/NiS$_2$/RGO | 2.90 | 16.55 | 0.749 | 0.69 | 8.55 | 864 |
| Bi$_2$S$_3$/RGO | FTO/TiO$_2$/N719/I$^-$/I$_3^-$/Bi$_2$S$_3$/RGO | 9.2 | 12.20 | 0.75 | 0.60 | 5.5 | 876 |
| RGO/NiO | FTO/TiO$_2$/N719/I$^-$/I$_3^-$/NiO/RGO | 1.93 | 15.57 | 0.763 | 0.624 | 7.42 | 865 |
| CuInS$_2$/graphene | FTO/TiO$_2$/N719/I$^-$/I$_3^-$/CuInS$_2$/graphene | 2.30 | 14.20 | 0.743 | 60.7 | 6.40 | 867 |
| CoS/RGO | FTO/TiO$_2$/N719/I$^-$/I$_3^-$/CoS/RGO | 3.59 | 19.42 | 0.764 | 0.633 | 9.39 | 868 |
| NiS/RGO | FTO/TiO$_2$/N719/I$^-$/I$_3^-$/NiS/RGO | 7.06 | 17.05 | 0.778 | 0.623 | 8.26 | 866 |
| MoS$_2$/graphite | FTO/TiO$_2$/dye/I$^-$/I$_3^-$//MoS$_2$/graphite | 8.05 | 15.64 | 0.685 | 0.67 | 7.18 | 870 |
| TiS$_2$/graphene | FTO/TiO$_2$/N719/I$^-$/I$_3^-$/TiS$_2$-graphene | 0.63 | 17.76 | 0.72 | 0.685 | 8.80 | 858 |
| WO$_x$/carbon | FTO/TiO$_2$/N719dye/I$^-$/I$_3^-$/WO$_x$/carbon/FTO | 12.70 | 14.30 | 0.705 | 0.591 | 6.00 | 871 |
| WO$_x$@WS$_2$/carbon | FTO/TiO$_2$/N719dye/I$^-$/I$_3^-$/WO$_x$@WS$_2$/carbon/FTO | 0.88 | 15.48 | 0.720 | 0.695 | 7.71 | 870 |
| MoS$_2$/graphene | FTO/TiO$_2$/N719dye/I$^-$/I$_3^-$/MoS$_2$/graphene/FTO | 34.02 | 20.5 | 0.800 | 0.42 | 8.1 | 862 |
| MoS$_2$/SnS$_2$ | FTO/TiO$_2$/N719/I$^-$/I$_3^-$/MoS$_2$/SnS$_2$/FTO | 0.32 | 15.99 | 0.73 | 0.65 | 7.6 | 872 |
| NiSe/GNS$_{0.50}$ | FTO/TiO$_2$/N719/I$^-$/I$_3^-$/NiSe$_2$-graphene (1:0.50) | 1.92 | 16.73 | 0.75 | 0.68 | 8.62 | 877 |
| WSe$_2$/MoS$_2$ | FTO/TiO$_2$/N719/I$^-$/I$_3^-$/WSe$_2$/MoS$_2$/FTO | 44.42 | 16.89 | 0.69 | 0.724 | 8.44 | 873 |
| Polyaniline/graphene | FTO/TiO$_2$/N719/I$^-$/I$_3^-$//polyaniline/graphene | 20.1 | 15.5 | 0.787 | 0.62 | 7.45 | 874 |
| NiO/NiS/graphene | FTO/TiO$_2$/N719/I$^-$/I$_3^-$/NiO/NiS/graphene | 23.2 | 4.86 | 0.76 | 0.56 | 2.10 | 875 |
| Co–Mo–S anchored on nitrogen-doped graphene (NG) | FTO/TiO$_2$/N719/I$^-$/I$_3^-$/Co–Mo–S anchored on nitrogen-doped graphene (NG) | — | 7.22 | 0.49 | 0.44 | 1.18 | 876 |

Murugadoss et al. grew NiSe NPs on GNSs with different mass ratios to obtain NiSe/GNS$_x$ ($x = 0.25$ to 1.00) nanohybrids by an in situ hydrothermal process.[878] This method takes advantages of the high SSA of GNSs to homogeneously immobilize NiSe NPs on top of them acting as the catalytic sites.[877] The nanohybrid with a mass ratio of 1:0.50 (NiSe/GNS$_{0.50}$) exhibited the highest electrocatalytic activity and electrolyte diffusion among the different hybrid compositions.[877] Thus, the DSSC with NiSe/GNS$_{0.50}$ CE exhibited an $\eta$ value of 8.62%, which is higher compared to a standard Pt-based DSSC ($\eta = 7.68\%$).[877] The NiSe/GNS$_{0.50}$ CE exhibited a superior PV performance compared to both Pt and pristine NiSe CEs.[877] Compared to the hybrid NiSe/GNS$_{0.50}$ CE, the lower performance of the NiSe electrode, when integrated in a DSSC ($\eta = 7.18\%$), was attributed to the aggregation of NiSe NPs and the poor connections between the NPs, which decrease the number of electrocatalytic active sites as well as the electrical conductivity of the CE.[877] The optimal performance of NiSe/GNS$_{0.50}$ CE is determined by the interfacial electron transfer pathways of GNSs and the exceptional catalytic activity of NiSe toward I$_3^-$ reduction at the CE/electrolyte interface.[877]

Table 9 summarizes the PV performance of DSSCs using 2D material-based hybrid CEs.

#### 5.2.4 Bifacial DSSCs using 2D material-based transparent CEs.
Research on transparent and cost-effective CEs is a persistent objective in the development of bifacial DSSCs. Transparent or semitransparent electrodes can be used for building-integrated photovoltaics (BIPVs) to make use of light from the interior of the building as well as the outside.[879–881] So far, efficient solution-processed transparent CEs have been reported using both PEDOT and binary-alloy transition metal chalcogenides (M–Se; M = Ni, Co, Fe, Cu, and Ru). For example, bifacial DSSCs with CEs composed by a metal selenide achieved front $\eta$ values of 8.3%, 7.8%, 6.4%, 7.6%, and 9.2% for Co$_{0.85}$Se, Ni$_{0.85}$Se, Cu$_{0.50}$Se, FeSe, and Ru$_{0.33}$Se, respectively, which were even higher than that of a cell with an electrode based on standard Pt (6.1%).[882] The corresponding rear $\eta$ values were 4.6%, 4.3%, 4.2%, 5.0%, 5.9%, and 3.5% for devices based on Co$_{0.85}$Se, Ni$_{0.85}$Se, Cu$_{0.50}$Se, FeSe, Ru$_{0.33}$Se, and Pt, respectively. Besides, PEDOT can provide an optimal conductivity–transparency trade-off, while being suitable for large-scale and cost-effective production.[883] However, the catalytic performance of PEDOT CE alone is inferior compared to those of Pt CE or other metal-based CEs. Therefore, PEDOT has been combined with other catalytic materials to improve the performance of bifacial DSSCs. In this context, Chen et al. developed transparent CEs of PEDOT/N-doped graphene (NG) for bifacial DSSCs, achieving an $\eta$ value of 8.3%, which is higher than that of DSSCs using Pt CE (8.17%).[884] In a recent study, Xia et al. prepared layered CoSe$_2$ nanorods with lengths of 70–500 nm and widths of 20–60 nm by a one-step hydrothermal reaction and used them as the CE material.[885] The resulting bifacial cells using a PVDF quasi-solid-state electrolyte revealed that the $\eta$ values for the front and rear irradiation of CoSe$_2$ CE-based devices reached 8.0% and 4.2%, which are higher than those achieved with Pt CE (7.4% and 4.0%, respectively). More recently, Xu et al. reported the preparation of transparent organic–inorganic hybrid composite films of MoS$_2$/PEDOT to take full advantage of the conductivity and electrocatalytic properties of the two components.[886] Researchers synthesized MoS$_2$ by the hydrothermal method. MoS$_2$ dispersions were spin coated to form an MoS$_2$ layer and subsequently prepared PEDOT films deposited on top of the MoS$_2$ film by the electrochemical polymerization to form composite CEs.[885] DSSCs using the MoS$_2$/PEDOT composite CE exhibited an $\eta$ value of 7% under front illumination and 4.82% under rear illumination.[885] Compared with other DSSCs based on PEDOT CE or Pt CE, DSSCs using MoS$_2$/PEDOT composite CE improved the front $\eta$ by 10.6% and

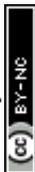

6.4%, respectively.[885] As discussed above, the DSSCs assembled with transparent CEs based on TMDs and graphene-like materials exhibit $\eta$ comparable or even higher than those of semitransparent Pt-based DSSCs. Such performances achieved with Pt-free CEs are promising to reduce the power-to-weight ratio and total cost of bifacial DSSCs, paving the way toward 2D material-based DSSCs for smart windows, power generators, and panel screens.

### 5.3 Summary and outlook

The global DSSC market size was valued at USD 90.5 million in 2019 and is expected to grow at a compound annual growth rate (CAGR) of 12.4% from 2020 to 2027.[887] The global DSSC market is segregated into portable charging, BIPV, embedded electronics, outdoor advertising, and automotive. In this scenario, BIPV represents a strategic sector, in which DSSCs reached an $\eta$ value higher than 25% (up to a record-high value of 32% at 1000 lux).[888-890] Regarding outdoor applications, the advent of PSCs has outclassed the use of DSSCs as single-junction devices, the $\eta$ value of DSSC being significantly inferior to that of PSCs.[897] However, DSSCs still represent an interesting PV technology to make cost-effective and efficient tandem systems, including those based on PSCs as sub-cells. Overall, the advancements achieved in DSSCs using 2D materials might be applied to the most promising configuration for convenient applications, as discussed above. In fact, the current record-high $\eta$ value of DSSCs was reached using GNPs as the CE[829] (previous record of 13% was also achieved using graphene-based CEs),[830] unequi $V_{OC}$ ally proving the potential of 2D materials for both improving the PV performance and decreasing the cost of traditional PV based on Pt CEs.[891,892] Prospectively, lifecycle assessment can provide a useful methodological framework to calculate the eco-profiles of solution-processed 2D material-based DSSCs with a future-oriented perspective. Importantly, DSSCs are the third-generation hybrid–organic technology that reached the highest maturity in terms of manufacturing, reporting several pilot-line semi-industrial production lines.[893,894] Overall, the synergistic use of 2D materials, novel dyes, and advanced anode structures are expected to play a major role in the further optimization of DSSC technology, both in the indoor and outdoor PV market.

## 6. PSCs

During the last few years, the rapid emergence of perovskite PV technology in the global energy scenario has strongly attracted efforts from the scientific community.[895-897]

Perovskite semiconductors have been used in SCs since the pioneering work of Kojima et al. who proposed the perovskite as a sensitizer in a DSSC, achieving an $\eta$ value of 3.8%.[894] Thereafter, much progress has been achieved in the last 12 years, achieving certified $\eta$ exceeding 25%,[898] which makes PSCs among the most promising class of devices in the broad context of 3rd-generation PV technologies.[897]

The term perovskite refers to a broad class of crystals sharing the crystalline structure of calcium titanate ($CaTiO_3$).[900]

Generally, the crystallographic structure is indicated with the chemical formula $ABX_3$, where A is a small organic or inorganic cation that occupies a cube-octahedral[901] site, B is a metal cation in the center, and X is a halogen anion in an octahedral site.[899]

The perovskite structure widely used in SCs is an organic–inorganic hybrid based on an organometallic halide material. In particular, A can be methylammonium $CH_3NH_3^+$ (MA), formamidinium (FA), or Cs; B is commonly the lead ion $Pb^{2+}$ (even though ions of Sn, Ge, Sb, Bi are also used), and X is a halide, typically iodide or bromide ($I^-$ or $Br^-$).[902] Thus, in the perovskite structure, A, located in a cage, is surrounded by four $BX_6$ octahedra, and can partially move inside the cage[901] (Fig. 17). Organic–inorganic halide perovskites offer attractive prospects in developing high-performance SCs,[903-905] low processing costs and facile fabrication processes.[906,907] In fact, due to their particular crystallographic structure and the peculiar choice of components,[908] organic–inorganic halide perovskites exhibit outstanding and unique optoelectronic properties, including large and broad absorption spectrum,[909] immediate charge generation within the bulk material (due to very low exciton binding energy),[910] optimal ambipolar charge transport with a long charge diffusion length ($\sim 100$ nm for $CH_3NH_3PbI_3$ and $\sim 1$ µm for $CH_3NH_3PbI_{3-x}Cl_x$),[911] efficient charge injection into the CTLs, and low trap density, which points them as "defect-tolerant" materials.[906,912-914] In addition, chemical modifications of the constituents enable the tuning of the material properties, for example, the $E_g$, which spans over a range wider than 1 eV. The architecture of a PSC consists of a perovskite active layer sandwiched between an HTL and an ETL. More specifically, two PSC structures have been mainly developed, i.e., mesoscopic[894,906] and planar[905] structures, and both of them can be found in n–i–p (direct) or p–i–n (inverted) configurations (Fig. 18).[915] The mesoscopic architecture is so called because a mesoporous oxide layer is used as the ETL in which the perovskite is infiltrated.

In detail, mesoscopic PSCs are composed of different layers.[894,906,916] The first one is a TCE (e.g., FTO or ITO),

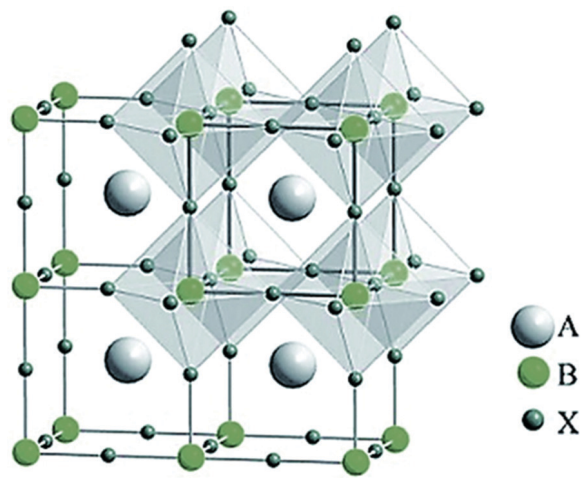

Fig. 17 The basic perovskite structure ($ABX_3$). Adapted from ref. 899.

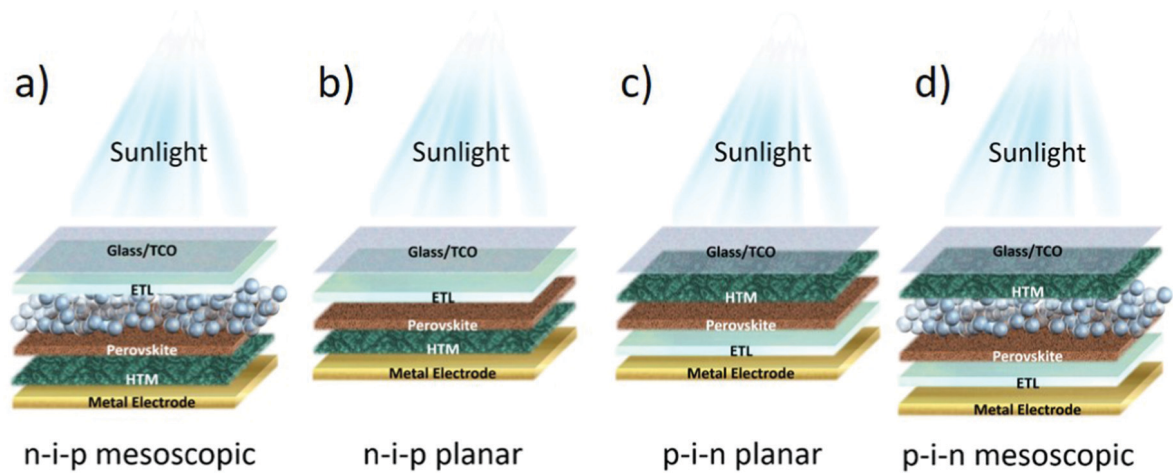

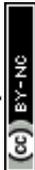

**Fig. 18** (a) Mesoscopic n–i–p, (b) planar n–i–p, and (c) planar p–i–n and mesoscopic p–i–n PSC structures.

deposited on the bottom of the glass surface and acting as the front transparent electrode. Subsequently, a compact ETL is deposited onto the TCE. The compact ETL is typically made of $TiO_x$[906] or $ZnO$[917,918] (thickness ranging from 10 to 400 nm, depending on the material), which can be deposited using different techniques (e.g., spray pyrolysis,[919] sol–gel,[920] DC magnetron sputtering,[921,922] electrodeposition,[923,924] electron-beam evaporation,[925] and pulsed laser deposition[926,927]). Then, a layer of mesoporous oxide is obtained by depositing a paste typically containing colloidal NPs (preferably $TiO_2$,[894,906] $ZrO_2$,[928,929] $Al_2O_3$,[930] and $SiO_2$[931,932]) using various techniques (e.g., sol–gel,[933,934] doctor blade,[935] spin coating,[936] spray coating,[937] electrospray deposition,[938] slot-die coating,[939,940] inkjet printing,[941] and pulsed laser deposition[733,942,943]). For glassy substrates, the device is subsequently heated (e.g., at 450–550 °C for 30 min for the calcination of $TiO_2$) in order to evaporate the organic binder and to create an electromechanical connection between the NPs.[902,944] For heat-sensitive plastic substrates, the mesoporous layer is alternatively formed by means of low-temperature processes, such as microwave sintering,[945] laser sintering,[946,947] intense pulsed light sintering,[948] and NIR sintering.[949,950] Thereafter, a perovskite layer (generally formed by one or more organic cations, such as methylammonium (MA) and formamidinium (FA), and one or more inorganic compounds, such as $PbI_2Cl$, $Pb_2$, or $PbI_2Br$) is infiltrated in the mesoporous electrodes by various techniques,[951] e.g., spin coating,[902,952–955] spray casting,[956–959] physical vapor deposition,[960,961] thermal evaporation,[962] dip coating,[963] slot-die coating,[964] roller coating,[965] and bar coating.[966] Subsequently, the resulting film is often crystallized through a heating step at 70–100 °C for 10–30 min, depending on the perovskite composition and deposition method.[950] On the so-formed photoelectrode, an HTL (e.g., P3HT,[967–969] spiro-OMeTAD,[970,971] PTAA,[972] PEDOT,[973] copper thiocyanate (CuSCN),[974,975] and triphenylamine-based molecules[976,977]) is deposited, typically via solution-based techniques, e.g., spin coating, spray casting, inkjet printing, and slot-die coating. Lastly, the mesoscopic PSC is completed by depositing a metal contact as the CE (e.g., Au, Ag, and Al). Alternatively, a planar architecture can be used as a simpler PSC structure.[905] The main difference between a planar direct configuration (n–i–p structure) and mesoscopic devices is the use of a single compact n-type metal oxide layer (e.g., $TiO_2$,[978] $ZnO$,[979–981] and $SnO_2$[982–986]) in the former rather than a combination of both compact and mesostructured scaffolds.[987,988] Differently, in an inverted planar configuration (p–i–n structure),[905] a p-type material (e.g., $NiO_x$,[989–991] PEDOT: PSS,[992,993] CuSCN,[994] $CuI$,[995] Cu oxides,[996,997] and $V_2O_x$[998]) is used as the bottom HTL, while PCBM is the typically used ETL.[999–1001] As a matter of fact, several potential material/structure combinations can be designed to implement new PSC structures. In fact, on one hand, the correct choice of materials is crucial for determining both optical and electronic properties (e.g., bandgap and commensurate absorption spectra, $\mu$, charge diffusion lengths, etc.). On the other hand, material arrangement in the different architectures plays a crucial role in the overall PSC performance.

Despite the undoubted interest in perovskites, mainly owing to their low cost and efficiency compared to the technologies currently in the market, PSCs currently suffer from low stability, a big challenge for their market uptake.[1002] Perovskite instability has been correlated to both intrinsic[1003–1006] and extrinsic factors,[1005] mainly associated with moisture[1007,1008] and oxygen exposure,[1006,1007] as well as UV radiation,[1009,1010] high-temperature exposure,[1011] and electrical biases.[1012,1013] In detail, under prolonged working conditions, PSCs often exhibit structural degradation of perovskites (as well as other component layers). For example, ion migration from the perovskite to metal and vice versa, as well as perovskite decomposition through the volatilization of perovskite species can cause the rapid failure of PSCs.[1005,1014]

Although some extrinsic degradation factors, such as oxygen, moisture, and UV exposure can be avoided by means of new strategies to encapsulate the assembled device,[1015–1017] the intrinsic structural instability can be addressed by designing more stable perovskites[1018–1021] or by adding crosslinking additives,[1022–1025] ionic liquids additives,[1026–1028] dopants,[1029–1031] and interlayers[1032–1035] that can stabilize the crystal structure.

Recently, pressure-tight polymer (polyisobutylene)/glass stack encapsulation has been shown to inhibit intake moisture, while preventing the outgassing of decomposition products of the perovskite.[1016] Consequently, the decomposition reaction for a prototypical Cs-containing triple-cation perovskite, namely, $Cs_{0.05}FA_{0.8}MA_{0.15}Pb(I_{0.85}Br_{0.15})_3$, were suppressed, permitting the devices to pass the International Electrotechnical Commission (IEC) 61215:2016 Damp Heat and Humidity Freeze tests.[1016]

In this context, (1) the exploitation of solution-processed GRMs to engineer the device interface[1036–1040] and (2) the introduction of 2D perovskites as active materials[111,1016,1041–1043] have recently been revealed as two main routes for the realization of highly efficient and durable PSCs. In the first case, the possibility to chemically or thermally modify GRMs allows their oxygen vacancy/defect/functional group concentration to be controlled to attain the desired optoelectronic properties.[1044] Recent advances in the production and processing of GRMs allowed their effective use in the PSC structures,[596,1037] for example, as ETLs[1045–1047] or HTLs[1048,1049] as well as interlayers at the perovskite/CTL interface.[1050,1051] In particular, GRM-induced improvement in charge transport and collection at the electrodes allows the $\eta$ value to be significantly enhanced.[1052,1053] Moreover, the use of GRMs in PSCs results in a remarkable increase in the device's stability under several stress conditions, by preventing interfacial perovskite degradation.[1039,1054–1059] In the second case, 2D perovskites have demonstrated superior moisture stability compared to the 3D counterparts, offering new approaches to stabilize PSCs.[111,1060] Owing to their versatile structure, 2D perovskites enable the *ad hoc* tuning of their optoelectronic properties through compositional engineering.[111,1061] Finally, recent achievements have demonstrated the possibility to combine 3D and 2D perovskites to simultaneously boost the device efficiency and stability, opening the route toward advanced mixed-dimensional PSCs.[111,1062]

In the subsequent sections, the use of GRMs and 2D perovskite layers into CTLs and perovskite, as well as their utilization as buffer layers and front/back electrodes, will be discussed.

## 6.1 ETLs

As shown in Fig. 18, the configurations of a PSC require a perovskite light-absorbing layer interposed between a wide-bandgap ETL and HTL, which assist charge carrier transport to the negative and positive electrodes, respectively.[1063,1064] Meanwhile, the CTLs must prevent charge transport toward the undesired electrode.[1062,1063]

In mesoscopic PSCs, the generated electrons have to travel through the mesoporous ETL, which plays a crucial role in terms of efficiency and stability of the whole device.[1065,1066] A large variety of materials has been used as ETLs to efficiently extract photoexcited electrons in the perovskite layer.[1067] It is worth pointing out that the mesoporous ETL can be conductive (*e.g.*, TiO$_2$, ZnO, and NiO) or insulating (*e.g.*, Al$_2$O$_3$, ZrO$_2$, and SiO$_2$). In the first case, the mesoporous layer acts as an ETL, while in the second one, it provides a scaffold functionality.

Due to the capacity to prevent leakage currents, and hence to prevent cell shunting, TiO$_2$ is the most established ETL for PSCs.[696,1068] Mesoscopic PSCs using anatase mesoporous TiO$_2$

(mTiO$_2$) with NPs in the 10–30 nm size range have shown $\eta$ often exceeding 21%.[902,1069] Nanomaterials based on ZnO could also be promising semiconducting metal oxides alternative to TiO$_2$, owing to their wide bandgap ($\sim$3.37 eV), large exciton binding energy ($\sim$60 meV), high $\mu_e$, unique photoelectric properties, optical transparency, electric conductivity, and piezoelectric properties.[1070–1072] To overcome the disruptive effect of moisture on the perovskite structure and to prolong the lifetime of the devices, ZnO QDs have also been exploited as an alternative scaffold to other ZnO nanostructures, owing to their tunable bandgap and chemical inertness.[1073] Lastly, ZnO offers the benefit to be easily processed at low temperatures, opening the possibility to develop efficient low-temperature fabricated mesoscopic PSCs.[1074]

In order to increase the $V_{OC}$ of PSCs, mesoporous Al$_2$O$_3$, ZrO$_2$, and SiO$_2$ have been investigated as a scaffold to alternative mTiO$_2$.[1075–1078] For example, a mesoporous Al$_2$O$_3$ (mAl$_2$O$_3$) layer was used to transport photoexcited electrons throughout the perovskite layer, allowing $\eta$ > 12% to be reached.[1079]

Ternary oxides, such as SrTiO$_3$,[1080,1081] Zn$_2$SnO$_4$,[1082] and BaSnO$_3$[1083] have also been used to obtain better performing devices in terms of $\eta$. For example, SrTiO$_3$ exhibits the same perovskite structure with $\mu_e$ and CB higher than those of TiO$_2$.[1084,1085] Thus, SrTiO$_3$-based PSCs showing $V_{OC}$ higher than 1 V have been reported.[1080]

In order to boost the electrical performance of PSCs, it is necessary to precisely control the charge carriers' pathway along the entire device,[1086,1087] by avoiding losses due to photon thermalization or carrier-trapping processes and improving faster electron injection. In fact, charge carrier injection at the perovskite/ETL interface is strongly limited by interfacial charge recombination when the interfaces are not properly engineered, such as in presence of nonoptimized energy-level alignment.[1088] Likewise, poor charge transport in the CTLs[1089,1090] severely limits charge collection toward the electrodes, reducing $I_{SC}$ and FF[1091–1093] and therefore the $\eta$ value.

Several strategies have been reported in the literature to enhance the charge transport and extraction properties at the interface between the perovskite/ETL. These strategies include TiO$_2$ doping[1094,1095] and the use of different TiO$_2$ nanostructures,[1096–1098] as well as the modification of interfacial energy-level alignment through the incorporation of appropriate buffer layers.[1036,1099]

In this framework, numerous solutions have been proposed to facilitate electron extraction and to increase the conductivity by taking advantage of GRMs.[1036,1100–1102]

The use of graphene with nanostructured ZnO or TiO$_2$ in PSCs results in a higher photocurrent density and consequently better device performance compared to the reference device without graphene.[1099] Wang *et al.* developed low-temperature-processed nanocomposites of pristine graphene nanoflakes and anatase–TiO$_2$ NPs to be used as the ETL in mesoscopic PSCs (Fig. 19a).[1100] The observed decrease in the series resistance, as well as a decrease in charge recombination, determined an improvement in the device performance, achieving an $\eta$ value of 15.6% (Fig. 19b).[1100] The use of graphene nanoflakes, with an appropriate $\phi_W$ (*i.e.*, 4.4 eV), reduced the energy barrier between

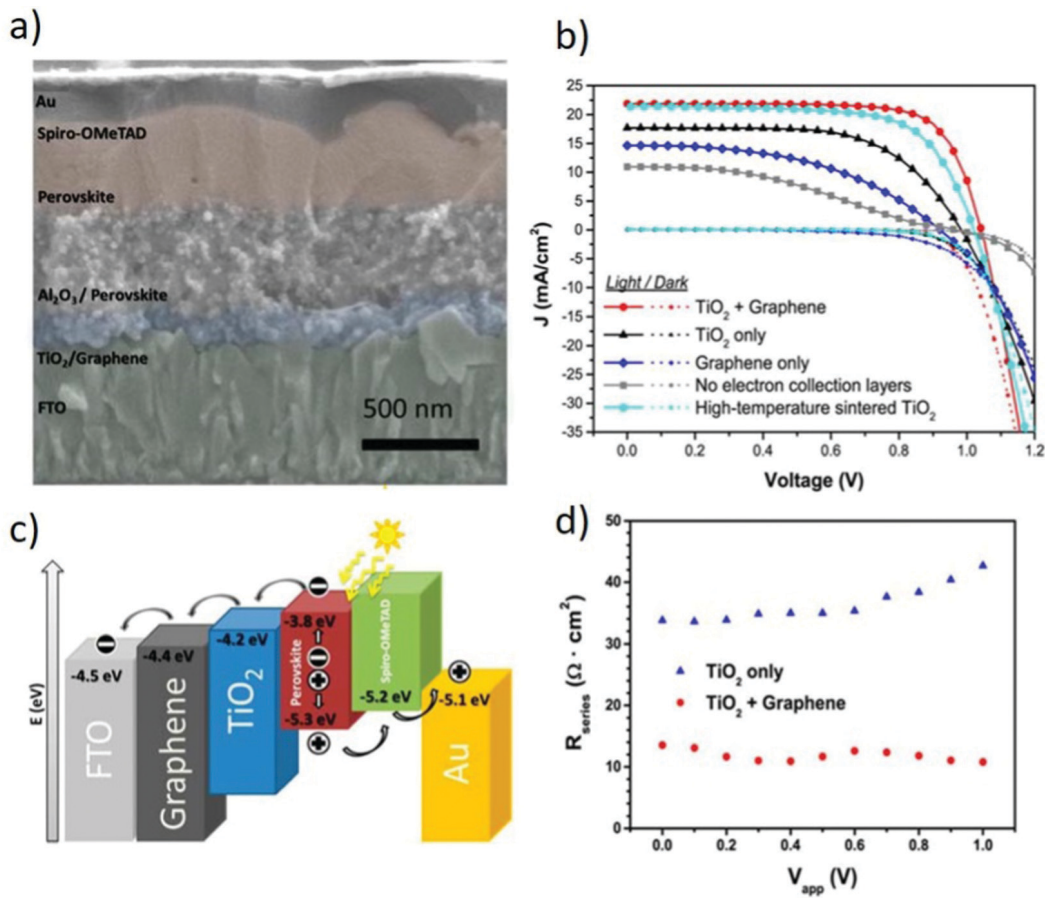

Fig. 19 (a) Cross-sectional SEM image of a PSC based on a graphene–TiO$_2$ composite as the ETL. (b) J–V curves measured for PSCs with different ETLs under solar irradiation and in the dark (c). Energy levels of a PSC based on a graphene–TiO$_2$ composite as the ETL. (d) Series resistances of PSCs using pristine TiO$_2$ and graphene–TiO$_2$ composite as the ETLs. Reprinted with permission from ref. 1100, Copyright 2013, American Chemical Society.

the LUMO of TiO$_2$ (4.2 eV) and $\Phi_w$ of FTO (4.5 eV), leading to better electron collection through the ETL (Fig. 19c).[1100] In addition, the $\mu$ value of graphene flakes increased the electrical conductivity of the graphene–TiO$_2$ ETL compared to bare TiO$_2$ (Fig. 19d).[1100] GRMs have also been used as the interlayer between perovskite and mTiO$_2$. Zhu et al. reported that the insertion of an ultrathin layer of GQDs between the perovskite and mTiO$_2$ impacts the PSC performance, increasing the $\eta$ value from 8.81% up to 10.15%.[1101] The insertion of a GQD interlayer causes strong quenching of perovskite photoluminescence at ~760 nm due to a reduced electron extraction time (90–106 ps) in the presence of GQDs.[1101] This means that the GQDs permit an efficient electron transfer from the photo-absorber to the acceptor, resulting in efficient electron extraction.[1101]

More recently, Tavakoli and co-workers developed a new and simple chemical process to synthesize a quasi-core–shell structure of ZnO NPs/RGO to be used as the ETL in PSCs.[1099] In fact, RGO passivates the ZnO NP surface, preventing degradation reactions at the perovskite/ETL interface caused by the presence of the hydroxide group.[1099] Furthermore, the ZnO/RGO ETL improves the electron transfer at the perovskite/ETL interface, increasing the EQE and photocurrent density.[1099] Thus, the use of RGO increased $\eta$ up to 15.2% on rigid PSCs using FTO-coated substrates, while flexible devices on ITO-coated PET achieved an $\eta$ value of 11.2%.[1099] In a subsequent work, Tavakoli et al. reported a very high performing PSC ($\eta$ = 17.2%) using a reduced-graphene scaffold (rGS) obtained through EpD.[1103] The authors fabricated a porous 3D scaffold of graphene with a large SSA that enabled a higher loading of perovskite materials.[1102] The addition of rGS improved the carrier transport of the PSC, yielding an $\eta$ value enhancement of ~27% compared to the conventional device.[1102] Besides, sealed rGS-based devices retained 80% of their initial $\eta$ for a time as long as a month under ambient conditions (~65% humidity).[1102] Ameen and co-workers used a graphene thin film as the barrier layer between O$_2$ plasma-treated ITO-PET and the ETL based on ZnO QDs.[1104] A subsequent atmospheric plasma jet (APjet) treatment of ZnO QD ETL improved the interfacial contacts, modifying the surface properties of the ITO-PET/Gr/ZnO QD structure.[1103] The use of a graphene interlayer and APjet treatment of ZnO QDs improved the carrier transport and collection efficiency.[1103] Moreover, modification with regard to the surface area, pore size, and porosity caused by the APjet treatment allowed perovskite infiltration to be optimized.[1103] Thus, the fabricated ITO-PET/Gr/ZnO QDs/CH$_3$NH$_3$PbI$_3$/spiro-OMeTAD/Ag flexible PSCs reached a high $\eta$ value of ~9.73%, along with a $J_{SC}$ value of ~16.8 mA cm$^{-2}$, $V_{OC}$ of ~0.935 V, and FF of ~0.62.[1103]

These performances outperformed those of the PSCs fabricated with ITO-PET/graphene and ZnO-QDs/graphene/ITO-PET structures.[1103] Graphene-based interface engineering through the incorporation of an additional buffer layer represents an effective strategy to improve the PV performance while overcoming oxygen- and moisture-induced instability. For example, Agresti et al. proposed a new, efficient PSC structure by including a GO-Li interlayer between TiO₂ and the perovskite.[1051] The main effect of the GO-Li interlayer is the enhancement—compared to the reference devices—of both $\eta$ and long-term stability.[1051] In particular, this work pointed out to improved charge extraction/injection at the negative photoelectrode when GO-Li was used as the ETL.[1051] In fact, the GO-Li interlayer favors the passivation of oxygen defects/vacancies in mTiO₂, eliminating reactive centers susceptible to moisture attack. Such an effect improved the stability of devices, which have shown an enlarged lifetime without encapsulation during aging tests. A similar interface strategy was used to further increase the $\eta$ and stability of PSCs using mTiO₂ doped with graphene flakes (mTiO₂ + G) and GO as an interlayer between the perovskite and HTL.[1036,1105] These PSCs achieved a remarkable $\eta$ value of 18.2% as a consequence of the improved charge-carrier injection/collection.[1036] In addition, the optimized PSCs improved their stability under several aging tests compared to the reference devices. In fact, when mTiO₂ + G was used, the PSCs retained more than 88% of their initial $\eta$ under prolonged 1 sun illumination at MPPT for a time as long as 16 h (Fig. 20a). Recently, time-of-flight secondary ion mass spectrometry (ToF-SIMS) 3D imaging and XPS depth profile analysis were used to evaluate the light-induced degradation of layers and interfaces both in mesoscopic PSCs with mTiO₂ + G and graphene-free PSCs (Fig. 20b).[1106] These results demonstrated that the incorporation of graphene within mTiO₂ improves the stability of PSCs by limiting the light-induced back-conversion of CH₃NH₃PbI₃ into PbIₓ and PbOₓ species.[1105] Therefore, the formation of iodine species was also reduced, impeding them to diffuse across the interface until modifying the Au electrode and mTiO₂ through the A–I and Ti–I bond formation.[1105] Even more recently, femtosecond transient absorption measurements proved that the incorporation of graphene can stabilize the PSCs owing to the potential exploitation of the contribution of hot carriers to the $\eta$ value of PSC (Fig. 20c).[1107] In particular, these results demonstrated that the insertion of graphene flakes into mTiO₂ leads to stable values of carrier temperature.[1106] In graphene-free PSCs aged over 1 week, the carrier temperature decreased from 1800 to 1300 K, while the graphene-based cell reported a reduction inferior to 200 K after the same aging time.[1106] The stability of carrier temperature was associated to the stability of the perovskite embedded in mTiO₂ + G. Overall, all these results involving mTiO₂ + G have opened the way for scalable large-area PSC production due to the use of GRMs in the form of dispersions and inks.[205] In fact, by means of graphene and GRMs, Agresti et al. realized large-area (50.6 cm²) perovskite-based solar modules (PSMs) with a remarkable $\eta$ value of 12.6%.[1108]

Recently, a similar approach was followed by Cho et al., which systematically investigated the role of RGO in PSCs by dispersing RGO into the mTiO₂ matrix to obtain highly efficient PSC ($\eta$ = 19.54%).[1109] Moreover, the role of RGO has been demonstrated to be crucial to improve the transport and injection of photoexcited electrons.[1108] Previously, several authors also used RGO as the dopant into a TiO₂ layer. Umeyama and co-workers doped both compact TiO₂ (cTiO₂) and mTiO₂ with RGO to increase the $\eta$ value from 6.6% to 9.3%.[1046] To maximize

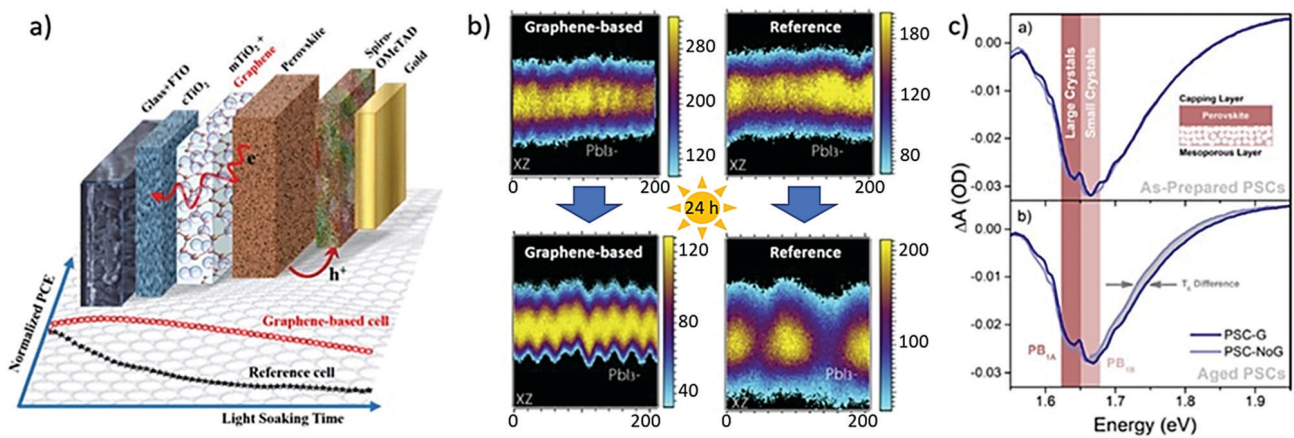

Fig. 20 (a) Graphene-based PSC energy-level alignment representation and $\eta$ stability trends under prolonged light-soaking condition (1 sun) for both standard and graphene-engineered devices.[1036] (b) ToF-SIMS 3D analysis showing the reconstructed XZ distribution of PbI₃⁻ (from the CH₃NH₃PbI₃ absorber) in the as-deposited (upper part) and 24 h light-aged (lower part) PSCs. The PbI₃ signal decay suggests the progressive decomposition of the perovskite absorber material, which was always more severe in the reference PSC structure. Adapted from ref. 1105, Copyright 2018, Elsevier. (c) Transient absorption spectra acquired at a pump–probe time delay of 0.75 ps for (a) as-prepared and (b) aged PSC with mTiO₂ + G (PSC-G) and graphene-free PSC (PSC-NoG). The photobleaching signal exhibits two peaks at 1.64 eV and 1.66 eV attributed to the absorption bleaching in large crystals of the capping layer and in small crystals of the mesoporous layer, respectively. Hot electron lifetime from the transient absorption measurements related to the degradation of small perovskite crystals wrapped in the mesoporous TiO₂ layer. When graphene is embedded into the mTiO₂ layer, the hot-carrier temperature is preserved over aging time by improving the device stability. Adapted from ref. 1106, Copyright 2019, American Chemical Society.

the charge transport properties in TiO$_2$ layers, the authors mixed TiO$_2$ with small quantities of GO, which was reduced to RGO during the subsequent calcination process.[1046]

Han *et al.* reported RGO/mTiO$_2$ nanocomposite ETL to reduce film resistivity and to increase the electron diffusion of pristine mTiO$_2$.[1044] Consequently, they achieved an $\eta$ value that was ~18% higher compared to the RGO-free reference PSC.[1044] Recently, GQDs have been similarly proposed for doping TiO$_2$.[1110] To improve the $I_{SC}$ value of SrTiO$_3$-based PSCs, Wang *et al.* successfully incorporated graphene in mesoporous SrTiO$_3$, reaching an $\eta$ value of 10%, which is 46.0% higher than that achieved by the reference device.[1080] Mali *et al.* proposed RGO-grafted porous zinc stannate (ZSO) scaffold-based PSCs, which achieved a $V_{OC}$ value of 1.046 V, $J_{SC}$ of 22.5 mA cm$^{-2}$, $\eta$ of 17.89%, and FF of 76%.[1111] The performance of the proposed PSCs was ascribed to the presence of RGO in the ZSO scaffold, where it served as a highway track for the photogenerated electrons, facilitating electron injection from the perovskite into the ZSO CB.[1110]

Recently, GRMs have been used to dope SnO$_2$ in planar PSCs. Zhu *et al.* proposed the incorporation of graphene into SnO$_2$ to improve the electron extraction efficiency, as well as to attenuate charge recombination at the ETL/perovskite interface.[1112] Consequently, PSCs based on graphene-doped SnO$_2$ ETL exhibited $\eta$ over 18% with negligible hysteresis.[1111] In addition, the use of graphene as the ETL dopant enhanced the stability of the device, which retained 90% of the initial $\eta$ after 300 h storage under the ambient condition with a relative humidity of 40 ± 5%.[1111]

Following a similar strategy, Zhao *et al.* incorporated naphthalene diimide–graphene into SnO$_2$ ETLs to increase the surface hydrophobicity and to generate van der Waals interaction between the surfactant and perovskite.[1113] These effects led to $\eta$ exceeding 20%.[1112] As the peculiar interface engineering of planar PSCs based on SnO$_2$ ETL, 2D g-C$_3$N$_4$ has been recently proposed as a heat-resisting n-type semiconductor to modify the interfaces of ETL/perovskite and perovskite/HTL, respectively.[1114] The g-C$_3$N$_4$ structure can passivate the surface trap states of the MAPbI$_3$ light absorber through the formation of Lewis adducts between N and the undercoordinated Pb, by reducing the grain boundaries between the perovskite crystal particles. The as-realized cells reached an $\eta$ value exceeding 19.6% with remarkable FF of over 80%.[1113] Moreover, new emerging 2D materials, including TiS$_2$ and SnS$_2$, were recently used as the ETL in the n–i–p planar architecture.[1044,1115,1116] For example, Huang and co-workers reported FTO/TiS$_2$/perovskite/spiro-OMeTAD/Au devices showing an $\eta$ value of 18.8% when the TiS$_2$-coated ITO film underwent UVO treatment.[1114] In fact, UVO can shift the TiS$_2$-coated ITO WF to 4.64 eV, thus speeding up electron collection.[1114] Moreover, UVO-treated TiS$_2$ ETL-based devices also exhibited excellent device stability, retaining 95.8% of their initial $\eta$ after 816 h of ambient storage (without any encapsulation).[1114] In addition, they maintained over 80% of their initial $\eta$ after exposure to a high humidity environment (45–60 RH) for 100 h.[1114] Lastly, highly efficient ($\eta$ = 21.73%) n–i–p planar PSCs were fabricated by Huang *et al.* in 2019, employing a double layer of SnO$_2$ and 2D TiS$_2$ as the ETL.[1115] Highly efficient ($\eta$ > 20%)

n–i–p planar PSCs were also recently demonstrated using SnS$_2$.[1044] Intermolecular Pb–S interactions between perovskite and SnS$_2$ were proposed to passivate the interfacial trap states.[1044] This effect can suppress charge recombinations and facilitate electron extraction, resulting in balanced charge transport at the ETL/perovskite and HTL/perovskite interfaces.[1044] Solution-processed BP quantum dots (BPQDs) with ambipolar conductivity were developed to be used as a dual-functional ESL material in plastic PSCs.[1117] BPQD-based ESL formed a cascade energy level for fast electron extraction and controlled the crystallization of the perovskite, thereby yielding compact high-quality (low-defect density) perovskite films with an ordered orientation.[1116]

The resulting plastic planar PSCs exhibited an $\eta$ value of 11.26%, owing to the efficient electron extraction and suppression of both radiative and trap-assisted nonradiative recombinations.[1116] More recently, phosphorene nanosheets, produced through vortex fluidic-mediated exfoliation under NIR pulsed laser irradiation, were also used as dopants for TiO$_2$ ETLs, resulting in low-temperature (100 °C) processed, planar n–i–p PSCs with a maximum $\eta$ value of 17.85%.[1122]

Tsikritzis *et al.* recently proposed a two-fold engineering approach for inverted PSCs, where ultrathin Bi$_2$Te$_3$ flakes were used (1) to dope the ETL and (2) to form a protective interlayer on top.[1123] This approach improved the electron extraction rate, increasing the overall $\eta$ by +6.6% compared to the reference cells. These effects were associated with an optimal alignment between the energy levels of the perovskite, cathode, and ETL. Furthermore, the interlayer of Bi$_2$Te$_3$ promoted efficient electron transport, while chemically protecting the underlying structure.[1122] By combining the two engineering approaches, the optimized PSCs reached an $\eta$ value as high as 19.46%, while retaining more than 80% of their initial $\eta$ value (after the burn-in phase) over 1100 h under continuous 1 sun illumination.[1122]

A complete replacement of the ETL with 2D materials was also presented for inverted planar PSCs by Castro *et al.* using functionalized GNRs.[1124] Compared to PC$_{61}$BM, the functionalized nanoribbons were hydrophobic and exhibited higher LUMO energy levels, thus providing superior $\eta$ and stability.[1123]

In addition, transition metal carbides, nitrides, and carbonitrides (*i.e.*, MXenes) have just started to be used for the design of high-performance ETLs.[1125,1126] Ti$_3$C$_2$ MXenes have been used as the dopant for SnO$_2$ ETLs to improve the $\eta$ value from 17.23% to 18.34%.[1124] The superior performance recorded for MXene-incorporated SnO$_2$-based PSCs was explained by both faster electron extraction and enhanced electrical conductivity compared to those exhibited in MXene-free ETLs.[1124] Very recently, Agresti *et al.* used Ti$_3$C$_2$ MXene-based ETLs to improve PSCs using perovskite absorbers modified with MXenes.[1125] The resulting cells exhibited a 26% increase in $\eta$ and hysteresis reduction compared with the reference cells without MXenes.[1125] Meanwhile, other less established 2D materials are also emerging as novel ETL material candidates. For example, Bi compounds, namely, Bi$_2$O$_2$Se nanoflakes, have recently been used as hydrophobic and smooth ETLs to improve the electron collection/transport while promoting the formation of large perovskite crystals, achieving $\eta$ value of up to 19.06%.[1120] Metallic group-5 TMDs, namely, $6R$-TaS$_2$ flakes, were exfoliated and incorporated as a buffer layer in inverted PSCs to simultaneously enhance their $\eta$, lifetime, and thermal stability.[1127] In detail, a thin buffer layer of $6R$-TaS$_2$ flakes on top of the ETL facilitated electron extraction, allowing the device to reach the maximum $\eta$ value of 18.45% (+12% *vs.* the reference cell).[1126] In addition, stability tests using ISOS-L2, ISOS-D1, ISOS-D1I, and ISOS-D2I protocols proved that the TaS$_2$ buffer layer retards the thermal degradation of PSCs, which retained more than 80% of their initial $\eta$ over 330 h under continuous 1 sun illumination at 65 °C.[1126]

Table 10 summarizes the main results achieved by PSCs using ETLs based on GRMs.

## 6.2 Perovskite layers

Beyond optoelectronic properties, the key factors influencing the performance of a perovskite absorber are the morphology and grain size.[1128–1130] Several works have highlighted the need to control the perovskite crystal morphology in order to obtain large grains that maximize charge photogeneration at the active layer.[1128,1129,1131–1133] In fact, on one hand, an accurate control of the crystallization process is an essential step to improve the perovskite film morphology for correct device operation. On the other hand, the perovskite grain interfaces play a crucial role to influence charge transport and recombination phenomena. In this context, the incorporation of graphene derivatives into a perovskite layer seems a practicable way to improve the quality of perovskite layer morphology.[1134] Hadadian *et al.* first reported the addition of N-doped reduced graphene oxide (N-RGO) into mixed organic–inorganic halide perovskites in order to increase the perovskite grain size.[1135] This effect was tentatively attributed to the slowing down of the crystallization process.[1135] Meanwhile, N-RGO decreased charge photogeneration owing to the surface passivation effect (Fig. 21a–e).[1134] Therefore, the presence of N-RGO in the perovskite layer improved $J_{SC}$ ($\sim 21$ mA cm$^{-2}$), $V_{OC}$ ($\sim 1.15$ V), and FF ($\sim 0.73\%$), increasing $\eta$ from 17.3% to 18.7%, compared to the reference PSC (Fig. 21f).[1134] Alternatively, GO was used as both HTL and additive in the perovskite absorber in an inverted PSC.[1136] The resulting PSC exhibited an $\eta$ value as high as 15%, which was attributed to the hole acceptor role of GO in the hybrid GO:perovskite composite.[1135] Moreover, the use of GQDs within the perovskite layer was reported as a promising strategy to passivate perovskite grain boundaries, improving the overall device performance.[1137] In fact, conductive GQDs were used to facilitate electron extraction and simultaneously passivate dangling bonds and eliminate electron traps at the perovskite grain boundaries.[1136] These effects enhanced the $\eta$ value up to 17.6%.[1136] Alternatively, 2D BP was proposed as an additive in the absorber precursor solution to obtain large ($>500$ nm) perovskite grains and to improve the $\eta$ value up to 20.65%.[1138]

The enhanced PV performance was attributed to the improved charge extraction and transport of MAPbI$_3$ perovskite in the presence of BP nanosheets.[1137] This approach was successfully applied to MAPbI$_3$-based n–i–p[1137] and p–i–n configurations,[1138] demonstrating the maximum $\eta$ value of 20.65% and 20.0%, respectively.

Moreover, the BP-doped n–i–p PSCs presented excellent photostability under prolonged light soaking, preserving 94%

of the initial $\eta$ after irradiation time of 1000 h,[1137] while p–i–n PSCs have shown encouraging thermal stability, maintaining over 80% of their initial $\eta$ value after aging for 100 h at 100 °C.[1139] As a further demonstration of the emergent role of BP in PSCs, X. Gong and co-workers demonstrated the use of BPQDs as an additive for inorganic CsPbI$_2$Br perovskite films.[1140] In that work, BPQDs were proposed as effective seed-like sites to modulate the nucleation and growth of CsPbI2Br perovskite crystals, affording device $\eta$ above 15%.[1139] Despite these results, the instability of few-layer phosphorene under ambient conditions[1141] still represents a major concern hampering its massive use in PSCs.

In terms of the intrinsic stability of perovskites, Ag NP-anchored reduced graphene oxide (Ag-RGO) was used as an additive in perovskite films to suppress ion migration by improving the thermal and light stability.[1142]

Recently, Guo *et al.* used MXenes as a perovskite additive in mesoscopic PSCs.[1143] In particular, the authors have shown that the termination groups of Ti$_3$C$_2$T$_x$ can retard the perovskite crystallization rate, thereby increasing the perovskite crystal size.[1142] After optimization, a 12% enhancement in $\eta$ compared to the reference PSCs was obtained with 0.03 wt% MXenes.[1142]

Agresti *et al.* used Ti$_3$C$_2$ MXenes as perovskite WF modifier to design n–i–p mesoscopic devices with $\eta$ exceeding 20%.[1125] Density functional theory calculations demonstrated that the formation of an interface dipole at the perovskite/Ti$_3$C$_2$T$_x$ interface strongly depends on T$_x$ terminations.[1125] For example, in the case of OH-terminated MXene, a larger interface dipole than O terminations was demonstrated.[1125] The overall reduction in perovskite WF upon MXene addition and the optimization of MXene-based ETL led to a 26% increase in $\eta$, together with hysteresis reduction, compared with the reference cells without MXenes.[1125] The possibility to vary the MXene WF on demand and control the band-energy alignments with other layers forming an electronic device represents a winning strategy to enlarge the design parameter space and improve the device performance.

The attempt from Hu *et al.* toward stabilizing the α-phase of Cs$_{0.1}$FA$_{0.9}$PbI$_3$ perovskite by using 2D phenyl ethyl ammonium lead iodide ((PEA)$_2$PbI$_4$) nanosheets as the additive deserves a separate discussion, see Section 6.5.[1144] Because of the 2D (PEA)$_2$PbI$_4$ nanosheets, the MA-free perovskite-based device reached a high $\eta$ value of 20.44% and retained 82% of its initial efficiency after 800 h of continuous white light (1 sun) illumination.

Table 11 summarizes the main results achieved by PSCs integrating GRMs as perovskite additives.

## 6.3 HTLs and back electrodes

The main role of HTLs is to extract positive charges from the perovskite layer, by minimizing charge recombination losses, and to efficiently transport them at the corresponding current collector.[1062,1063] Depending on the device's structure, HTLs have additional functions: in direct planar structures, HTLs also act as a perovskite protective layers against environmental factors (*e.g.*, moisture and oxygen) and can even contribute to heat dissipation,[1145,1146] improving the long-term stability of devices. In inverted planar structures, HTLs are often used as a scaffold layers for the growth of perovskites. Therefore, specific

**Table 10** Summary of the PV performance of PSCs using ETLs based on GRMs[ab]

| Material | Usage | Device structure | Cell performance | | | | Ref. |
|---|---|---|---|---|---|---|---|
| | | | $J_{SC}$ [mA cm⁻²] | $V_{OC}$ [V] | FF | η [%] | |
| Liquid phase exfoliated (LPE) graphene nanoflakes | Dopant for ETL | FTO/cTiO₂ + G/mesoporous Al₂O₃/MAPbI₍₃₋ₓ₎Clₓ/spiro/Au | 21.9 | 1.04 | 0.73 | 15.5 | 1100 |
| RGO quantum dots (QD) | ETL (quasi-core–shell structure of ZnO-RGO QDs) | FTO/ZnO-RGO QD/MAPbI₃/spiro-OMeTAD/Au | 21.7 | 1.03 | 0.68 | 15.2 | 1099 |
| Graphene quantum dots (GQDs) | Interlayer at ETL/perovskite interface | FTO/cTiO₂/mTiO₂/GQDs/MAPbI₃/spiro-OMeTAD/Au | 17.06 | 0.94 | 0.64 | 10.15 | 1101 |
| Reduced-graphene scaffold (rGS) | Scaffold for perovskite growth | FTO/cTiO₂/rGS/MAPbI₃/spiro-OMeTAD/Au | 22.8 | 1.05 | 0.72 | 17.2 | 1102 |
| Graphene | Interlayer at front electrode/ETL interface | ITO-PET/graphene/ZnO-QDs/MAPbI₃/spiro-OMeTAD/Ag | 16.8 | 0.94 | 0.62 | 9.73 | 1103 |
| Graphene nanoflakes and GO-Li | Dopant for ETL; interlayer at ETL/perovskite interface | FTO/cTiO₂/mTiO₂:G/GO-Li/MAPbI₃/spiro-OMeTAD/Ag | 18.16 (50 cm² active area) | 8.57 (50 cm² active area) | 0.65 (50 cm²) | 12.6 (50 cm² active area) | 1107 |
| RGO | Dopant for ETL | FTO/cTiO₂/mTiO₂:RGO/(FAPbI₃)₀.₈₅(MAPbBr₃)₀.₁₅/spiro-OMeTAD/Au | 21.98 | 1.11 | 0.8 | 19.54 | 1108 |
| RGO | RGO grafted highly porous zinc stannate (Zn₂SnO₄-zSO) scaffold nanofibers | (2) FTO/cZSO/RGO-ZSO₀.₇₅/(FAPbI₃)₀.₈₅ (MAPbBr₃)₀.₁₅/PTAA/Au | 22.51 | 1.04 | 0.75 | 17.83 | 1110 |
| Functionalized Graphene nanoribbon | ETL | Glass/ITO/PEDOT:PSS/MAPbI₃/functionalized-graphene nanoribbon | 22.66 | 0.93 | 0.78 | 16.5 | 1123 |
| Naphthalene diimide (NDAI)-graphene | ETL | ITO/NDI-Graphene:SnO₂/FA₀.₇₅MA₀.₁₅Cs₀.₁P-bI₂.₆₅Br₀.₃₅/spiro-OMeTAD/Au | 22.66 | 1.084 | 0.82 | 20.16 | 1112 |
| Graphene QD | Dopant for ETL | FTO/cTiO₂/GQD/MAPbI₃/spiro-OMeTAD/Au | 22.47 | 1.12 | 0.76 | 19.11 | 1109 |
| GO and MoS₂ | Interlayer at HTL/perovskite interface; interlayer at ETL/perovskite interface | ITO/PEDOT:PSS/GO/MAPbI₃/PCBM/MoS₂/Ag | 22.834 | 1.135 | 0.74 | 19.14 | 1163 |
| GO | HTL and additive for perovskite | ITO/GO/MAPbI₃:cGO/PCBM/Ag | 20.71 | 0.96 | 0.76 | 15.2 | 1135 |
| N-Doped graphene (NG) | Dopant for ETL | FTO/NiMgLiO/MAPbI/NG:PCBM/CQDs/Ag | 19.68 | 1.07 | 0.75 | 15.57 | 1209 |
| TiS₂ | ETL | FTO/TiS₂/perovskite/spiro-OMeTAD/Au | 24.75 | 1.00 | 0.75 | 18.79 | 1114 |
| TiS₂ | Interlayer at ETL/perovskite interface | ITO/SnO₂/TiS₂/perovskite/Spiro-OMeTAD/Ag | 24.57 | 1.11 | 0.79 | 21.73 | 1115 |
| Black phosphorous quantum dots (BPDQs) | ETL | ITO/BPQDs/FA₀.₈₃MA₀.₁₇PbBr₀.₅I₂.₅/Spiro-OMeTAD/Au | 16.77 | 1.03 | 0.65 | 11.26 | 1116 |
| Phosiporene nanosheets | Dopant for ETL | FTO/(TiO₂:phosporene)/FA₀.₉MA₀.₂PbBr₀.₃I₂.₅/Spiro-OMeTAD/Au | 23.32 | 1.08 | 0.71 | 17.85 | 1124 |
| Ti₃C₂ (MXene) | Interlayer at ETL/cathode interface | ITO/SnO₂-Ti₃C₂/perovskite/Spiro-OMeTAD/Ag | 23.14 | 1.06 | 0.75 | 18.34 | 1124 |
| SnS₂ | ETL | ITO/SnS₂/perovskite/spiro-OMeTAD/Au | 23.55 | 1.16 | 0.73 | 20.12 | 1044 |
| Graphene ink | Dopant for ETL | ITO/SnO₂:G/perovskite/spiro-OMeDAT/Au | 23.06 | 1.09 | 0.72 | 18.11 | 1111 |
| Ti₃C₂ (MXene) | ETL, interlayer at ETL/perovskite interface, dopant for perovskite | FTO/cTiO₂/MXenes/mTiO₂:MXenes/Mxenes/Cs₅(MA₀.₁₇FA₀.₈₃)₉₅₋ₓPb(I₀.₈₃Br₀.₁₇)₃:MXenes/spiro-OMeTAD/Au | 23.82 | 1.09 | 0.78 | 20.14 | 675 |
| Bi₂Te₃ | Interlayer at ETL/perovskite interface | ITO/PTAA/(tbPbI₃)₀.₀₄[(CsPbI₃)₀.₀₅[(FAPbI₃)₀.₈₅MAPbBr₃)₀.₁₅]₀.₉₅/PCBM:Bi₂Te₃/Bi₂Te₃/BCP/Ag | 22.52 | 1.096 | 0.788 | 19.46 | 1122 |
| g-C₃N₄ | Interlayer at ETL/perovskite interface | FTO/SnO₂/g-C₃N₄/MAPbI₃/spiro-OMeTAD/Au | 21.45 | 1.14 | 0.81 | 19.69 | 1113 |
| Thiazole-modified-C₃N₄ nanosheets | Interlayer at ETL/cathode interface | ITO/PTAA/MAPbI₃/PC₆₁BM/thiazole-modified:g-C₃N₄ nanosheets/AZO/Ag | 20.17 | 1.090 | 0.78 | 17.15 | 284 |
| Sn₂-PCBM | ETL | ITO/NiOx/perovskite/PCBM-SnS₂/ZnO/Ag | 22.70 | 1.06 | 0.83 | 19.95 | 1118 |
| TiS₂ | Interlayer at ETL/perovskite interface | FTO/TiO₂ nanograss/TiS₂/Cs₀.₀₅[MA₀.₁₃FA₀.₈₇]₀.₉₅Pb[I₀.₈₇Br₀.₁₃]/spiro-OMeTAD/Au | 22.05 | 1.13 | 0.725 | 18.73 | 1119 |
| Ti₃C₂Tₓ | Dopant for ETL, and additive for perovskite | ITO/NiO/MAPbI₃ + MXenes/PCBM + MXenes/BCP/Ag | 22.88 | 1.09 | 0.77 | 19.20 | 1120 |

**Table 10** (continued)

| Material | Usage | Device structure | Cell performance | | | | Ref. |
|---|---|---|---|---|---|---|---|
| | | | $J_{SC}$ [mA cm$^{-2}$] | $V_{OC}$ [V] | FF | $\eta$ [%] | |
| g-C$_3$N$_4$ QDs | Dopant For ETL | ITO/g-C$_3$N$_4$ QDs:SnO$_2$ QDs:SnO$_2$/Cs$_{0.5}$FA$_{0.4}$MA$_{0.1}$PbI$_3$/Spiro-OMeTAD/Au | 24.03 | 1.18 | 0.78 | 22.13 | 283 |
| g-C$_3$N$_4$ QDs | Interlayer at ETL/perovskite interface | FTO/SnO$_2$/g-C$_3$N$_4$ QDs/Cs$_{0.5}$FA$_{0.4}$MA$_{0.1}$PbI$_3$/Spiro-OMeTAD/Au | 23.41 | 1.13 | 0.77 | 20.30 | 285 |
| Bi$_2$O$_2$Se nanoflakes | ETL | ITO/Bi$_2$O$_2$Se/MAPbI$_3$/Spiro-OMeTAD/MoO$_3$/Ag | 16.16 | 0.99 | 0.57 | 9.12 | 1121 |
| SnO$_2$/Bi$_2$O$_2$Se nanoflakes | ETL | ITO/SnO$_2$/2D Bi$_2$O$_2$Se/MAPbI$_3$/MoO$_x$/Ag | 23.48 | 1.07 | 0.76 | 19.06 | 1120 |
| 6R-TaS$_2$ | Interlayer at ETL/cathode interface | ITO/PTAA/[(RbPbI$_3$)$_{0.05}$[(CsPbI$_3$)$_{0.05}$](FAPbI$_3$)$_{0.85}$(MAPbBr$_3$)$_{0.13}$]$_{0.91}$/PC$_{70}$BM/TaS$_2$/BCP/Ag | 21.45 | 1.08 | 0.796 | 18.45 | 1126 |

$^a$ MA = CH$_3$NH$_3$, $^b$ FA = HC(NH$_2$)$_2$.

morphology and structural properties are required, especially for low-temperature solution-processed device fabrication on flexible substrates.[1147] Moreover, because in such a configuration, sunlight comes from the p-type electrode, HTLs must be thin to prevent optical losses, while impeding short-circuiting between the conductive oxide (FTO or ITO for rigid and flexible substrates, respectively) and perovskite active layer.[1148–1150] In addition, the HTLs need to ensure efficient hole transport toward the electrode by minimizing the series resistance, as well as charge recombination processes.[1147–1149]

Finally, HTL-covered substrates should exhibit optimal wettability and compatibility with the solvent used for the perovskite deposition step. In this context, PEDOT:PSS has been the most frequently used HTL material in inverted PSCs, due to the following properties: (1) energy-level (HOMO level at 5.25 eV)[1151] matching with ITO $\varphi_W$ (4.9 eV)[1152] and perovskite HOMO level (5.4 eV);[1153] (2) excellent $\mu$; (3) simple solution processability.[1154] The doping with GRMs has been used to improve the physical, mechanical, and electrical features of PEDOT:PSS. For example, RGO was added into PEDOT:PSS.[1155,1156] An $\eta$ improvement of $\sim$22% was observed in RGO-doped PEDOT:PSS (RGO:PEDOT:PSS)-based device compared to the nondoped HTL-based reference due to the suppression of leakage current.[1155] Giuri and co-workers investigated the cooperative effect of GO and glucose inclusion in the PEDOT:PSS matrix.[1157] Chemically functionalized GO with the glucose molecule was used to modify the chemical properties of the PEDOT:PSS surface, changing the wettability, as well as improving the electrical conductivity of PEDOT:PSS.[1156] Concurrently, glucose molecules favored the reduction of GO[1158] and enhanced the wettability of the PEDOT:PSS substrate due to the presence of numerous hydroxyl terminations. Consequently, the GO-doped glucose/PEDOT:PSS HTL increased the $V_{OC}$ value compared to the PEDOT:PSS-based devices, indicating minimal losses, high hole selectivity, and reduced trap density at the optimized HTL/perovskite coverage.[1156] The use of a chemical approach to control the optical and electrical properties of GO was also reported by Liu et al., who used silver trifluoromethanesulfonate (AgOTf) as an inorganic dopant for single-layer GO.[1159] In particular, the spin coating of AgOTf in a nitromethane solution over a GO-doped PEDOT:PSS layer allowed the HTL $\varphi_W$ to be finely tuned, thus lowering the energy barrier for hole transfer at the PEDOT:PSS:AgOTf-doped GO/perovskite interface.[1158] This effect led to an $\eta$ improvement for both flexible and rigid PSCs in comparison to the reference devices based on PEDOT:PSS.

Li et al. demonstrated that GO can also be used as an efficient interlayer between the conductive layer and PEDOT:PSS HTL.[1160] In fact, the high conductivity of PEDOT:PSS combined with the electron-blocking capability of GO suppressed current leakage in the PSC structure, while improving the carrier injection and perovskite film morphology.[1159] The as-realized device has shown a maximum $\eta$ value of 13.1%, which was higher than that reached by the reference PSC based on PEDOT:PSS ($\eta$ = 10%).[1159]

The insertion of a buffer layer between ITO and PEDOT:PSS has been demonstrated to significantly increase the long-term stability of nonencapsulated devices under atmospheric conditions

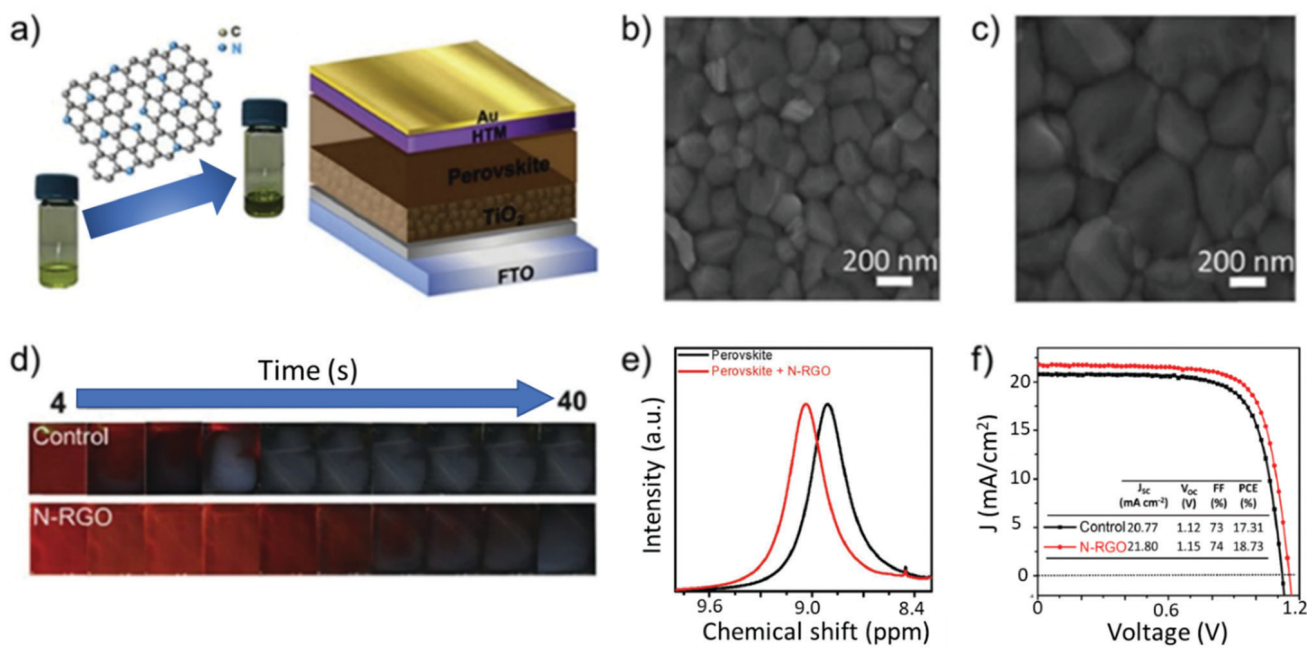

Fig. 21 (a) Schematic showing the N-RGO-doped perovskite solution and PSC with a structure of FTO/TiO$_2$/N-RGO/perovskite/spiro-OMeTAD/Au. (b and c) SEM images of perovskite films before and after the incorporation of N-RGO. (d) Photographs of perovskite films during the annealing process at 100 °C. (e) $^1$H NMR spectra of perovskite and N-RGO/perovskite solution. (f) J–V characteristics of the control device and N-RGO-incorporated device. Adapted from ref. 1134.

Table 11 Summary of the PV performance of PSCs incorporating GRMs in the perovskite active layer[ab]

| Material | Usage | Device structure | Cell performance | | | | Ref. |
|---|---|---|---|---|---|---|---|
| | | | $J_{SC}$ [mA cm$^{-2}$] | $V_{OC}$ [V] | FF | $\eta$ [%] | |
| N-Doped RGO nanosheets | Additive in perovskite | (FTO)/cTiO$_2$/mTiO$_2$/FA$_{0.85}$MA$_{0.15}$Pb(I$_{0.85}$Br$_{0.15}$)$_3$/ spiro-OMeTAD/Au | 21.8 | 1.15 | 0.74 | 18.73 | 1134 |
| Graphene quantum dots (GQDs) | Additive for perovskite | FTO/cTiO$_2$/mTiO$_2$/MAPbI$_3$/spiro-OMeTAD/Au | 22.91 | 1.05 | 0.76 | 18.34 | 1136 |
| 2D BP | Additive for perovskite | FTO/cTiO$_2$/SnO$_2$/perovskite:2D BP/Spiro-OMeTAD/Ag, | 1.82 | 23.31 | 0.82 | 20.65 | 1137 |
| Silver nanoparticle-anchored reduced graphene oxide (Ag-rGO) | Additive in perovskite | FTO/bl-TiO$_2$/m-TiO$_2$/Al$_2$O$_3$/MAPbI$_{3-x}$Cl$_x$/Ag-rGO/ spiro-OMeTAD/Au | 0.929 | 23.501 | 0.74 | 16.101 | 1141 |
| Ti$_3$C$_2$ MXenes | Additive in perovskite | FTO/SnO$_2$/MAPbI$_3$:MXenes/spiro-OMeTAD/Au | 1.03 | 22.26 | 0.76 | 17.41 | 1142 |
| Ti$_3$C$_2$ (MXene) | ETL, interlayer at ETL/ perovskite, additive in perovskite | FTO/cTiO$_2$:MXenes/mTiO$_2$:MXenes/Mxenes/ Cs$_x$(MA$_{0.17}$FA$_{0.83}$)$_{(1-x)}$Pb(I$_{0.83}$Br$_{0.17}$)$_3$:MXenes/ spiro-OMeTAD/Au | 1.09 | 23.82 | 0.78 | 20.14 | 675 |
| Black phosphorus quantum dots (BPQDs) | Additive in perovskite | ITO/PTAA/MAPbI$_3$-BPQDs/PCBM/BCP/Ag | 21.9 | 1.10 | 0.83 | 20.0 | 1138 |
| BPQDs | Additive in perovskite | FTO/SnO$_2$/BPQDs + CsPbI$_2$Br/spiro-OMeTAD/Au | 15.86 | 1.25 | 0.78 | 15.47 | 1139 |
| 2D (PEA)$_2$PbI$_4$ nanosheets | Additive in perovskite | FTO/cTiO$_2$/mTiO$_2$/Cs$_{0.1}$FA$_{0.9}$PbI$_3$ + 2D (PEA)$_2$PbI$_4$/spiro-OMeTAD/Au | 24.8 | 1.05 | 0.78 | 20.27 | 1143 |
| g-C$_3$N$_4$ | Additiver in perovskite | FTO/cTiO$_2$/MAPbI$_3$:g-C$_3$N$_4$/spiro-OMeTAD/Au | 24.31 | 1.07 | 0.74 | 19.49 | 282 |

[a] MA = CH$_3$NH$_3$. [b] FA = HC(NH$_2$)$_2$.

(temperature of 21–24 °C and humidity of 38–55%).[1058] In fact, the GO buffer layer can prohibit direct contact between the ITO and highly acidic PEDOT:PSS by slowing down photoelectrode degradation.[1058] The improved $\eta$ and stability achieved with 2D interlayers was clearly demonstrated by Kakavelakis and co-workers, who used a MoS$_2$ interlayer between PTAA (HTL) and perovskite.[1039] The introduction of MoS$_2$ flakes afforded a device with an $\eta$ value of 16.42% and a prolonged lifetime, corresponding to an 80% retention of their initial performance after 568 h of continuous illumination.[1039] By following this approach, Tang and co-workers demonstrated planar inverted PSCs with the glass/ITO/PTAA/MoS$_2$/perovskite/PCBM/BCP/Ag structure exceeding $\eta$ of 20%.[1161] In this case, the in-plane coupling between epitaxially grown CH$_3$NH$_3$PbI$_3$ and MoS$_2$ crystal lattices led to perovskite films with a large grain size, low trap density, and preferential growth orientation along the

(110) direction normal to the MoS$_2$ surface. Very recently, Zhang *et al.* replaced the MoS$_2$ interlayer with 2D antimonene.[1162] This new 2D material exhibits a thickness-dependent bandgap that can be advantageous in PV and other optoelectronic devices.[1161] In particular, antimonene-based PSCs displayed an outstanding $\eta$ value of 20.11% with a remarkable $V_{OC}$ value of 1.114 V, while the reference device has shown an $\eta$ value of 17.60% with a $V_{OC}$ value of 1.065 V.[1161] Antimonene provided sufficient nucleation sites, promoting perovskite crystallization and therefore speeding-up hole extraction at the photoelectrode.[1161]

Moreover, Cao *et al.* proposed the use of WS$_2$ flakes as an efficient interlayer at the PTAA/perovskite interface, acting as a template for the van der Waals epitaxial growth of mixed perovskite films.[1163] The WS$_2$/perovskite heterojunction has shown an engineered energy alignment, boosting charge extraction and reducing interfacial recombination.[1162] Inverted PSCs with WS$_2$ interlayers reached $\eta$ values up to 21.1%, which is among the highest value reported for inverted planar PSCs.[1162] A further evolution in the use of 2D interlayers for planar PSCs was proposed by Wang and co-workers using a double interlayer approach: GO was used at the interface between PEDOT:PSS and perovskite, while MoS$_2$ was used at the interface between PCBM and the Ag electrode.[1164] The PSC with GO and MoS$_2$ layers has shown an increase in $V_{OC}$ from 0.962 to 1.135 V, and $\eta$ from 14.15% to 19.14%.[1163] However, despite the extensive use of PEDOT:PSS in PSCs, PEDOT:PSS suffers from hygroscopicity and acidic properties, which cause faster degradation of both organic layers and organolead halide perovskites layer.[1165] Thus, several efforts have been made in order to replace PEDOT:PSS with the most stable HTL based on graphene or other 2D materials, including TMDs.[1166,1167] In fact, the lone pair of electrons of the carbon and chalcogen atoms in the structure of graphene (and derivative) and TMDs, respectively, improves the $\mu$ value of HTL, due to the demonstrated ballistic transport.[1168–1170] Moreover, chemical doping and simple surface treatment allow the easy modulation of the energy levels of both (R)GO[1171,1172] and TMD films.[1173,1174] For example, Kim and co-workers demonstrated the feasibility to replace PEDOT:PSS with polycrystalline structure of MoS$_2$ and WS$_2$ layers, which were synthesized through a chemical deposition method.[1150] The devices with a planar inverted architecture of ITO/MoS$_2$ or WS$_2$/perovskite/PCBM/BCO/LiF/Al exhibited $\eta$ of 9.53% and 8.02% for MoS$_2$ and WS$_2$ cases, respectively, comparable to that measured for PEDOT:PSS-based devices (9.93%).[1150] In the same work, the use of GO as the HTL resulted in an $\eta$ of 9.62%, demonstrating the effectiveness of GRMs in PSCs.[1150] The use of TMDs for HTL in inverted PSCs has been recently optimized by Huang and co-workers, who achieved an $\eta$ of 14.35% and 15.00% for MoS$_2$- and WS$_2$-based DSCs, also demonstrating enhanced stability compared to the PEDOT:PSS-based reference devices.[1175]

Following the aforementioned studies, GO and RGO in both pristine and functionalized forms have been extensively tested as HTLs, yielding valuable results in terms of $\eta$ and stability. For example, Wu *et al.* improved the $\eta$ of ITO/HTL/CH$_3$NH$_3$PbI$_{3-x}$Cl$_x$/PCBM/ZnO/Al PSCs from 9.2% to 12.4% by replacing PEDOT:PSS with a GO layer.[1050] In particular, by tuning the concentrations of GO in neutral aqueous suspensions from 0.25 mg mL$^{-1}$ to 4 mg mL$^{-1}$, the authors were able to deposit a GO layer with a thickness in the range of $\sim$2–20 nm, finely controlling the PV performance of the devices.[1050]

More recently, GO has been exploited as the HTM even in the form of nitrogen-doped nanoribbons (NGONRs) by Kim and co-workers.[1176] The NGONRs were synthetized starting from MWCNTs and subsequently doped by pyrolyzing nanoribbons/polyaniline (PANI) composites at 900 °C for 1 h in an Ar atmosphere.[1175] Different from the case of PEDOT:PSS, the deposition of perovskite films onto NGONRs allowed the perovskite film to grow into large textured domains, yielding an almost complete coverage.[1175] Thus, small-area devices reached an $\eta$ value of 12.41%, which was higher than that of the PEDOT:PSS-based reference (9.70%).[1175] Notably, NGONR-based cells demonstrated negligible current hysteresis along with improved stability under ambient conditions (average temperature and humidity of 20 °C and 47%, respectively), since the absence of the PEDOT:PSS layer prevented perovskite degradation caused by the acidic nature of the polymer.[1175] A significant $\eta$ of 16.5% and extraordinary stability was also achieved by using GO as the HTL in an inverted PSC.[1177] Long-term aging test under ambient humidity with a relative humidity of 60% was carried out on the encapsulated devices.[1176] After initial $J$–$V$ measurements, the devices were continuously illuminated and then stored in the dark under standard laboratory conditions.[1176] The GO-based devices reported long-term stability compared to PEDOT:PSS-based PSCs.[1176] In particular, their $\eta$ decreased by only 10% after nearly 2000 h.[1176]

The improvement in device stability was demonstrated using a RGO nanosheet as the HTL in the inverted structure of ITO/RGO/perovskite/PCBM/BCP/Ag.[1049] The RGO/perovskite junction induced faster charge transfer across its interface, resulting in reduced charge recombination compared to the PEDOT:PSS-based PSCs.[1049] Furthermore, the perovskite grains (100–200 nm grains) of the perovskite film grown on the RGO layer reduced the total number of grain boundaries, increasing the cell FF, compared to the PEDOT:PSS-based reference (perovskite with grain size of <100 nm).[1049] RGO-based devices have shown promising stability, retaining 62% of the initial $\eta$ even after 140 h of light exposure, while PEDOT:PSS-based devices failed.[1049] The stability of cells with RGO stemmed from the quasi-neutral properties of RGO with few surface oxygen functionalities and the inherent passivation ability of RGO against moisture and oxygen.[1049] With the aim to further enhance the stability of PEDOT:PSS-based inverted PSCs, GO[1178] and ammonia-modified GO (GO:NH$_3$)[1179] were reported as efficient interlayers between the HTL and perovskite active layer. In the latter case, a thin GO:NH$_3$ layer of $\sim$2 nm was spin coated onto the PEDOT:PSS surface and subsequently annealed at 120 °C for 10 min.[1178] Similar to the results reported using the RGO nanosheet,[1049] the perovskite film realized onto the PEDOT:PSS/GO:NH$_3$ substrate displayed improved crystallization with a preferred orientation order and nearly complete coverage, improving its optical absorption.[1178] Furthermore, the optimal energy-level matching between the PEDOT:PSS-GO:NH$_3$ HTL and perovskite led to an $\eta$ value up to 16.11%, which was significantly superior compared to the values measured for bare

PEDOT:PSS-based PSCs ($\eta$ = 12.5%).[1178] Notably, the highly ordered perovskite structure led to a marked improvement in the structural stability of the active film, extending the device lifetime in ambient conditions.[1178] Organo-sulfonate graphene (oxo-G) was reported to replace PEDOT:PSS in inverted PSCs, significantly enlarging the device's lifetime.[1180] In fact, the use of oxo-G as the HTL effectively prevented the access of water vapor into the device stack, without penalizing the overall $\eta$ of the devices, which reached valuable $\eta$ of 15.6%.[1179] Noteworthily, the unencapsulated devices retained ~60% of the initial $\eta$ after ~1000 h light soaking under 0.5 sun and ambient condition.[1179] The obtained results confirmed the use of functionalized graphene-based materials as a viable route to stabilize inverted PSCs.[1179]

In the archetypical mesoscopic structure of cTiO$_2$/mTiO$_2$ (or Al$_2$O$_3$)/perovskite/spiro-OMeOTAD/Au, GRMs have been widely used to replace traditional HTMs, as well as interlayers, mainly aiming to solve certain issues related to the spiro-OMeOTAD HTL, such as instability and high cost ($170–475/g).[1181] In particular, spiro-OMeTAD needs to be doped to increase the intrinsic low electrical conductivity of its pristine amorphous form.[1182] To this end, lithium bis(trifluoromethanesulfonyl)imide (Li-TFSI)[1183,1184] and tert-butylpyridine (TBP)[1185] are the most frequently used dopants to increase $\mu_h$ and to improve contact at the spiro-OMeTAD/perovskite interface, respectively. However, major instability drawbacks must be considered when the aforementioned dopants are used. In fact, Li-TFSI exposed to ambient conditions is deliquescent[1186] and tends to dissociate from the spiro-OMeTAD, negatively affecting its performances.[1185] In addition, TBP corrodes the perovskite layer due to its polar nature.[1187]

With the aim to replace common dopants of the spiro-OMeTAD layer, Luo and co-authors recently proposed the use of GO reduced by a ferrous iodide acidic solution as an alternative HTM dopant.[1188] The devices prepared using iodine-RGO/spiro-OMeTAD HTL displayed an $\eta$ of 10.6%, which was lower compared to that of doped spiro-OMeTAD-based devices (13.01%).[1187] However, the cell stability was significantly improved.[1187] In particular, the $\eta$ value of the RGO-based devices retained above 85% of the initial value even after 500 h of storage in air, while the $\eta$ value of

the device fabricated with doped spiro-OMeTAD decreased to 35% under the same aging conditions.[1187]

An alternative strategy to enhance cell $\eta$ by preventing perovskite/spiro-OMeTAD interface degradation involves the use of an interfacial layer based on 2D materials. As a recent example, phosphorene has been used at both mTiO$_2$/perovskite and perovskite/spiro-OMeTAD interfaces in mesoscopic n–i–p PSCs, achieving a remarkable $\eta$ of 19.83%.

Solution-processed phosphorene has shown ambipolar carrier transport behavior and can be considered as a viable route for enabling great advances in PSC performance via judicious interfacial positioning of phosphorene in the cell structure.[1189] Regarding the perovskite/spiro-OMeTAD interface stability, Capasso et al. showed that the insertion of few-layers MoS$_2$ retarded PSC degradation with higher lifetime stability, of over 550 h, compared to the reference MoS$_2$-free PSC ($\Delta\eta/\eta$ = −7% vs. $\Delta\eta/\eta$ = −34%).[1190] The authors justified the extended lifetime to the role of MoS$_2$ flakes that act as a protective layer, preventing the formation of shunt contacts between the perovskite and Au electrode.[1189] The enhanced stability of the mesoscopic device using an MoS$_2$ interlayer was recently demonstrated even under prolonged light soaking condition at 1 sun illumination.[1105,1191] However, the MoS$_2$ VB does not perfectly match with the perovskite HOMO level, possibly forming an energy barrier for the hole extraction process that causes $V_{OC}$ reduction. In order to fully exploit the potential of MoS$_2$ as an interlayer, Najafi et al. produced MoS$_2$ quantum dots (MoS$_2$ QDs), derived by LPE-produced MoS$_2$ flakes and hybridized with functionalized reduced graphene oxide (f-RGO), to provide both hole extraction and electron blocking properties (Fig. 22).[1192] In fact, the intrinsic n-type doping of the MoS$_2$ flakes introduce intraband gap states that can extract holes through an electron injection mechanism.[1191] Meanwhile, quantum confinement effects increase the $E_g$ of MoS$_2$ (from 1.4 eV for flakes to more than 3.2 eV for QDs), rising its CB minimum energy from −4.3 eV to −2.2 eV. The latter value is above the CB of CH$_3$NH$_3$PbI$_3$. Therefore MoS$_2$ QDs exhibit electron-blocking properties.[1191] In addition, the hybridization of MoS$_2$ QDs with f-RGO, obtained by chemical silanization-induced linkage between RGO and (3-mercaptopropyl)trimethoxysilane, promoting the deposition of a homogeneous interlayer onto the

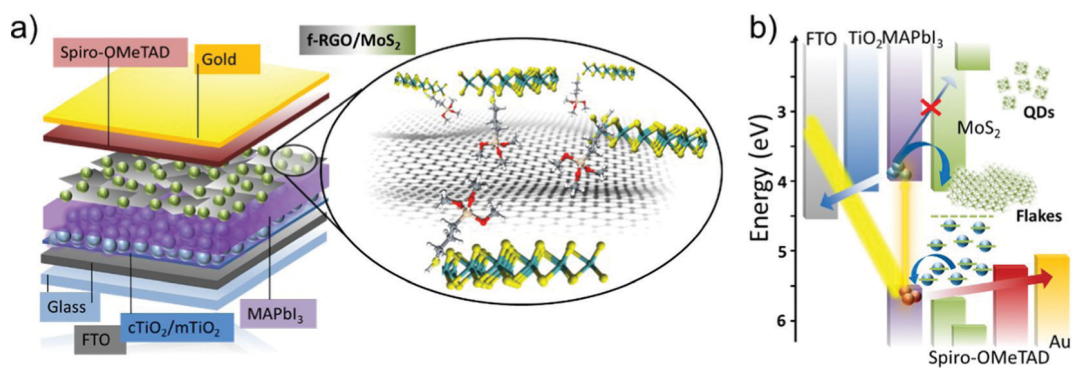

Fig. 22 (a) Schematic of mesoscopic MAPbI$_3$-based PSC using MoS$_2$ QDs:f-RGO hybrids as both HTL and active buffer layer. (b) Schematic of the energy band edge positions of the materials used in different components of the assembled mesoscopic MAPbI$_3$-based PSCs. Reprinted with permission from ref. 1191, Copyright 2018, American Chemical Society.

perovskite film.[1191] In fact, the f-RGO flakes plug the pinholes of MoS$_2$ QD films.[1191] The as-prepared PSCs achieved $\eta$ values of up to 20.12% (average $\eta$ of 18.8%).[1191] As an alternative to MoS$_2$ QDs, Agresti *et al.* proposed a chemical functionalization of the MoS$_2$ flakes (fMoS$_2$), by linking a thiol group of 3-mercaptopropionic acid (MPA) moieties to the MoS$_2$ surface *via* S–S van der Waals physisorption and/or S-vacancy passivation (Fig. 23).[1193] Apart from chemically and electronically repairing the defective lattice of the MoS$_2$ flakes, MPA-based functionalization is effective to upshift the MoS$_2$ energy bands.[1192] The upshift of the MoS$_2$ energy bands aligns the VB edge of MoS$_2$ with the HOMO level of the perovskite, improving the hole extraction process.[1192] In addition, the MPA-based functionalization shifts the CB edge of MoS$_2$ above the LUMO level of the perovskite, hindering undesired electron transfer (*i.e.*, providing electron blocking properties).[1192] Owing to these effects, the MPA-based functionalization of MoS$_2$ flakes, when integrated in PSCs, improved the $\eta$ value of the reference devices without MoS$_2$-based interlayer by +11.6%.[1192]

Recently, another mechanism has been proposed by Shi *et al.* to explain the hole extraction properties of 2D MoS$_2$.[1194] This mechanism relies on the presence of intrinsic S vacancies at the MoS$_2$ edges that stabilize halide vacancies at the perovskite/MoS$_2$ interface.[1193] This process induces an interface dipole moment, which reverses the offset of the VB maxima.[1193] Overall, this effect can lead to an ultrafast (picosecond timescale) hole transport from the perovskite to the current collector, boosting the performance of MoS$_2$ HTL-based PSCs.

Beyond MoS$_2$, both GO[1195] and functionalized GO (fGO)[1196] were used as an efficient buffer layer at the perovskite/spiro-OMeTAD interface. In particular, when GO was deposited onto the perovskite surface, it performed as a base that absorbed spiro-OMeTAD onto its surface.[1194] Moreover, parts of the O atoms in GO were demonstrated to connect with unsaturated Pb atoms in the perovskite, improving adhesion between spiro-OMeTAD and the active layer.[1194] Furthermore, the surface defect states of the perovskite were dramatically reduced, leading to an $\eta$ increase of 45.5%, from 10.0% in the case of a standard PSC structure to 14.5% when GO was inserted as the interlayer.[1194]

Amino-functionalized N-doped graphene (NG) was tested as an interlayer between perovskite and undoped spiro-OMeTAD in the standard mesoscopic structure (FTO/cTiO$_2$/mTiO$_2$/perovskite/dopant-free spiro-OMeTAD/Au), reaching higher $\eta$ (14.6%) compared to the reference device ($\eta$ = 10.7%).[1195] These results were explained by the absorption of spiro-OMeTAD from NG-treated perovskite surface *via* $\pi$–$\pi$ interactions, ensuring electron-rich molecules. In fact, N atoms interact with undercoordinated Pb$_2^+$ ions by donating electron density.[1195] Thus, the perovskite surface is passivated and the charge extraction toward the HTL is optimized.[1195,1197] Consequently, $\eta$ enhancement was associated with the increase in $J_{SC}$ and FF due to reduced charge recombination at the perovskite/HTM interface.[1195] Functionalized RGO was also used at the perovskite/spiro-OMeTAD interface in the planar configuration to reduce interfacial recombination and enhance hole extraction.[1198] An alternative strategy to

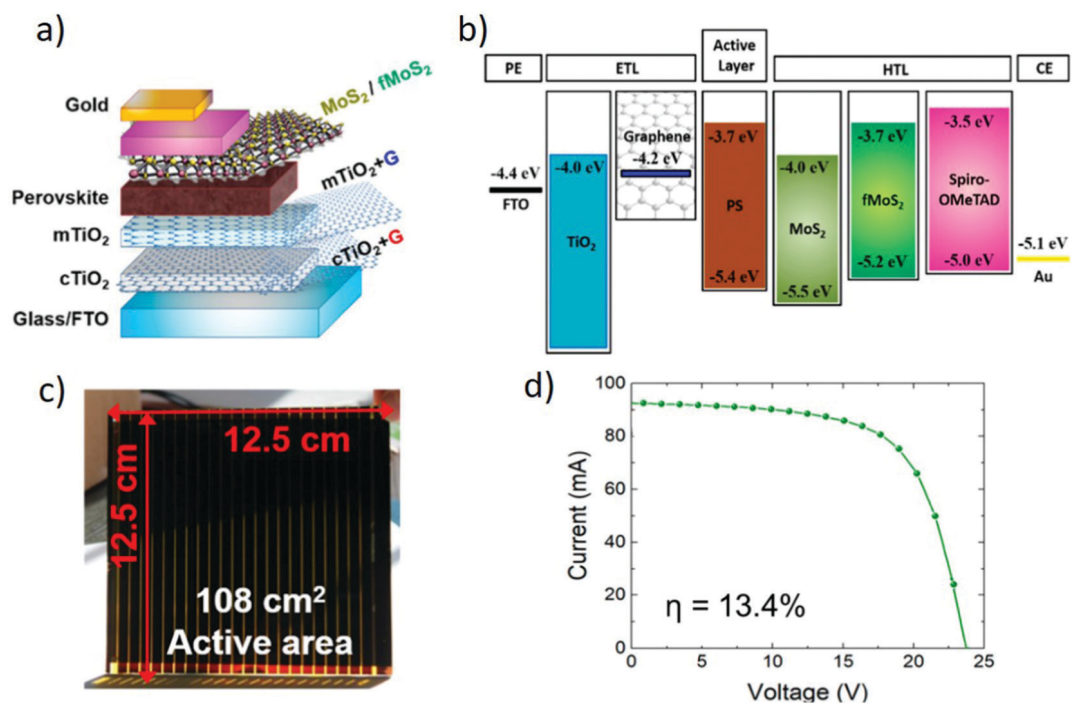

**Fig. 23** (a) Engineered PSC architecture using chemically functionalized molybdenum disulfide (fMoS$_2$) as the interlayer at the perovskite/HTL interface for improving the hole injection/collection at the CE and (b) its energy band diagram. (c) Photograph of a representative large-area PSM (108 cm$^2$ active area, 156.25 cm$^2$ substrate area), showing an $\eta$ value of 13.4% under 1 sun illumination, as shown by (d) its *J–V* characteristic. Adapted with permission from ref. 1192, Copyright 2019, American Chemical Society.

replace spiro-OMeTAD is to use more stable and dopant-free HTMs. To this end, Nouri et al. demonstrated that the use of copper phthalocyanine (CPC) as an alternative HTM can produce valuable $\eta$ only if an interlayer of GO was used between the perovskite and HTL.[1199] In a recent work, You et al. proposed the use of solution-processed high-mobility 2D materials, namely, $MoS_2$ and BP, to conduct holes from the grain boundary of the perovskite layer to the HTL, proposing a novel strategy to passivate defects in PSC grain boundaries.[1200]

An attempt to replace spiro-OMeTAD with sprayed RGO was reported by Palma et al.[1052] Despite the $\eta$ values of RGO-based PSCs ($\eta = 5\%$) were lower than that obtained with devices based on doped spiro-OMeTAD as the HTL ($\eta = 11\%$), the authors have reported an impressive improvement in device stability.[1052] In fact, PSC stability was demonstrated in an endurance test carried out under both shelf-life conditions (in air, in dark, at ambient temperature (RT = 23 °C) and relative humidity (RH = 50%)) and open-circuit load conditions for prolonged light-soaking stress test (1 sun at 65 °C and ambient RH).[1052] In particular, 1987 h of shelf-life testing revealed that $\eta$ increased by more than 30% in RGO-based-PSCs, while spiro-OMETAD-based PSCs evidenced a drastic $\eta$ reduction (−44%).[1052] Notably, the consecutive light-soaking tests induced a further $\eta$ decrease of only 26% in RGO-based PSCs, while spiro-OMeTAD-based device completely failed the test.[1052]

RGO was also used as a dopant for poly-3(hexylthiophene) (P3HT) polymer, a valid HTL alternative to spiro-OMeTAD when doped with LiTFSI salts and TBP.[1201] The enhanced $\eta$ (∼9%) of PSC using RGO-doped P3HT compared to devices using bare P3HT (6.5%) was complemented by improved shelf-life stability.[1200] The authors demonstrated that RGO doping introduced additional charge percolation pathways in P3HT and enhanced the interfacial contacts with the underlying perovskite layer and Au back electrode.[1200] Thus, improved hole depletion from the perovskite layer limits charge recombination effects by avoiding trapped charges at the perovskite/HTM interface.[1202] The increase in $\mu_h$ of P3HT was reported as the key point to improve the PSC performance by Ye and co-workers, who proposed an imidazole-functionalized GO (IGO) as the HTL dopant.[1053] PSCs using IGO-doped P3HT achieved an $\eta$ value of 13.82%, which was among the highest reported for P3HT-based PSCs.[1053] Apart from the increase in $\mu_h$, IGO doping allowed the P3HT HOMO level to be shifted from −5.0 eV to −5.2 eV.[1053] Moreover, the hydrophobicity of the P3HT/graphene layer resulted in excellent stability of the PSCs, which retained more than 70% of their initial performance after 8 weeks of storage in ambient conditions (25 °C, 20–40% RH).[1053]

The hydrophobicity of GRMs is a peculiar property that drives their exploitation in the development of new and more robust HTMs and/or protecting layers for PSCs. Very recently, Cao and co-workers successfully replaced spiro-OMeTAD with a perthiolated tri-sulfur-annulated hexa-peri-hexabenzocoronene (TSHBC)/graphene layer, achieving an $\eta$ value exceeding 14% on small-area devices.[1055] Such a tested compound combined the hydrophobicity of both graphenes and thiols, providing an effective molecular sealing approach to improve the stability of complete devices.[1055] Moreover, the TSHBC/graphene layer exhibited an excellent hole extraction capability, ensuing from the Pb–S coordination bond between TSHBC and perovskite, together with enhanced $\mu_h$ due to the presence of GNSs in the HTL.[1055] A similar approach was also reported by Wang and co-authors using a multilayered buffer layer with the aim to replace spiro-OMeTAD and to protect perovskite from moisture.[1056] The realized SWNT/GO/polymethyl methacrylate (PMMA) layer conjugated the SWNT capability in assisting photogenerated carrier extraction/transport with the electron blocking property of GO.[1056] Moreover, the built-in potential across the device drastically increased upon the insertion of the GO layer, which prevented carrier recombination losses.[1056] With regard to stability, the $\eta$ of spiro-OMeTAD-based PSC significantly reduced from 10.5% to 5.8% during a stable day test, while the SWNT/GO/PMMA-based cells exhibited stable performance, showing a decrease of $\eta$ from 10.5% to 10.0% in the same timeframe.[1056] This result was attributed to the PMMA layer, which acts as an effective barrier to moisture and oxygen penetration, preventing the degradation of the perovskite layer.[1056] We should mention that HTM doping was also realized using 2D materials other than GO.[1203] Indeed, effective HTMs were produced by means of BP,[1204,1205] graphene,[1206] and functionalized $MoS_2$[1207] dopants. For example, BP/spiro-OMeTAD blend-based PSCs have shown a remarkable increase in $\eta$ (more than 20%) compared to PSCs without BP.[1203] Lastly, solution-processed 2D-conjugated polymers have also been proposed as effective dopant-free HTL materials alternative to spiro-OMeTAD, confirming that the design of novel 2D materials can prospectively offer advanced strategies to further boost the $\eta$ and the stability of PSCs.[1208] In detail, planar n–i–p-structured PSCs based on 2DP-TDB as a dopant-free HTM recently achieved champion $\eta$ as high as 22.17%, while showing improved stability under continuous light soaking in an inert atmosphere compared to control devices.[1207]

Recently, graphene-based dopants for HTLs have been demonstrated to have a crucial role in CE replacement.[1209] In fact, one of the main hurdles of PSC technology is that the hole transporting materials established for state-of-the art Au-based devices are not compatible with carbon pastes used for the fabrication of carbon-based PSCs. Thus, Chu et al. proposed the use of HTL based on solution-processed P3HT/graphene composites, exhibiting outstanding $\mu_h$ and thermal tolerance.[1208] In fact, after annealing at 100 °C, the $\mu_h$ value of this HTL increased from $8.3 \times 10^{-3}$ to $1.2 \times 10^{-2}$ cm$^2$ V$^{-1}$ s$^{-1}$, which was two orders of magnitude larger than that of pure P3HT.[1208] As a result, the authors reported carbon-based PSCs with a record $\eta$ value of 17.8% (certified by Newport).[1208] This cell was the first PSC to be certified under a stabilized testing protocol.[1208] The P3HT/graphene composite-based HTL device yielded a champion device with $\eta$ of 18.2%.[1208] In comparison, the use of sole P3HT as the HTL resulted in a device with inferior performance, i.e., $\eta = 11.1\%$.[1208] The outstanding stability of a unencapsulated device based on P3HT/graphene HTL was demonstrated by only 3% drop after 1680 h storage in ambient conditions with a relative humidity of ∼50%.[1208] After encapsulation, the device retained ∼89% of its initial $\eta$ under continuous 1 sun illumination at RT for 600 h in a N₂ environment.[1208] In comparison, the device using P3HT HTL exhibited rapid degradation, reaching ∼25% of its original $\eta$ after ∼75 h.[1208] Device stability improvement using GRMs was also demonstrated by Bi and co-workers using a nanostructured carbon layer into the device structure.[1210] In particular, an ETL based on PCBM containing N-doped graphene coupled with a carbon quantum dot (CQD) interlayer before Ag CE effectively suppressed the diffusion of ions/molecules within PSCs, preventing perovskite degradation.[1209] In fact, the stable $\eta$ of a CQDs/G-PCBM-based device over 15% was measured when the device was kept in the dark at RT for 5000 h or under AM 1.5G simulated solar light for 1000 h.[1209] In particular, during the thermal aging test at 85 °C for 500 h, the devices retained 98% of the initial $\eta$.[1209]

Aurora *et al.* recently reported a breakthrough in the race for the design and realization of stable PSCs.[1038] The authors demonstrated the possible replacement of expensive spiro-OMeTAD with CuSCN as the HTL by achieving a remarkable $\eta$ above 20%.[1038] The addition of a conductive RGO spacer layer between CuSCN and Au allowed the PSCs to retain more than 95% of the initial $\eta$ after aging at MPP for 1000 h under 1 sun illumination at 60 °C (Fig. 24).[1038]

Graphene was demonstrated to play a major role in reducing the high sheet resistance of PEDOT:PSS used in form of adhesive CEs.[1211] In fact, PEDOT:PSS was easily spin coated on graphene/PMMA/poly(dimethylsiloxane) (PDMS) substrates to realize the CE, which was subsequently laminated on the perovskite substrate.[1210] When 4-layer graphene was embedded in the CE, a remarkable $\eta$ value of 12.4% was achieved under light illumination from the FTO side, while device semi-transparency was demonstrated by reporting an $\eta$ of 4.37% during illumination from the CE side.[1210] However, this work was conducted using CVD graphene, and a similar approach based on solution-processed graphene must be consolidated. Notably, the number of graphene layers was key for the device performance optimization.[1210] In fact, even though a large number of layers decreases the series resistance, a number of graphene layers higher than 5 compromises the adhesion between graphene and spiro-OMeTAD.[1210]

An effective approach to improve PSC stability is represented by the replacement of the metal CE with a carbon-based back electrode, to form the so-called carbon perovskite solar cells (C-PSCs).[1212–1219] In fact, Au is a well-known cause of instability, since it suffers from metal-ion migration phenomenon degrading the perovskite and HTLs when the device experiences an operating temperature above 70 °C.[1220] So far, three types of C-PSCs have been proposed, namely, mesoporous,[1221] embedment,[1222–1224] and paintable C-PSCs.[1125,1126,1225] In mesoporous C-PSCs, a porous carbon electrode is first deposited and then the perovskite is infiltrated within it to complete the structure.[1218,1220] To produce embedment C-PSCs, a porous carbon electrode is deposited onto a perovskite precursor (*e.g.*, PbI₂), followed by the conversion of the precursor to perovskite by infiltrating a reaction solution.[1221–1223] Lastly, the carbon CE can be directly deposited onto the

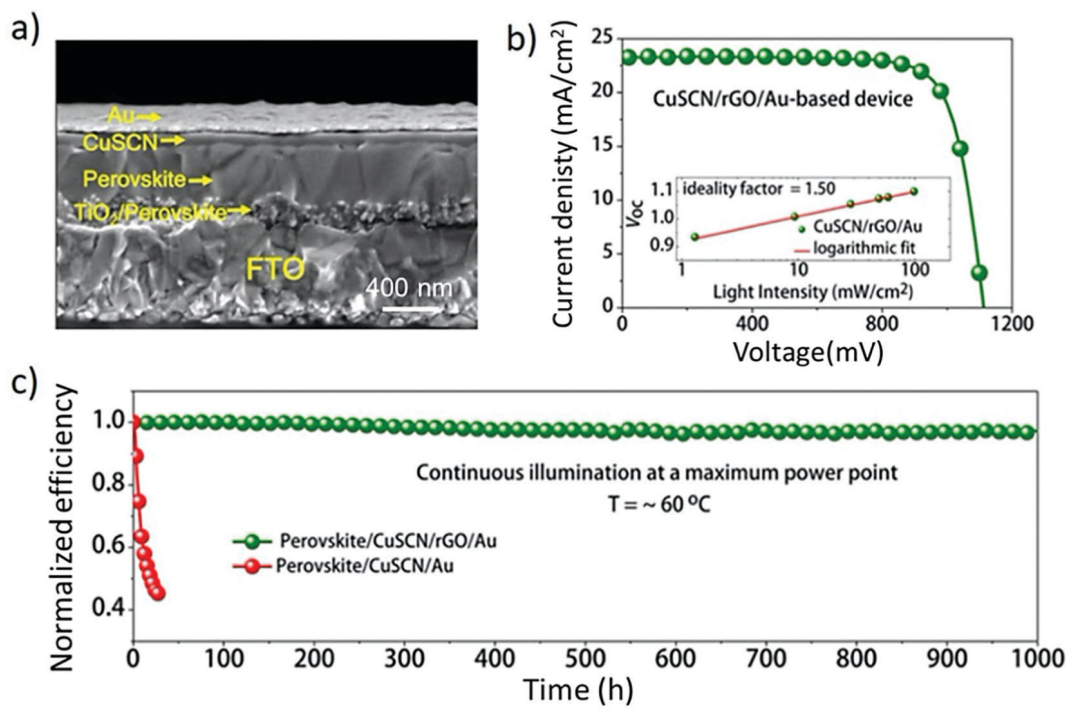

Fig. 24 (a) Cross-sectional SEM micrograph displaying the thickness of different layers in a complete mesoscopic n–i–p PSC employing rGO as the buffer layer between CuSCN-based HTL and Au CE, (b) J–V curve of the CuSCN-based device showing $\eta$ = 20.4%; the inset shows $V_{OC}$ as a function of illumination intensity with an ideality factor of 1.50. (c) Stabilities of an unencapsulated device based on CuSCN HTL and an unencapsulated CuSCN-based device incorporating a thin layer of RGO between the Au and CuSCN layers, evaluated at the MPPs under continuous simulated sunlight illumination at 60 °C in a N₂ atmosphere. Adapted from ref. 1038.

perovskite layer, HTL, or ETL depending on the device configuration (*i.e.*, CTL-free devices, n–i–p, and p–i–n configurations, respectively) to obtain paintable C-PSCs. Recent reviews on C-PSCs summarized the advantages of such technology compared to conventional PSCs,[1211–1217] including low cost, chemical inertness of the carbon-based material to halide ions, and hydrophobic characteristics. Therefore, we refer the reader to these earlier reviews, while, here, we will specifically focus on the progresses achieved in C-PSCs by using solution-processed 2D materials. In particular, Grancini *et al.* used a 2D/3D (HOOC(CH$_2$)$_4$NH$_3$)$_2$PbI$_4$/CH$_3$NH$_3$PbI$_3$ perovskite junction as the active layer to develop $10 \times 10$ cm$^2$ solar modules by a fully printable industrial-scale process, delivering a stable $\eta$ value of 11.2% for more than 10 000 h with zero loss in performance measured under controlled standard conditions.[1218] However, due to poor perovskite layer uniformity, the perovskite infiltration process still represents a critical step, and a facile carbon paste deposition process onto the perovskite is highly pursued. For this purpose, Chen *et al.* recently applied carbon CE over all-inorganic PSCs based on a CsPbBr$_3$ absorber.[1227] In this work, Ti$_3$C$_2$-MXene nanosheets were used as the interlayer to eliminate energy-level mismatches, accelerate hole extraction, and reduce the recombination at the interface of perovskite/carbon electrode.[1226] Following this approach, PSCs showing an initial $\eta$ of 9.0% and long-term stability in a moisture environment over 1900 h (over 600 h under thermal conditions) have been demonstrated.[1226] As alternatives to conventional carbon paste, SLG, FLG, and multilayer graphene (MLG) have been reported for the realization of metal-free CEs in mesoscopic PSCs.[1228] In particular, an $\eta$ value of 11.5% was achieved using reduced multilayered graphene oxide (MGO) at 1000 °C under an Ar atmosphere.[1227] In comparison to SLG, the better hole extraction of MGO was ascribed to the as-formed Schottky barrier,[1227] while an ohmic contact was established for the case of SLG.[1227] Furthermore, larger transport coefficient, longer photocarrier lifetime, and twice the diffusion length have been demonstrated for MGO in comparison with SLG, opening a new route toward the low-cost production of Au-free PSCs.[1227] N-Doped graphene frameworks (N-GFs), forming covalently bonded 3D structures, were also used as excellent CEs in HTL-free PSCs, achieving an $\eta$ of 10.32%.[1229]

Recently, Mariani *et al.* reported low-temperature graphene-based carbon pastes in alcoholic solvents compatible with prototypical PSC materials used in standard configurations, namely, the triple cation Cs$_{0.05}$(FA$_{0.85}$MA$_{0.15}$)$_{0.95}$Pb(I$_{0.85}$Br$_{0.15}$)$_3$ perovskite and spiro-OMeTAD HTL.[1230] The corresponding graphene-based CEs have been applied to large-area (1 cm$^2$) mesoscopic devices and low-temperature-processed planar n–i–p devices that reached $\eta$ values of 13.85% and 14.06%, respectively.[1229] Moreover, proof-of-concept metallized mini-wafer-like C-PSCs over a substrate area of 6.76 cm$^2$ (aperture area = 4.00 cm$^2$) afforded an $\eta$ of 13.86%, which corresponded to a record-high $\eta$ value of 12.10% on the aperture area. These results proved, for the first time, the metallization compatibility with such paintable C-PSC configurations.[1229]

Carbon back electrode mechanically stacked with another carbon-coated FTO glass under pressure was also proposed to realize an innovative modular flexible C-PSC design.[1231] Among the different carbon nanomaterials (*i.e.*, carbon black, graphite sheet, and RGO), RGO has shown $\eta$ as high as 18.65%, which was the record-high value reported for C-PSCs.[1230] Furthermore, graphene-based C-PSCs retained 90% of their initial $\eta$ after aging at an elevated temperature of 85 °C for 1000 h without any encapsulation.[1230] Very recently, Ti$_3$C$_2$ MXenes have been used as the back electrode for mesoscopic n–i–p PSCs in HTL-free configurations. In particular, Ti$_3$C$_2$ has been directly deposited over a MAPbI$_3$ perovskite layer by doctor-blade coating[1232] or alternatively by using a simple hot-pressing method.[1233] Despite the highest $\eta$ value for PSCs using MXene-based back electrode is 13.83%, the as-produced devices have shown improved stability in the ambient atmosphere at RT (humidity: 30%) compared to gold-based PSCs.[1232] Table 12 summarizes the main results achieved by PSCs using HTLs and back electrodes based on GRMs.

## 6.4 Front electrodes (ITO replacement)

GRMs have been recently explored in PSCs with the aim to replace TCOs (*e.g.*, ITO and FTO) used for the TCE in PSC architecture. In fact, ITO and FTO are difficult to fabricate *via* low-temperature solution processes and exhibit poor mechanical flexibility, hindering the development of solution-processed flexible PSCs.[1235–1237] PEDOT:PSS has also been tested as alternative transparent electrodes.[1238] However, as a consequence of its hygroscopicity, it can absorb moisture, which decomposes the perovskite layers and rapidly degrades the device performance.[1239] In this context, graphene and graphene-based nanocomposites have been demonstrated to be reliable alternatives for TCO replacement in both rigid and flexible PSCs. The highest ever reported $\eta$ (17.1%) on TCO-free rigid devices was claimed in 2016 by Sung and co-workers using CVD graphene substrates.[1240] The tested inverted planar structure reported a graphene/MoO$_3$/PEDOT:PSS photoelectrode, in which a 2 nm-thick MoO$_3$ layer improved the PEDOT:PSS deposition onto the hydrophobic graphene surface.[1239] The lower conductivity of the graphene electrode compared to that of ITO was compensated by the higher transparency and lower surface roughness, resulting in comparable $J_{SC}$, higher $V_{OC}$, and improvement of $\eta$ from 16.9% in ITO/MoO$_3$-based device up to 17.1%.[1239] In 2019, Yao and co-workers proposed the use of solution-processed graphene:ethyl cellulose (G:EC) as a transparent electrode for both rigid and flexible substrates using a planar inverted PSC architecture.[1241] Apart from the remarkable results achieved on rigid substrates ($\eta$ = 16.93%), the highly dispersed graphene composite-based transparent electrode satisfied the requirements in terms of $\sigma$ and $T_r$ for flexible PSCs, resulting in a champion device with an $\eta$ value of 15.71%.[1240]

A different way to replace the TCO layer can be the deposition of Ag conductive grids on rigid or flexible transparent substrates.[1242] However, during initial attempts, the reaction between Ag and halide ions in the perovskite caused rapid, permanent device degradation.[1241] To overcome such a limitation, Lu and co-workers proposed a protective GO coating for the Ag grids, and remarkable $\eta$ of 9.23% and 7.92% were reported for rigid and flexible substrates, respectively.[1241] The optimal $\Phi_w$ alignment and

Table 12  Summary of the PV performance of PSCs using HTLs or back electrodes based on GRMs[a,b]

| Material | Usage | Device structure | $J_{SC}$ [mA cm$^{-2}$] | $V_{OC}$ [V] | FF | $\eta$ [%] | Ref. |
|---|---|---|---|---|---|---|---|
| (a) GO or (b) WS$_2$ or (c) MoS$_2$ | HTL | ITO/GO or WS$_2$ or MoS$_2$/MAPbI$_{3-x}$Cl$_x$/PCBM/BCO/LiF/Al | (a) 14.51 (b) 15.91 (c) 14.89 | (a) 0.92 (b) 0.82 (c) 0.96 | (a) 0.72 (b) 0.64 (c) 0.67 | (a) 9.62 (b) 8.02 (c) 9.53 | 1150 |
| RGO | Dopant for HTL | ITO/RGO-PEDOT:PSS/MAPbI$_3$/PC$_{61}$BM/Al | 17.1 | 0.95 | 0.64 | 10.6 | 1155 |
| RGO | Dopant for HTL | ITO/PEDOT:PSS@glucose/MAPbI$_3$/PC$_{61}$BM/LiF/Al | 17.6 | 1.05 | 0.69 | 12.8 | 1156 |
| AgOTf-doped GO | Dopant for HTL | ITO/PEDOT:PSS:AgOTf-doped GO/MAPbI$_{3-x}$Cl$_x$/PCBM/Au | 19.18 | 0.88 | 0.7 | 11.9 | 1158 |
| GO | Interlayer at front electrode/HTL interface | ITO/GO-PEDOT:PSS/MAPbI$_{3-x}$Cl$_x$/PC$_{61}$BM/Au | 17.96 | 0.96 | 0.76 | 13.1 | 1159 |
| GO | HTL | ITO/GO/MAPbI$_{3-x}$Cl$_x$/PC$_{61}$BM/ZnO/Al | 15.59 | 0.99 | 0.72 | 11.11 | 1050 |
| NGONRs | HTL | ITO/NGONR/MAPbI$_3$/ZnO NPs/Al | 17.93 | 1.00 | 0.72 | 12.94 | 1175 |
| Ammonia modified GO (GO:NH$_3$) | Interlayer at HTL/perovskite interface | ITO/PEDOT:PSS-GO:NH$_3$/MAPbI$_{3-x}$Cl$_x$/PC$_{61}$BM/Bphen/Ag. | 22.06 | 1.03 | 0.71 | 16.11 | 1178 |
| Organosulfate-G (oxo-G) | HTL | Oxo-G1/MAPbI$_3$/PC$_{60}$BM/Zn/Al | 18.06 | 1.08 | 0.78 | 15.2 | 1179 |
| GO | Interlayer at HTL/perovskite interface | FTO/cTiO$_2$/mTiO$_2$/MAPbI$_3$/GO/spiro-OMeTAD/Au | 20.2 | 1.04 | 0.73 | 15.1 | 1186 |
| Iodine reduced GO | Dopant for HTL | FTO/cTiO$_2$/mTiO$_2$/MAPbI$_3$/GO/spiro-OMeTAD/Au | 19.6 | 1.00 | 0.68 | 13.33 | 1187 |
| MoS$_2$ | Interlayer at HTL/perovskite interface | FTO/c TiO$_2$/mTiO$_2$/MAPbI$_3$/MoS$_2$/spiro-OMeTAD/Au | 21.5 | 0.93 | 0.67 | 13.3 | 1192 |
| Amino-functionalized GO (NGO) | Interlayer at HTL/perovskite interface | FTO/c TiO$_2$/mTiO$_2$/MAPbI$_3$/NGO/spiro-OMeTAD/Au | 23.6 | 0.94 | 0.66 | 14.6 | 1195 |
| (p-Methoxyphenyl -single walled carbon nanotube) SWCNT-PhOMe and RGO | Dopants for HTL | FTO/cTiO$_2$/mTiO$_2$/MAPbI$_3$/P3HT:SWCNT-PhOMe:RGO)/Au | 18.8 | 0.87 | 0.62 | 10.0 | 1200 |
| (a) (SLG) or (b) MLG | HTL and back electrode | (a) FTO/cTiO$_2$/mTiO$_2$/MAPbI$_3$/SLG (b) FTO/cTiO$_2$/mTiO$_2$/MAPbI$_3$/MLG | (a) 14.2 (b) 16.7 | (a) 0.88 (b) 0.94 | (a) 0.54 (b) 0.73 | (a) 6.7 (b) 11.5 | 1227 |
| RGO | Interlayer at HTL/back electrode interface |  | 23.24 | 1.112 | 0.782 | 20.4 | 1038 |
| MoS$_2$ | Interlayer at HTL/perovskite | ITO/PTAA/MoS$_2$/MAPbI$_{3-x}$Cl$_x$/PCBM/PFN/Al. | 20.71 | 1.011 | 0.784 | 16.89 | 1039 |
| Black phosphorous (BP) nanosheets | HTL | ITO/TiO$_2$/MoS$_2$/(spiro-OMeTAD:BP nanosheets)/Au | 20.22 | 1.06 | 0.761 | 16.4 | 1203 |
| Black phosphorous QDs (BPQDs) | Interlayer at HTL/perovskite | ITO/(PEDOT:PSS:PBQDs)/MAPbI$_3$/ZrAcAc-modified PCBM/Ag | 20.56 | 1.01 | 80.0 | 16.69 | 1204 |
| Ti$_3$C$_2$ MXenes | Interlayer at perovskite/CE interface | FTO/TiO$_2$/(Cs$_2$PbBr$_4$/MXenes/Carbon | 8.54 | 1.44 | 0.73 | 9.01 | 1226 |
| Graphene | HTL | FTO/TiO$_2$/FAMAPbI$_3$/(PEDOT:graphene)/Au | 21.32 | 0.79 | 0.518 | 8.79 | 1205 |
| Graphene | HTL | FTO/TiO$_2$/(FAPbI$_3$)$_{0.85}$(MAPbBr$_3$)$_{0.15}$/(PTh:graphene)/Au | 20.07 | 0.57 | 0.43 | 4.95 | 1205 |
| Phenyl acetylene silver [PAS]-functionalized MoS$_2$ | Dopant for HTL | FTO/(PEDOT:PSS:MoS$_2$)/MAPbI$_3$/PCBM/Ag | 24.035 | 0.998 | 0.686 | 16.47 | 1206 |
| RGO functionalized with 4-fluorophenyl-hydrazine hydrochloride (4FPH) | Interlayer at HTL/perovskite interface | ITO/TiO$_2$/MAPbI$_3$Cl$_{3-x}$/RGO-4FPH/spiro-OMeTAD/Au | 21.5 | 1.11 | 0.786 | 18.75 | 1197 |
| GO | Dopant for HTL | ITO/GO-doped PEDOT:PSS/(FAPbI$_3$)$_{0.85}$(MAPbBr$_3$)$_{0.15}$/PC$_{61}$BM/BCP/Ag | 20.01 | 0.90 | 0.79 | 14.20 | 1202 |
| GO | Interlayer at HTL/perovskite interface | FTO/TiO$_2$/MAPbI$_3$/GO/CuBuPC/Au | 20.9 | 1.04 | 0.66 | 14.4 | 1198 |
| GO | Interlayer at HTL/perovskite interface | FTO/PEDOT:PSS/GO/MAPbI$_3$/PCBM/Ag. | 21.92 | 0.94 | 0.748 | 15.34 | 1177 |
| MoS$_2$ | HTL | ITO/MoS$_2$/MAPbI$_3$/PCBM/Al | 12.60 | 0.84 | 0.57 | 6.01 | 1166 |
| RGO | Dopant for HTL | ITO/RGO:PEDOT:PSS/MAPbI$_3$/PCBM/BCP/Ag. | 16.75 | 0.87 | 0.75 | 10.7 | 1234 |
| MoS$_2$ | HTL | ITO/MOS$_2$/MAPbI$_3$/C$_{60}$/BCP/Al | 20.94 | 0.88 | 0.779 | 14.35 | 1174 |
| WS$_2$ | HTL | ITO/WS$_2$/MAPbI$_3$/C$_{60}$/BCP/Al | 21.22 | 0.97 | 0.73 | 15.00 | 1174 |
| GO | HTL | ITO/GO/MAPbI$_3$/C$_{60}$/BCP/Au | 21.6 | 1.00 | 0.762 | 16.5 | 1176 |
| N-Doped graphene frameworks (N-GFs) | Interlayer at HTL/perovskite interface | FTO/TiO$_2$/MAPbI$_3$/N-GF | 20.02 | 0.87 | 0.593 | 10.32 | 1176 |
| GO | Interlayer at HTL/perovskite interface | ITO/PEDOT:PSS/GO/MAPbI$_3$/PCBM/MoS$_2$/Ag | 22.834 | 1.135 | 0.738 | 19.14 | 1163 |
| GO | HTL and additive for perovskite | ITO/GO/MAPbI$_3$:GO/PCBM/Ag | 20.71 | 0.96 | 0.76 | 15.2 | 1135 |
| Carbon quantum dots (CQDs) | Interlayer at HTL/back electrode interface | FTO/NiMgLiO/MAPbI/NG:PCBM/CQDs/Ag | 19.68 | 1.07 | 0.75 | 15.57 | 1209 |
| MoS$_2$ flakes suspended in 2-propanol (LPE) | Interlayer at HTM/perovskite interface | glass/ITO/PTAA/MoS$_2$/perov/PCBM/BCP/Ag | 1.13 | 22.66 | 0.80 | 20.55 | 1160 |
| Antimonene | Interlayer at HTM/perovskite interface | ITO/PTAA/antimonene/CH$_3$NH$_3$PbI$_3$/PCBM/Bphen/Al | 1.114 | 23.52 | 0.768 | 20.11 | 1161 |
| Graphene | Dopant for HTL | FTO/SnO$_2$/TiO$_2$/perov/P3HT:graphene/Carbon | 1.09 | 22.5 | 0.74 | 18.2 | 1208 |
| MoS$_2$ QDs:RGO | Interlayer at HTM/perovskite interface | FTO/TiO$_2$/MAPbI$_3$/MoS$_2$ QDs:RGO/spiro-OMeTAD/Au | 1.11 | 22.81 | 0.79 | 20.12 | 1191 |

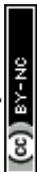

**Table 12 (continued)**

| Material | Usage | Device structure | Cell performance | | | | Ref. |
|---|---|---|---|---|---|---|---|
| | | | $J_{SC}$ [mA cm$^{-2}$] | $V_{OC}$ [V] | FF | $\eta$ [%] | |
| fMoS$_2$ | Interlayer at HTM/perovskite interface | FTO/TiO$_2$/Cs$_6$(MA$_{0.17}$FA$_{0.83}$)$_{(1-x)}$Pb(I$_{0.83}$Br$_{0.17}$)$_3$/fMoS$_2$/spiro-OMeTAD/Au | 1.14 | 22.76 | 0.74 | 19.2 | 720 |
| Graphene | Back electrode | FTO/SnO$_2$/Cs$_{0.05}$MA$_{0.16}$FA$_{0.79}$Pb(I$_{0.83}$Br$_{0.16}$)$_3$/spiro-OMeTAD/graphene/ITO | 1.05 | 22.78 | 0.75 | 18.65 | 1230 |
| WS$_2$ flakes | Interlayer at HTL/perovskite interface | FTO/cTiO$_2$/mTiO$_2$/MAPbI$_3$/Ti$_3$C$_2$ | 1.15 | 22.75 | 0.81 | 21.1 | 1162 |
| Ti$_3$C$_2$ MXenes | Back electrode | FTO/cTiO$_2$/mTiO$_2$/MAPbI$_3$/Ti$_3$C$_2$ | 0.89 | 13.78 | 0.64 | 7.78 | 1231 |
| Ti$_3$C$_2$ MXenes | Back electrode | FTO/cTiO$_2$/mTiO$_2$/MAPbI$_3$/Ti$_3$C$_2$ | 0.95 | 22.97 | 0.63 | 13.83 | 1232 |
| 2D black phosphorene (BP) | ETL/perovskite and HTL/perovskite Interlayers | FTO/cTiO$_2$/mTiO$_2$/BP/Cs$_{0.05}$MA$_{0.16}$FA$_{0.79}$Pb(I$_{0.83}$Br$_{0.17}$)$_3$/BP/spiro-OMeTAD/Au | 1.12 | 23.86 | 0.74 | 19.83 | 1188 |
| Graphene | Back electrode And ETL dopant | FTO/cTiO$_2$/graphene/mTiO$_2$/graphene/Cs$_{0.05}$(FA$_{0.83}$MA$_{0.15}$)$_{0.95}$Pb(I$_{0.85}$Br$_{0.15}$)$_3$ | 1.04 | 21.8 | 0.70 | 15.81 | 1229 |
| GO | Interlayer at HTL/perovskite interface | FTO/cTiO$_2$/mTiO$_2$/(CsPbI$_3$)$_{0.05}$(FAPbI$_3$)$_{0.05}$(MAPbBr$_3$)$_{0.05}$/spiro-OMeTAD/Au | 1.085 | 22.72 | 0.75 | 18.45 | 1199 |
| BPQDs | Interlayer at HTL/perovskite interface | FTO/cTiO$_2$/mTiO$_2$/(CsPbI$_3$)$_{0.05}$(FAPbI$_3$)$_{0.05}$(MAPbBr$_3$)$_{0.05}$/spiro-OMeTAD/Au | 1.085 | 23.12 | 0.75 | 18.86 | 1199 |
| MoS$_2$ | Interlayer at HTL/perovskite interface | FTO/cTiO$_2$/mTiO$_2$/(CsPbI$_3$)$_{0.05}$(FAPbI$_3$)$_{0.05}$(MAPbBr$_3$)$_{0.05}$/spiro-OMeTAD/Au | 1.095 | 23.05 | 0.75 | 19.03 | 1199 |
| BP nanosheets | Interlayer at HTL/perovskite interface | FTO/cTiO$_2$/mTiO$_2$/(CsPbI$_3$)$_{0.05}$(FAPbI$_3$)$_{0.05}$(MAPbBr$_3$)$_{0.05}$/spiro-OMeTAD/Au | 1.11 | 23.40 | 0.78 | 20.32 | 1199 |
| 2DP-TDB | HTL | ITO/SnO$_2$/FA$_{0.9}$MA$_{0.1}$PbI$_3$ + 4-fluorobenzamide hydrochloride (p-FPhFACl)/2DP-TDP/MoO$_x$/Ag | 1.16 | 24.02 | 0.796 | 22.17 | 1207 |

$^a$ MA = CH$_3$NH$_3$, $^b$ FA = HC(NH$_2$)$_2$.

surface wetting of Ag nanonetwork compared to PEDOT:PSS HTL were finely controlled *via* the reduction degree of GO flakes by means of a self-assembly approach at room temperature.[1241]

It is worth noting that the feasible realization of solution-processed conductive front electrode opens the route toward large-scale, low-cost fabrication of PSCs exclusively through R2R technologies.

Table 13 summarizes the main results achieved by PSCs using graphene-based front electrodes. Some representative results achieved using nonsolution-processed graphene (*i.e.*, CVD graphene) are also reported to facilitate comparison.

## 6.5 Two-dimensional/three-dimensional PSCs

As discussed in the introduction of Section 6, careful interface engineering between the perovskite active layer and CTL (ETL/HTL) is pivotal to push device development and optimization. Beyond interface engineering with graphene and other GRMs, the use of a layered perovskite, namely, 2D perovskites, has recently attracted a lot of interest.[1243,1244] This is motivated by their superior stability against moisture, far exceeding those of their standard 3D parent structures.[111,1242] Layered perovskites usually possess the general structure R$_2$A$_{n-1}$B$_n$X$_{3n+1}$, where A, B, and X are the organic cation, metal cation, and halide anion typically forming the 3D perovskites, while R is a large organic cation (for example, aliphatic or aromatic alkylammonium), functioning as a spacer between the inorganic layers. In the structure, $n$ determines the number of inorganic sheets held together (Fig. 25a). By controlling the A/R ratio, the $n$ value could be adjusted from $n$ = 1 (2D), $n$ > 1 (quasi-2D), and $n$ = $\infty$ (3D).[111,1016,1040–1042,1245] For low $n$, 2D perovskites have large $E_g$ and stable excitons with large binding energy and limited transport through the organic spacer.[110,111] Such properties limit the PV effect, leading to poor performances in SCs.[111,1246] By increasing $n$, the device $\eta$ improves (Fig. 25b) in concomitance with a decrease in bandgap and improvement in charge transport across the inorganic layers, reaching values up to 17%.[111] A large family of R cations can be inserted forming a layered 2D material, as shown in Fig. 25c. This family includes, for example, an organic cation designed *ad hoc* with additional functional groups or atom (such as fluorine moieties) to enhance the water repellent characteristics of the material.[1247,1248] Importantly, compared to 3D perovskites, 2D perovskites show remarkably higher moisture resistance, due to the hydrophobic nature of the R cation, as well as the highly oriented crystalline structure and dense packing.[1246,1247] These properties reduce the possibility of direct contact of water or oxygen molecules within the perovskite grain boundaries.[1246,1247]

The integration of 2D perovskites into PSCs as a stabilizer component has become increasingly popular in the last few years as a tool to increase the lifetimes of PSCs. Beyond that, a 2D perovskite also functions as a surface passivation layer, significantly improving the device $V_{OC}$.[1249,1250] In detail, the most common approach intends to combine the high efficiency of 3D perovskites with the superior stability of 2D perovskites by means of synergistic interface functionalization. This has been demonstrated by either mixing the 2D and 3D precursors together[110,111] or by engineering a layer-by-layer deposition method to obtain a clean 2D/3D vertical bilayer architecture.[966,1041,1042,1251,1252] The top 2D perovskite layers can simultaneously act as surface passivators, improving the surface robustness and hydrophobic character of the active layer, while also reducing surface charge recombination, ultimately improving the device open-circuit voltage.[1040,1041,1059,1247,1253] Cho et al. developed a method for the deposition of a 3D/2D bilayer composed of mixed halide perovskites and (PEAI)$_2$PbI$_4$ (PEAI = phenethylammonium iodide).[1251] The layer-by-layer growth is induced by the spin coating of PEAI in an isopropanol solution on the mixed halide 3D perovskite with PbI$_2$ excess.[1251] The PbI$_2$ excess has been demonstrated to segregate on top of the 3D perovskite, reacting in situ with PEAI at the top surface and forming a thin 2D layer on top of the 3D material (the model of the device architecture is shown in Fig. 25d and e).

**Table 13** Summary of the PV performance of PSCs using graphene-based front electrodes[ab]

| Material | Usage | Device structure | Cell performance | | | | Ref. |
|---|---|---|---|---|---|---|---|
| | | | $J_{SC}$ | $V_{OC}$ | FF | $\eta$ | |
| CVD graphene (CVD-G)(not solution-processed) | Transparent back electrode | FTO/cTiO$_2$/MAPbI$_{3-x}$Cl$_x$/spiro/PEDOT:PSS:sorbitol/PDMS-PMMA-CVD-G | 19.17 | 0.96 | 0.67 | 12.37 | 1210 |
| CVD-graphene (CVD-G)(not solution-processed) | Front electrode | CVD-G/molybdenum trioxide (MoO$_3$)/PEDOT:PSS/MAPbI$_3$/C$_{60}$/BCP/LiF/Al | 21.9 | 1.03 | 0.72 | 17.1 | 1239 |
| Single-layer graphene (SLG)(not solution-processed) | Bottom contact for top cell in tandem configuration | Bottom cell:crystalline silicon top cell:SLG/PEDOT:PSS/spiro-OMeTAD/MAPbI$_{3-x}$Cl$_x$/TiO$_2$/FTO/glass | 21.9 (Top cel from FTO side) | 0.96 (Top cel from FTO side) | 0.56 (Top cel from FTO side) | 11.8 (Top cel from FTO side) | 1272 |
| Nano-composite of silver nano-network and GO | Front electrode | Substrate/Ag nanonetwork/GO/PEDOT:PSS/MAPbI$_3$/PCBM/PFN-P1/Ag | 13.78 | 0.94 | 0.71 | 9.23 | 1241 |
| Graphene:ethyl cellulose (G:EC) | Front electrode | Substrate/G:EC transparent electrode/perovskite/PCBM/Ag | 1.06 | 20.68 | 0.77 | 16.93 | 1240 |

[a] MA = CH$_3$NH$_3$. [b] FA = [HC(NH$_2$)$_2$].

Since the 2D perovskite lies on the top surface at the interface with the HTM, the interfacial charge carrier recombination is reduced, increasing $\eta$ to values higher than 20%.[1251] More recently, Jung et al. reported a double-layered halide architecture incorporating an ultrathin wide-bandgap halide stacked onto a narrow-bandgap halide light-absorbing layer. This layer effectively reduced charge recombinations at the perovskite/P3HT interface, resulting in an $\eta$ value of around 23% and long-term operational stability.[966] In addition to improving the surface robustness, imparting hydrophobicity, and passivating the surface, it has been recently demonstrated that the 2D overlayer is also crucial in preventing ion diffusion at the interface with the HTM.[1040,1041] Sutanto et al. indeed observed a slower evolution (timescale of months) of the PV characteristics of 2D/3D PSCs using thiophene alkylammonium-based

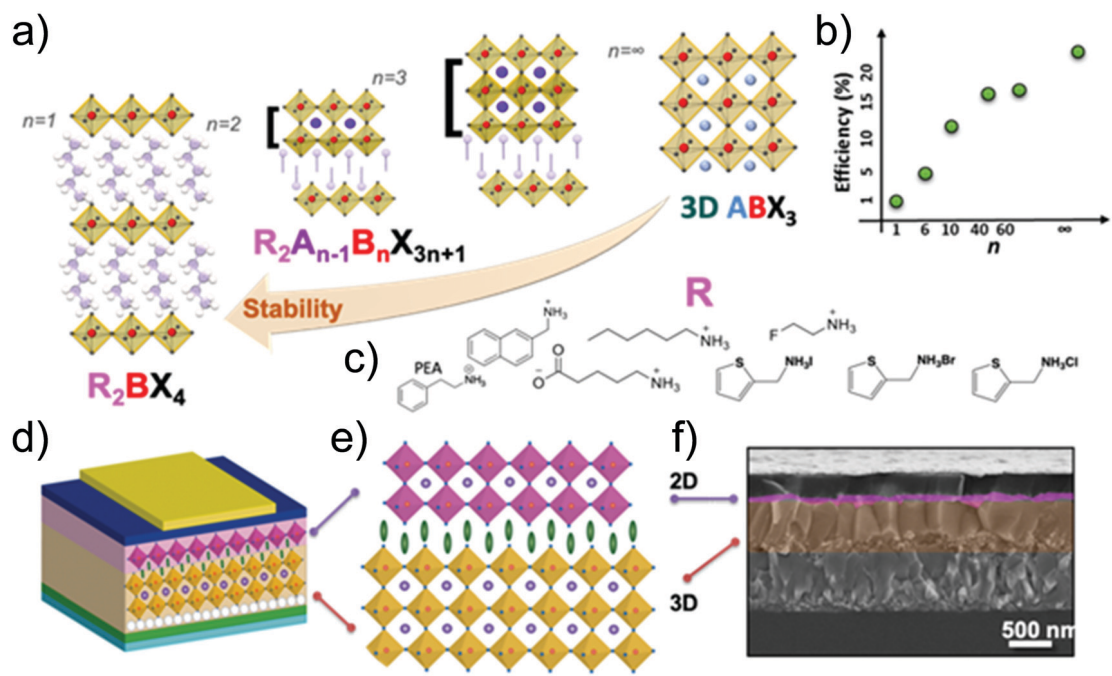

**Fig. 25** (a) Structure of perovskite from 2D to 3D forms. (b) $\eta$ versus $n$. C) Typical cations used as R for the 2D perovskite formulation. (d) Structure of an n–i–p PSC based on FTO; cTiO$_2$, mTiO$_2$, 3D/2D perovskite, organic HTM, and Au CE. (e) Schematic of the 3D/2D active perovskite layer. (f) Cross-sectional SEM image of the 2D/3D device.

organic cations as the building blocks for the 2D perovskite (Fig. 25c).[1040,1041] A boost in $\eta$ has been associated with the slow structural rearrangement of the 2D/3D interface, which depends on the "softness" of the 2D perovskite overlayer that can act as an ion scavenger.[1040] Because of the movement of ions in the 3D perovskite, small MA cations accumulate at the interface.[1040] The 2D structure can incorporate the MA cations by altering its pristine layered structure into a mixed (or quasi-2D) phase.[1040] In addition, a "more robust" 2D layer can prevent such structural changes, mechanically blocking the movement of ions.[1040] This ion blockage leads to a dramatic increase in device stability, while maintaining high device $\eta$.[1040] In addition, these 2D modifiers also dramatically improve the thermal stability of PSCs,[1040] demonstrating that a conscious choice of proper 2D components can control the structural, physical, and energetic properties of the 2D/3D interfaces, a key element to be controlled for the design and realization of efficient and stable devices.[11]While defining the interface structure–function relationship is of utmost importance to control ion and charge accumulation and dynamical effects, the exact knowledge on interface energetics is also pivotal. To this end, it has been recently demonstrated that such thiophene-based cations form a p–n junction at the 2D/3D interface, which is the key to enable efficient charge transfer. As a consequence, electron accumulation at the interface is reduced, nullifying interfacial recombinations. This beneficial effect is reflected in the device $V_{OC}$, which reached 1.19 V, among the highest reported so far in the literature.[1254] As an illustrative example, an intact 2D/3D heterojunction, realized by growing a stable and highly crystalline 2D ($C_4H_9NH_3)_2PbI_4$ film on top of a 3D perovskite (using a solvent-free solid-phase in-plane growth), reached a certified steady-state $\eta$ of 24.35%, while retaining 94% of its initial $\eta$ after 1056 h under the damp heat test (85 °C/85% relative humidity) and 98% after 1620 h under 1 sun illumination (without any encapsulation).[1255] Meanwhile, substantial progresses have been recently achieved in controlling the film formation of 2D perovskites.[1256] For example, an $\eta$ value of 15.81% has been achieved using hot-cast Dion–Jacobson 2D perovskite ($(PDMA)(MA)_{n-1}Pb_nI_{3n+1}$ (PDMA = 1,4-phenylenedimethanammonium; $\langle n \rangle = 4$)) as the photoactive layer.[1255] Moreover, by elucidating the critical role of additives in regulating the nucleation and crystallization kinetics of 2D $(PEA)_2(MA)_3Pb4I_{13}$ films with low trap states and desired carrier transport/collection properties, Yang et al. recently achieved an $\eta$ value up to 18.5%, together with FF of 83.4%.[1257]

## 6.6 Tandem SCs based on PSCs

The tunability of the $E_g$ value of perovskites via halide replacement[1258] or cation exchange[1259] and their high absorption coefficient across the entire visible range[1260] make these materials attractive for tandem SCs, particularly in combination with Si sub-cells.

In a Si/perovskite tandem configuration, higher-energy photons are absorbed by the perovskite sub-cell, while infrared photons are transmitted through the perovskite top cell and absorbed by the Si sub-cell, covering a wide absorption spectral range defined by the $E_g$ value of Si.[1261,1262]

Therefore, the perovskite-based tandem configurations require the stacking of constituent sub-cells, with the perovskite top cell having two transparent electrodes, one of them directly processed on top of the charge selective layer (e.g., spiro-OMeTAD).[1263,1264] Both high $\sigma$ and optimal $T_r$ of the top-cell transparent electrode are the key requirements for the successful design/realization of tandem devices. Conventional TCOs optically optimized for single-junction devices cannot be easily deposited onto the perovskite top cell due to the ion bombardment-induced degradation of the underlying materials during TCO sputtering.[1265,1266] A strategy to minimize the underlying material damage is the deposition of additional buffer layers, which can absorb the energy impact of ions crashing on the device. Either thermally evaporated sub-stoichiometric molybdenum oxide ($MoO_x$) buffer layers[1267,1268] or ultrathin layers of Au[1269] have been reported to protect spiro-OMeTAD during TCO sputtering. However, the aforementioned strategies inevitably add complexity to the perovskite top-cell fabrication process or cause additional optical losses. Moreover, the simplest solution offered by the $MoO_x$ buffer raises concerns on long-term stability, since the iodide of the perovskite layer can chemically react with $MoO_x$, resulting in an unfavorable interface energy-level alignment for hole extraction.[1270]

In order to address these challenges, transparent graphene-based electrodes are promising for the realization of efficient and stable bifacial PSCs. Although graphene-based electrodes for PSC-based tandem SCs produced by solution-processed methods are still missing, several groups already reported their practical implementation through other techniques, such as CVD. For example, Lang and co-workers addressed this challenge by implementing large-area CVD-graphene as a highly transparent photoelectrode in a perovskite top-cell.[1271] In fact, the electrodes based on graphene combined an excellent $T_r$ (97.4%) with $R_s$ of 100 Ω □$^{-1}$.[1272] Zhou and co-workers demonstrated two-layer CVD graphene as a transparent contact for a top cell based on a Cl-doped perovskite film with a bandgap of 1.59 eV.[1273] The graphene electrodes permitted to achieve a top-cell $\eta$ of 11.8%, resulting in a tandem SC with an $\eta$ of 18.1%.[1272]

Even though solution-processed 2D material-based recombination layers or transparent conductive contacts have not been reported in tandem devices yet, the use of graphene in the ETL of the perovskite top-cell was recently reported in a two-terminal (2T) mechanically stacked Si/perovskite tandem SCs.[1274] With this approach, the sub-cells were fabricated and independently optimized and subsequently coupled by contacting the back electrode of the mesoscopic perovskite top-cell with the texturized and metallized front contact of the silicon bottom cell.[1273] Then, the graphene-doped mesoporous ETL used in the perovskite top-cell allowed the tandem SCs to improve their $\eta$ up to 26.3% over an active area of 1.43 cm$^2$.[1273] Overall, the "mechanical approach," based on the independent optimization and fabrication of sub-cells, as well as graphene-based top cell, is ready to synergistically exploit the most recent progress achieved in both PSCs and Si cells in order to boost perovskite/silicon tandem SCs beyond current PV technology established in the market.[1273]

### 6.7 Summary and outlook

PSCs are an exciting PV technology aiming to enter a massive market. In fact, they can be produced through scalable and cost-effective solution-based techniques compatible with R2R and S2S manufacturing processes,[1275] while reaching outstanding $\eta$ up to certified values of 25.2%.[897] This value approach to the record-high $\eta$ of monocrystalline and HIT Si SCs (26.1 and 26.7%, respectively),[897] even superior to those of thin-film PV technologies, such as CdTe and copper indium gallium selenide (CIGS) SCs (22.1% and 23.4%, respectively).[897] Furthermore, perovskite-based tandem SCs, namely, perovskite-Si tandem SCs, have reached certified $\eta$ up to 29.1%,[897] which is as high as the value of costly GaAs SCs (that holds the record-high certified $\eta$ for single-junction cells).[897] Prospectively, the LCOE of perovskite solar panels has been estimated to be lower than 5 US cents kW h$^{-1}$.[131–133] This value is competitive with the LCOEs of fossil fuels,[134,135] thus enabling the achievement of grid parity. However, the instability of photoactive perovskites[1001,1276–1278] and CTLs[1001,1277] represents the main technical barrier for PSC technology. In this scenario, the use of solution-processed 2D materials in PSCs demonstrated exciting results in resolving current PSC issues, boosting both stability and $\eta$ by means of scalable and cost-effective strategies.[1037,1279,1280] These advances can be ascribed to the progresses achieved in the preparation of 2D material inks and their large-area (i.e., wafer-scale) printing,[205,297,1281,1282] addressing controllable optoelectronic properties to be exploited in PSC structures.[1192] By formulating 2D material-based inks in solvents compatible with materials composing the PSCs, GRMs have been successfully integrated as both CTLs and interlayers,[1283] improving the charge collection (while providing effective barriers against humidity) and migration of ions within the PSC structures.[1037,1284,1285] Graphene and its derivatives have also been investigated to form efficient back electrodes (CE) as an alternative to Au or Ag.[1217,1227] Beside reaching relevant $\eta$ up to 18.65%,[1230] the use of metal-free back electrodes eliminates the degrading reaction between the perovskite layer and Au and Ag, which are the cause of device instability.[1219,1286,1287] Recent advances in low-temperature processable graphene inks are promising for the realization of paintable C-PSCs based on structures that achieved record-high $\eta$ using metal-based back electrodes.[1217] Importantly, graphene and other GRMs (e.g., MoS$_2$) can also regulate the perovskite crystal over both mesoscopic scaffolds and planar CTLs,[1104,1160] increasing the reproducibility of high-$\eta$ devices. Besides, pristine FLG flakes were used to stabilize perovskite films, slowing down charge thermalization.[1106] The realization of 2D material-enabled hot-carrier extraction and collection paves the way for the creation of advanced SC concepts, which are still unexplored.[1106]

Despite the implementation of GRMs has not been reported yet for PSCs showing state-of-the-art $\eta$, outstanding results have been achieved in large-area PSCs and PSMs.[1192] The deposition of 2D material-based inks by means of printing techniques, such as slot-dye coating, blade coating, spray coating, and screen printing have been established in a wide range of applications, including energy storage and conversion systems beyond SCs.[5,207,1288] For the case of PSCs, the printing processes of 2D materials can be easily customized and optimized in combination with a protective layer on top of the perovskite absorber, such as 2D perovskite[1289] or polymeric interlayer (e.g., PMMA).[1290] As a striking example of PSC scale-up, some authors of the present review (belonging to University of Rome Tor Vergata, Hellenic Mediterranean University, Italian Institute of Technology, and BeDimensional S.p.A.), in collaboration with the industrial partner GreatCell Solar, realized the first example of real-time characterized, standalone 2D-material-enabled perovskite solar farm. The latter was installed in 2020 in Heraklion (Crete), a site with favorable climate conditions. Initially, it comprised 9 solar panels, each one with an active area of 0.32 m$^2$ (Fig. 26). According to the planned activity, other panels will be integrated into the solar farm, and the output of the solar farm will be continuously monitored, providing a clear understanding of (1) the correlation of environmental conditions with the outdoor performance of solar panels and (2) the benchmarking of 2D material-based perovskite solar panels against conventional PV technologies (Si, CdTe, and CIGS). The preliminary data, provided to the Commission of the European Union (project founders),[1291] revealed the key advantages of 2D materials in providing PV FoMs competitive in the market, bringing PSC commercialization closer to reality. It is noteworthy that under the umbrella of European Graphene Flagship, the solar farm project has been recently extended to a graphene-integrated perovskite–silicon tandem SC technology, involving a key player of the PV industry, namely, Enel Green Power and Siemens.[1292,1293] Not by chance, the results achieved using 2D materials on single-junction PSCs have already been exploited in perovskite-based tandem SCs, namely, a perovskite–Si tandem device.[1273] In particular, the doping of TiO$_2$-based ETLs of PSCs with graphene flakes enabled the tandem devices to reach $\eta$ over 26%.[1273] Nevertheless, the incorporation of GRMs in perovskite-based tandem devices is still at a premature stage. Prospectively, solution-processed graphene and other metallic 2D materials can play a major role in developing advanced interconnecting layers with a satisfactory trade-off between optical transparency and electrical conductivity.

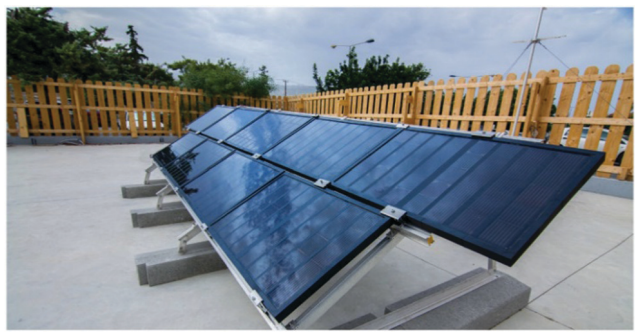

**Fig. 26** 2D materials–enabled perovskite solar farm installed in Heraklion (Crete).

Lastly, it is worth noting that 2D materials can play a relevant role in developing new encapsulation strategy for perovskite devices, which are particularly sensitive to oxygen, moisture, and volatilization of internal species (*i.e.*, decomposition products and dopants).[1294,1295] For example, a recent work demonstrated a cost-effective and scalable flexible transparent lamination encapsulation method based on graphene films with a PDMS buffer on a PET substrate.[1296] Moreover, the impermeability of graphene or other related materials can be successfully exploited to create novel encapsulants or edge sealers, decreasing the water vapor transmission rate (WVTR) or oxygen transmission rate (OTR) of the current encapsulants used in PSCs, as well as other PV technologies.

Overall, 2D materials are expected to play protagonists in the optimization of perovskite-based PV technology, which could represent a game changer in the PV market for the near future.

# 7. Other SCs

## 7.1 QDSCs

QDSCs are an attractive PV technology owing to various advantages,[1297] such as cost-effectiveness and simple device manufacturing processes.[57,1298–1307] As comprehensively discussed in recent review articles (for example, ref. 1296), such SCs are based on photoactive semiconductors (organic, inorganic, or hybrid) QD films, which act as both absorbers and charge transporting media. Different types of QDSC architectures have been proposed: (1) Schottky QDSCs, which consist of a heterojunction between a planar film of p-type colloidal QDs and a shallow-$\Phi_w$ metal, which produce a Schottky barrier generating a depletion region for carrier separation;[1308–1313] (2) depleted heterojunction QDSCs, which use a highly doped n-type metal oxide (typically, $TiO_2$ or ZnO, but even metal chalcogenides, *e.g.*, CdS) in a p–n heterojunction with a p-type QD film;[1314–1316] (3) heterojunction QDSCs, also referred to as QD-sensitized solar cells (QDSSCs), in which the n-type wideband-gap semiconductor and QD film form an interpenetrating layer.[1296,1301,1317–1325] This structure is usually obtained by infiltrating QDs into the structured n-type semiconductors. Since this architecture resembles that of DSSCs, such cells are often referred to as QD-based DSSCs (QDDSSCs) (see Section 5); (4) quantum junction QDSCs, which consist of a homojunction-like architecture where both p- and n-type materials of the junction are composed of QDs;[1326] (5) bulk nanoheterojunction SCs in which an n-type material and p-type QDs are mixed similar to a BHJ architecture.[1327]

The optoelectronic properties of semiconductor QDs, *e.g.*, $E_g$, optical absorption coefficient ($\alpha$), and charge carrier transport, can be effectively tuned by modulating their size and shape,[1298–1304,1328] offering versatile systems to be used in graded doping architectures[1329–1331] and multijunction (tandem) SCs.[1332,1333] Initially, chalcogenide semiconductors, such as CdX and PbX (X = S, Se, and Te), have been used for QDSCs due to their ability to harvest light in the visible and IR regions and their low cost.[1300,1304–1306,1327,1334] However, the limited $\eta$ achieved with these inorganic QDs drove researchers to design novel QDs, including inorganic alloys, organic, and organic–inorganic hybrid QDs with superior PV capabilities.[1304–1306,1327] Therefore, over the past decade, QDSCs have seen rapid improvements, until reaching a certified $\eta$ value of 16.6% with mixed Cs and formamidinium lead triiodide perovskite system[1335] (previous record was 13.4%).[1336] These important results are the fruits of progress achieved in both control of the QD surface chemistry and the understanding of device physics,[1305,1306,1337] and they are now leading QDSCs toward commercialization.[1305,1306]

Despite recent progresses, the record-high $\eta$ of the QDSCs is still far from their theoretical maximum $\eta$, which is as high as 33%[1338] (or 44%, depending on whether or not multiple exciton generation of the QDs is considered).[1337,1339] Actually, the major issue in QDSCs is the presence of structural defects or unpassivated states on the QD surface, which leads to recombination reactions limiting the overall performance of the devices.[1340–1347] To resolve this issue, several strategies, including the implementation of atomic ligand/anionic passivation schemes,[1348–1350] use of passivation layers over QD films,[1351–1355] and design of core–shell structures,[1356–1363] have been developed in various type of QDSCs. For example, a hybrid passivation scheme, which introduces halide anions during the end stages of the QD synthesis process, was used to realize depleted heterojunction QDSCs with a certified $\eta$ value of 7.0%.[1347] Sequential inorganic ZnS/$SiO_2$ double-layer treatment onto the QD-sensitized photoanode strongly inhibited the interfacial recombination processes in QDSSCs, which reached a certified $\eta$ value of 8.21%.[1354] CdSeTe/CdS type-I core–shell QDSSCs, obtained by overcoating CdS shells around CdSeTe-core QDs, achieved an $\eta$ value of 9.48%.[1357] Binary QD films have also been investigated in heterojunction QDSCs in order to improve the charge separation using p–n junctions at the nanoscale, while passivating possible surface defects of QDs.[1326,1364,1365] In addition, such junctions enabled the dissociation of excitons in free carriers, drastically reducing bimolecular recombination processes.[1326,1363,1364] The use of mixed QD films was targeted to extend the carrier diffusion length, allowing thicknesses of the photoactive films to become comparable to the optical absorption length.[1363,1364] However, binary QD systems have limitations in simultaneously controlling the $E_g$ value as well as CB and VB edges for both charge photogeneration and collection. Hence, to overcome the limitations of binary QDs, alloy QDs[1366,1367] and hybrid organic–inorganic QDs[1334,1368,1369] have been successfully proposed together with the abovementioned strategy to passivate surface defects. For example, Du *et al.* reported a Zn–Cu–In–Se-alloyed QD sensitizer to construct Pb- and Cd-free QDSSCs with a certified $\eta$ value of 11.61%.[1370] Very recently, the $Cs_{1-x}FA_xPbI_3$ system in the form of QDs enabled the realization of QDSSCs with a certified record $\eta$ of 16.6%, together with superior stability (94% of the original $\eta$ under continuous 1 sun illumination for 600 h) compared with their thin-film counterpart.[1334]

In addition to the aforementioned strategies, the engineering of various QDSC configurations through the introduction of interfacial layers and doping of components is crucial to improve the charge extraction and transport from the photoactive layer to the metal contacts, thereby achieving performance rivaling those of other PV technologies. Especially in this context, solution-processed graphene and other 2D materials have attracted a primary interest for QDSCs. Tavakoli *et al.* reported an *in situ* solution-based process to prepare hybrid ZnO/graphene QDs (Fig. 27a and b), where the graphene shell quenches the PL intensity of ZnO nanocrystals (size of NPs: 5 nm) by ~72%, primarily due to charge transfer and static quenching.[1371] This nanocomposite was used as a CE material in a PbS/TiO₂ depleted heterojunction QDSCs, which achieved an $\eta$ value of up to 4.5%.[1370] Fig. 27c shows a schematic of the architecture of the device, in which fast electron extraction is achieved by means of ZnO–graphene CE (Fig. 27d).[1370] In particular, the band diagram of device shows the electron extraction process from PbS to ZnO–graphene-coated TiO₂ (Fig. 27e).[1370]

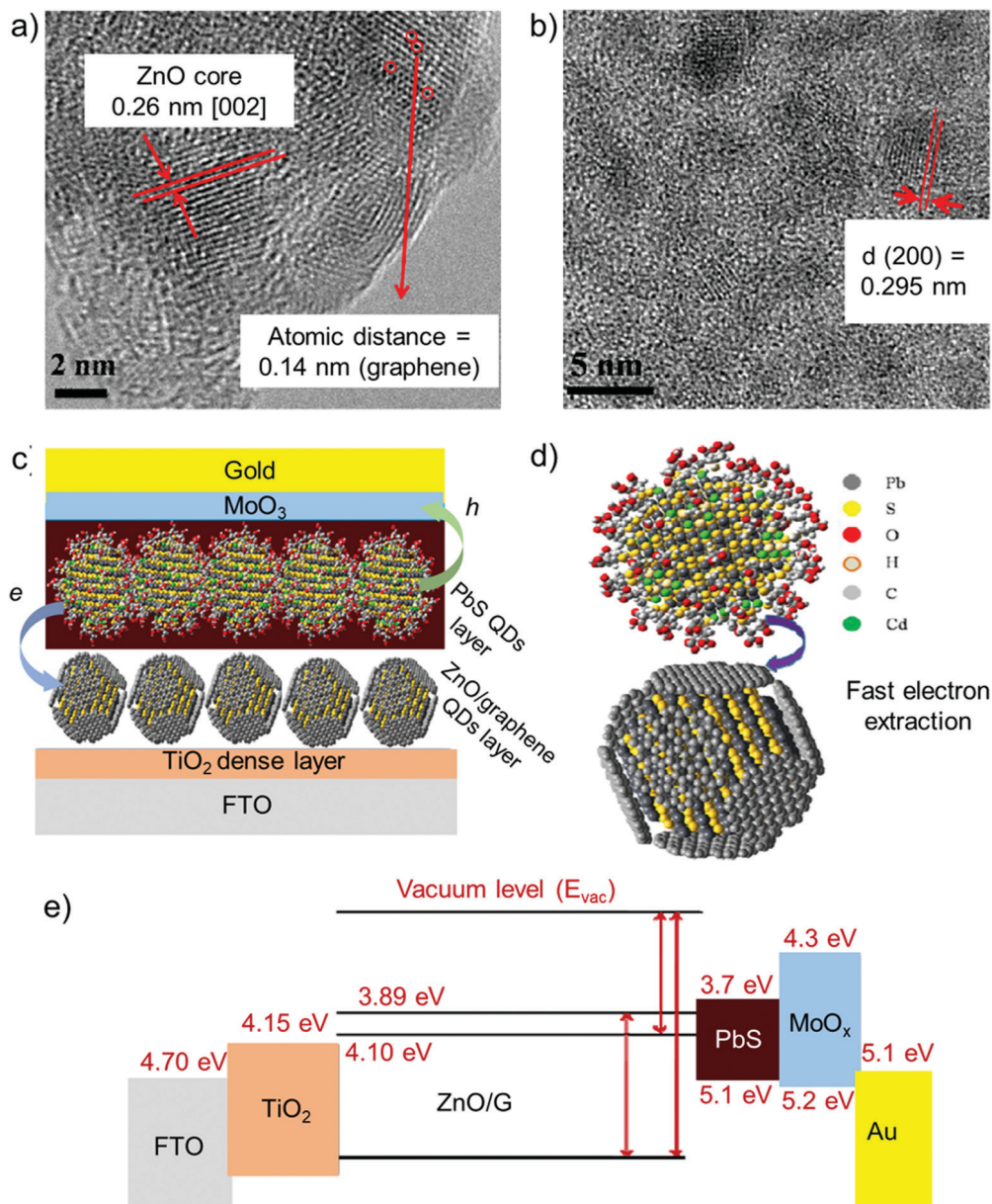

**Fig. 27** High-resolution TEM images of (a) graphene QDs and (b) ZnO/graphene QDs. (c) PbS QD and ZnO/graphene QD-based depleted heterojunction QDSCs. (d) Schematic of the electron extraction process from PbS QD to ZnO/graphene QD. (e) Energy-level diagram of PbS QD- and ZnO/graphene QD-based depleted heterojunction QDSCs (G = graphene). Adapted from ref. 1370.

The authors explained their results by suggesting efficient electron injection from the CB of ZnO QDs to the LUMO levels of graphene, which occurs through Zn–O–C bonding, and slow electron recombination in the presence of ZnO-graphene buffer layer.[1370] Graphene frameworks were incorporated into the TiO₂ photoanode as an electron transport medium to improve the PV performance of QDSSCs (up to an $\eta$ value of 4.2%) owing to their excellent conductivity and isotropic framework structure.[1372]

Kim *et al.* reported the use of a hierarchical ZnO nanostructure array, produced by a two-step solution reaction and composed of nanosheet branched ZnO nanorods as an efficient anode for QDSSCs.[1373] This 2D (nanosheet)–1D (nanorod) combined hierarchical ZnO nanostructure considerably enhanced light capture compared with ZnO thin films and ZnO nanorods, allowing the corresponding CdSe/CdS-based QDSSCs to achieve an $\eta$ value of 4.4%.[1372]

Recently, 2D MoS$_2$ nanosheets were used as an efficient HTL for PbS-based depleted heterojunction QDSCs.[1374] All-solution-processed n–p–p$^+$ architecture was fabricated by sequentially depositing ZnO NPs, PbS QDs, and 2D MoS$_2$ nanosheets acting as n-, p-, and p$^+$-type layers, respectively.[1373] The incorporation of MoS$_2$ HTL improved the $\eta$ value from 0.92% (in the free-MoS$_2$ reference) to 2.48%.[1373]

Noteworthily, 2D MoS$_2$ has recently been coupled to Sn-doped In$_2$O$_3$ nanocrystals to collect holes from the latter and driving permanent charge separation across a novel type of ultrathin solid-state 0D/2D hybrid interface that can store light in the contactless mode.[1375] Therefore, these results further prove the potential of MoS$_2$ as the HTL in QD-based optical devices.

Jin *et al.* reported graphdiyne, which is a π-conjugated structure consisting of sp$^2$- and sp-hybridized carbons in a typical 2D configuration, as a potential solution-processed hole transporter for PbS-based QDSCs, which reached an $\eta$ value of 10.64%.[1376] The use of graphdiyne-based anode buffer layer improved hole extraction from the QDs to Au anodes, while providing long-term shelf-life stability over 120 days.[1375]

Dangling bond-free 2D h-BN with self-terminated atomic planes, produced through LPE in 2-propanol, was used to passivate the TiO$_2$ surface in CdSe-based QDSCs.[1377] By decreasing the recombination rate at the TiO$_2$/CdSe interface, the resulting QDSCs achieved an $\eta$ value of 7%, corresponding to a 46% improvement in $\eta$ exhibited by the h-BN-free reference.[1376]

In addition to the aforementioned roles of 2D materials in QDSCs, liquid-phase synthetized antimonene QDs have been applied as the photoactive layer in QDSSCs.[1378] Owing to their strong light–matter interaction, moderate $E_g$ for an optimal absorption in the visible spectrum, and antioxidation properties, antimonene QDs enabled the realization of QDSSCs with an $\eta$ value up to 3.07%.[1377] Moreover, the as-fabricated SCs have shown long-term stability, retaining more than 90% of the initial $\eta$ after 1000 h.[1377] Therefore, antimonene QDs, as well as other 2D material-derived QDs, may provide a new pathway for a novel kind of cost-effective solution-processed QDSCs.[1377]

Although the above examples clearly indicate that 2D materials can play a significant role in further improving the performance of QDSCs, their use in this type of SCs is not strongly established as those for OSCs, DSSCs, and PSCs. However, both the advent of a novel type of efficient QDSCs and successful implementation of 2D materials in other PV devices can provide the fundamentals for the future establishment of 2D material-enabled efficient QDSCs.

## 7.2 Organic–inorganic hybrid SCs

Organic–inorganic hybrid SCs combine organic and inorganic materials as the photoactive material. As discussed in previous reviews in the literature,[1379–1385] the rationale of this combination is to implement the advantages offered by both OSCs and inorganic components. As proposed for OSCs (Section 3), organic materials are solution-processable and thus compatible with low-cost and high-throughput deposition methods, including R2R printing techniques. Moreover, they have high $\sigma$ in the visible spectrum. Thus, they allow their thin (thickness of a few hundred nanometers) films to efficiently absorb solar light. Meanwhile, inorganic materials can be formulated in the form of solution-processable nanocrystals with tunable optoelectronic properties, as shown in Section 7.1 for QDSCs. Furthermore, they have a large dielectric constant (*e.g.*, ∼10.4 for CdSe),[1386] which decreases the Coulombic attraction between electrons and holes, facilitating their separation in free charges. Thus, when mixed with organic photoactive components, they can provide an interfacial force driving the dissociation of excitons generated in the organic materials in free charge.[1387–1390] Therefore, inorganic nanocrystals can act as ideal acceptor materials in BHJ OSC-like devices using either organic polymers or conjugated small molecules as the donors.[1378–1384,1391]

The first hybrid SC was reported in 1996 using CdSe nanodots as the acceptor and poly[2-methoxy-5-(2′-ethylhexyloxy)-p-phenylene vinylene] (MEH-PPV) as the donor.[1392] However, the corresponding $\eta$ was low (<0.5%) as a consequence of the poor charge transport through the CdSe nanodots.[1391] Thereafter, much effort has been devoted toward improving charge transport by tuning the nanocrystal shapes.[1393–1395] Studies on QDSCs also helped to rapidly advance hybrid SCs.[1396] In 2011, Ren *et al.* reached an $\eta$ value of 4.1% with hybrid SCs based on P3HT and CdS nanocrystals as the donor and acceptor, respectively.[1397] More recently, hybrid SCs based on Si as the inorganic component have drawn relevant attention due to their room-temperature, facile, and cost-effective fabrication processes, which is promising to lower the cost of conventional Si SCs.[1398–1400] Owing to advances in the synthesis of organic materials and design of novel device structures, hybrid SCs based on n-type Si substrate achieved an $\eta$ higher than 16%,[1401,1402] (record value of 17.4%).[1403] Despite the aforementioned results, the $\eta$ value of hybrid SCs is still insufficient to compete with conventional inorganic Si SCs and PSCs (Section 6). Moreover, the stability of hybrid SCs is also limited compared to conventional inorganic PV technology.[1404–1407] These drawbacks are strongly hindering the commercialization of hybrid SCs at a large scale. In this context, the incorporation of GRMs can help resolve both $\eta$ and stability limits of hybrid SCs. For example, RGO has been proposed to produce a buffer layer in hybrid SCs to improve the light-induced charge extraction of ∼50%, as well as to replace the PEDOT:PSS contact.[1408] Recently, GQDs were mixed with PEDOT to be used in hybrid SCs using PEDOT:GQDs/porous Si/n-Si/TiO$_x$ structure.[1409] In detail, GQDs improved the conductivity of PEDOT, porous Si reduced the overall reflectivity, and TiO$_x$ acted as a passivation layer to reduce the recombination layer.[1408] The as-produced devices reached a maximum $\eta$ value of 10.49%, retaining 78% of the initial $\eta$ under ambient conditions for 15 days.[1408] GQDs, produced through a top-down strategy based on laser fragmentation in a post-hydrothermal treatment, were also used as a buffer layer between TiO$_2$ and P3HT to form a cascade energy-level scheme in hybrid SCs.[1410] The introduction of GQDs into a BHJ hybrid SC led to the enhancement of $\eta$ from 2.04% to 3.16%.[1409]

Although the aforementioned examples demonstrated the potential of GRMs in hybrid SCs, further studies are needed to formulate 2D materials in overcoming the fundamental issues exhibited by such type of SCs. Both progresses of 2D material science and hybrid SC-related technology could help for an in-depth reconsideration of 2D material-enabled hybrid SCs.

# 8. Outlook and conclusions

Several progresses have been achieved in the use of graphene and related 2D materials (GRMs) in solution-processed PVs. Regarding TCE applications, the implementation of solution-processed 2D materials is still at a premature stage. In fact, solution-processed graphene-based films typically exhibit sheet resistance ($R_s$) values in the order of kΩ sq$^{-1}$ (for $T_r \geq$ 80%),[1411] which are significantly higher than typical benchmarks (e.g., less than kΩ sq$^{-1}$ for ITO and FTO films). The origin of such low performance is mainly ascribed to the low lateral size of the liquid-phase exfoliated graphene flakes (typically in the order of few micrometers for high-quality graphene flakes)[359] and the high contact resistance between the graphene flakes composing the electrode. However, the development of hybrids between solution-processed graphene and metal nanowires or CNTs, as well as the use of micromesh structures on top of the graphene-based films, represent promising approaches to overcome the current limitations. Prospectively, they could allow the design/realization of TCEs compatible with R2R large-area manufacturing. However, the high cost of metal nanowires[1412] (several hundreds of dollars per kilogram),[1413] CNTs (even more than 1000 \$ kg$^{-1}$ for single-walled CNTs)[1414] and microscale metal grids (\$30–40 m$^{-2}$)[1411] is not lower than the cost of ITO (\$5 m$^{-2}$ for a film with $R_s$ of 150 Ω sq$^{-1}$ films and higher than \$20 m$^{-2}$ for films with $R_s$ of 10 Ω sq$^{-1}$,[1415] or 600 \$ kg$^{-1}$),[1416] making currently available graphene-based TCEs not competitive for massive use in large-area PV devices. Recently, transparent electrodes have also been demonstrated by spin coating 2D $Ti_3C_2$ from an aqueous dispersion for photodetector applications.[1417] However, as for the case of solution-processed graphene, such a transparent electrode shows high sheet resistance, still being ineffective in collecting current density in the order of tens of milliamperes, as those displayed by PV devices. In addition, it should be noted that 2D materials have been used to develop efficient, transparent CEs for bifacial DSSCs, which emerged as interesting systems for both BIPVs and tandem SCs.[878–880,889]

The most successful applications of GRMs in PV technologies rely on their use in the form of CTLs for both holes and electrons (or interlayers in tandem PV architectures). For example, GRMs effectively act as dopants to improve the properties of traditional CTLs. The amount of GRMs needed for this purpose is often minimal, in the order of few weight (volume) percentages of the overall material (dispersion). For example, just 1.6 mL of graphene flakes dispersion at a concentration of 1 g L$^{-1}$ is sufficient to realize 1 m$^2$ of advanced ETLs for PSCs.[1107,1192,1273] By considering $\eta$ higher than 18% in single-junction SCs,[1107,1192] and even higher than 25% in tandem SCs,[1273] only a few grams of graphene flakes are needed for the realization of a 1 MWp PV plant. This amount of graphene flakes corresponds to a negligible added marginal cost, in the order of tens of dollars per megawatts-peak.[205,328,359] Thus, the integration of graphene and other metallic 2D materials,[1418] including group-5 TMDs (e.g., $TaS_2$, $TaSe_2$, $NbS_2$, $NbS_2$, $VS_2$, $VSe_2$, etc.), group-6 TMDs (e.g., the 1T polytype of $MoS_2$ and $WS_2$), topological insulators (e.g., $Bi_2S_3$, $Bi_2Se_3$, and $Bi_2Te_3$), and MXenes, as dopants in the CTL is an approach that can be immediately implemented on different solution-processed PV technologies at the industrial scale, without increasing the overall costs. Beyond their use as dopants, GRMs have been successfully used for the realization of a thin buffer layer (or interlayer) to improve the extraction/collection of the charge photogenerated in the photoactive layer of the cells toward the CTLs and current collectors. In this context, several studies have focused on the formulation of 2D material dispersions in solvents compatible with other materials composing the SC structure. For example, 2D TMD inks have been formulated in 2-propanol to be deposited as a buffer layer over the perovskite layer for the realization of PSCs, showing $\eta$ exceeding 20%. Therefore, the incorporation of 2D material-based buffer layers into the most advanced SC architectures is highly promising to further boost the $\eta$ value of PV technologies beyond the current record-high values. In addition, 2D material-based buffer layers can have a tangible impact on the enhancement of the long-term stability of SCs, particularly for OSCs and PSCs. In fact, 2D materials intrinsically act as shielding layers against humidity, offering promising potential as oxygen/moisture barriers. Moreover, they can also provide effective barriers against ion migration, stabilizing the photoactive perovskite layers or blocking metal/ion migration effects, which determines the degradation of PV devices. With regard to dopants, the amount of 2D materials required for the realization of thin films of 2D materials can be minimal, allowing almost zero additional costs. Not by chance, TMD-based buffer layers (e.g., $MoS_2$) have been used by research groups comprising authors of this work to build a 2D material-enabled solar farm (Fig. 26), without any significant impact over the technology lifecycle assessment (LCA) (data unpublished but reviewed by the European Commission in the context of the Graphene Flagship project).[1290]

Another prospective application of 2D materials in solution-processed SCs are their use as additives in photoactive layers. In particular, the use of GRMs as energy cascade materials can increase the solar-light absorption, whilst eliminating charge recombination pathways occurring in the native materials. In addition, 2D materials can alter the interfacial properties of the photoactive material in contact with other materials composing the SC structure. Such effects can be used to improve charge transfer toward the CTLs (or current collectors), as recently shown with MXenes.[1125] Therefore, the implementation of 2D material-based buffer layers has higher potential for boosting the PV performance of 3rd-generation SCs toward commercially competitive values. To accomplish these, the chemical functionalization of GRMs can be a key step to tune on-demand their optoelectronic properties, thereby adequately matching their energy levels with those of the active materials and CTLs. In addition to GRMs, 2D perovskites have been recently established to improve the thermal stability of PSCs,[1040] demonstrating that rational perovskite engineering can advantageously regulate the structural, physical, and energetic properties of 2D/3D interfaces for the realization of efficient and stable PSCs.[111] Thus, the impact of 2D materials on the structural and optoelectronic properties of the photoactive layer represents a current "hot topic" for the future optimization of current state-of-the-art SCs. Even though the success of solution-processed 2D materials has been established in several PV technologies, we notice that major efforts are currently focused on PSCs, probably because of their attracting $\eta$ exceeding 25%, together with their advantageous combination with Si SCs in tandem systems. In this context, the use of solution-processed 2D materials combined with advanced strategies proposed for optimizing the photoactive layer formulation and processing, as well as for device structure engineering, is promising to boost the $\eta$ of SCs beyond the current state-of-the-art values. The same approach is also viable in enabling similar performance over large-area systems (from a module up to a solar farm). Moreover, the outcomes consolidated for PV technologies discussed in this work could also be extended to other types of thin-film SCs and Si SCs, in which the implementation of solution-processed 2D materials is still premature. Overall, we do believe that the conscious use of the ever-growing 2D materials portfolio can renew the expectation for the rapid establishment of advanced PV technologies worldwide. To accomplish these advances, the standardization of the morphological and structural characterization of 2D materials is crucial for the establishment of industrial-scale technologies, which also requires the setting up of reliable 2D material suppliers with a massive production capability. In this context, the recent standardization sequence of methods for characterizing the structural properties of graphene, bilayer graphene, and graphene nanoplatelets (SO/TS 21356-1:2021) represents a step forward toward the upscaling of solution-processed 2D material-enabled SCs. Meanwhile, emerging solution-processed 2D materials, such as nonlayered materials, carbon nitrides ($C_xN_y$), 2D c-MOFs, layered double hydroxides, and other poorly investigated GRMs (e.g., metal monochalcogenides, group-4 and group-5 TMDs, and polar and/or ferroelectric non-centrosymmetric materials) represent a playground for the realization of cutting-edge concepts of SCs.

## Abbreviations

| | |
|---|---|
| 2D c-MOF | Two-dimensional conjugated metal–organic framework |
| AFM | Atomic force microscopy |
| AgNWs | Silver nanowires |
| ALD | Atomic layer deposition |
| ANIGQDs | Aniline graphene quantum dots |
| APjet | Atmospheric plasma jet |
| a-Si | Amorphous silicon |
| $\alpha$ | Optical absorption coefficient |
| $\alpha_{vis}$ | Optical absorption coefficient in the visible spectrum |
| BCP | Bathocuproine |
| BHJ | Bulk heterojunction |
| BIPVs | Building-integrated photovoltaics |
| BP | Black phosphorus |
| BPNFs | Black phosphorus nanoflakes |
| BPQDs | Black phosphorous quantum dots |
| CCG | Chemically converted graphene |
| CB | Conduction band |
| CE | Counter electrode |
| CIGS | Copper indium gallium diselenide |
| CIGSSe | Copper indium gallium selenide sulfide |
| C-PSC | Carbon perovskite solar cell |
| CQDs | Carbon quantum dots |
| CZTSe | Copper zinc tin sulfur-selenide alloy |
| CNTs | Carbon nanotubes |
| c-Si | Crystalline silicon |
| CTAB | Cetyl-trimethyl-ammonium-bromide |
| CV | Cyclic voltammetry |
| CVD | Chemical vapor deposition |
| $D_n$ | Electron diffusion coefficient |
| DGU | Density gradient ultracentrifugation |
| DSSC | Dye-sensitized solar cell |
| $\delta_p$ | Optical penetration depth |
| ECS | Energy conversion and storage |
| EDNB | Ethylenediamine dinitrobenzoyl |
| $e$ | Elementary charge |
| $e^-$ | Electron |
| $E_F$ | Fermi energy level |
| $E_g$ | Optical bandgap |
| $E_{ph}$ | Photon energy |
| EIS | Electrochemical impedance spectroscopy |
| EpD | Electrophoretic deposition |
| EQE | External quantum efficiency |
| ETLs | Electron transporting layers |
| e-graphene | Electrochemically exfoliated graphene |
| FA | $HC(NH_2)_2$ |
| FF | Fill factor |
| FGSs | Functionalized graphene sheets |
| $fMoS_2$ | Functionalized molybdenum disulfide |
| FoM | Figures of merit |
| FRGO | Fluorinated reduced graphene oxide |
| FTO | Fluorine-doped tin oxide |
| $\phi_W$ | Work function |
| GMo | Graphene-molybdenum disulfide heterostructure |
| GNPs | Graphene nanoplatelets |
| GNSs | Graphene nanosheets |
| GNRs | Graphene nanoribbons |
| GO | Graphene oxide |
| GO-EDNB | Graphene oxide functionalized with ethylenediamine's amino groups |
| GO-TPP | Graphene oxide linked with porphyrin moieties |
| GO-Cl | Chlorinated graphene oxide |
| GQDs | Graphene quantum dots |
| GRMs | Graphene-related materials |
| GSs | Graphene sheets |
| h-BN | Hexagonal boron nitride |
| HIT | Heterojunction with intrinsic thin layer |
| HOMO | Highest occupied molecular orbital |

| | | | |
|---|---|---|---|
| HTLs | Hole transporting layers | PEAI | Phenethylammonium iodide |
| HSN | Hierarchically structured nanoparticles | (PEA)$_2$PbI$_4$ | Phenyl ethyl ammonium lead iodide |
| $\eta$ | Solar-to-electrical energy conversion efficiency | $P_{in}$ | Power of incident light |
| h$^+$ | Hole | PC$_{61}$BM | [6,6]-Phenyl-C$_{61}$-butyric acid methyl ester |
| $h$ | Planck's constant | PC$_{71}$BM | [6,6]-Phenyl-C$_{71}$-butyric acid methyl ester |
| $\hbar$ | Reduced Planck's constant | PCDTBT | PC$_{71}$BM:poly[$N$-9′-heptadecanyl-2,7-carbazole-$alt$-5,5-(4′,7′-di-2-thienyl-2′,1′,3′benzothiadiazole)]]:[6,6]-phenyl-C71 butyric acid methyl ester |
| $I$ | Electrical current | | |
| ICBA | Indene–C$_{60}$ bisadduct | | |
| ICLs | Interconnection layers | | |
| IGO | Imidazole-functionalized GO | PDINO | $N,N$-Dimethyl-ammonium $N$-oxide)propyl perylene diimide |
| $I_{MPP}$ | Current at the maximum power point | | |
| $I_{SC}$ | Short-circuit current | PDINO-G | Graphene doped with $N,N$-dimethyl-ammonium $N$-oxide)propyl perylene diimide |
| IQE | Internal quantum efficiency | | |
| ITO | Indium–tin oxide | PDMS | Polydimethylsiloxane |
| $k_B$ | Boltzmann's constant | PEDOT:PSS | Poly(3,4-ethylenedioxythiophene) polystyrene sulfonate |
| $\kappa$ | Molar extinction coefficient | | |
| LCOE | Levelized cost of energy | PET | Poly(ethylene terephthalate) |
| Li-TFSI | Lithium bis(trifluoromethanesulfonyl)imide | PFN | Poly((9,9-bis(3′-($N,N$-dimethylamino)propyl)-2,7-fluorene)-$alt$-2,7-(9,9-dioctylfluorene)) |
| LPE | Liquid phase exfoliation | | |
| LRGO | Laser-treated reduced graphene oxide | PFN-Br | Poly9,9-bis6-($N,N,N$-trimethylammonium)hexyl-fluorene-$alt$-co-phenylenebromide) |
| LUMO | Lowest unoccupied molecular orbital | | |
| $\lambda$ | Photon wavelength | P3HT | Poly(3-hexylthiophene) |
| MA | CH$_3$NH$_3^+$ | P3HT:PC$_{61}$BM | Poly(3-hexylthiophene):[6,6]-phenyl C61-butyric acid methyl ester |
| MBE | Molecular beam epitaxy | | |
| MEH-PPV | Poly[2-methoxy-5-(2′-ethylhexyloxy)-p-phenylene vinylene] | P3OT | Poly(3-octylthiophene-2,5-diyl) |
| | | PH1000 | Poly(3,4-ethylenedioxythiophene): poly-(styrenesulfonate) |
| MGO | Multilayer graphene oxide | | |
| MIR | Mid-infrared | PMMA | Polymethyl methacrylate |
| MLG | Multilayer graphene | PM6 | Poly[[4,8-bis[5-(2-ethylhexyl)-4-fluoro-2-thienyl]benzo[1,2-$b$:4,5-$b'$]dithiophene-2,6-diyl]-2,5-thiophenediyl[5,7-bis(2-ethylhexyl)-4,8-dioxo-4$H$,8$H$-benzo[1,2-$c$:4,5-$c'$]dithiophene-1,3-diyl]-2,5-thiophenediyl] |
| MOF | Metal–organic framework | | |
| MPA | 3-Mercaptopropionic acid | | |
| MPPT | Maximum power point | | |
| mTiO$_2$ | Mesoporous TiO$_2$ | | |
| MWCNTs | Multiwalled carbon nanotubes | | |
| $\mu$ | Charge carrier mobility | PTB7:PCB$_{71}$M | Thieno[3,4-$b$]thiophene/benzodithiophene: phenyl-C$_{71}$-butyric acid methyl ester |
| $\mu_e$ | Electron mobility | | |
| $\mu_h$ | Hole mobility | PTAA | Poly(triaryl)amine |
| $N_d$ | Charge carrier density | pRGO | Partially reduced graphene oxide |
| $n_{film}$ | Film refractive index | PSCs | Perovskite solar cells |
| $n_{sub}$ | Substrate refractive index | PSMs | Perovskite solar modules |
| $\eta_{th}$ | Theoretical solar-to-electrical energy conversion efficiency | PV | Photovoltaic |
| | | QDs | Quantum dots |
| NFA | Non-fullerene acceptors | QDDSSCs | Quantum dot-based dye sensitized solar cells |
| NG/NiO | N-Doped graphene@nickel oxide | QDSCs | Quantum dot solar cells |
| N-GFs | N-Doped graphene frameworks | QDSSCs | Quantum dot-sensitized solar cells |
| NGNP | N-Doped graphene nanoplatelets | R2R | Roll-to-roll |
| NP | Nanoparticles | $R_{CT}$ | Charge transfer resistance |
| NG | Amino-functionalized graphene | $R_S$ | Sheet resistance |
| NIR | Near-infrared | $R_{TiO_2}$ | Transport resistance of electrons in the TiO$_2$ film |
| NR | Nanorods | $R_{TCO–TiO_2}$ | Resistance at transparent conductive oxide/TiO$_2$ contact |
| NRGO | Nitrogen-doped reduced graphene oxide | | |
| OLED | Organic light-emitting diodes | $R_{rec}$ | Charge transfer resistance of the charge recombination between electrons in the TiO$_2$ film and I$_3^-$ in the electrolyte |
| O-MoS$_2$ | Oxygen-incorporated molybdenum disulfide | | |
| OSCs | Organic solar cells | | |
| OTR | Oxygen transmission rate | $R_{CT}$ | Charge transfer resistance at the counter electrode/electrolyte interface |
| oxo-G | Organo-sulfonate graphene | | |
| | | $R_{TCO–electr.}$ | Charge transfer resistance at the TCO/electrolyte interface |

| RGO | Reduced graphene oxide |
| RGOMM | Reduced graphene oxide micromesh |
| rGS | Reduced graphene scaffold |
| RH | High humidity environment |
| RT | Room temperature |
| S–Q | Shockley–Queisser |
| SBS | Sedimentation-based separation |
| SCs | Solar cells |
| SLG | Single-layer graphene |
| Spiro-OMeTAD | 2,2′,7,7′-Tetrakis[$N,N$-di(4-methoxyphenyl)amino]-9,9′-spirobifluorene |
| SSA | Specific surface area |
| SWCNT | Single-walled carbon nanotube |
| $\sigma$ | Electrical conductivity |
| $\sigma_{dc}$ | d.c. conductivity |
| $\sigma_{opt}$ | Optical conductivity |
| $t$ | Photoactive material thickness |
| TBP | Tert-butylpyridine |
| TCEs | Transparent conductive electrodes |
| TCOs | Transparent conductive oxides |
| TCPP | Tetrakis(4-carboxyphenyl)porphyrin |
| TEGr | Thermally exfoliated graphene |
| TEM | Transmission electron microscopy |
| TFSCs | Thin-film solar cells |
| TFSI | Trifluoromethanesulfonyl imide |
| TMD | Transition metal dichalcogenide |
| ToF-SIMS | Time-of-flight secondary ion mass spectrometry |
| $T_r$ | Optical transmittance |
| TRGO | Thermally reduced GO |
| TSHBC | Perthiolated tri-sulfur-annulated hexa-peri-hexabenzocoronene |
| $\tau$ | Electron lifetime |
| UVO | UV-ozone |
| VB | Valence band |
| $V_{MPP}$ | Voltage at the maximum power point |
| $V_{OC}$ | Open-circuit voltage |
| WVTR | Water vapor transmission rate |
| WJM | Wet-jet milling |
| XPS | X-ray photoelectron spectroscopy |
| XRD | X-ray diffraction |
| $Z$ | Vacuum impedance |
| $Z_d$ | Warburg impedance |
| ZnP | Zn–porphyrin |
| ZSO | Zinc stannate |

## Conflicts of interest

There are no conflicts to declare.

## Acknowledgements

This project has received funding from the European Union's Horizon 2020 research and innovation program under grant agreements No. 785219 and 881603-GrapheneCore2 and GrapheneCore3, the MSCA-ITN ULTIMATE project under grant agreement No. 813036, European Union's SENSIBAT project under grant agreement No. 957273, and the Bilateral project GINSENG between NSFC (China) and MAECI (Italy) (2018–2020), by the Natural Science Foundation of Shandong Province (ZR2019QEM009). ADC acknowledge the financial support of the Ministry of Education and Science of the Russian Federation in the framework of Megagrant No. 14.Y26.31.0027. G. G. acknowledges the ''HY-NANO'' project that has received funding from the European Research Council (ERC) Starting Grant 2018 under the European Union's Horizon 2020 research and innovation programme (Grant agreement No. 802862) and the project Cariplo Economia Circolare 2021 FLHYPER (num. 2020-1067).

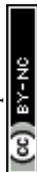